\documentclass[12pt]{article}  %fuer beidseitige Version

\usepackage{a41}
\usepackage{cite}
\usepackage{epsfig}
\usepackage{graphicx}
\usepackage{amsmath}
\usepackage{amssymb}
\usepackage{ulem}
\usepackage{mdwlist}
\usepackage{color}

\usepackage[rflt]{floatflt}
\usepackage{float}
\usepackage{slashed}

\usepackage{array}

\usepackage[english]{babel}
\usepackage[latin1]{inputenc}
\usepackage[T1]{fontenc}
\usepackage{ae}

\usepackage{url}

\usepackage{amsmath, amsthm, amssymb}
\newtheorem{thm}{Theorem}[section]

\newtheorem{definition}[thm]{Definition}

\newtheorem{remark}[thm]{Remark}

\newcommand{\IIntegral}{I}

\newcommand{\HA}{{\rm H}^*}
\newcommand{\HHH}{{\rm H}}

\newcommand{\abs}[1]{\left|{#1}\right|}

\newcommand{\sign}[1]{\textnormal{sign}\hspace{-0.1em}\left(#1\right)}

\newcommand{\ds}{\displaystyle}
\newcommand{\Li}{{\rm Li}}
\newcommand{\Mvec}{{\rm \bf M}}

\newcommand{\ie}{i.e.,\ }

\newcommand{\ep}{\varepsilon}

\let\set\mathbb

\usepackage{rotating}

\usepackage{graphicx}
\newcounter{mmacnt}
\def\restartmma{\setcounter{mmacnt}{0}}
\restartmma \catcode`|=\active
\def|#1|{\mathrm{#1}}
\catcode`|=12
\newenvironment{mma}{
 \par\smallskip
 \catcode`|=\active
 \parskip=0pt\parindent=0pt % locally
 \small
 \def\In##1\\{%
   \def\linebreak{\hfill\break\null\qquad}%
   \refstepcounter{mmacnt}
   \hangindent=2.5em\hangafter=0
   \leavevmode
   \llap{\tiny\sffamily In[\arabic{mmacnt}]:=\kern.5em}%
   \mathversion{bold}\footnotesize$\displaystyle##1$\normalsize
   \mathversion{normal}\par
 }%
 \def\Print##1\\{%
   \def\linebreak{\hfill\break}%
   \hangindent=2.5em\hangafter=0
   \leavevmode ##1\par}%
 \def\Out##1\\{%
   \def\linebreak{$\hfill\break\null\hfill$}%
   \kern\abovedisplayskip\par
   \hangindent=2.5em\hangafter=0
   \leavevmode
   \llap{\tiny\sffamily Out[\arabic{mmacnt}]=\kern.5em}
   \footnotesize$\displaystyle##1$\normalsize\hfill\null\par
   \kern\belowdisplayskip
 }%
 \def\Warning##1##2\\{%
   \def\linebreak{\hfill\break}%
   \hangindent=2.5em\hangafter=0
   \leavevmode
   {\scriptsize##1 : ##2}\par}%
}{%
 \par\smallskip
}

\usepackage{color}

\newenvironment{fshaded}{%
\MakeFramed {\FrameRestore}
}%
{\endMakeFramed}

\allowdisplaybreaks[4]

\begin{document}
\setlength{\baselineskip}{0.515cm}
\sloppy
\thispagestyle{empty}
\begin{flushleft}
DESY 15--049
%\hfill {\tt arXiv:xxx}
\\
DO--TH 15/06\\
MITP/15-080 \\
September 2015
\end{flushleft}

\mbox{}
\vspace*{\fill}
\begin{center}

{\Large\bf Calculating Three Loop Ladder and}

\vspace*{2mm}
{\Large\bf \boldmath{$V$}-Topologies for Massive Operator Matrix Elements}

\vspace*{2mm}
{\Large\bf by Computer Algebra}

\vspace{3cm}
{\large
J.~Ablinger$^a$,
A.~Behring$^b$,
J.~Bl\"umlein$^b$,
A.~De Freitas$^b$,
A.~von~Manteuffel$^c$, 

\vspace*{2mm}
and C.~Schneider$^a$}

\vspace{1cm}
\normalsize
{\it $^a$~Research Institute for Symbolic Computation (RISC),\\
                          Johannes Kepler University, Altenbergerstra\ss{}e 69,
                          A--4040, Linz, Austria}\\

\vspace*{3mm}
{\it  $^b$ Deutsches Elektronen--Synchrotron, DESY,}\\
{\it  Platanenallee 6, D-15738 Zeuthen, Germany}
\\

\vspace*{3mm}
{\it  $^c$ PRISMA Cluster of Excellence,  Institute of Physics, J. Gutenberg
University,}\\ 
{\it D-55099 Mainz, Germany.}
\\

%%\today

\end{center}
\normalsize
\vspace{\fill}
\begin{abstract}
\noindent
Three loop ladder and {\it V}-topology diagrams contributing to the massive operator matrix element $A_{Qg}$ are calculated. 
The corresponding objects can all be expressed in terms of nested sums and recurrences depending on the Mellin variable $N$ 
and the dimensional parameter $\ep$. Given these representations, the desired Laurent series expansions in $\ep$ can be obtained 
with the help of our computer algebra toolbox. Here we rely on generalized hypergeometric functions and Mellin-Barnes 
representations, on difference ring algorithms for symbolic summation, on an optimized version of the multivariate 
Almkvist-Zeilberger algorithm for symbolic integration, and on new methods to calculate Laurent series solutions of coupled systems 
of differential equations. The solutions can be computed for general coefficient matrices directly for any basis also performing 
the expansion in the dimensional parameter in case it is expressible in terms of indefinite nested product-sum expressions. This 
structural result is based on new results of our difference ring theory. In the cases discussed we deal with iterative sum- 
and integral-solutions over general alphabets. The final results are expressed in terms of special sums, forming quasi-shuffle algebras, 
such as nested harmonic sums, generalized harmonic sums, and nested binomially weighted (cyclotomic) sums. Analytic continuations to 
complex values of $N$ are possible through the recursion relations obeyed by these quantities and their analytic asymptotic expansions. 
The latter lead to a host of new constants beyond the multiple zeta values, the infinite generalized harmonic and cyclotomic sums 
in the case of $V$-topologies.
\end{abstract}

\vspace*{\fill}
\noindent
\numberwithin{equation}{section}
%%%%%%%%%%%%%%%%%%%%%%%%%%%%%%%%%%%%%%%%%%%%%%%%%%%%%%%%%%%%%%%%%%%%%%%

\newpage
\tableofcontents

%%%%%%%%%%%%%%%%%%%%%%%%%%%%%%%%%%%%%%%%%%%%%%%%%%%%%%%%%%%%%%%%%%%%%%%
\section{Introduction}
\label{sec:2}
%%%%%%%%%%%%%%%%%%%%%%%%%%%%%%%%%%%%%%%%%%%%%%%%%%%%%%%%%%%%%%%%%%%%%%%

\vspace*{1mm}
\noindent
The heavy flavor corrections to deep-inelastic scattering at 3-loop order to the structure function
$F_2(x,Q^2)$ can be expressed in terms of massive operator matrix elements (OMEs) and massless Wilson 
coefficients at large enough virtualities $Q^2$ \cite{Buza:1995ie}. These corrections will complete 
all terms requested to perform the complete next-to-next-to-leading order (NNLO) analyses for precision
determinations of the strong coupling constant $\alpha_s(M_Z^2)$, the parton distribution functions
and the charm quark mass $m_c$ from deep-inelastic data \cite{Bethke:2011tr,Moch:2014tta,Alekhin:2013nda,Alekhin:2012vu}.
At the level of a series of fixed Mellin moments the problem has been solved in Ref.~\cite{Bierenbaum:2009mv}. 
The experimental analysis, however, requests the corresponding expressions for the general Mellin variable $N$, 
the calculation of which is much more demanding and relies on very different  calculation techniques, which are
much more involved. 

The Feynman integrals for single-scale quantities like the QCD splitting functions
or Wilson coefficients for hard scattering processes \cite{Z1} depending on one characteristic variable
$x \in [0,1]$ can be expressed in terms of special functions out of certain classes, which
grow with the loop order, see Refs.~\cite{Bierenbaum:2007qe,Ablinger:2010ty,Ablinger:2014nga,Ablinger:2014uka}.
Here the variable $x$ may denote a momentum fraction or another characteristic ratio of Lorentz invariants like $x = t/s$ \cite{BHA}. 
Because it is defined for $x \in [0,1]$ one may apply the Mellin transform
%---------------------------------------------------------------------------------------------------
\begin{eqnarray}
\Mvec\left[f(x)\right](N) = \int_0^1 dx~x^N~f(x),~~N \in \mathbb{N}
\end{eqnarray}
%---------------------------------------------------------------------------------------------------
to obtain discrete representations for the corresponding physical problem. These may also be given
directly in the case of the light cone expansion for deep-inelastic scattering \cite{LCE}. The Mellin transforms
obey recurrence relations and modern symbolic summation techniques \cite{Karr:81,Schneider:01,Schneider:05a,
Schneider:07d,Schneider:10b,Schneider:10c,Schneider:15a} can be used to calculate 
the corresponding Feynman integrals, whenever the recurrences are solvable in  $R\Pi\Sigma^*$-difference rings~\cite{Schneider:08c}. There are 
corresponding algorithms \cite{SIG1,SIG2} to determine whether indeed this is possible or not. 

In the calculation of the massive operator matrix elements to 3-loop order we have obtained so far only representations
in terms of nested sums. Here the simplest structures are nested harmonic sums \cite{Blumlein:1998if,Vermaseren:1998uu}, 
followed by generalized sums \cite{Moch:2001zr,Ablinger:2013cf}, and binomial sums \cite{Ablinger:2014bra} in the final 
representations. Intermediary results and the final representations for $V$-topology diagrams also request generalized 
cyclotomic  sums being binomially weighted \cite{Ablinger:2011te}.\footnote{For recent surveys on these structures see
Refs.~\cite{SURV}.}

The simplest topologies at 3-loop order are those with only one massive loop with up to 3 propagators 
\cite{Ablinger:2010ty,Behring:2014eya} which can be calculated using hypergeometric representations and applying summation 
methods \cite{SIG1,SIG2}, extending the methods which have been applied in the 2-loop case 
\cite{Bierenbaum:2007qe,Bierenbaum:2007pn,Bierenbaum:2008yu,Bierenbaum:2009zt,Blumlein:2014fqa} and for the structure 
function $F_L(x,Q^2)$ before \cite{Blumlein:2006mh,Behring:2014eya}. The next more involved topologies are those with two internal fermion 
loops of equal mass \cite{Ablinger:2014uka}, followed by those of Benz-topologies with up to 3 massive propagators 
\cite{Ablinger:2014lka,Ablinger:2014vwa,Ablinger:2014nga}.

In the present paper we study the 3-loop ladder and $V$-topologies with operator insertions on the massive lines. Using 
integration-by-parts (IBP) techniques \cite{IBP} encoded in {\tt Reduze 2} \cite{Reduze1,Reduze2}\footnote{The 
package {\tt Reduze2} uses the codes {\tt Fermat} \cite{FERMAT} and {\tt GiNac} \cite{Bauer:2000cp}.}, we reduce 
the diagrams 
to master integrals. We apply different methods to calculate the contributing master integrals. In the simpler cases their 
representation through generalized hypergeometric series, up to Appell functions and their integrals 
\cite{HYP,Slater,Appell,Schlosser}, or more generally, Mellin-Barnes \cite{MB} integrals, are possible. Exploiting our symbolic summation toolbox mentioned above, their expansion in 
the dimensional parameter $\ep = D - 4$ leads to multiple nested sum representations. A second method consists in applying 
the multivariate Almkvist-Zeilberger algorithm \cite{AZ} to establish a difference equation in the Mellin variable $N$. A 
third method relies on the solution of systems of differential equations \cite{DEQ}, which are cast into difference 
equation systems. These are uncoupled using the package {\tt Oresys} \cite{ORESYS} leading to individual difference 
equations. The method presented in the present paper does not require a specific form of the coupled system of 
first order differential equations like requested in \cite{Henn:2013pwa}, but works for the general case, also 
performing the $\ep$-expansion automatically. Here we aim on solutions in difference rings, i.e. indefinite nested sum 
representations in Mellin space.\footnote{Synonymous solutions in terms of iterated integrals in $x$-space are possible in an 
equivalent manner, see Refs.~\cite{Blumlein:2009ta,Ablinger:2011te,Ablinger:2013cf,Ablinger:2014bra}.}

The solution of the difference equations and the summation of the nested sums is finally performed using modern summation 
techniques \cite{Karr:81,Schneider:01,Schneider:05a,Schneider:07d,Schneider:08c,Schneider:10b,Schneider:10c,Schneider:15a} in 
$R\Pi\Sigma$-difference rings using the packages {\tt Sigma} \cite{SIG1,SIG2}, {\tt EvaluateMultiSum},  {\tt SumProduction} 
\cite{EMSSP}, and {\tt HarmonicSums} \cite{HARMONICSUMS,Ablinger:PhDThesis,Ablinger:2011te,Ablinger:2013cf,Ablinger:2014bra}. In 
previous studies 
\cite{Ablinger:2012qm,Ablinger:2014yaa} we have considered examples of scalar diagrams of the ladder and $V$-topology type 
containing no numerator structure besides the operator insertion. The calculation of these diagrams turns out to be far 
easier than that of the complete Feynman diagrams because of both, the number of terms and the singularity structure in the 
dimensional parameter. While scalar diagrams usually need not to be decomposed using the IBP-relations, this is unavoidable 
in the case of QCD diagrams. In this way the number of final integrals to be calculated is reduced. However, some of the master 
integrals have to be evaluated up to $O(\ep^{3})$ in case no complete representation has been available\footnote{Simpler 
master integrals being known for general values of $\ep$ had to expanded to $O(\ep^{5})$.}, given the way they enter
the diagram representation. Using 12 representative cases, we will show in the following sections that this can indeed be 
achieved applying a chain of algorithms. Finally, we apply the algebraic relations 
\cite{Blumlein:2003gb,Ablinger:2011te,Moch:2001zr,Ablinger:2013cf} for the nested sums obtained, in order to 
represent all expressions by the most elementary sums. In the case of the binomial sums, this operation turns out to be 
insufficient, as the reduction over the $R\Pi\Sigma$-difference rings yields a shorter basis. It could be shown, however, that in this 
case one has to apply additionally their contiguous relations, which finally leads to the desired result. The analytic 
continuation to complex values of $N$ is performed as has been outlined previously in the case of harmonic 
\cite{Blumlein:2009ta,Blumlein:2009fz}, cyclotomic \cite{Ablinger:2011te} and generalized harmonic sums \cite{Ablinger:2013cf}, 
giving the corresponding asymptotic representations. Furthermore, integer steps can be performed from any value in the analyticity 
region due to the recurrence relations for these quantities.

The paper is organized as follows. In Section~\ref{sec:CAChallenge} a general outline of the approach to compute
individual Feynman diagrams is given. The representation of the ladder and $V$-graphs in terms of master integrals
is described in Section~\ref{Sec:LadderGraphs}. In Section~\ref{sec:Calc-MIs} we outline the calculation techniques for the 
master integrals. These rely on representations in terms of generalized hypergeometric functions, in terms of Mellin-Barnes integrals
and in terms of differential and difference equation systems. We describe how to deal with the $\ep$-expansion and 
outline the solution of systems of coupled difference equations and the use of the multivariate Almkvist-Zeilberger 
algorithm. In Section~\ref{sec:GetMoment} we describe how the solution for a Feynman diagram is found from the
master integrals. The results obtained using the presented algorithms are illustrated for a physical graph with ladder 
topology in Section~\ref{sec:LDIA} and 
examples of the planar and non-planar parts of  $V$-diagrams  in Section~\ref{sec:VDIA}. Section~\ref{sec:Conc} contains the 
conclusions. In the Appendix we present the results for the remaining ladder and $V$-diagrams.
%%%%%%%%%%%%%%%%%%%%%%%%%%%%%%%%%%%%%%%%%%%%%%%%%%%%%%%%%%%%%%%%%%%%%%%
\section{General outline of the computer algebra approach}
\label{sec:CAChallenge}
%%%%%%%%%%%%%%%%%%%%%%%%%%%%%%%%%%%%%%%%%%%%%%%%%%%%%%%%%%%%%%%%%%%%%%%

\vspace{1mm}
\noindent
The massive three loop Feynman diagrams with a local operator insertion can be written in terms of multiple integrals, which depend 
on a discrete variable $N \in \mathbb{N}$ and the dimensional parameter $\varepsilon = D - 4$, 
where $D \in \mathbb{R}$ denotes the space-time dimension, see e.g. see~\cite{Blumlein:2010zv,Weinzierl:13}. We denote one of these 
diagrams by $D(N)$. Given such an expression, we are interested in the first coefficients of its Laurent series expansion 
w.r.t.\ $\ep$ (in short, $\ep$-expansion):
%------------------------------------------------------------------------------------------------------------------------------
\begin{equation}\label{Equ:epExpansion}
D(N)=F_{o}(N)\,\ep^o+F_{o+1}(N)\,\ep^{o+1}+F_{o+2}(N)\,\ep^{o+2}+\dots \,\, .
\end{equation}
%------------------------------------------------------------------------------------------------------------------------------
In the 3--loop case we usually have $o=-3$ and we are interested in the coefficients $F_{-3}(N)$, $F_{-2}(N), F_{-1}(N), 
F_{0}(N)$.

More precisely, we want to express the coefficients in terms of special functions that can be represented by indefinite 
nested sums and products, which can be defined as follows. Let $f(N)$ be an expression that evaluates at non-negative 
integers (from a certain point on) to elements of a field $\set K$ containing the rational numbers $\set Q$. Then $f(N)$ 
is called an expression in terms of indefinite nested sums and products w.r.t.\ $N$ (or in short, an indefinite nested 
product-sum expression) if it is composed by elements from the rational function field $\set K(N)$, the three operations 
($+,-,\cdot$), and sums and products of the type $\sum_{k=l}^Nh(k)$ or $\prod_{k=l}^Nh(k)$ where $l\in\set N$ and where 
$h(k)$ is an indefinite nested product-sum expression w.r.t.\ $k$ which is free of $N$. In this article we restrict this 
class further such that any product is a hypergeometric expression, i.e., for any occurring product $\prod_{i=l}^{k}h(i)$ 
we have that $h\in\set K(i)$ is a rational function in $i$. Note that we can write $(\prod_{i=l}^kh(i))^{-1} = 
\prod_{i=l}^kh(i)^{-1}$ if $h(i)\neq0$ for all $i\geq l$. This class of special functions covers as special cases: 
\begin{enumerate}
\item
The harmonic sums~\cite{Blumlein:1998if,Vermaseren:1998uu}
%--------------------------------------------------------------------------------------------
\begin{eqnarray}\label{Equ:HarmonicSumsIntro}
	S_{a_1,\ldots ,a_k}(N)= \sum_{N\geq i_1 \geq i_2 \geq \cdots \geq i_k \geq 1} \frac{\sign{a_1}^{i_1}}{i_1^{\abs {a_1}}}\cdots
	\frac{\sign{a_k}^{i_k}}{i_k^{\abs {a_k}}}	
\end{eqnarray}
%--------------------------------------------------------------------------------------------
for non-negative integers $N$ and nonzero integers $a_i$ $(1 \leq
i \leq k)$. 
\item
The generalized harmonic sums (or $S$-sums)~\cite{Moch:2001zr,Ablinger:2013cf}
%--------------------------------------------------------------------------------------------
\begin{equation}\label{Equ:SSumsIntro}
	S_{a_1,\ldots ,a_k}(x_1,\ldots ,x_k;N)= \sum_{n\geq i_1 \geq i_2 \geq
        \cdots \geq i_k \geq 1} \frac{x_1^{i_1}}{i_1^{a_1}}\cdots
	\frac{x_k^{i_k}}{i_k^{a_k}},
\end{equation}  
%--------------------------------------------------------------------------------------------
with positive integers $a_i$ and (usually) real numbers $x_i\neq0$ $(1\leq i\leq k)$.
\item
The (inverse) binomial sums~\cite{Fleischer:1998nb,Davydychev:2003mv,Weinzierl:2004bn,Ablinger:2014bra}
%--------------------------------------------------------------------------------------------
\begin{equation}\label{Equ:BinomSum}
\sum_{n\geq i_1 \geq i_2 \geq
        \cdots \geq i_k \geq 1}a_1(i_1)\dots a_k(i_k),
\end{equation}
%--------------------------------------------------------------------------------------------
where the summands are of the form
%--------------------------------------------------------------------------------------------
$$a_j(n)=\binom{2 n}{n}^{b_j}\frac{c_j^n}{n^{m_j}},$$
%--------------------------------------------------------------------------------------------
with $b_j \in \{-1,0,1\}$, $c_j \in \mathbb{R}\setminus\{0\}$ and $m_j \in \mathbb{N}$. 
\end{enumerate}

\noindent
The weight of the sums~\eqref{Equ:HarmonicSumsIntro} and~\eqref{Equ:SSumsIntro} is defined by $|a_1|+\dots+|a_k]$, and the 
(nesting) depth of the sums~\eqref{Equ:HarmonicSumsIntro}, \eqref{Equ:SSumsIntro} and ~\eqref{Equ:BinomSum} 
is $k$. Also the class of cyclotomic sums~\cite{HARMONICSUMS,Ablinger:PhDThesis} arises in all the above cases. They 
are characterized by denominators
of the kind
%--------------------------------------------------------------------------------------------
\begin{equation}
\frac{1}{(k i + l)^n},~~~~k,l,n \in \mathbb{N} \backslash \{0\}
\end{equation}
%--------------------------------------------------------------------------------------------
and $i$ denotes the summation quantifier. For $k = 2, l = n = 1$ terms of this kind emerge in binomial sums
in the present project. In what follows, we will omit sometimes the explicit $N$ dependence of the 
(generalized) harmonic sums, so for example, $S_{2,1}(N)$ will be simply written as $S_{2,1}$, etc.

For many scalar ladder graphs direct symbolic summation techniques in the setting of difference 
rings~\cite{SIG1,SIG2,EMSSP,Schneider:07d,Karr:81,Schneider:08c,DFTheory,Schneider:10c,Schneider:15a,Schneider:10b,RecSolver,Schneider:01,Schneider:05a} and in combination with 
special 
function 
algorithms~\cite{Vermaseren:1998uu,Blumlein:2009ta,Blumlein:2009fz,HARMONICSUMS,Ablinger:2013cf,Ablinger:2014bra}
have been very effective~\cite{Ablinger:2012qm,Ablinger:2014yaa}. Note that these symbolic summation algorithms generalize and 
enhance substantially the \hbox{($q$--)}hypergeometric and holonomic toolbox~\cite{AlternativeSummation,WegschaiderWZ} in order 
to rewrite definite sums to indefinite nested sums. In addition, we applied successfully  
extensions~\cite{Ablinger:2014yaa,FWTHESIS} 
of Brown's hyperlogarithm algorithm~\cite{Brown:2008um} treating massive graphs with resummed local operator insertions
for several scalar ladder and $V$-diagrams. Here, even non-linear variable dependences were allowed in the last integration steps,
partly using variable transformations. In case of Feynman integrals with poles, not being covered by the algorithm \cite{Brown:2008um}, 
one may map these for usual (massive) Feynman diagrams, cf.~\cite{Panzer:2014gra,vonManteuffel:2014qoa}. Yet, whether the 
constant part can be calculated using the method of hyperlogarithms is not clear a priori. This depends on whether or not the 
multi-linearity of the integration variables can be kept, including suitable mappings of non-linear terms, 
cf.~\cite{Ablinger:2014yaa,FWTHESIS}, in case.

Due to the size and structure of the diagrams to be dealt with, essential preparatory work is necessary before one can apply 
our toolbox: we will reduce the diagrams, written in terms of scalar integrals, to simpler integrals, the so called master integrals. In order to 
accomplish this task, we introduce a new variable $x$ and switch to its power series representation
%--------------------------------------------------------------------------------------------
\begin{equation}\label{Equ:PowerSeries}
\hat{D}(x)=\sum_{N=0}^{\infty}D(N)x^N.
\end{equation}
%--------------------------------------------------------------------------------------------
Given this encoding, we can now use the whole host of
integration by parts identities \cite{IBP} for the corresponding diagrams delivered by the package {\tt Reduze 2} \cite{Reduze1,Reduze2}, 
which is a {\tt C++} program 
based on Laporta's algorithm \cite{Laporta:2001dd}. Namely, using this technology we can express $\hat{D}(x)$ in terms of 
master integrals, say $\hat{M}_i(x)$. More precisely, it can be written as a linear combination 
%--------------------------------------------------------------------------------------------
\begin{equation}\label{Equ:DExp}
\hat{D}(x)=c_1(x,\ep)\hat{M}_1(x)+c_2(x,\ep)\hat{M}_2(x)+\dots+c_r(x,\ep)\hat{M}_r(x),
\end{equation}
%--------------------------------------------------------------------------------------------
where the coefficients $c_i(x,\ep)$ are rational functions in $x$ and $\ep$. The number of master integrals is 
significantly smaller than the number of Feynman parameter integrals to be computed in general. Often the master integrals 
are simpler than the input integrals. 
%%One complication implied is, however, that higher order 
%%expansions in the dimensional parameter $\ep$ are requested also in a series of non-trivial cases.

In order to derive the $\ep$-expansion in~\eqref{Equ:epExpansion}, we could apply the following tactic: calculate the 
$\ep$-expansion for all master integrals $\hat{M}_i(x)$ in terms of special functions up to the necessary order, assemble 
the results to obtain an expression $\hat{D}(x)$ in terms of the found special functions, and finally extract the $N$th-coefficient 
in~\eqref{Equ:PowerSeries} in order to derive the desired coefficients $F_i(N)$ of the $\ep$-expansion~\eqref{Equ:epExpansion}.
However, currently we do not yet have the necessary computer algebra tools at hand to carry out this approach. Hence we apply a 
slightly different methodology. The master integrals $\hat{M}_i(x)$ produced by the integration by parts method can be directly 
translated to a formal power series representation
%--------------------------------------------------------------------------------------------
\begin{equation}\label{Equ:MxRep}
\hat{M}_i(x)=\sum_{N=0}^{\infty} M_i(N)x^N,
\end{equation}
%--------------------------------------------------------------------------------------------
where the coefficients $M_i(N)$ can be given explicitly in terms of multiple integrals depending on the discrete parameter $N$ 
and the dimensional parameter $\ep$. 
%Note that this representation suits again our available computer algebra toolbox.
To obtain the $\ep$-expansion~\eqref{Equ:epExpansion} of $D(N)$, we will proceed as follows. First, we aim at doing the 
$\ep$-expansions
%--------------------------------------------------------------------------------------------
\begin{equation}\label{Equ:epExpansionMI}
M_i(N)=M_{i,-3}(N)\ep^{-3}+M_{i,-2}(N)\ep^{-2}+M_{i,-1}(N)\ep^{-1}+M_{i,0}(N)\ep^{0}+\dots,
\end{equation}
%--------------------------------------------------------------------------------------------
i.e., we compute the coefficients $M_{i,j}(N)$  in terms of indefinite nested product-sum expressions with our computer algebra 
technologies. This task is highly non-trivial and a big proportion of this article will be devoted to new computer algebra tools 
to tackle this challenge. Besides our symbolic summation technologies mentioned already earlier, we will use new 
ideas~\cite{NewUncouplingMethod} to solve coupled systems of difference/differential equations based on uncoupling 
algorithms~\cite{Zuercher:94,ORESYS}\footnote{For further references on uncoupling algorithms see e.g. \cite{UNCOUPL}. For a recent 
idea on uncoupling of homogenous systems of IBP relations see \cite{Tancredi:2015pta}.} and a 
recurrence solver for Laurent series~\cite{Blumlein:2010zv}. Our algorithm will find the solution of the (inhomogeneous) recurrence in 
$\Pi\Sigma^*$-fields and $R\Pi\Sigma^*$-rings or return that the problem has no solution in these fields (rings).
Within our algorithm, we will 
rely heavily on a fine-tuned version~\cite{Ablinger:PhDThesis,LL12:Technolgy} of the multivariate Almkvist-Zeilberger 
algorithm~\cite{AZ} for symbolic integration.\footnote{For an alternative integration algorithm that can tackle 
definite integrals see  \cite{RAAB} which generalizes Risch's indefinite integration algorithm \cite{RISCH} in the
setting differential fields.}

After the calculation of the $\ep$-expansions of the master integrals $M_i(N)$, we will write them again as functions in $x$. 
More precisely, we insert the truncated $\ep$-expansion into~\eqref{Equ:MxRep}. Finally, we plug these expressions 
into~\eqref{Equ:epExpansion}, expand the coefficients $c_i(x,\ep)$ and obtain an $\ep$-expansion of the form
%--------------------------------------------------------------------------------------------
\begin{equation}\label{equ:DxExp}
\hat{D}(x)=\ep^{-3}\hat{F}_{-3}(x)+\ep^{-2}\hat{F}_{-2}(x)+\ep^{-1}\hat{F}_{-1}(x)+\ep^{0}\hat{F}_{0}(x)+\dots,
\end{equation}
%--------------------------------------------------------------------------------------------
where the $\hat{F}_{i}(x)$ are given in terms of a linear combination of  formal power series in $x$ with coefficients 
being rational functions in $x$. Finally, we are in the position to calculate the $N$th coefficient $F_{i}(N)$  of 
$\hat{F}_{i}(x)$. This yields precisely the $\ep$-expansion of~\eqref{Equ:epExpansion}.

We emphasize that the proposed strategy has a subtle bottleneck: since the calculated coefficients $c_i(x,\ep)$ 
in~\eqref{Equ:DExp} may have poles in $\ep$, we have to compute the corresponding master integrals to higher orders 
in $\varepsilon$ in order to get the diagram $\hat{D}(x)$ to the constant term. Precisely these extra coefficients 
in the $\ep$-expansions are one of the true challenges to carry out the described tactic with our computer algebra toolbox.
In the next Sections we will present the key technologies to tackle these heavy calculations. 
%--------------------------------------------------------------------------------------------
\section{From ladder and \boldmath $V$-graphs to expressions in terms of master integrals}
\label{Sec:LadderGraphs}
%--------------------------------------------------------------------------------------------

\begin{figure}[ht]
%%%%%%%%%%%%%%%%%%%%%%%%%%%%%%%%%%%%%%%%%% 1 %%%%%%%%%%%%%%%%%%%%%%%%%%%%%%%%%%%%%%%%%%%%%%%%%%%%
\begin{minipage}[c]{0.31\linewidth}
     \includegraphics[width=1\textwidth]{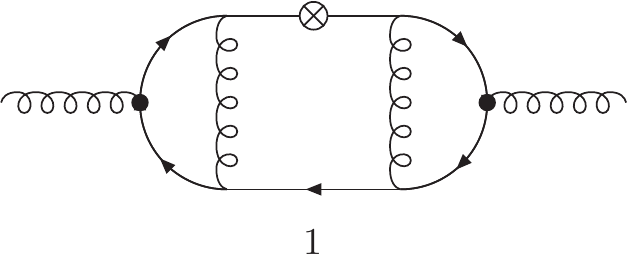}
%\vspace*{-8mm}
\end{minipage}
\hspace*{1mm}
\begin{minipage}[c]{0.31\linewidth}
     \includegraphics[width=1\textwidth]{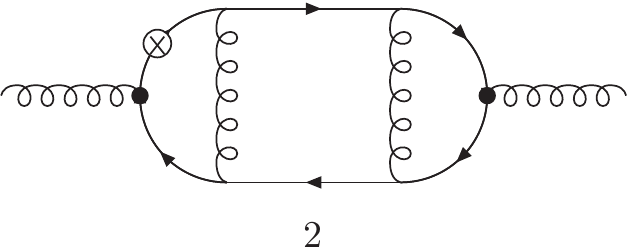}
%\vspace*{-8mm}
\end{minipage}
\hspace*{1mm}
\begin{minipage}[c]{0.31\linewidth}
     \includegraphics[width=1\textwidth]{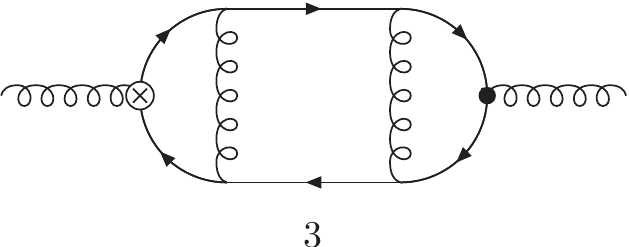}
%\vspace*{-8mm}
\end{minipage}
\hspace*{1mm}
%\begin{minipage}[c]{0.23\linewidth}
%     \includegraphics[width=1\textwidth]{axo4.pdf}
%\vspace*{-8mm}
%\end{minipage}

\vspace*{4mm}
%%%%%%%%%%%%%%%%%%%%%%%%%%%%%%%%%%%%%%%%%%%%% 2 %%%%%%%%%%%%%%%%%%%%%%%%%%%%%%%%%%%%%%%%%%%%%%%%%%%%%%%
\begin{minipage}[c]{0.31\linewidth}
     \includegraphics[width=1\textwidth]{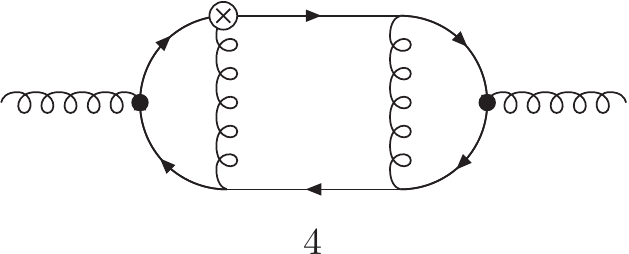}
%\vspace*{-8mm}
\end{minipage}
\hspace*{1mm}
\begin{minipage}[c]{0.31\linewidth}
     \includegraphics[width=1\textwidth]{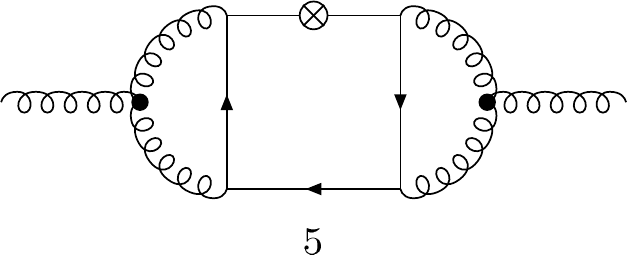}
%\vspace*{-8mm}
\end{minipage}
\hspace*{1mm}
\begin{minipage}[c]{0.31\linewidth}
     \includegraphics[width=1\textwidth]{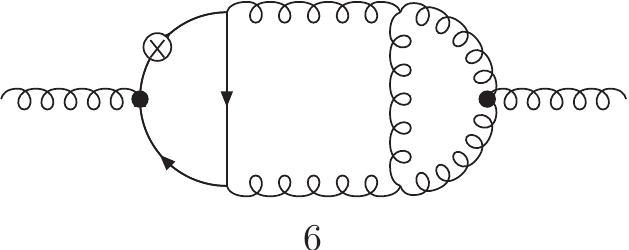}
%\vspace*{-8mm}
\end{minipage}
\hspace*{1mm}
%\begin{minipage}[c]{0.23\linewidth}
%     \includegraphics[width=1\textwidth]{axo8.pdf}
%\vspace*{-8mm}
%\end{minipage}

\vspace*{4mm}
%%%%%%%%%%%%%%%%%%%%%%%%%%%%%%%%%%%%%%%%%%%%% 3 %%%%%%%%%%%%%%%%%%%%%%%%%%%%%%%%%%%%%%%%%%%%%%%%%%%%%%%
\begin{minipage}[c]{0.31\linewidth}
     \includegraphics[width=1\textwidth]{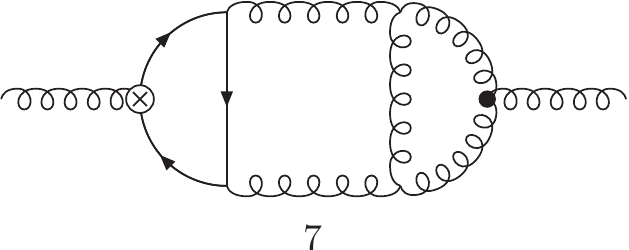}
%\vspace*{-8mm}
\end{minipage}
\hspace*{1mm}
\begin{minipage}[c]{0.31\linewidth}
     \includegraphics[width=1\textwidth]{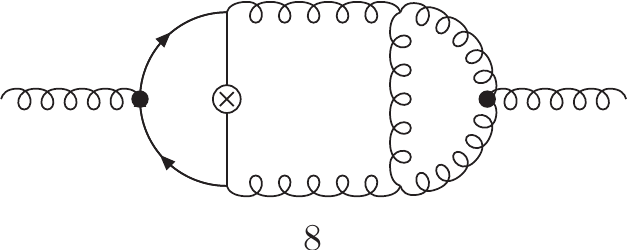}
%\vspace*{-8mm}
\end{minipage}
\hspace*{1mm}
\begin{minipage}[c]{0.31\linewidth}
     \includegraphics[width=1\textwidth]{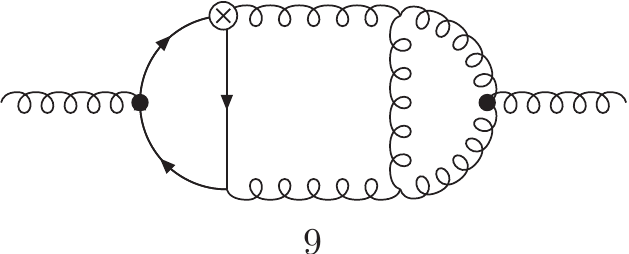}
%\vspace*{-8mm}
\end{minipage}
\hspace*{1mm}
%\begin{minipage}[c]{0.23\linewidth}
%     \includegraphics[width=1\textwidth]{axo12.pdf}
%\vspace*{-8mm}
%\end{minipage}

\iffalse
\vspace*{4mm}
%%%%%%%%%%%%%%%%%%%%%%%%%%%%%%%%%%%%%%%%%%%%% 4 %%%%%%%%%%%%%%%%%%%%%%%%%%%%%%%%%%%%%%%%%%%%%%%%%%%%%%%
\begin{minipage}[c]{0.31\linewidth}
     \includegraphics[width=1\textwidth]{axo10.pdf}
%\vspace*{-8mm}
\end{minipage}
\hspace*{1mm}
\begin{minipage}[c]{0.31\linewidth}
     \includegraphics[width=1\textwidth]{axo11.pdf}
%\vspace*{-8mm}
\end{minipage}
\hspace*{1mm}
\begin{minipage}[c]{0.31\linewidth}
     \includegraphics[width=1\textwidth]{axo12.pdf}
%\vspace*{-8mm}
\end{minipage}
\hspace*{1mm}
%\begin{minipage}[c]{0.23\linewidth}
%     \includegraphics[width=1\textwidth]{Diag_Aqq59.eps}
%\vspace*{-8mm}
%\end{minipage}
\fi
\caption{\sf \small Sample of ladder diagrams.}
\label{samplediagrams}
\end{figure}

\vspace{1mm}
\noindent
In Figure~\ref{samplediagrams} we show a sample of ladder diagrams, which will serve as examples to demonstrate the calculation 
techniques outlined in the present paper. All of these diagrams contribute to the operator matrix element $A_{Qg}^{(3)}$, but the 
methods described here can 
be applied to the ladder diagrams in $A_{gg,Q}^{(3)}$ as well, cf.~\cite{AGG}. The OMEs $A_{Qg}^{(3)}$ and $A_{gg,Q}^{(3)}$, 
together with  $A_{qg,Q}^{(3)}$, are the only operator matrix elements where ladder diagrams appear. In the latter 
case one of the closed fermion lines is massless \cite{Ablinger:2010ty,Blumlein:2012vq,Behring:2014eya}. Those graphs can 
be computed, however, by applying hypergeometric function techniques \cite{HYP,Slater}. 
For all other OMEs, 
such as $A_{Qq}^{(3), \rm PS}$ and $A_{qq,Q}^{(3), \rm NS}$, neither ladder nor non-planar topologies arise. 
The most complicated topologies in case of $A_{gq}^{(3),\rm S}, A_{qq,Q}^{(3), \rm NS}$ 
and $A_{Qq}^{(3), \rm PS}$ correspond to Benz diagrams with three massive internal lines 
\cite{Ablinger:2014lka,Ablinger:2014vwa,Ablinger:2014nga}.

%--------------------------------------------------------------------------------------------
%--------------------------------------------------------------------------------------------
%\subsection{The application of Feynman rules}
%--------------------------------------------------------------------------------------------

The diagrams are assembled using the standard Feynman rules of QCD \cite{YND}, together with the Feynman rules for operator 
insertions, cf.~\cite{Bierenbaum:2009mv}. After the corresponding Green functions are projected as described in 
\cite{Bierenbaum:2009mv,Klein:2009ig}, the operator matrix elements end up being expressed in terms of a linear combination 
of scalar integrals. For example, diagrams 1 and 2 in Figure~\ref{samplediagrams} can be written as a linear combination of 
the following type of integrals,
%--------------------------------------------------------------------------------------------
\begin{equation}
 \int \frac{d^Dk_1}{(2 \pi)^D}  \frac{d^Dk_2}{(2 \pi)^D} \frac{d^Dk_3}{(2 \pi)^D} \,\,
\frac{(\Delta.k_1)^{a_1} (\Delta.k_2)^{a_2} (\Delta.k_3)^{a_3}}{P_1^{\nu_1} P_2^{\nu_2} \cdots P_9^{\nu_9}} (\Delta.k_3)^{N-1} \, ,
\label{int1}
\end{equation}
%--------------------------------------------------------------------------------------------
while diagram 3 in Figure~\ref{samplediagrams} can be expressed in terms of the integrals
%--------------------------------------------------------------------------------------------
\begin{equation}
 \int \frac{d^Dk_1}{(2 \pi)^D}  \frac{d^Dk_2}{(2 \pi)^D} \frac{d^Dk_3}{(2 \pi)^D} \,\,
\frac{(\Delta.k_1)^{a_1} (\Delta.k_2)^{a_2} (\Delta.k_3)^{a_3}}{P_1^{\nu_1} P_2^{\nu_2} \cdots P_9^{\nu_9}}
\sum_{j=0}^{N-2} (\Delta.k_3)^j (\Delta.k_3-\Delta.p)^{N-2-j} \, ,
\label{int2}
\end{equation}
%--------------------------------------------------------------------------------------------
where $\nu_1, \ldots, \nu_9$, $a_1$, $a_2$ and $a_3$ are integers, and the inverse propagators $P_1, \ldots, P_9$ are
given in both cases by
%--------------------------------------------------------------------------------------------
\begin{eqnarray}
&& P_1 = k_1^2-m^2, \quad P_2 = (k_1-p)^2-m^2, \quad P_3 = k_2^2-m^2 \phantom{()}, \quad P_4 = (k_2-p)^2-m^2,  \nonumber \\
&& P_5 = k_3^2-m^2, \quad P_6 = (k_1-k_3)^2, \phantom{-m^2} \quad P_7 = (k_2-k_3)^2, \quad P_8 = (k_1-k_2)^2 \quad {\rm and} 
\nonumber \\
&& P_9 = (k_3-p)^2-m^2\, .  
\label{props1}
\end{eqnarray}
%--------------------------------------------------------------------------------------------
Here, the momentum of the external gluon is given by $p$, with $p^2 = 0$, the mass of the heavy quark is $m$, and $\Delta$ is 
a light-like vector. The integrals contributing to the diagrams in Figure~\ref{samplediagrams} have at most eight propagators. The 
additional ninth 
propagator is an auxiliary one that allows us to express any dot product of momenta as a linear combination of (inverse) 
propagators. In this way, any integral with irreducible numerators consisting of such dot products can be uniquely expressed 
in terms of integrals where the indices $\nu_1, \ldots, \nu_9$ are allowed to be negative. Later, we will also introduce 
artificial propagators that will allow us to express the dot products of $\Delta$ with internal momenta in a similar way. 

Notice that in the case of diagrams 
where the operator insertion is connected to an external gluon, such as in diagram~3, the
sum in $j$ can be performed, leading to two terms, each of which has the form of a line insertion. For example, the operator 
insertion in Eq. (\ref{int2}) gives
%--------------------------------------------------------------------------------------------
\begin{equation}
\sum_{j=0}^{N-2} (\Delta.k_3)^j (\Delta.k_3-\Delta.p)^{N-2-j} =
\frac{1}{\Delta.p} \left[(\Delta.k_3)^{N-1} - (\Delta.k_3-\Delta.p)^{N-1}\right] \, .
\label{summedj}
\end{equation}
%--------------------------------------------------------------------------------------------
The form of Eq.~(\ref{summedj}) will play a role later when we try to reduce the complexity of the problem using integration by 
parts identities.

\begin{figure}
\begin{minipage}[c]{0.31\linewidth}
     \includegraphics[width=1\textwidth]{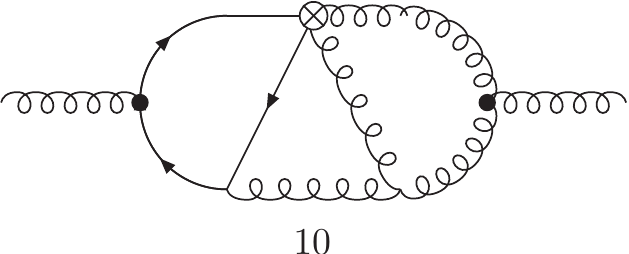}
%\vspace*{-8mm}
\end{minipage}
\hspace*{1mm}
\begin{minipage}[c]{0.31\linewidth}
     \includegraphics[width=1\textwidth]{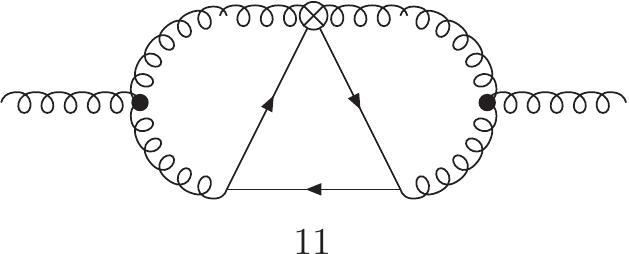}
%\vspace*{-8mm}
\end{minipage}
\hspace*{1mm}
\begin{minipage}[c]{0.31\linewidth}
     \includegraphics[width=1\textwidth]{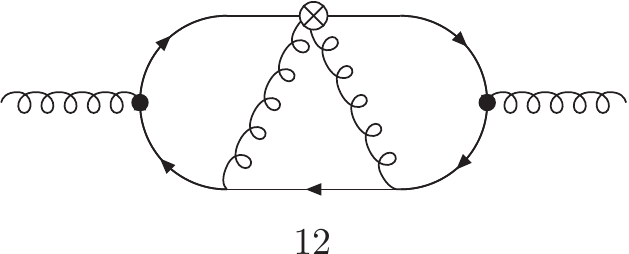}
\end{minipage}
\caption{\sf \small Diagrams with a central triangle.}
\label{samplediagrams2}
\end{figure}

In Figure~\ref{samplediagrams2} a few examples of diagrams with a central triangle ($V$-diagrams) are shown. These are 
ladder diagrams where the top middle propagator is missing, and the operator vertex insertion has two gluons coming out of the 
vertex. These diagrams will in general need to be expressed in terms of two types of integrals, each one corresponding to 
the two terms appearing in the $q\bar{q}gg$ operator insertion. For example, diagram 12 in Figure~\ref{samplediagrams2} can be 
expressed in terms of the integrals
%--------------------------------------------------------------------------------------------
\begin{equation}
\int \frac{d^Dk_1}{(2 \pi)^D}  \frac{d^Dk_2}{(2 \pi)^D} \frac{d^Dk_3}{(2 \pi)^D} \,\,
\frac{(\Delta.k_1)^{a_1} (\Delta.k_2)^{a_2} (\Delta.k_3)^{a_3}}{P_1^{\nu_1} P_2^{\nu_2} \cdots P_9^{\nu_9}}
\sum_{j=0}^{N-3} \sum_{l=j+1}^{N-2} (\Delta.k)^j (\Delta.k_2)^{N-2-l} (\Delta.k_3)^{l-j-1} 
\label{int3}
\end{equation}
%--------------------------------------------------------------------------------------------
and 
%--------------------------------------------------------------------------------------------
\begin{equation}
\int \frac{d^Dk_1}{(2 \pi)^D}  \frac{^Ddk_2}{(2 \pi)^D} \frac{d^Dk_3}{(2 \pi)^D} \,\,
\frac{(\Delta.k)^{a_1} (\Delta.k_2)^{a_2} (\Delta.k_3)^{a_3}}{P_1^{\nu_1} P_2^{\nu_2} \cdots P_7^{\nu_7} {P'}_8^{\nu_8} 
P_9^{\nu_9}} 
\sum_{j=0}^{N-3} \sum_{l=j+1}^{N-2} (\Delta.k)^j (\Delta.k_2)^{N-2-l} (\Delta.k_3)^{l-j-1}  \, ,
\label{int4}
\end{equation}
%--------------------------------------------------------------------------------------------
where the propagators are the ones defined in Eq. (\ref{props1}), and in the case of the integral in Eq. (\ref{int4}), the 
eight's propagator gets modified by
%--------------------------------------------------------------------------------------------
\begin{equation}
P'_8 = (k_1+k_2-k_3-p)^2-m^2\, .
\end{equation}
%--------------------------------------------------------------------------------------------
%%In Eqs. (\ref{int3}--\ref{int4}), we have used the shorthand notation
%--------------------------------------------------------------------------------------------
%%\begin{equation}
%%dk \rightarrow \frac{d^D k}{(2 \pi)^D}  \frac{d^D k_2}{(2 \pi)^D} \frac{d^D k_3}{(2 \pi)^D} \, .
%%\end{equation}
%--------------------------------------------------------------------------------------------

%--------------------------------------------------------------------------------------------
%--------------------------------------------------------------------------------------------

\begin{figure}[H]
\begin{center}
\includegraphics[width=\textwidth]{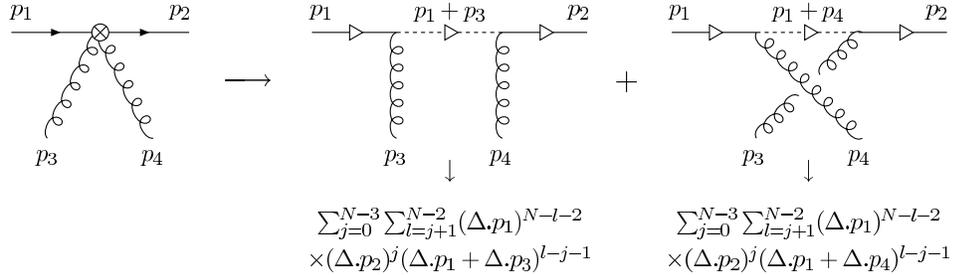}
%%qqggVrtxDecomp.eps}
\end{center}
\vspace*{-17cm}
\caption{\sf \small Representation showing the origin of the two terms appearing in the $q\bar{q}gg$-operator insertion. 
The gluon momenta is taken incoming.}
\label{qqggvrtxdecomp}
\end{figure}

The two terms appearing in the $q\bar{q}gg$-vertex operator insertion are directly related to the two ways in which an 
additional line can be inserted in between the gluons, as represented in Figure~\ref{qqggvrtxdecomp}. 
%--------------------------------------------------------------------------------------------
%--------------------------------------------------------------------------------------------

\begin{figure}
\begin{center}
%%%%%%%%%%%%%%%%%%%%%%%%%%%%%%%%%%%%%%%%%% 1 %%%%%%%%%%%%%%%%%%%%%%%%%%%%%%%%%%%%%%%%%%%%%%%%%%%%
\begin{minipage}[c]{0.31\linewidth}
     \includegraphics[width=\textwidth]{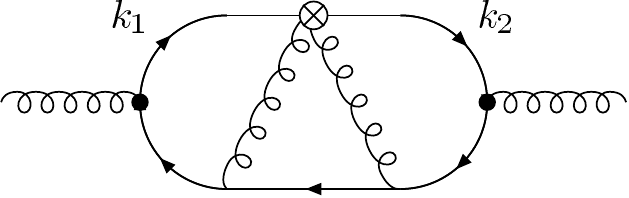}
\end{minipage}
\begin{minipage}[c]{0.04\linewidth}
\vspace*{-1mm}
{\Large $\nearrow$}

$\phantom{a}$

{\Large $\searrow$}
\end{minipage}
\begin{minipage}[c]{0.31\linewidth}
     \includegraphics[width=1\textwidth]{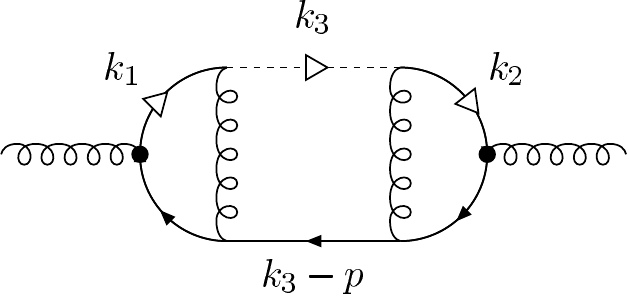}

$\phantom{a}$

     \includegraphics[width=1\textwidth]{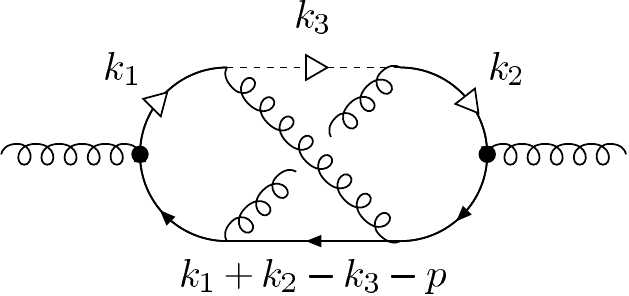}
\end{minipage}
%\vspace*{1mm}
%%%%%%%%%%%%%%%%%%%%%%%%%%%%%%%%%%%%%%%%%%%%% 2 %%%%%%%%%%%%%%%%%%%%%%%%%%%%%%%%%%%%%%%%%%%%%%%%%%%%%%%
\end{center}
\caption{\sf \small A depiction of the two terms appearing in diagram 12 of Figure~\ref{samplediagrams}. The internal momenta 
have been chosen so that the operator insertion term is the same in both diagrams on the right hand side. The bottom middle 
propagator then differs in each case.}
\label{Vexplain}
\end{figure}

\noindent
Here, a dashed line represents an additional fermion line, which does not appear as a propagator in a diagram. There are three 
dot products of $\Delta$ with momenta appearing in the double sum of the two terms of the operator insertion. These are the 
momenta going through the fermion lines with the triangles in Figure~\ref{qqggvrtxdecomp}, in the direction the triangles 
indicate. Two of these momenta are $p_1$ and $p_2$ (the momenta of the fermion lines actually present), and the third one is the 
one going through the dashed line. The corresponding terms in the case of diagram 12 are depicted in Figure~\ref{Vexplain}. 
Notice that one of the diagrams on the right hand side of Figure~\ref{Vexplain} has a non-planar structure, even though the 
original diagram is a planar one. In other words, the operator insertion is introducing non-planarity to an otherwise planar 
diagram, which is the main reason why this diagram turns out to be particularly difficult. As we will see, the planar part 
of diagram~12 (the top diagram on the right hand side of Figure~\ref{Vexplain}) is quite simple and involves only standard 
harmonic sums. The non-planar part (the bottom diagram on the right hand side of Figure~\ref{Vexplain}), on the other hand, 
requires sophisticated computation techniques, leading to generalized cyclotomic binomial sums.

In the case of diagrams~10 and 11 in Figure~\ref{samplediagrams2}, the operator insertion again has two parts. However, in 
these cases the non-planar part is not present by virtue of having a vanishing color factor. Only the planar part survives, 
making the calculation of these diagrams much simpler than it would be otherwise.

%%%%%%%%%%%%%%%%%%%%%%%%%%%%%%%%%%%%%%%%%%%%%%%%%%%%%%%%%%%%%%%%%%%%%%%
\subsubsection*{Reduction to master integrals}
%\label{sec:Reduction2MIs}
%%%%%%%%%%%%%%%%%%%%%%%%%%%%%%%%%%%%%%%%%%%%%%%%%%%%%%%%%%%%%%%%%%%%%%%

\vspace*{1mm}
\noindent
As usual in QCD perturbative calculations, the number of scalar integrals required to express the diagrams is quite large. 
This demands the use of integration by parts identities \cite{IBP} in order to express all scalar integrals in terms of a 
much smaller set of so called master integrals. For this purpose we use the package {\tt Reduze 2} \cite{Reduze1,Reduze2}. 
It can be applied whenever the integrals to be reduced can be uniquely identified by a set of propagators with definite powers. In 
our case, the presence of operator insertions, leading to terms such as $(\Delta.k_3)^{N-1}$ in Eq. (\ref{int1}) or 
$\sum_{j=0}^{N-2} (\Delta.k_3)^j (\Delta.k_3-\Delta.p)^{N-2-j}$ in Eq. (\ref{int2}), represents a complication for the 
application of this method, since the dot products of $\Delta$ with internal momenta are raised to indefinite powers.
We deal with this complication by transforming the operator insertion terms into artificial 
propagators~\cite{Ablinger:2014uka,Ablinger:2014vwa,Ablinger:2014nga}. As mentioned in Section \ref{sec:CAChallenge}, this is 
achieved by 
introducing
a new variable $x$, multiplying the integrals by $x^{N-i}$ (where $i=1,2,3$, depending on the type of insertion), and then 
summing from $N=i$ to infinity\footnote{We can always, of course, shift $N$ by $i$ so that the sum starts at $N=0$ as 
in~Eq.~(\ref{Equ:MxRep}).}. In this way, the term $(\Delta.q)^{N-1}$ appearing in the operator insertion on a fermion line becomes
%---------------------------------------------------------------------------------------------------------------------------
\begin{equation}
(\Delta.q)^{N-1} \rightarrow \sum_{N=1}^{\infty} x^{N-1} (\Delta.q)^{N-1} = \frac{1}{1-x \Delta.q} \, .
\label{N2x1}
\end{equation}
The vertex operator insertion with one gluon coming out of the vertex gets transformed into
\begin{eqnarray}
\sum_{j=0}^{N-2} (\Delta.q_1)^j (\Delta.q_2)^{N-2-j} &\rightarrow& \sum_{N=2}^{\infty} x^{N-2} \sum_{j=0}^{N-2} (\Delta.q_1)^j (\Delta.q_2)^{N-2-j} \nonumber \\
&=& \frac{1}{(1-x \Delta.q_1) (1-x \Delta.q_2)} \, .
\label{N2x2}
\end{eqnarray}
%---------------------------------------------------------------------------------------------------------------------------
Each term in the vertex operator insertion with two gluons in the vertex turns into
%---------------------------------------------------------------------------------------------------------------------------
\begin{equation}
\sum_{j=0}^{N-3} \sum_{l=j+1}^{N-2} (\Delta.q_1)^j (\Delta.q_2)^{N-2-l} (\Delta.q_3)^{l-j-1} \rightarrow 
\frac{1}{(1-x \Delta.q_1) (1-x \Delta.q_2) (1-x \Delta.q_3)} \, .
\label{N2x3}
\end{equation}
%---------------------------------------------------------------------------------------------------------------------------
The transformed operator insertions can be treated as additional propagators that will now have definite powers, and Laporta's 
algorithm can then be applied without problems. One may introduce additional artificial propagators on top of the ones 
generated by Eqs.~(\ref{N2x1}--\ref{N2x3}), in such a way that $\Delta.k_1$, $\Delta.k_2$ and $\Delta.k_3$ can all be uniquely 
expressed in terms of these propagators, which requires three such artificial propagators. Together with the original nine 
propagators, we end up with twelve propagator integrals, where the dot products of any vector with internal momenta can be 
uniquely expressed as a linear combination of these twelve (inverse) propagators. A set of propagators satisfying this 
property defines an {\it integral family}. Different integrals belonging to a given integral family are uniquely identified 
by the powers of the propagators, which can be negative, denoting the presence of irreducible numerators.

We have found that the calculation of {\it all} operator matrix elements can be achieved using a total of 24 integral families, 
although only a few of them are needed for the examples shown in this paper. In Table~\ref{families1}, we show the integral 
families required in order to express all diagrams appearing in Figures~\ref{samplediagrams} and \ref{samplediagrams2} in 
terms of scalar integrals. The names {\tt B1a}, {\tt B3a} and {\tt C3a} label the different families (the rationale behind 
these names was explained in Ref.~\cite{Ablinger:2014nga}). Diagrams 1 to 4 can be expressed in terms of scalar integrals in the {\tt 
B3a} 
family, diagrams~5 to 11 are written in terms of {\tt B1a} integrals, and diagram~12 can be written in terms of {\tt B3a} and 
{\tt C3a} integrals. It turns out that when we perform the reduction of these integrals using {\tt Reduze 2}, most of the 
master integrals remain within these three integral families, but in our setup, a few master integrals are mapped to
the simpler families {\tt B5a}, {\tt B5b} and {\tt B5c} shown in Table \ref{families2}. This occurs when we reduce the {\tt C3a} 
integrals for diagram~12.

%%%%%%%%%%%%%%%%%%%%%%%%%%%%%%%%%%%%%%%%%%%%%%%%%%%%%%%%%%%%%%%%%%%%%%%%%%%%%%%%%%%%%%%%%%%%%%%%%%%%%%%%%%%%%%%%%%%%%%%%%%%%%%%%%%%%%%%%%
\begin{table}[ht]
\begin{center}
\begin{tabular}{|l|l|l|l|l|}
\hline
          &  {\tt B1a}                     &  {\tt B3a}               &  {\tt C3a}                 \\
\hline
$P_1$     &  $k_1^2$                       &  $k_1^2-m^2$             &  $k_1^2-m^2$               \\
$P_2$     &  $(k_1-p)^2$                   &  $(k_1-p)^2-m^2$         &  $(k_1-p)^2-m^2$           \\
$P_3$     &  $k_2^2$                       &  $k_2^2-m^2$             &  $k_2^2-m^2$               \\
$P_4$     &  $(k_2-p)^2$                   &  $(k_2-p)^2-m^2$         &  $(k_2-p)^2-m^2$           \\
$P_5$     &  $k_3^2-m^2$                   &  $k_3^2-m^2$             &  $k_3^2-m^2$               \\
$P_6$     &  $(k_3-k_1)^2-m^2$             &  $(k_3-k_1)^2$           &  $(k_3-k_1)^2$             \\
$P_7$     &  $(k_3-k_2)^2-m^2$             &  $(k_3-k_2)^2$           &  $(k_3-k_2)^2$             \\
$P_8$     &  $(k_1-k_2)^2$                 &  $(k_1-k_2)^2$           &  $(k_1+k_2-k_3-p)^2-m^2$   \\
$P_9$     &  $(k_3-p)^2-m^2$               &  $(k_3-p)^2-m^2$         &  $(k_3-p)^2-m^2$           \\
$P_{10}$  &  $1-x \Delta.(k_3-k_1)$        &  $1-x \Delta.k_1$        &  $1-x \Delta.k_1$          \\
$P_{11}$  &  $1-x \Delta.k_3$              &  $1-x \Delta.k_3$        &  $1-x \Delta.k_3$          \\
$P_{12}$  &  $1-x \Delta.(k_3-k_2)$        &  $1-x \Delta.k_2$        &  $1-x \Delta.k_2$          \\
\hline
\end{tabular}
\end{center}
\caption{\sf \small Propagators of the integral families used to express all ladder diagrams in Figures~\ref{samplediagrams}
and \ref{samplediagrams2} in terms of scalar integrals.}
\label{families1}
\end{table}
%%%%%%%%%%%%%%%%%%%%%%%%%%%%%%%%%%%%%%%%%%%%%%%%%%%%%%%%%%%%%%%%%%%%%%%%%%%%%%%%%%%%%%%%%%%%%%%%%%%%%%%%%%%%%%%%%%%%%%%%%%%%%%%%%%%%%%%%%

%%%%%%%%%%%%%%%%%%%%%%%%%%%%%%%%%%%%%%%%%%%%%%%%%%%%%%%%%%%%%%%%%%%%%%%%%%%%%%%%%%%%%%%%%%%%%%%%%%%%%%%%%%%%%%%%%%%%%%%%%%%%%%%%%%%%%%%%%
\begin{table}[ht]
\begin{center}
\begin{tabular}{|l|l|l|l|}
\hline
         &  {\tt B5a}            &  {\tt B5b}                 &  {\tt B5c}               \\
\hline
$P_1$    &  $k_1^2-m^2$          &  $k_1^2-m^2$               &  $k_1^2-m^2$             \\
$P_2$    &  $(k_1-p)^2-m^2$      &  $(k_1-p)^2-m^2$           &  $(k_1-p)^2-m^2$         \\
$P_3$    &  $k_2^2-m^2$          &  $k_2^2-m^2$               &  $k_2^2-m^2$             \\
$P_4$    &  $(k_2-p)^2-m^2$      &  $(k_2-p)^2-m^2$           &  $(k_2-p)^2-m^2$         \\
$P_5$    &  $k_3^2$              &  $k_3^2$                   &  $k_3^2$                 \\
$P_6$    &  $(k_3-k_1)^2-m^2$    &  $(k_3-k_1)^2-m^2$         &  $(k_3-k_1)^2-m^2$       \\
$P_7$    &  $(k_3-k_2)^2-m^2$    &  $(k_3-k_2)^2-m^2$         &  $(k_3-k_2)^2-m^2$       \\
$P_8$    &  $(k_1-k_2)^2$        &  $(k_1-k_2)^2$             &  $(k_1-k_2)^2$           \\
$P_9$    &  $(k_3-p)^2$          &  $(k_3-p)^2$               &  $(k_3-p)^2$             \\
$P_{10}$ &  $1-x \Delta.k_1$     &  $1-x \Delta.(k_3-k_1)$    &  $1-x \Delta.k_1$        \\
$P_{11}$ &  $1-x \Delta.k_3$     &  $1-x \Delta.k_3$          &  $1-x \Delta.(k_1-k_3)$  \\
$P_{12}$ &  $1-x \Delta.k_2$     &  $1-x \Delta.(k_3-k_2)$    &  $1-x \Delta.(k_2-k_3)$  \\
\hline
\end{tabular}
\end{center}
\caption{\sf \small Additional integral families used for the reduction to master integrals.}
\label{families2}
\end{table}
%%%%%%%%%%%%%%%%%%%%%%%%%%%%%%%%%%%%%%%%%%%%%%%%%%%%%%%%%%%%%%%%%%%%%%%%%%%%%%%%%%%%%%%%%%%%%%%%%%%%%%%%%%%%%%%%%%%%%%%%%%%%%%%%%%%%%%%%%

\noindent
It must be pointed out that, in order to perform the reductions, for each one of the families mentioned above one needs to consider
also the version of the family with all momenta reversed ($k_i \rightarrow -k_i$, $p \rightarrow -p$). We refer to 
these families as 
the crossed version of the families. Reversing the momenta is equivalent to multiplying the integrals by $(-1)^N$ before we introduce 
the variable $x$, or performing $x \rightarrow -x$.

Diagrams with an operator insertion connected to an external gluon require special treatment. For example, in the case of 
diagram 3 in Figure~\ref{samplediagrams}, when we introduce the variable $x$ and perform the sum in $N$, we obtain
%-------------------------------------------------------------------------------------------------------------------------
\begin{equation}
\sum_{j=0}^{N-2} (\Delta.k_3)^j (\Delta.k_3-\Delta.p)^{N-2-j} \rightarrow 
\frac{1}{(1-x \Delta.k_3) (1-x (\Delta.k_3-\Delta.p))} \, .
\end{equation}
%-------------------------------------------------------------------------------------------------------------------------
The artificial propagators $1/(1-x \Delta.k_3)$ and $1/(1-x (\Delta.k_3-\Delta.p))$ cannot be both part of an integral 
family, since we would then need more than three artificial propagators in order to be able to express the dot products of 
$\Delta$ with internal momenta as linear combinations of such (inverse) propagators. Moreover, at least in the case of 
$\Delta.k_3$, such a linear combination would not be unique. Therefore, to deal with this case we first perform the partial 
fraction decomposition of the product of artificial propagators
%-------------------------------------------------------------------------------------------------------------------------
\begin{equation}
\frac{1}{(1-x \Delta.k_3) (1-x (\Delta.k_3-\Delta.p))} = \frac{1}{x \Delta.p} \left(\frac{1}{1-x \Delta.k_3} 
- \frac{1}{1-x (\Delta.k_3 - \Delta.p)}\right) \, .
\label{externalvrtxEq}
\end{equation}
%-------------------------------------------------------------------------------------------------------------------------
This is related to what we did before in Eq. (\ref{summedj}). Now each term looks like a line insertion and can be treated 
separately. The term $1/(1-x \Delta.k_3)$ already corresponds to one of the propagators in the integral families in 
Table~\ref{families1}. In the case of the term $1/(1-x (\Delta.k_3 - \Delta.p))$, we can perform the shift $k_i \rightarrow 
-k_i+p$ ($i=1,2,3)$, and we end up with the crossed version of the same integral family as in the case of the first term. This shift 
will, of course, 
reshuffle the powers of the propagators by $\nu_1 \leftrightarrow \nu_2$, $\nu_3 \leftrightarrow \nu_4$ and $\nu_5 
\leftrightarrow \nu_9$. The overall factor of $1/x$ on the right-hand side of (\ref{externalvrtxEq}) just implies a shift by 1 
in $N$ in the $N$-representation of the integrals.

%%%%%%%%%%%%%%%%%%%%%%%%%%%%%%%%%%%%%%%%%%%%%%%%%%%%%%%%%%%%%%%%%%%%%%%
\section{Calculation of the master integrals}
\label{sec:Calc-MIs}
%%%%%%%%%%%%%%%%%%%%%%%%%%%%%%%%%%%%%%%%%%%%%%%%%%%%%%%%%%%%%%%%%%%%%%%

\vspace*{1mm}
\noindent
In the previous sections we worked out the following reduction: Given a ladder diagram $D(N)$, we can switch to its formal 
power series representation~\eqref{Equ:PowerSeries} and can express it in terms of a linear combination~\eqref{Equ:DExp} of 
master integrals using the transformations~(\ref{N2x1}--\ref{N2x3}). For the current considerations we will set $m=1$ and 
$\Delta.p=1$. Also the color factors are suppressed and will be hidden within the ground field $\set K$. Furthermore, we will 
omit an overall factor of $i$ and the spherical factor $S_\ep = \exp[\tfrac{\ep}{2} (\gamma_E - \ln(4\pi))]$, with $\gamma_E$ the 
Euler-Mascheroni constant.

Once we perform the reduction to master integrals $\hat{M}_i(x)$ with~\eqref{Equ:MxRep}, the next step is their calculation as 
functions of the Mellin variable $N$. More precisely, based on the output of the integration by parts methods, the integral 
representation of the $N$th coefficient $M(N)$ can be read off. For the ladder diagrams considered in this paper, we 
obtained the coefficients of their $\ep$-expansion~\eqref{Equ:epExpansionMI} with the following techniques:
%-------------------------------------------------------------------------------------------------------------------------
\begin{enumerate}
\item Beta functions, generalized hypergeometric and associated functions \cite{HYP,Slater,Appell} and symbolic summation 
methods~\cite{Schneider:15a,Schneider:07d,Schneider:10c,Schneider:10b,RecSolver,Schneider:01,Schneider:05a} based on difference ring/field theory~\cite{Karr:81,Schneider:08c,DFTheory}, 
implemented in 
Schneider's 
{\tt Mathematica} packages {\tt Sigma} \cite{SIG1,SIG2}, {\tt EvaluateMultiSums}, {\tt SumProduction} \cite{EMSSP}, and 
special function tools implemented in Ablinger's {\tt Mathematica} package {\tt HarmonicSums} 
\cite{HARMONICSUMS,Ablinger:PhDThesis,Ablinger:2011te,Ablinger:2013cf,Ablinger:2014bra}.
\item Mellin-Barnes integral representations \cite{MB} in combination with the symbolic summation packages mentioned above.
\item An enhanced Almkvist-Zeilberger algorithm~\cite{AZ} implemented within Ablinger's package 
\texttt{MultiIntegrate}~\cite{Ablinger:PhDThesis} to compute recurrences for multiple integrals and a recurrence solver 
for Laurent series solutions~\cite{Blumlein:2010zv} implemented in \texttt{Sigma}.
\item A coupled differential (difference) equations solver \cite{DEQ} implemented in \texttt{SumProduction} which is based 
on the symbolic summation packages above and which requires Z\"urcher's uncoupling algorithm~\cite{Zuercher:94} implemented 
in Gerhold's package {\tt OreSys}~\cite{ORESYS}.
\end{enumerate}
%-------------------------------------------------------------------------------------------------------------------------

\noindent 
Below we will explain the different technologies and illustrate them with concrete examples. 

%%%%%%%%%%%%%%%%%%%%%%%%%%%%%%%%%%%%%%%%%%%%%%%%%%%%%%%%%%%%%%%%%%%%%%%
\subsection{Beta functions, hypergeometric functions, and symbolic summation}
\label{sec:Sigma-MIs}
%%%%%%%%%%%%%%%%%%%%%%%%%%%%%%%%%%%%%%%%%%%%%%%%%%%%%%%%%%%%%%%%%%%%%%%

Many of the master integrals were calculated using standard Feynman parameterization, and in many cases
the resulting Feynman parameter integrals could be done immediately in terms of Beta functions, if it were not
for the term coming from the operator insertion. Consider for example the following master integral
%-------------------------------------------------------------------------------------------------------------------------
\begin{equation}
J_1 = \int \frac{d^Dk_1}{(2 \pi)^D}  \frac{d^Dk_2}{(2 \pi)^D} \frac{d^Dk_3}{(2 \pi)^D} \,\,
      \frac{\sum_{j=0}^N (\Delta.k_3)^j (\Delta.k_3-\Delta.k_1)^{N-j}}{P_1 P_2 P_5 P_7 P_8 P_9},
\end{equation}
%-------------------------------------------------------------------------------------------------------------------------
where the propagators are again those of family {\tt B1a}, i.e.
%-------------------------------------------------------------------------------------------------------------------------
\begin{eqnarray}
&& P_1 = k_1^2, \,\,\phantom{-m^2} \quad P_2 = (k_1-p)^2, \phantom{3-m^2} \quad P_3 = k_2^2, \,\,\,\, \phantom{-k_3)^2-m^2} 
\quad P_4 = (k_2-p)^2, \nonumber \\
&& P_5 = k_3^2-m^2,            \quad P_6 = (k_1-k_3)^2-m^2,               \quad P_7 = (k_2-k_3)^2-m^2,             \quad P_8 
= (k_1 - k_2)^2 \quad {\rm and} \nonumber \\
&& P_9 = (k_3-p)^2-m^2\, .
\label{B1aProps}
\end{eqnarray}
%-------------------------------------------------------------------------------------------------------------------------
Introducing five Feynman parameters we obtain
%-------------------------------------------------------------------------------------------------------------------------
\begin{eqnarray}
J_1 &=& - \int_0^1 dx_1 \cdots \int_0^1 dx_5 \,\, \Gamma\left(-{\textstyle \frac{3}{2}} \varepsilon\right)
x_2^{-1+\varepsilon/2} (1-x_2)^{\varepsilon/2}
x_4^{-1-\varepsilon} (1-x_4)^{-\varepsilon}
x_5^{\varepsilon} (1-x_5)^{-\varepsilon/2} \nonumber \\ 
&& \phantom{- \int_0^1 dx_1 \cdots} \times
\sum_{j=0}^N \left[ x_3 (1-x_4)-x_1 (1-x_2)+(x_4-x_2) (x_1 (1-x_5)+x_3 x_5) \right]^j \nonumber \\
&& \phantom{- \int_0^1 dx_1 \cdots \times \sum_{j=0}^N} \times
\left[ x_3 (1-x_4)+x_4 (x_1 (1-x_5)+x_3 x_5) \right]^{N-j} \, .
\end{eqnarray}
%-------------------------------------------------------------------------------------------------------------------------
Notice that in the case $N=0$, this integral can be done directly in terms of Beta functions. We can now try to split
the polynomials in the operator insertion term using binomial expansions, leading to a result in terms of multi-sums that
can then be integrated in terms of Beta functions. This can be done in multiple ways, and in terms of the number of sums arising,
some ways are more efficient than others. Here we rewrite the original sum coming from the operator insertion using
%-------------------------------------------------------------------------------------------------------------------------
\begin{equation}
\sum_{j=0}^N A^j B^{N-j} = \frac{A^{N+1}-B^{N+1}}{A-B} = \sum_{j=0}^N \binom{N+1}{j} (A-B)^{N-j} B^j \, .
\end{equation}
%-------------------------------------------------------------------------------------------------------------------------
In this way, our integral becomes
%---------------------------------------------------------------------------------------------------------------------------------
\begin{eqnarray}
J_1 &=& - \int_0^1 dx_1 \cdots \int_0^1 dx_5 \,\, \Gamma\left(-{\textstyle \frac{3}{2}} \varepsilon\right)
x_2^{-1+\varepsilon/2} (1-x_2)^{\varepsilon/2}
x_4^{-1-\varepsilon} (1-x_4)^{-\varepsilon}
x_5^{\varepsilon} (1-x_5)^{-\varepsilon/2} \nonumber \\ 
&& \phantom{- \int_0^1 } \times
\sum_{j=0}^N \binom{N+1}{j} \left[ x_4 (1-x_5) (x_1-x_3)+x_3 \right]^j 
\left[ x_2 x_5 (x_1-x_3)-x_1 \right]^{N-j} \nonumber \\
&=& - \int_0^1 dx_1 \cdots \int_0^1 dx_5 \,\, \Gamma\left(-{\textstyle \frac{3}{2}} \varepsilon\right)
\sum_{j=0}^N \sum_{k=0}^j \sum_{i=0}^{N-j} 
(-1)^{N-j-i} \binom{N+1}{j} \binom{j}{k} \binom{N-j}{i} \nonumber \\
&& \phantom{- \int_0^1 dx_1 \cdots \int_0^1 dx_5} \times
x_1^{N-j-i} x_2^{i-1+\varepsilon/2} (1-x_2)^{\varepsilon/2} x_3^{j-k} x_4^{k-1-\varepsilon} (1-x_4)^{-\varepsilon} \nonumber \\
&& \phantom{- \int_0^1 dx_1 \cdots \int_0^1 dx_5} \times
x_5^{i+\varepsilon} (1-x_5)^{k-\varepsilon/2} (x_1-x_3)^{i+k} \, .
\end{eqnarray}
%---------------------------------------------------------------------------------------------------------------------------------
And now we can perform the integrals in $x_1$ and $x_3$ using
%---------------------------------------------------------------------------------------------------------------------------------
\begin{equation}
\int_0^1 dx \int_0^1 dy \,\, x^{a-1} y^{b-1} (x-y)^{c-1} =
\frac{\Gamma(c)}{a+b+c-1} \left(\frac{\Gamma(b)}{\Gamma(b+c)}-(-1)^c \frac{\Gamma(a)}{\Gamma(a+c)} \right) \, ,
\end{equation}
%---------------------------------------------------------------------------------------------------------------------------------
which leads to
%---------------------------------------------------------------------------------------------------------------------------------
\begin{eqnarray}
J_1 &=& -\sum_{j=0}^N \sum_{k=0}^j \sum_{i=0}^{N-j} \, \Gamma\left(-{\textstyle \frac{3}{2}} \varepsilon\right)
(-1)^{N-j-i} \binom{N+1}{j} \binom{j}{k} \binom{N-j}{i} \frac{(i+k)!}{N+2} \nonumber \\
&& \phantom{\sum_{j=0}^N \sum_{k=0}^j \sum_{i=0}^{N-j}} \times
\left[(-1)^{i+k} \frac{(N-j-i)!}{(N-j+k+1)!}+\frac{(j-k)!}{(i+j+1)!}\right]
\Gamma(i+\varepsilon/2) \nonumber \\
&& \phantom{\sum_{j=0}^N \sum_{k=0}^j \sum_{i=0}^{N-j}} \times
\Gamma(1+\varepsilon/2)
\frac{\Gamma(k-\varepsilon) \Gamma(1-\varepsilon)}{\Gamma(k+1-2 \varepsilon)}
\frac{\Gamma(k+1-\varepsilon/2)}{\Gamma(i+k+2+\varepsilon/2)} \, .
\end{eqnarray}
%---------------------------------------------------------------------------------------------------------------------------------
We end up with a triple sum that can be performed using {\tt EvaluateMultiSums} in combination with \texttt{Sigma} and 
\texttt{HarmonicSums}; for further details on the interplay of the different difference ring/field algorithms we refer, e.g., to~\cite{EMSSP}.  
The calculation up to order $\ep^0$ gives
%---------------------------------------------------------------------------------------------------------------------------------
\begin{eqnarray}
J_1 &=& \frac{1}{N+2} \Biggl\{
\frac{1}{\varepsilon^3} \Biggl[
-\frac{4}{3} S_1
-\frac{4 (2 N+3)}{3 (N+1) (N+2)}
-\frac{4 (-1)^N}{3 (N+2)}
\Biggr]
+\frac{1}{\varepsilon^2} \Biggl[
\frac{2 \big(4 N^2+7 N+1\big)}{3 (N+1) (N+2)^2}
\nonumber \\ &&
+(-1)^N \biggl(
\frac{2 \big(2 N^3+6 N^2+9 N+7\big)}{3 (N+1)^2 (N+2)^2}
-\frac{2 S_1}{3 (N+2)}
\biggr)
-S_1^2
+\frac{2 (N-2) (2 N+3)}{3 (N+1) (N+2)} S_1
\nonumber \\ &&
+S_2
-\frac{4}{3} S_{-2}
\Biggr]
+\frac{1}{\varepsilon} \Biggl[
-\biggl(
\frac{4 N^2-6 N-1}{3 (N+1)^2}
+\frac{5}{6} S_2
+\frac{\zeta_2}{2}
\biggr) S_1 
+\frac{6 N^2+19 N+6}{6 (N+1) (N+2)} S_1^2
\nonumber \\ &&
+\biggl(
\frac{2 \big(2 N^2-5\big)}{3 (N+1) (N+2)}
-2 S_1
\biggr) S_{-2} 
-\frac{\big(6 N^2+17 N+16\big) S_2}{6 (N+1) (N+2)}
-S_{2,1}
+\frac{4}{3} S_{-2,1}
\nonumber \\ &&
+(-1)^N \biggl(
-\frac{4 N^5+24 N^4+57 N^3+60 N^2+16 N-11}{3 (N+1)^3 (N+2)^3}
-\frac{S_1^2+9 S_2+3 \zeta_2}{6 (N+2)}
%-\frac{S_1^2}{6 (N+2)}
%-\frac{3 S_2}{2 (N+2)}
%-\frac{\zeta_2}{2 (N+2)}
%
\nonumber \\ &&
+\frac{2 N^3+12 N^2+21 N+13}{3 (N+1)^2 (N+2)^2} S_1
\biggr)
-\frac{8 N^5+46 N^4+100 N^3+104 N^2+63 N+27}{3 (N+1)^3 (N+2)^3}
\nonumber \\ &&
-\frac{7}{18} S_1^3
+\frac{2}{9} S_3
-\frac{2}{3} S_{-3}
-\frac{(2 N+3) \zeta_2}{2 (N+1) (N+2)}
\Biggr]
+\biggl(
-\frac{12 N^3+50 N^2+59 N+62}{12 (N+1)^2 (N+2)}
\nonumber \\ &&
-\frac{25 S_2}{24}
-\frac{3 \zeta_2}{8}
\biggr) S_1^2
+\biggl(
\frac{8 N^4+12 N^3-22 N^2-53 N-66}{6 (N+1)^3 (N+2)}
+\frac{\big(10 N^2-N-36\big) S_2}{12 (N+1) (N+2)}
\nonumber \\ &&
-2 S_3
-\frac{3}{2} S_{2,1}
+2 S_{-2,1}
+\frac{(N-2) (2 N+3) \zeta_2}{4 (N+1) (N+2)}
+\frac{7 \zeta_3}{6}
\biggr) S_1
+\frac{65}{48} S_2^2
+\frac{7 S_4}{24}
-\frac{1}{3} S_{-4}
\nonumber \\ &&
+\frac{16 N^7+140 N^6+508 N^5+979 N^4+1065 N^3+623 N^2+130 N-35}{6 (N+1)^4 (N+2)^4}
+\frac{2}{3} S_{-2,2}
\nonumber \\ &&
+\biggl(
\frac{2 N^2-5}{3 (N+1) (N+2)}
-S_1
\biggr) S_{-3} 
-\frac{\big(4 N^2+31 N+63\big) S_3}{18 (N+1) (N+2)}
+\frac{\big(2 N^2+4 N-1\big) S_{2,1}}{2 (N+1) (N+2)}
\nonumber \\ &&
+\biggl(
-\frac{3}{2} S_1^2
+\frac{2 N^2+2 N-3}{(N+1) (N+2)} S_1
-\frac{4 N^4+12 N^3+13 N^2+21 N+25}{3 (N+1)^2 (N+2)^2}
-\frac{3}{2} S_2
\nonumber \\ &&
-\frac{\zeta_2}{2}
\biggr) S_{-2}
+\biggl(
\frac{12 N^4+70 N^3+151 N^2+178 N+102}{12 (N+1)^2 (N+2)^2}
+\frac{3 \zeta_2}{8}
\biggr) S_2
-\frac{4}{3} S_{-2,1,1}
\nonumber \\ &&
+\frac{4 N^2+7 N+1}{4 (N+1) (N+2)^2} \zeta_2
+\frac{3}{2} S_{3,1}
+\frac{2}{3} S_{-3,1}
-\frac{2 \big(2 N^2-5\big)}{3 (N+1) (N+2)} S_{-2,1}
-\frac{1}{2} S_{2,1,1}
\nonumber \\ &&
+\frac{7 (2 N+3) \zeta_3}{6 (N+1) (N+2)}
+(-1)^N \biggl[
-\frac{S_1^3}{36 (N+2)}
+\frac{2 N^3+12 N^2+21 N+13}{12 (N+1)^2 (N+2)^2} S_1^2
\nonumber \\ &&
-\biggl(
\frac{4 N^5+36 N^4+129 N^3+240 N^2+226 N+79}{6 (N+1)^3 (N+2)^3}
+\frac{3 S_2}{4 (N+2)}
+\frac{\zeta_2}{4 (N+2)}
\biggr) S_1
\nonumber \\ &&
+\frac{8 N^7+72 N^6+274 N^5+562 N^4+656 N^3+434 N^2+183 N+67}{6 (N+1)^4 (N+2)^4}
-\frac{S_{2,1}}{N+2}
\nonumber \\ &&
+\frac{2 N^3+6 N^2+9 N+7}{4 (N+1)^2 (N+2)^2} \zeta_2
%+\frac{1}{4} \big(2 N^3+6 N^2+9 N+7\big) \frac{\zeta_2}{(N+1)^2 (N+2)^2}
+\frac{6 N^3+28 N^2+47 N+31}{4 (N+1)^2 (N+2)^2} S_2
+\frac{21 \zeta_3-25 S_3}{18 (N+2)}
%-\frac{25 S_3}{18 (N+2)}
%+\frac{7 \zeta_3}{6 (N+2)}
\biggr]
\nonumber \\ &&
+\frac{14 N^2+59 N+30}{36 (N+1) (N+2)} S_1^3
-\frac{5}{48} S_1^4
\Biggr\} \, .
\label{eq:4.8}
\end{eqnarray}
%---------------------------------------------------------------------------------------------------------------------------------
Here and in the following we use the shorthand notation $S_{\vec{a}}(N) \equiv S_{\vec{a}}$ for the harmonic sums. The calculation 
of (\ref{eq:4.8}) took 4828 seconds. In order to assemble, e.g. diagram~10 in Figure~\ref{samplediagrams2}, we  needed in 
addition the $\ep^1$ term which took another 50339 seconds. This extra coefficient can be expressed in terms of harmonic sums up to 
weight {\sf w = 5}. 
%B1a:950

In other occasions, the Feynman parameter integrals lead to hypergeometric functions. For example, in the case of the following
master integral, 
%---------------------------------------------------------------------------------------------------------------------------------
\begin{equation}
J_2 = \int \frac{d^Dk_1}{(2 \pi)^D}  \frac{d^Dk_2}{(2 \pi)^D} \frac{d^Dk_3}{(2 \pi)^D} \,\,
\frac{(\Delta.k_3-\Delta.k_1)^N}{P_2 P_3 P_5 P_6 P_8 P_9} \, ,
\end{equation}
%---------------------------------------------------------------------------------------------------------------------------------
where the propagators are those of family {\tt B1a} as given in Table \ref{families1}, we obtain
%---------------------------------------------------------------------------------------------------------------------------------
\begin{eqnarray}
J_2 &=& - \int_0^1 dx_1 \cdots \int_0^1 dx_5 \,\, \Gamma\left(-{\textstyle \frac{3}{2}} \varepsilon\right)
x_1^{\varepsilon/2} (1-x_1)^{\varepsilon/2}
(1-x_2)^{-1-\varepsilon/2}
x_3^{\varepsilon} (1-x_3)^{N-\varepsilon}
\nonumber \\ && \phantom{\int_0^1 dx_1 \cdots \int_0^1 dx_5 \,\,} \times
x_5^{N+1} (1-x_5)^{-1-\varepsilon} 
(x_4-x_2)^N (1-x_3 x_5)^{\frac{3}{2} \varepsilon} \, .
\end{eqnarray}
%---------------------------------------------------------------------------------------------------------------------------------
The integral in $x_1$ gives a Beta function, the integral in $x_4$ is trivial, and the integral in $x_5$ gives a ${}_2F_1$ 
hypergeometric function.
%---------------------------------------------------------------------------------------------------------------------------------
\begin{eqnarray}
J_2 &=& - \int_0^1 dx_2 \int_0^1 dx_3 \,\, \Gamma\left(-{\textstyle \frac{3}{2}} \varepsilon\right)
\frac{\Gamma \left(1+\varepsilon/2\right)^2}{\Gamma(2+\varepsilon)}
\left((1-x_2)^{N+1}+(-1)^N x_2^{N+1}\right)
\nonumber \\ && \phantom{- \int_0^1 dx_2 \int_0^1 dx_3 \,\,} \times
\frac{1}{N+1}
(1-x_2)^{-1-\varepsilon/2} 
x_3^{\varepsilon} (1-x_3)^{N-\varepsilon}
\nonumber \\ && \phantom{- \int_0^1 dx_2 \int_0^1 dx_3 \,\,} \times
\frac{\Gamma(N+2) \Gamma(-\varepsilon)}{\Gamma(N+2-\varepsilon)} 
{}_2F_1\left(-{\textstyle \frac{3}{2}} \varepsilon, N+2; N+2-\varepsilon; x_3\right) \, .
\end{eqnarray}
%---------------------------------------------------------------------------------------------------------------------------------
Now the integral in $x_3$ leads to 
%---------------------------------------------------------------------------------------------------------------------------------
\begin{eqnarray}
{}_3F_2\left(-\frac{3}{2} \varepsilon,1+\varepsilon,N+2;N+2,N+2-\varepsilon;1\right) =
{}_2F_1\left(-\frac{3}{2} \varepsilon,1+\varepsilon;N+2-\varepsilon;1\right), 
\end{eqnarray}
%---------------------------------------------------------------------------------------------------------------------------------
which can then be written in terms of $\Gamma$-functions using Gau\ss{}' summation theorem. We get 
%---------------------------------------------------------------------------------------------------------------------------------
\begin{eqnarray}
J_2 &=& -\Gamma\left(-{\textstyle \frac{3}{2}} \varepsilon\right)
\frac{\Gamma \left(1+\varepsilon/2\right)^2 \Gamma(1+\varepsilon) \Gamma(N+1-\varepsilon) \Gamma(-\varepsilon) \Gamma \left(N+1-\varepsilon/2\right)}{\Gamma(2+\varepsilon) \Gamma(N+1-2 \varepsilon) \Gamma \left(N+2+\varepsilon/2\right)}
\nonumber \\ && \times
\left(\frac{1}{(N+1) (N+1-\varepsilon/2)}+(-1)^N \frac{\Gamma(N+1) \Gamma \left(-\varepsilon/2\right)}{\Gamma \left(N+2-\varepsilon/2\right)}\right) \, .
\label{J2alleps}
\end{eqnarray}
%---------------------------------------------------------------------------------------------------------------------------------
The result can be expanded in $\varepsilon$ directly. The result up to the constant term is
%---------------------------------------------------------------------------------------------------------------------------------
\begin{eqnarray}
J_2 &=& \frac{1}{(N+1)^2} \Biggl\{
\frac{4}{3 \varepsilon^3} (-1)^N
+\frac{2}{3 \varepsilon^2} \Biggl[
(-1)^N \big(
S_1
-2
\big)
-\frac{1}{N+1}
\Biggr]
+\frac{1}{\varepsilon} \Biggl[
\frac{2}{3 (N+1)}
\nonumber \\ &&
+(-1)^N \biggl(
\frac{4 N^2+8 N+5}{3 (N+1)^2}
+\frac{1}{6} S_1^2
-\frac{2}{3} S_1
+\frac{13}{6} S_2
+\frac{\zeta_2}{2}
\biggr)
\Biggr]
-\frac{4 N^2+8 N+5}{6 (N+1)^3}
\nonumber \\ &&
-\frac{4 S_2+\zeta_2}{4 (N+1)}
%-\frac{S_2}{N+1}
%-\frac{\zeta_2}{4 (N+1)}
+(-1)^N \biggl[
\biggl(
\frac{4 N^2+8 N+5}{6 (N+1)^2}
+\frac{13}{12} S_2
+\frac{\zeta_2}{4}
\biggr) S_1 
+\frac{1}{36} S_1^3
-\frac{1}{6} S_1^2
\nonumber \\ &&
-\frac{4 N^2+8 N+5}{3 (N+1)^2}
-\frac{13}{6} S_2
+\frac{55}{18} S_3
-\frac{\zeta_2}{2}
-\frac{7}{6} \zeta_3
\biggr]
+\varepsilon \Biggl[
\frac{4 N^2+8 N+5}{6 (N+1)^3}
+\frac{S_2}{N+1}
\nonumber \\ &&
+(-1)^N \biggl[
\frac{4 N^2+8 N+5}{24 (N+1)^2}
\big(
S_1^2-4 S_1+13 S_2+3 \zeta_2
\big)
%%%%
+\biggl(
%\frac{4 N^2+8 N+5}{24 (N+1)^2}
\frac{13}{48} S_2
+\frac{\zeta_2}{16}
\biggr) S_1^2 
-\frac{13}{12} S_1 S_2
\nonumber \\ &&
+\biggl(
%-\frac{4 N^2+8 N+5}{6 (N+1)^2}
\frac{55}{36} S_3
-\frac{\zeta_2}{4}
-\frac{7}{12} \zeta_3
\biggr) S_1 
+\frac{13}{16} \zeta_2 S_2
%+\biggl(
%\frac{13 \big(4 N^2+8 N+5\big)}{24 (N+1)^2}
%+\frac{13}{16} \zeta_2
%\biggr) S_2 
%+\frac{\big(4 N^2+8 N+5\big)}{8 (N+1)^2} \zeta_2
+\frac{\big(4 N^2+6 N+3\big) \big(4 N^2+10 N+7\big)}{12 (N+1)^4}
\nonumber \\ &&
+\frac{1}{288} S_1^4
-\frac{1}{36} S_1^3
+\frac{169}{96} S_2^2
-\frac{55}{18} S_3
+\frac{241}{48} S_4
-\frac{173}{160} \zeta_2^2
+\frac{7}{6} \zeta_3
\biggr]
%+\frac{7 \zeta_3+3 \zeta_2-18 S_3}{12 (N+1)}
-\frac{3 S_3}{2 (N+1)}
\nonumber \\ &&
+\frac{\zeta_2}{4 (N+1)}
+\frac{7 \zeta_3}{12 (N+1)}
\Biggr]
\Biggr\}.
\end{eqnarray}
%---------------------------------------------------------------------------------------------------------------------------------
This master integral is part of diagram~9 in Figure~\ref{samplediagrams}.
Notice that, based on the expression in Eq. (\ref{J2alleps}), in principle, we can obtain $J_2$ to any order in $\varepsilon$.
Of course, things will not always be so smooth, and in many cases we will end up with a hypergeometric function that will
lead to multi-sums, requiring the use of symbolic summation. Let us consider, for example, the integral
%---------------------------------------------------------------------------------------------------------------------------------
\begin{equation}
J_3 = \int \frac{d^Dk_1}{(2 \pi)^D}  \frac{d^Dk_2}{(2 \pi)^D} \frac{d^Dk_3}{(2 \pi)^D} \,\,
\frac{\sum_{j=0}^N (\Delta.k)^j 
(\Delta.k_1-\Delta.k_3)^{N-j}}{P_1 P_2 P_4 P_7 P_8} \, ,
\end{equation}
%---------------------------------------------------------------------------------------------------------------------------------
where the propagators are now those of family {\tt B5c} shown in Table \ref{families2}. Introducing three Feynman parameters we obtain
%---------------------------------------------------------------------------------------------------------------------------------
\begin{eqnarray}
J_3 &=& \int_0^1 dx \int_0^1 dy \int_0^1 dz \,\, 
\Gamma(-1-\varepsilon/2) \Gamma(-\varepsilon) \sum_{j=0}^N (-x)^{N-j} y^{N-j+\varepsilon/2} (1-y)^{-\varepsilon/2}
\nonumber \\ && \phantom{\int_0^1 dx \int_0^1 dy \int_0^1 dz \,\,} \times
z^{N-j+1} (1-z)^{-1-\varepsilon/2} (1-x z)^j (1-y z)^{\varepsilon} \, .
\end{eqnarray}
%---------------------------------------------------------------------------------------------------------------------------------
After we express the term $(1-x z)^j$ in terms of its binomial expansion, the integral in $x$ becomes trivial, and the integrals
in $y$ and $z$ lead to a ${}_3F_2$-hypergeometric function evaluated at 1:
%---------------------------------------------------------------------------------------------------------------------------------
\begin{eqnarray}
J_3 &=& \Gamma(-1-\varepsilon/2) \Gamma(-\varepsilon) \sum_{j=0}^N \sum_{k=0}^j 
\frac{(-1)^{N-j+k}}{N-j+k+1} \binom{j}{k} 
\nonumber \\ && \times
\frac{\Gamma(N-j+k+2) \Gamma(-\varepsilon/2)}{\Gamma(N-j+k+2-\varepsilon/2)}
\frac{\Gamma(N-j+1+\varepsilon/2) \Gamma(1-\varepsilon/2)}{\Gamma(N-j+2)}
\nonumber \\ && \times
{}_3F_2\left(-\varepsilon,N-j+1+\frac{\varepsilon}{2},N-j+k+2;N-j+2,N-j+k+2-\frac{\varepsilon}{2};1\right) \, .
\end{eqnarray}
%---------------------------------------------------------------------------------------------------------------------------------
The series representation of this hypergeometric function is convergent. We obtain
%---------------------------------------------------------------------------------------------------------------------------------
\begin{eqnarray}
J_3 &=&
\Gamma(-1-\varepsilon/2) \Gamma(-\varepsilon) \Gamma(1-\varepsilon/2) \sum_{j=0}^N \sum_{k=0}^j 
\frac{(-1)^{N-j+k}}{N-j+k+1} \binom{j}{k} 
\nonumber \\ && \times
\sum_{n=0}^{\infty} \frac{\Gamma(n-\varepsilon) \Gamma(n+N-j+1+\varepsilon/2) \Gamma(n+N-j+k+2)}{n! \Gamma(n+N-j+2) \Gamma(n+N-j+k+2-\varepsilon/2)} \, .
\end{eqnarray}
%---------------------------------------------------------------------------------------------------------------------------------
This can now be given to {\tt EvaluateMultiSums} based on \texttt{Sigma}, which performs the $\varepsilon$ expansion and solves the multi-sum in terms of standard harmonic sums.
The result up to the constant term 
%---------------------------------------------------------------------------------------------------------------------------------
\begin{eqnarray}
J_3 &=& 
\frac{1}{\varepsilon^3} \Biggl[
        -\frac{4}{(N+1)^2}
        +\frac{8 (-1)^N}{(N+1)^2}
        -4 S_2
        -8 S_{-2}
\Biggr]
+\frac{1}{\varepsilon^2} \Biggl[
        \frac{2 (N-2)}{(N+1)^3}
        -\frac{4 (-1)^N N}{(N+1)^3}
        -4 S_1 S_2
\nonumber \\ &&
        +\frac{2 (N-1)}{N+1} S_2
        -6 S_3
        +\biggl(
                \frac{4 (N-1)}{N+1}
                -8 S_1
        \biggr) S_{-2}
        -4 S_{-3}
        +4 S_{2,1}
        +8 S_{-2,1}
\Biggr]
\nonumber \\ &&
+\frac{1}{\varepsilon} \Biggl[
        \frac{-N^2+N+9}{(N+1)^4}
        +\frac{2 (-1)^N \big(
                N^2+N-3\big)}{(N+1)^4}
        +\biggl(
                -\frac{3}{2 (N+1)^2}
                +\frac{3 (-1)^N}{(N+1)^2}
                -\frac{3}{2} S_2
\nonumber \\ &&
                -3 S_{-2}
        \biggr) \zeta_2
        +\biggl(
                \frac{2 (N-1)}{N+1} S_2
                -6 S_3
                +4 S_{2,1}
                +8 S_{-2,1}
        \biggr) S_1
        -2 S_1^2 S_2
        +\frac{3 (N-1)}{N+1} S_3
\nonumber \\ &&
        +\biggl(
                -\frac{4 (-1)^N}{(N+1)^2}
                +\frac{3-N^2}{(N+1)^2}
        \biggr) S_2
        +3 S_2^2
        +2 S_4
        +\biggl(
                -\frac{8 (-1)^N}{(N+1)^2}
                -\frac{2 (N-1)}{N+1}
                -4 S_1^2
\nonumber \\ &&
                +\frac{4 (N-1) S_1}{N+1}
                +4 S_2
        \biggr) S_{-2}
        +4 S_{-2}^2
        +\biggl(
                \frac{2 (N-1)}{N+1}
                -4 S_1
        \biggr) S_{-3}
        +2 S_{-4}
        +6 S_{3,1}
\nonumber \\ &&
        -\frac{2 (N-1)}{N+1} S_{2,1}
        -\frac{4 (N-1)}{N+1} S_{-2,1}
        +4 S_{-2,2}
        +4 S_{-3,1}
        -4 S_{2,1,1}
        -8 S_{-2,1,1}
\Biggr]
\nonumber \\ &&
+\frac{(N-4) \big(
        N^2+4 N+6\big)}{2 (N+1)^5}
+\biggl[
        \frac{3 (N-2)}{4 (N+1)^3}
        -\frac{3 (-1)^N N}{2 (N+1)^3}
        +\frac{3 (N-1) S_2}{4 (N+1)}
        -\frac{3}{2} S_1 S_2
\nonumber \\ &&
        -\frac{9}{4} S_3
        +\biggl(
                \frac{3 (N-1)}{2 (N+1)}
                -3 S_1
        \biggr) S_{-2}
        -\frac{3}{2} S_{-3}
        +\frac{3}{2} S_{2,1}
        +3 S_{-2,1}
\biggr] \zeta_2
+\biggl(
        \frac{2 (-1)^N-1}{2 (N+1)^2}
%        -\frac{1}{2 (N+1)^2}
%        +\frac{(-1)^N}{(N+1)^2}
\nonumber \\ &&
        -\frac{1}{2} S_2
        -S_{-2}
\biggr) \zeta_3
-\frac{(-1)^N}{(N+1)^5} \big(
        N^3+2 N^2-2 N-10\big)
+\biggl[
        \frac{4 (-1)^N}{(N+1)^2} S_2
        +3 S_2^2
\nonumber \\ &&
        +\frac{N-1}{N+1} \left(3 S_3-2 S_{2,1}-4 S_{-2,1}-S_2\right)
%        +\frac{3 (N-1) S_3}{N+1}
%        -\frac{2 (N-1) S_{2,1}}{N+1}
%        -\frac{4 (N-1) S_{-2,1}}{N+1}
%        -\frac{(N-1) S_2}{N+1}
        +2 S_4
        +6 S_{3,1}
        +4 S_{-2,2}
        +4 S_{-3,1}
        -4 S_{2,1,1}
\nonumber \\ &&
        -8 S_{-2,1,1}
\biggr] S_1
+\biggl(
        \frac{N-1}{N+1} S_2
        -3 S_3
        +2 S_{2,1}
        +4 S_{-2,1}
\biggr) S_1^2
+\biggl(
        \frac{2 (-1)^N (N+3)}{(N+1)^3}
\nonumber \\ &&
        +\frac{N^3+N^2-3 N-9}{2 (N+1)^3}
        -\frac{4}{3} S_3
        -4 S_{2,1}
        -12 S_{-2,1}
\biggr) S_2
-\frac{2}{3} S_1^3 S_2
-\frac{9}{2} S_5
+4 S_1 S_{-2}^2
\nonumber \\ &&
+\biggl(
        \frac{6 (-1)^N}{(N+1)^2}
        +\frac{5-3 N^2}{2 (N+1)^2}
\biggr) S_3
+\frac{N-1}{N+1} \biggl[
-2 S_{-2}^2
-\frac{3}{2} S_2^2
- S_4
+\left(2 S_1-1\right) S_{-3}
\nonumber \\ &&
+\left(1-2 S_1+2 S_1^2-2 S_2\right) S_{-2}
-S_{-4}
+S_{2,1}
+2 S_{-2,1}
-3 S_{3,1}
-2 S_{-2,2}
-2 S_{-3,1}
\nonumber \\ &&
+2 S_{2,1,1}
+4 S_{-2,1,1}
\biggr]
%-\frac{2 (N-1)}{N+1} S_{-2}^2
%-\frac{3 (N-1) S_2^2}{2 (N+1)}
%-\frac{(N-1) S_4}{N+1}
%+\frac{N-1}{N+1} \left(1-2 S_1+2 S_1^2-2 S_2\right) S_{-2}
%+\frac{N-1}{N+1} \left(2 S_1-1\right) S_{-3}
%-\frac{N-1}{N+1} S_{-4}
%+\frac{N-1}{N+1} S_{2,1}
%+\frac{2 (N-1)}{N+1} S_{-2,1}
%-\frac{3 (N-1) S_{3,1}}{N+1}
%-\frac{2 (N-1) S_{-2,2}}{N+1}
%-\frac{2 (N-1) S_{-3,1}}{N+1}
%+\frac{2 (N-1) S_{2,1,1}}{N+1}
%+\frac{4 (N-1) S_{-2,1,1}}{N+1}
+\biggl[
%        \frac{N-1}{N+1}
        \frac{4 (-1)^N (N+3)}{(N+1)^3}
        +\biggl(
                \frac{8 (-1)^N}{(N+1)^2}
%                -\frac{2 (N-1)}{N+1}
                +4 S_2
        \biggr) S_1
%        +\frac{2 (N-1) S_1^2}{N+1}
        -\frac{4}{3} S_1^3
%        -\frac{2 (N-1) S_2}{N+1}
        -\frac{8}{3} S_3
\nonumber \\ &&
        +8 S_{2,1}
        -4 S_{-2,1}
\biggr] S_{-2}
%+\biggl(
%        -\frac{2 (N-1)}{N+1}
%        +4 S_1
%\biggr) S_{-2}^2
+\biggl(
        \frac{4 (-1)^N}{(N+1)^2}
%        +\frac{1-N}{N+1}
%        +\frac{2 (N-1) S_1}{N+1}
        -2 S_1^2
        +2 S_2
\biggr) S_{-3}
+2 S_1 S_{-4}
-S_{-5}
+4 S_{2,3}
\nonumber \\ &&
-\frac{4 (-1)^N}{(N+1)^2} \left(S_{2,1}+2 S_{-2,1}\right)
%-\frac{4 (-1)^N}{(N+1)^2} S_{2,1}
%-\frac{8 (-1)^N}{(N+1)^2} S_{-2,1}
+8 S_{2,-3}
-3 S_{4,1}
-6 S_{-2,3}
+8 S_{-2,-3}
-2 S_{-4,1}
+2 S_{2,2,1}
\nonumber \\ &&
-8 S_{2,1,-2}
-6 S_{3,1,1}
-12 S_{-2,1,-2}
-4 S_{-2,2,1}
-4 S_{-3,1,1}
+4 S_{2,1,1,1}
+8 S_{-2,1,1,1} 
\end{eqnarray}
%---------------------------------------------------------------------------------------------------------------------------------
is needed for diagram~12 in Figure~\ref{samplediagrams2}. This calculation took 3011 seconds.

%%%%%%%%%%%%%%%%%%%%%%%%%%%%%%%%%%%%%%%%%%%%%%%%%%%%%%%%%%%%%%%%%%%%%%%
\subsection{Mellin-Barnes integral representations and symbolic summation}
\label{sec:MB-MIs}
%%%%%%%%%%%%%%%%%%%%%%%%%%%%%%%%%%%%%%%%%%%%%%%%%%%%%%%%%%%%%%%%%%%%%%%

\vspace*{1mm}\noindent
The introduction of Feynman parameters does not always lead to integrals that can be solved in terms of Beta functions or (generalized)
hypergeometric functions or associated extensions, cf.~\cite{HYP,Slater,Appell,Schlosser}. It can also happen that a result in 
terms of hypergeometric functions can be obtained, but no
convergent series representation corresponding to it can be found. A possible way to deal with these cases is to split a
denominator in the integrand by introducing a Mellin-Barnes contour integral, allowing the integration of the Feynman parameters.
For example, let us consider the following master integral
%B1a:1460
%---------------------------------------------------------------------------------------------------------------------------------
\begin{equation}
J_4 = \int \frac{d^Dk_1}{(2 \pi)^D}  \frac{d^Dk_2}{(2 \pi)^D} \frac{d^Dk_3}{(2 \pi)^D} \,\,
\frac{(\Delta.k_3)^N}{P_3 P_5 P_6 P_8 P_9} 
\, ,
\end{equation}
%---------------------------------------------------------------------------------------------------------------------------------
where the propagators are those of family {\tt B1a}. This integral is needed up to the quadratic term in $\ep$. The Feynman 
parameterization yields
%---------------------------------------------------------------------------------------------------------------------------------
\begin{eqnarray}
J_4 &=& \int_0^1 dx \int_0^1 dy \int_0^1 dz \,\, \frac{\Gamma\left(-1-{\textstyle \frac{3}{2}} \varepsilon\right)}{N+1} 
x^{\varepsilon/2} (1-x)^{\varepsilon/2} y^{1+\varepsilon} (1-y)^{-1-\varepsilon} \nonumber \\ && \phantom{\int_0^1 dx \int_0^1 dy \int_0^1 dz \,\,}
\times z^{N+1} (1-z)^{-2-\varepsilon} \left(1-yz\right)^{1+\frac{3}{2} \varepsilon}
\nonumber \\ 
&=& \frac{\Gamma\left(-1-{\textstyle \frac{3}{2}} \varepsilon\right)}{N+1} \Gamma(1+\varepsilon/2)^2 \Gamma(-\varepsilon)
\frac{\Gamma(N+2) \Gamma(-1-\varepsilon)}{\Gamma(N+1-\varepsilon)} 
\nonumber \\ && \times
{}_3F_2\left(N+2,-1-{\textstyle \frac{3}{2}} \varepsilon,2+\varepsilon; N+1-\varepsilon,2; 1\right) \phantom{\frac{A}{B}}
\nonumber \\
&=& \frac{\Gamma(1+\varepsilon/2)^2 \Gamma(-\varepsilon) \Gamma(-1-\varepsilon)}{(N+1) \Gamma(2+\varepsilon)}
\sum_{k=0}^{\infty} \frac{\Gamma(k+N+2) \Gamma\left(k-1-{\textstyle \frac{3}{2}} 
\varepsilon\right) \Gamma(k+2+\varepsilon)}{k! \Gamma(k+N+1-\varepsilon) \Gamma(k+2)} \, .
\end{eqnarray}
%---------------------------------------------------------------------------------------------------------------------------------
Here the latter infinite sum turns out to be divergent, and thus does not form an adequate representation. We therefore  proceed as 
follows. 
First, for convenience, we perform the change $z \rightarrow 1-z$ in the first two lines of the previous equation, which leads to
%---------------------------------------------------------------------------------------------------------------------------------
\begin{equation}
J_4 = \int_0^1 dy \int_0^1 dz \,\, \frac{\Gamma\left(-1-{\textstyle \frac{3}{2}} \varepsilon\right)}{N+1}
\frac{\Gamma(1+\varepsilon/2)^2}{\Gamma(2+\varepsilon)}
\frac{y^{1+\varepsilon} (1-y)^{-1-\varepsilon} z^{-2-\varepsilon} (1-z)^{N+1}}{(1-y+zy)^{-1-{\textstyle \frac{3}{2}} \varepsilon}} \, .
\end{equation}
%---------------------------------------------------------------------------------------------------------------------------------
We can now split the denominator $1/(1-y+zy)^{-1-{\textstyle \frac{3}{2}} \varepsilon}$ using
%---------------------------------------------------------------------------------------------------------------------------------
\begin{equation}
\frac{1}{(A+B)^{\nu}} = \frac{1}{2 \pi i} \int_{\gamma-i \infty}^{\gamma+i \infty} d\sigma \,\, 
\frac{\Gamma(-\sigma) \Gamma(\sigma+\nu)}{\Gamma(\nu)}
A^{\sigma} B^{-\sigma-\nu} \, .
\end{equation}
%---------------------------------------------------------------------------------------------------------------------------------
The integrals in $y$ and $z$ can then be done in terms of Beta functions, and we end up with the following contour integral
%---------------------------------------------------------------------------------------------------------------------------------
\begin{eqnarray}
J_4 &=& \frac{1}{2 \pi i} \int_{\gamma-i \infty}^{\gamma+i \infty} d\sigma \,\, \Gamma(-\sigma) \Gamma\left(\sigma-1-{\textstyle \frac{3}{2}} \varepsilon\right)
\frac{\Gamma\left(1+\varepsilon/2\right)^2}{\Gamma(2+\varepsilon)}
\frac{\Gamma(\sigma+2+\varepsilon) \Gamma\left(-\sigma+1+\varepsilon/2\right)}{\Gamma\left(3+{\textstyle \frac{3}{2}} \varepsilon\right)}
\nonumber \\ && \phantom{\frac{1}{2 \pi i} \int_{-i \infty}^{+i \infty} d\sigma}
\times
\frac{\Gamma(\sigma-1-\varepsilon) \Gamma(N+1)}{\Gamma(\sigma+N+1-\varepsilon)}.
\label{MBint1}
\end{eqnarray}
%---------------------------------------------------------------------------------------------------------------------------------
At this point, we can use the {\tt Mathematica} package {\tt MB} \cite{Czakon:2005rk} in order to find a value for $\gamma \in 
\mathbb{R}$ and $\varepsilon$ such
that the integral in Eq. (\ref{MBint1}) is well defined. This package can then do the analytic continuation to $\varepsilon \rightarrow 0$, by
moving $\varepsilon$ towards zero, and taking a residue every time a pole is crossed in the integrand. 
In this case, we find that $\gamma=-\frac{1}{8}$, and we have to take residues at $\sigma=1+\frac{3}{2} \varepsilon$,
$\sigma=\frac{3}{2} \varepsilon$, $\sigma=1+\varepsilon$ and $\sigma=\varepsilon$.
After this, the expansion in $\varepsilon$ can be performed. We get the following expression
%---------------------------------------------------------------------------------------------------------------------------------
\begin{eqnarray}
J_4 &=& J'_4
+\frac{\Gamma \left(1+\frac{\varepsilon}{2}\right)^2 \Gamma (-\varepsilon)}{\Gamma (2+\varepsilon) \Gamma \left(3+\frac{3 \varepsilon}{2}\right)} \Biggl[
-
\Gamma (2+2 \varepsilon) \left(\frac{2 \Gamma \left(-\frac{\varepsilon}{2}\right)^2}{N+1}+\Gamma \left(-1-\frac{\varepsilon}{2}\right) \Gamma \left(1-\frac{\varepsilon}{2}\right)\right)
\nonumber \\ &&
+
\Gamma (N+1) \Gamma \left(\frac{\varepsilon}{2}\right)
\left(\frac{\Gamma \left(-1-\frac{3}{2} \varepsilon\right) \Gamma \left(3+\frac{5}{2} \varepsilon\right)}{\Gamma \left(N+2+\frac{\varepsilon}{2}\right)}
-\frac{4 \Gamma \left(1-\frac{3}{2} \varepsilon\right) \Gamma \left(2+\frac{5}{2} \varepsilon\right)}{3 (\varepsilon-2) \Gamma \left(N+1+\frac{\varepsilon}{2}\right)}\right)
\Biggr],
\end{eqnarray}
%---------------------------------------------------------------------------------------------------------------------------------
where the second term comes from taking the four residues mentioned above, which we have left still in unexpanded form.
%---------------------------------------------------------------------------------------------------------------------------------
$J'_4$ is given by
%---------------------------------------------------------------------------------------------------------------------------------
\begin{eqnarray}
J'_4 &=& \frac{1}{2 \pi i} \int_{\gamma-i \infty}^{\gamma+i \infty} d\sigma \,\, 
\frac{\Gamma (N+1) \Gamma (1-\sigma) \Gamma (\sigma-1)^2 \Gamma (-\sigma) \Gamma (\sigma+2)}{\Gamma (\sigma+N+1)}
\Biggl[ \frac{1}{2}
\nonumber \\ && \phantom{\frac{1}{2 \pi i}}
+\varepsilon \left(\frac{1}{2} \psi(\sigma+N+1) 
+\frac{1}{4} \psi(1-\sigma)-\frac{5}{4} \psi(\sigma-1)+\frac{1}{2} \psi(\sigma+2)-\frac{13}{8}\right)
\nonumber \\ && \phantom{\frac{1}{2 \pi i}}
+\varepsilon^2 \left(
\frac{1}{4} \psi(1-\sigma) \psi(\sigma+N+1)+\frac{1}{4} \psi(\sigma+N+1)^2
+\frac{1}{4} \psi'(\sigma+2)
\right. \nonumber \\ && \phantom{\frac{1}{2 \pi i} +\varepsilon^2 (} \left.
+\frac{1}{2} \psi(\sigma+2) \psi(\sigma+N+1)-\frac{13}{8} \psi(\sigma+N+1)-\frac{1}{4} \psi'(\sigma+N+1)
\right. \nonumber \\ && \phantom{\frac{1}{2 \pi i} +\varepsilon^2 (} \left.
+\frac{1}{16}\psi(1-\sigma)^2-\frac{5}{8} \psi(\sigma-1) \psi(1-\sigma)+\frac{1}{4} \psi(\sigma+2) \psi(1-\sigma)
\right. \nonumber \\ && \phantom{\frac{1}{2 \pi i} +\varepsilon^2 (} \left.
-\frac{13}{16} \psi(1-\sigma)+\frac{25}{16} \psi(\sigma-1)^2+\frac{1}{4} \psi(\sigma+2)^2+\frac{65}{16} \psi(\sigma-1)
\right. \nonumber \\ && \phantom{\frac{1}{2 \pi i} +\varepsilon^2 (} \left.
-\frac{5}{4} \psi(\sigma-1) \psi(\sigma+2)-\frac{13}{8} \psi(\sigma+2)+\frac{1}{16} \psi'(1-\sigma)+\frac{13}{16} \psi'(\sigma-1)
\right. \nonumber \\ && \phantom{\frac{1}{2 \pi i} +\varepsilon^2 (} \left.
-\frac{5}{4} \psi(\sigma-1) \psi(\sigma+N+1)
-\frac{11}{96} \pi^2+\frac{115}{32}
\right)
+O(\varepsilon^3)
\Biggr] \, .
\end{eqnarray}
%---------------------------------------------------------------------------------------------------------------------------------
Here $\psi^{(k)}(x),~k \geq 0,~k \in \mathbb{N}$, denotes the polygamma function 
%---------------------------------------------------------------------------------------------------------------------------------
\begin{eqnarray}
\psi(x) = \frac{1}{\Gamma(x)} \frac{d}{dx} \Gamma(x)~.
\end{eqnarray}
%---------------------------------------------------------------------------------------------------------------------------------
We can perform  the $J'_4$ integral by closing the contour either to the right or to the left, and taking residues. 
If we close the contour to the left, then the poles will be located at $\sigma=-k$, with $k$ an integer
greater than 1. In this case, we have to distinguish the cases $k \leq N$ and $k>N$, due to the presence
of $\Gamma (\sigma+N+1)$ in the integrand. It is simpler to close the contour to the right, where we
have to sum the residues at $\sigma=k$, with $k \geq 0$. We get
%---------------------------------------------------------------------------------------------------------------------------------
\begin{equation}
J'_4 = a_0+a_1+\sum_{k=2}^{\infty} \frac{(k+1) \Gamma (k-1) \Gamma (N+1)}{2 (k-1) \Gamma (k+N+1)}
\left(b_0+\varepsilon b_1+\varepsilon^2 b_2\right) \, ,
\end{equation}
%---------------------------------------------------------------------------------------------------------------------------------
where $a_0$ and $a_1$ are the residues of $J'_4$ at $\sigma=0$ and $\sigma=1$, which have to be taken separately, and
%---------------------------------------------------------------------------------------------------------------------------------
\begin{eqnarray}
b_0 &=& S_1(k+N)-S_1(k)+\frac{2 k^2+3 k-1}{(k-1) k (k+1)} \, ,
\\
b_1 &=& 
\left(\frac{13 k^3-26 k^2-25 k+10}{4 (k-1) k (k+1)}\right) \left(S_1(k)-S_1(k+N)\right)
-\frac{3}{2} S_1(k+N) S_1(k)
-\frac{3}{4} S_2(k)
\nonumber \\ && 
+\frac{3}{4} S_2(k+N)
+\frac{3}{4} S_1(k+N)^2
+\frac{3}{4} S_1(k)^2
-\frac{26 k^4-47 k^3-64 k^2+51 k-14}{4 (k-1)^2 k^2 (k+1)} \, , 
\\
b_2 &=&
\frac{39 k^3-70 k^2-63 k+26}{16 (k-1) k (k+1)} \bigl[2 S_1(k) S_1(k+N)-S_2(k+N)-S_1(k)^2-S_1(k+N)^2\bigr]
\nonumber \\ &&
-\frac{7}{8} S_1(k+N)^2 S_1(k)
+\frac{7}{8} S_1(k+N) S_1(k)^2 
+\frac{39 k^3-118 k^2-135 k+50}{16 (k-1) k (k+1)} S_2(k)
\nonumber \\ &&
+\frac{115 k^5-453 k^4+471 k^3+341 k^2-310 k+84}{16 (k-1)^2 k^2 (k+1)} \bigl[S_1(k+N)-S_1(k)\bigr]
\nonumber \\ &&
+\left(\frac{7}{8} S_2(k+N)
-\frac{19}{8} S_2(k)
+\frac{3}{8} \zeta_2\right) \bigl[S_1(k+N)-S_1(k)\bigr]
+\frac{7}{24} S_1(k+N)^3
\nonumber \\ &&
+\frac{7}{12} S_3(k+N)
-\frac{7}{24} S_1(k)^3
-\frac{43}{12} S_3(k)
+\frac{3 \left(2 k^2+3 k-1\right)}{8 (k-1) k (k+1)} \zeta_2
+3 \zeta_3
\nonumber \\ &&
+\frac{230 k^6-895 k^5+937 k^4+725 k^3-1071 k^2+598 k-124}{16 (k-1)^3 k^3 (k+1)} \, .
\end{eqnarray}
%---------------------------------------------------------------------------------------------------------------------------------
The sums in $k$ can be performed using {\tt Sigma}. Adding all the pieces, we obtain
%---------------------------------------------------------------------------------------------------------------------------------
\begin{eqnarray}
J_4 &=& \frac{1}{N+1} \Biggl\{
\frac{8}{3 \varepsilon^3}
+\frac{1}{\varepsilon^2} \Biggl[
\frac{N^2-9 N-8}{3 (N+1)}
+\frac{2}{3} S_1
\Biggr]
+\frac{1}{\varepsilon} \Biggl[
\frac{N^2-2}{3 (N+1)} S_1
-\frac{1}{6} S_1^2
-\frac{1}{6} S_2
+\zeta_2
\nonumber \\ &&
-\frac{19 N^3+18 N^2-13 N-8}{12 (N+1)^2}
\Biggr]
+S_{2,1}
+\frac{2 N^2+9 N+2}{12 (N+1)} \left(S_1^2+S_2\right)
+\frac{N^2-9 N-8}{8 (N+1)} \zeta_2
\nonumber \\ &&
+\biggl(
-\frac{19 N^3+60 N^2+59 N+16}{12 (N+1)^2}
-\frac{5}{12} S_2
+\frac{\zeta_2}{4}
\biggr) S_1 
-\frac{5}{36} S_1^3
-\frac{5}{18} S_3
-\frac{7}{3} \zeta_3
\nonumber \\ &&
+\frac{229 N^4+835 N^3+1119 N^2+665 N+160}{48 (N+1)^3}
+\varepsilon \Biggl[
\frac{N^2-2}{2 (N+1)} S_{2,1}
-\frac{1}{2} S_{3,1}
+\frac{5}{2} S_{2,1,1}
\nonumber \\ &&
+\biggl(
\frac{N^2-2}{8 (N+1)} \zeta_2
+\frac{229 N^4+1039 N^3+1749 N^2+1289 N+346}{48 (N+1)^3}
-\frac{13}{36} S_3
-\frac{7}{12} \zeta_3
\nonumber \\ &&
-\frac{1}{2} S_{2,1}
%+\frac{4 N^2+27 N+10}{24 (N+1)} S_2
\biggr) S_1 
+\frac{4 N^2+27 N+10}{72 (N+1)} \big(3 S_1 S_2+S_1^3+2 S_3\big)
-\frac{7 \big(N^2-9 N-8\big)}{24 (N+1)} \zeta_3
\nonumber \\ &&
-\frac{19 N^3+18 N^2-13 N-8}{32 (N+1)^2} \zeta_2
%+\frac{4 N^2+27 N+10}{72 (N+1)} S_1^3
%+\frac{4 N^2+27 N+10}{36 (N+1)} S_3
-\biggl(
\frac{19 N^3+81 N^2+95 N+28}{24 (N+1)^2}
+\frac{13}{48} S_2
+\frac{\zeta_2}{16}
\biggr) S_1^2 
\nonumber \\ &&
-\biggl(
\frac{19 N^3+81 N^2+95 N+28}{24 (N+1)^2}
+\frac{\zeta_2}{16}
\biggr) S_2 
-\frac{13}{288} S_1^4
-\frac{37}{96} S_2^2
-\frac{25}{48} S_4
-\frac{173}{80} \zeta_2^2
\nonumber \\ &&
-\frac{2239 N^5+11312 N^4+22366 N^3+21592 N^2+10171 N+1888}{192 (N+1)^4}
\Biggr]
\nonumber \\ &&
+\varepsilon^2 \Biggl[
-\biggl(
%%%%%%%%%\frac{38 N^3+183 N^2+226 N+68}{144 (N+1)^2}
\frac{29}{288} S_2
+\frac{5}{96} \zeta_2
\biggr) S_1^3
-\frac{38 N^3+183 N^2+226 N+68}{144 (N+1)^2} \big(S_1^3+3 S_1 S_2+2 S_3\big)
\nonumber \\ &&
-\frac{29}{2880} S_1^5
+\frac{2 N^2+9 N+2}{32 (N+1)} \left(\zeta_2 S_1^2+8 S_1 S_{2,1}+8 S_{3,1}\right)
-\frac{173 \big(N^2-9 N-8\big)}{640 (N+1)} \zeta_2^2
\nonumber \\ &&
+\biggl(
\frac{7}{48} \zeta_3
-\frac{29}{144} S_3
-\frac{5}{8} S_{2,1}
%+\frac{2 N^2+9 N+2}{32 (N+1)} \zeta_2
%+\frac{229 N^4+1141 N^3+2064 N^2+1601 N+439}{96 (N+1)^3}
%+\frac{8 N^2+63 N+26}{96 (N+1)} S_2
\biggr) S_1^2
+\frac{56 N^2+279 N+74}{192 (N+1)} S_2^2
-\frac{179}{120} S_5
+\frac{7}{4} S_{2,3}
\nonumber \\ &&
+\frac{8 N^2+63 N+26}{576 (N+1)} \big(S_1^4+6 S_1^2 S_2+8 S_1 S_3\big)
%+\frac{8 N^2+63 N+26}{576 (N+1)} S_1^4
+\biggl(
-\frac{5}{4} S_{3,1}
-\frac{5}{32} \zeta_2 S_2
%+\biggl(
%%%%%%%%%-\frac{38 N^3+183 N^2+226 N+68}{48 (N+1)^2}
%-\frac{5}{32} \zeta_2
%\biggr) S_2
-\frac{7 \big(N^2-2\big)}{24 (N+1)} \zeta_3
\nonumber \\ &&
-\frac{2239 N^5+12560 N^4+28096 N^3+31330 N^2+17401 N+3838}{192 (N+1)^4}
-\frac{149}{192} S_2^2
\nonumber \\ &&
-\frac{173}{320} \zeta_2^2
%-\frac{19 N^3+60 N^2+59 N+16}{32 (N+1)^2} \zeta_2
%+\frac{8 N^2+63 N+26}{72 (N+1)} S_3
-\frac{89}{96} S_4
%+\frac{2 N^2+9 N+2}{4 (N+1)} S_{2,1}
+\frac{7}{4} S_{2,1,1}
\biggr) S_1
-\frac{19 N^3+60 N^2+59 N+16}{32 (N+1)^2} \left(\zeta_2 S_1+4 S_{2,1}\right)
\nonumber \\ &&
-\frac{N^2+18 N+10}{4 (N+1)} S_{2,1,1}
+\frac{229 N^4+1141 N^3+2064 N^2+1601 N+439}{96 (N+1)^3} \big(S_1^2+S_2\big)
\nonumber \\ &&
+\frac{7 \big(19 N^3+18 N^2-13 N-8\big)}{96 (N+1)^2} \zeta_3
+\frac{32 N^2+171 N+50}{96 (N+1)} S_4
%+\frac{2 N^2+9 N+2}{4 (N+1)} S_{3,1}
+\frac{1}{2} S_{4,1}
-\frac{1}{4} S_{2,2,1}
\nonumber \\ &&
+\biggl(
\frac{229 N^4+835 N^3+1119 N^2+665 N+160}{128 (N+1)^3}
-\frac{7}{8} \zeta_3
\biggr) \zeta_2 
+\frac{7}{4} S_{3,1,1}
+\frac{1}{4} S_{2,1,1,1}
\nonumber \\ &&
-\frac{5}{48} \zeta_2 S_3
%+\biggl(
%%%%%%%%-\frac{38 N^3+183 N^2+226 N+68}{72 (N+1)^2}
%-\frac{5}{48} \zeta_2
%\biggr) S_3
+ \frac{3}{8} \zeta_2 S_{2,1}
%+\biggl(
%\frac{3 \zeta_2}{8}
%-\frac{19 N^3+60 N^2+59 N+16}{8 (N+1)^2}
%\biggr) S_{2,1} 
+\biggl(
%\frac{229 N^4+1141 N^3+2064 N^2+1601 N+439}{96 (N+1)^3}
\frac{7}{48} \zeta_3
-\frac{209}{144} S_3
+\frac{9}{8} S_{2,1}
+\frac{2 N^2+9 N+2}{32 (N+1)} \zeta_2
\biggr) S_2 
-\frac{239}{20} \zeta_5
\nonumber \\ &&
+\frac{19405}{768} N-\frac{455}{64 (N+1)}-\frac{41}{32 (N+1)^2}-\frac{1}{16 (N+1)^3}+\frac{1}{12 (N+1)^4}+\frac{5407}{192}
\nonumber \\ &&
+\frac{1}{24 (N+1)^5}
%+\frac{19405 N^6+118653 N^5+296730 N^4+387506 N^3+277545 N^2+102721 N+15232}{768 (N+1)^5}
\Biggr]
\Biggr\}~.
\end{eqnarray}
%---------------------------------------------------------------------------------------------------------------
Again this master integral is needed to assemble diagram~12 in Figure~\ref{samplediagrams2}. It took 160 seconds to execute the 
underlying symbolic summation algorithms within~\texttt{EvaluateMultiSums}.

%%%%%%%%%%%%%%%%%%%%%%%%%%%%%%%%%%%%%%%%%%%%%%%%%%%%%%%%%%%%%%%%%%%%%%%
\subsection{Symbolic integration and a recurrence solver for \boldmath{$\ep$}-expansions}
\label{sec:Zeilberger}
%%%%%%%%%%%%%%%%%%%%%%%%%%%%%%%%%%%%%%%%%%%%%%%%%%%%%%%%%%%%%%%%%%%%%%%
%B5b-729

Let us consider the following class of integrals
%---------------------------------------------------------------------------------------------------------------
\begin{equation}
J_5(\nu_1,\nu_4,\nu_5) = \int \frac{d^Dk_1}{(2 \pi)^D}  \frac{d^Dk_2}{(2 \pi)^D} \frac{d^Dk_3}{(2 \pi)^D} \,\,
 \frac{(\Delta.k_1 - \Delta.k_3)^N}{P_1^{\nu_1} P_4^{\nu_4} P_5^{\nu_5} P_7 P_8} \, ,
\end{equation}
%---------------------------------------------------------------------------------------------------------------
where the propagators are those of family {\tt B5b}, and the operator insertion corresponds to the crossed version of this family. We remark 
that these master integrals were needed to treat diagram~12 in Figure~\ref{samplediagrams2}. 
Introducing four Feynman parameters we obtain
%---------------------------------------------------------------------------------------------------------------
\begin{eqnarray}
J_5(\nu_1,\nu_4,\nu_5) &=& \int_0^1 dx \int_0^1 dy \int_0^1 dz_1 \int_0^{1-z_1} dz_2 \,\,
(-1)^{\nu+1} \frac{\Gamma \left(\nu-4-{\textstyle \frac{3}{2}} \varepsilon\right)}{\Gamma(\nu_1) \Gamma(\nu_4) \Gamma(\nu_5)}
(x+y-1)^N
\nonumber \\ && \phantom{\int_0^1} \times
x^{\varepsilon/2} (1-x)^{-\nu_1+1+\varepsilon/2} 
y^{-\nu_5+1+\varepsilon/2} (1-y)^{\varepsilon/2}
z_1^{\nu_1-2-\varepsilon/2} z_2^{\nu_5-2-\varepsilon/2} 
\nonumber \\ && \phantom{\int_0^1} \times
(1-z_1-z_2)^{N+\nu_4-1}
\left(1-z_1 \frac{x}{x-1}-z_2 \frac{y}{y-1}\right)^{-\nu+4+\frac{3}{2} \varepsilon} \, ,
\end{eqnarray}
where $\nu = \nu_1+\nu_4+\nu_5$.
The $z_1$ and $z_2$ integrals can be done in terms of an Appell hypergeometric function \cite{Slater,Appell}
\begin{eqnarray}
J_5(\{\nu_i\}) &=& \int_0^1 dx \int_0^1 dy \,\,
(-1)^{\nu+1} \frac{\Gamma \left(\nu-4-{\textstyle \frac{3}{2}} \varepsilon\right)}{\Gamma(\nu_1) \Gamma(\nu_4) \Gamma(\nu_5)}
(x+y-1)^N
x^{\varepsilon/2} (1-x)^{-\nu_1+1+\varepsilon/2} 
\nonumber \\ &&  \times
y^{-\nu_5+1+\varepsilon/2} (1-y)^{\varepsilon/2}
%\nonumber \\ &&  \times
\frac{\Gamma\left(\nu_1-1-\frac{\varepsilon}{2}\right) \Gamma\left(\nu_5-1-\frac{\varepsilon}{2}\right) \Gamma(N+\nu_4)}{\Gamma(N+\nu-2-\varepsilon)}
\nonumber \\ &&  \times
F_1\left(\nu-4-{\textstyle \frac{3}{2}} \varepsilon; \nu_1-1-\frac{\varepsilon}{2}, \nu_5-1-\frac{\varepsilon}{2}; 
N+\nu-2-\varepsilon; \frac{x}{x-1}, \frac{y}{y-1}\right)~.
\nonumber\\
\end{eqnarray}
%---------------------------------------------------------------------------------------------------------------
Now we can use the following analytic continuation relation \cite{Slater}
%---------------------------------------------------------------------------------------------------------------
\begin{equation}
F_1\left(a;b,b';c;\frac{x}{x-1},\frac{y}{y-1}\right) = (1-x)^b (1-y)^{b'} F_1(c-a;b,b';c;x,y) \, ,
\end{equation}
%---------------------------------------------------------------------------------------------------------------
which in this case leads to
%---------------------------------------------------------------------------------------------------------------
\begin{eqnarray}
J_5(\{\nu_i\}) &=& \int_0^1 dx \int_0^1 dy \,\,
(-1)^{\nu+1} \frac{\Gamma \left(\nu-4-{\textstyle \frac{3}{2}} \varepsilon\right)}{\Gamma(\nu_1) \Gamma(\nu_4) \Gamma(\nu_5)}
(x+y-1)^N
x^{\varepsilon/2} 
\nonumber \\ &&  \times
y^{-\nu_5+1+\varepsilon/2} (1-y)^{\nu_5-1}
\frac{\Gamma\left(\nu_1-1-\frac{\varepsilon}{2}\right) \Gamma\left(\nu_5-1-\frac{\varepsilon}{2}\right) \Gamma(N+\nu_4)}{\Gamma(N+\nu-2-\varepsilon)}
\nonumber \\ &&  \times
F_1\left(N+2+\frac{\varepsilon}{2}; \nu_1-1-\frac{\varepsilon}{2}, \nu_5-1-\frac{\varepsilon}{2}; N+\nu-2-\varepsilon; x, y\right) \, .
\end{eqnarray}
%---------------------------------------------------------------------------------------------------------------
At this point we use the binomial expansion of the term $(x+y-1)^N$,
and apply the series representation of the Appell hypergeometric function, which in this case is convergent in the integration
region of $x$ and $y$. This allows us to do the integrals in terms of Beta functions. E.g, we obtain
%---------------------------------------------------------------------------------------------------------------
\begin{equation}\label{Equ:J5ForInitialValues}
J_5(1,1,1)=\frac{2 \left[
        \big(
                -1
                -N
                -\frac{\ep}{2}
        \big)\,F_1
        -2 F_2
        +2 F_3
\right]}{3 \ep +2},
\end{equation}
%---------------------------------------------------------------------------------------------------------------
with
%---------------------------------------------------------------------------------------------------------------
\begin{eqnarray}
F_1&=&\sum_{j=0}^N 
        \sum_{n=0}^{\infty } 
                \sum_{m=0}^{\infty } \frac{(-1)^{-j
                +N
                } N! \Gamma (
                        -\frac{3 \ep}{2}) \Gamma (1+N) \Gamma (
                        -\frac{\ep}{2}
                        +m
                ) \Gamma (
                        -\frac{\ep}{2}
                        +n
                ) }
{(
                        1
                        +\frac{\ep}{2}
                        +j
                        +m
                ) j! m! n! (-j
                +N
                )! } 
\nonumber\\ && \times \frac{
 \Gamma (
                        1
                        +\frac{\ep}{2}
                        +n
                ) \Gamma (1
                -j
                +N
                )\Gamma (
                        1
                        +\frac{\ep}{2}
                        +m
                        +n
                        +N
                )}
{\Gamma (
                        2
                        +\frac{\ep}{2}
                        +N
                ) \Gamma (
                        2
                        +\frac{\ep}{2}
                        -j
                        +n
                        +N
                )\Gamma (1
                -\ep
                +m
                +n
                +N
                )},
\\
F_2&=& \sum_{k=0}^{\infty } \frac{
\Gamma (-\ep) \Gamma (
        -\frac{\ep}{2}) \Gamma (1+N) \Gamma (
        -\frac{\ep}{2}
        +k
) \Gamma (
        1
        +\frac{\ep}{2}
        +k
        +N
)^2}{k! \Gamma (
        2
        +\frac{\ep}{2}
        +N
) \Gamma (
        1
        -\frac{\ep}{2}
        +k
        +N
) \Gamma (
        2
        +\frac{\ep}{2}
        +k
        +N
)},\\
F_3&=&\sum_{j=0}^N 
        \sum_{n=0}^{\infty } 
                \sum_{m=0}^{\infty } \frac{(-1)^{-j
                +N
                } (
                        1
                        +\frac{\ep}{2}
                        +j
                ) N! \Gamma (
                        -\frac{3 \ep}{2}) \Gamma (1+N) \Gamma (
                        -\frac{\ep}{2}
                        +m
                ) \Gamma (
                        -\frac{\ep}{2}
                        +n
                ) \Gamma (
                        1
                        +\frac{\ep}{2}
                        +n
                )}{(
                        1
                        +\frac{\ep}{2}
                        +j
                        +m
                ) j! m! n! (-j
                +N
                )! \Gamma (
                        2
                        +\frac{\ep}{2}
                        +N
                ) \Gamma (
                        2
                        +\frac{\ep}{2}
                        -j
                        +n
                        +N
                )} 
\nonumber\\ && \times \frac{\Gamma (1
                -j
                +N
                ) \Gamma (
                        1
                        +\frac{\ep}{2}
                        +m
                        +n
                        +N
                )}
{\Gamma (1
                -\ep
                +m
                +n
                +N
                )}.
\end{eqnarray}
%---------------------------------------------------------------------------------------------------------------
Using \texttt{EvaluateMultiSums} together with \texttt{Sigma} enabled us to calculate the expansion up the 
linear term in $\ep$. However, we need the expansions up to $\ep^4$ in later calculations. At the moment, such a high order in 
$\varepsilon$ seems beyond what {\tt Sigma} can compute in a
reasonable amount of time.

Fortunately, there is an alternative. Using \cite{Slater}
%---------------------------------------------------------------------------------------------------------------
\begin{equation}
F_1(a;b,b';c;x,y) = \frac{\Gamma(c)}{\Gamma(a) \Gamma(c-a)} \int_0^1 dz \,\, z^{a-1} (1-z)^{c-a-1} (1-zx)^{-b} (1-zy)^{-b'} \, ,
\end{equation}
%---------------------------------------------------------------------------------------------------------------
we obtain
%---------------------------------------------------------------------------------------------------------------
\begin{eqnarray}
J_5(\{\nu_i\}) &=& \int_0^1 dx \int_0^1 dy \int_0^1 dz \,\,
(-1)^{\nu+1} 
\frac{\Gamma\left(\nu_1-1-\frac{\varepsilon}{2}\right) \Gamma\left(\nu_5-1-\frac{\varepsilon}{2}\right) \Gamma(N+\nu_4)}{\Gamma(\nu_1) \Gamma(\nu_4) \Gamma(\nu_5) 
\Gamma\left(N+2+\frac{\varepsilon}{2}\right)}
\nonumber \\ && \phantom{\int_0^1 dx \int_0^1 dy \int_0^1} \times
(x+y-1)^N
x^{\varepsilon/2} 
y^{-\nu_5+1+\varepsilon/2} (1-y)^{\nu_5-1}
z^{N+1+\varepsilon/2}
\nonumber \\ && \phantom{\int_0^1 dx \int_0^1 dy \int_0^1} \times
(1-z)^{\nu-5-\frac{3}{2} \varepsilon} (1-zx)^{-\nu_1+1+\varepsilon/2} (1-zy)^{-\nu_5+1+\varepsilon/2}. \label{Equ:J5Integral}
\end{eqnarray}
%---------------------------------------------------------------------------------------------------------------
Let us stress again that for specific, not too large values of $N$ such integrals are relatively easy to solve, while 
obtaining the general $N$ dependence is much more involved. We will attack this problem with symbolic integration methods, 
see Subsection~\ref{Sec:AlmkvistZeil}, and recurrence solving, see Subsection~\ref{Sec:RecSolver}.

%%%%%%%%%%%%%%%%%%%%%%%%%%%%%%%%%%%%%%%%%%%%%%%%%%%%%%%%%%%%%%%%%%%%%%%%%%%%%%%%%%%%%%%%%%%%%%%%%%%%%%%%%%%%%%%%%%%%%%%%%%%%%%%%%%%%%%%%%%%%%%
%%%%%%%%%%%%%%%%%%%%%%%%%%%%%%%%%%%%%%%%%%%%%%%%%%%%%%%%%%%%%%%%%%%%%%%%%%%%%%%%%%%%%%%%%%%%%%%%%%%%%%%%%%%%%%%%%%%%%%%%%%%%%%%%%%%%%%%%%%%%%%

\subsubsection{A fine-tuned multi-variate Almkvist-Zeilberger algorithm}\label{Sec:AlmkvistZeil}

\vspace*{1mm}
\noindent
First, we will calculate linear recurrences in $N$ for several instances of~\eqref{Equ:J5Integral} using a slight modification of the 
multi-variable Almkvist-Zeilberger algorithm~\cite{AZ} which is implemented in the package \ttfamily{MultiIntegrate} \rmfamily  
and described in \cite{Ablinger:PhDThesis}.

In general, consider the hyperexponential integrand
%---------------------------------------------------------------------------------------------------------------
\begin{equation}
F(N;x_1, \dots , x_d)=P(N;x_1, \dots, x_d) \cdot H(N;x_1, \dots, x_d), \label{mAZintegrand}
\end{equation}
%---------------------------------------------------------------------------------------------------------------
with  the multivariate polynomials $P(N;x_1, \dots, x_d) \in \set K[N,x_1, \dots, x_d]$ and
%---------------------------------------------------------------------------------------------------------------
$$
H(N;x_1, \dots, x_d)=
e^{a(x_1, \dots,x_d)/b(x_1, \dots, x_d)} \cdot
\left ( \prod_{p=1}^P {S_p(x_1, \dots, x_d)}^{\alpha_p} \right )\cdot
\left ( { \frac{s(x_1, \dots, x_d)}{t(x_1, \dots, x_d)} } \right )^N,
$$
%---------------------------------------------------------------------------------------------------------------
where $a(x_1, \dots, x_d),b(x_1, \dots, x_d)$,
$s(x_1, \dots, x_d),t(x_1, \dots, x_d)$ and
$S_p(x_1, \dots, x_d) \in \set K[x_1, \dots, x_d]$, and $\alpha_p\in \set K$. Note that this class of integrands 
covers as a special case the integral family~\eqref{Equ:J5Integral} with exact $\ep$-dependence by choosing the 
rational function field $\set K=\set Q(\ep)$. More generally, a big 
class of Feynman integrals that contains at most one mass can be represented in this form~\cite{Blumlein:2010zv,Weinzierl:13}.\\
Then due to~\cite{AZ} there exists a non-negative integer $L$, 
there exist $e_0(N),e_1(N), \dots , e_L(N)\in\set K[N]$ (or equivalently from $\set K(N)$),
{\it not all zero}, and  there also exist $R_i(N;x_1, \dots, x_d)\in\set K(N,x_1, \dots, x_d)$
%($i=1, \dots ,d$) 
such that
%---------------------------------------------------------------------------------------------------------------
\begin{equation}
G_i(N;x_1, \dots, x_d):=R_i(N;x_1, \dots,x_d)F(N;x_1, \dots, x_d)  \label{Gs}
\end{equation}
%---------------------------------------------------------------------------------------------------------------
satisfy the integrand recurrence
%---------------------------------------------------------------------------------------------------------------
\begin{equation}
\sum_{i=0}^L e_i(N) F(N+i;x_1, \dots, x_d)= \sum_{i=1}^d D_{x_i} G_i(N;x_1, \dots, x_d), \label{mAZrec}
\end{equation}
%---------------------------------------------------------------------------------------------------------------
where $D_{x_i}$ stands for the derivative w.r.t $x_i$.

%---------------------------------------------------------------------------------------------------------------
\subsubsection*{The general method} 
%---------------------------------------------------------------------------------------------------------------

\vspace*{1mm}\noindent
The proof of the existence, and in particular a method to compute such an integrand recurrence~\eqref{mAZrec}, 
is based on the following observation~\cite{AZ}.
Fix a non-negative integer $L$ (with the role given above), define
%---------------------------------------------------------------------------------------------------------------
$$
\overline{H} (N;x_1, \dots, x_d)
    :=e^{a(x_1, \dots,x_d)/b(x_1, \dots, x_d)} \cdot\left (\prod_{p=1}^P {S_p(x_1, \dots, x_d)}^{\alpha_p} \right )\cdot{ \frac{s(x_1, \dots, x_d)^N}{t(x_1, \dots, x_d)^{N+L}}},
$$
%---------------------------------------------------------------------------------------------------------------
and make for $i=1, \dots, d$ the general Ansatz
%---------------------------------------------------------------------------------------------------------------
\begin{equation}
G_i(N;x_1, \dots, x_d)=\overline{H}(N;x_1, \dots, x_d) \cdot r_i(x_1,\dots, x_d ) \cdot X_i(N; x_1, \dots, x_d). \label{mAZansatz}
\end{equation}
%---------------------------------------------------------------------------------------------------------------
Then it turns out that for $L$ chosen sufficiently large\footnote{There exist upper bounds for a particular input. But usually, these bounds are too high and one tries smaller values.} there exist polynomials
%---------------------------------------------------------------------------------------------------------------
$X_i(N; x_1, \dots , x_d)\in \set K[N][x_1, \dots, x_d]$ with $1\leq i\leq L$ and polynomials $e_i(N)\in\set K[N]$ (not all zero) such that~\eqref{mAZrec} holds. Motivated by this fact, one searches for these unknowns $X_i$ and $e_i$ as follows. By construction the logarithmic derivative of $\overline{H} (N;x_1, \dots, x_d)$ is a rational function in the $x_i$, i.e., we have that
$$
{ \frac{D_{x_i} \overline{H}(N;x_1, \dots, x_d)}{\overline{H}(N;x_1, \dots, x_d)} }={\frac{q_i(x_1, \dots, x_d)}{r_i(x_1, \dots, x_d)}}
$$
%---------------------------------------------------------------------------------------------------------------
for explicitly given $q_i(x_1, \dots, x_d),r_i(x_1, \dots, x_d)\in\set K[x_1,\dots,x_d]$. 
Hence the Ansatz (\ref{mAZrec})
is equivalent to~\cite{AZ}
%---------------------------------------------------------------------------------------------------------------
\begin{eqnarray}
&&\sum_{i=1}^d [D_{x_i}r_i(x_1, \dots, x_d)+q_i(x_1, \dots, x_d)] \cdot X_i(N;x_1, \dots, x_d)+r_i(x_1, \dots, x_d)\cdot D_{x_i}X_i(x_1, \dots, x_d)
\nonumber \\
&&=\sum_{i=0}^L e_i(N)\,P(N;x_1,\dots,x_d)\,s(x_1,\dots,x_d)^i\,t(x_1,\dots,x_d)^{L-i}. \label{mAZrec2}
\end{eqnarray}
%---------------------------------------------------------------------------------------------------------------
Finally, we chose appropriate degree bounds w.r.t.\ the $x_1,\dots,x_d$ for the $X_i$ ($1\leq i\leq d$) and plug the polynomials 
with unknown coefficients from $\set K[N]$ (from $\set K(N)$) into~\eqref{mAZrec2}. By coefficient comparison this yields 
a linear system in $\set K(N)$ with the unknowns $e_i(N)$ and the unknown coefficients of the polynomials $X_i$. Finally, we 
can seek a non-trivial solution for~\eqref{mAZrec2} and thus for~\eqref{mAZrec}. In the end, we clear 
denominators in $N$ such that the $e_i(N)$ turn to polynomials.

If $F(N;\dots,x_{i-1},u_i,x_{i+1},\dots)=0$ and $F(N;\dots,x_{i-1},o_i,x_{i+1},\dots)=0$ 
%(and hence $G(n;\dots,x_{i-1},u_i,x_{i+1},\dots)=0$ and $G(n;\dots,x_{i-1},o_i,x_{i+1},\dots)=0$ )
then
%---------------------------------------------------------------------------------------------------------------
$$
a(N):=\int_{u_d}^{o_d} \dots\int_{u_1}^{o_1}F(N;x_1, \dots, x_d) dx_1 \dots dx_d,
$$
%---------------------------------------------------------------------------------------------------------------
satisfies the homogeneous linear recurrence equation with polynomial coefficients 
%---------------------------------------------------------------------------------------------------------------
\begin{equation}\label{AZlinrec}
\sum_{i=0}^L e_i(N) a(N+i)= 0.
\end{equation}
%---------------------------------------------------------------------------------------------------------------

%---------------------------------------------------------------------------------------------------------------
\subsubsection*{Dealing with non-standard boundary conditions}
%---------------------------------------------------------------------------------------------------------------

\vspace*{1mm} \noindent
Unfortunately, in many cases the integrand (\ref{mAZintegrand}) does not vanish at the integration bounds and we end up in a linear 
recurrence with a non-trivial inhomogeneous part which can be written as a linear combination of integrals with at least one integral 
operator less. 

%---------------------------------------------------------------------------------------------------------------
\begin{remark}
\rm
In order to solve such recurrences using the methods presented below in Subsection~\ref{Sec:RecSolver}, also the new arising 
integrals need to be simplified to indefinite nested product-sum expressions. In a nutshell, a recursive application of the 
presented method has to be carried out. We remark that this approach works nicely, if the initial values of the integrals in 
the inhomogeneous part can be calculated efficiently. Further details on this approach are given 
in~\cite{Ablinger:PhDThesis,LL12:Technolgy}. We remark further that similar approaches have been explored in~\cite{Blumlein:2010zv}
and~\cite{LL12:Technolgy} based on~\cite{WegschaiderWZ}
and~\cite{NewSigmaApproach}, respectively, in order to derive recurrences for hypergeometric
multi-sums.
\end{remark}

%---------------------------------------------------------------------------------------------------------------

In the following we will avoid these difficulties by adapting the above Ansatz. Namely, we can always obtain a homogeneous recurrence of the form~\eqref{AZlinrec}
by changing (\ref{mAZansatz}) to
%---------------------------------------------------------------------------------------------------------------
\begin{equation}
G_i(n;x_1, \dots, x_d)=\overline{H}(n;x_1, \dots, x_d) \cdot r_i(x_1,\dots, x_d ) \cdot X_i(x_1, \dots, x_d)(x_i-u_i)(x_i-o_i), \label{mAZansatz2}
\end{equation}
%---------------------------------------------------------------------------------------------------------------
\ie the $G_i$ are forced to vanish at the integration bounds. 
Then with this Ansatz (\ref{mAZrec2}) the underlying linear system turns into
%---------------------------------------------------------------------------------------------------------------
\begin{eqnarray}
&&\sum_{i=1}^d [D_{x_i}r_i(x_1, \dots, x_d)+q_i(x_1, \dots, x_d)] \cdot X_i(x_1, \dots, x_d)(x_i-u_i)(x_i-o_i) \nonumber \\
&&\hspace{1cm}+r_i(x_1, \dots, x_d)\cdot D_{x_i}X_i(x_1, \dots, x_d)(x_i-u_i)(x_i-o_i)\nonumber\\
&&\hspace{1cm}=\sum_{i=0}^L e_i(N)\,P(N;x_1,\dots,x_d)\,s(x_1,\dots,x_d)^i\,t(x_1,\dots,x_d)^{L-i}.\label{mAZrec3}
\end{eqnarray}
%---------------------------------------------------------------------------------------------------------------

The general method now is straightforward: Given an integrand of the form (\ref{mAZintegrand}), we can set $L=0,$ look for degree bounds for $X_i(x_1, \dots , x_d)$ and try to find a 
solution of (\ref{mAZrec3}) by coefficient comparison. If we do not find a solution of (\ref{mAZrec3}) with not all $e_i(n)$'s equal to zero, we increase $L$ by one, look for new degree bounds 
for $X_i(x_1, \dots , x_d)$ and try again to find a solution of (\ref{mAZrec3}). Again, if we do not find a solution  with not all $e_i(n)$'s equal to zero, we increase $L$ by one and repeat 
the process.

As an example we apply this modification of the Almkvist-Zeilberger algorithm to  
%---------------------------------------------------------------------------------------------------------------
\begin{eqnarray}
\tilde{J}_5(N,\varepsilon) &=& \frac{\Gamma \left(\frac{\varepsilon}{2}+N+2\right)}{\Gamma \left(-\frac{\varepsilon}{2}\right)^2 \Gamma (N+1)}J_5(N,\varepsilon)\nonumber\\
		   &=&\int_0^1 dx \int_0^1 dy \int_0^1 dz \,\,x^{\varepsilon/2} y^{\varepsilon/2} (1-z)^{-\frac{3 \varepsilon}{2}-2}z^{\frac{\varepsilon}{2}+N+1} (1-x z)^{\varepsilon/2}\nonumber
		    \\ && \phantom{\int_0^1 dx \int_0^1 dy \int_0^1} \times (1-y z)^{\varepsilon/2} (x+y-1)^N\, ,
\label{eq:J5a}
\end{eqnarray}
%---------------------------------------------------------------------------------------------------------------
with
%---------------------------------------------------------------------------------------------------------------
\begin{eqnarray}
 J_5(N,\varepsilon):=J_5(1,1,1).
\end{eqnarray}
%---------------------------------------------------------------------------------------------------------------
Using the package \ttfamily{MultiIntegrate} \rmfamily we find a recurrence of order 5:
%---------------------------------------------------------------------------------------------------------------
\begin{eqnarray}
\sum_{k = 0}^5 \tilde{A}_k(N,\varepsilon) \tilde{J}_5(N+k,\varepsilon) = 0,
\label{Equ:tileARec}
\end{eqnarray}
%---------------------------------------------------------------------------------------------------------------
where
%---------------------------------------------------------------------------------------------------------------
\begin{align*}
\tilde{A}_0&(N,\varepsilon)=(N+1) (N+2)\big(
        8 \varepsilon^{10}
        +104 \varepsilon^9 (N+3)
	+4 \varepsilon^8  \big(96 N^2+601 N+887\big)
        \nonumber\\&\ \ +4 \varepsilon^7  \big(12 N^3+414 N^2+1583 N+1393\big)
        \nonumber\\&\ \ -8 \varepsilon^6  \big(264 N^4+2436 N^3+8643 N^2+14518 N+9947\big)
        \nonumber\\&\ \ -16 \varepsilon^5 \big(156 N^5+1690 N^4+6847 N^3+12661 N^2+9537 N+717\big)
        \nonumber\\&\ \ +32 \varepsilon^4 \big(68 N^6+1158 N^5+8155 N^4+30114 N^3+61712 N^2+67616 N+31693\big)
        \nonumber\\&\ \ +64 \varepsilon^3 \big(40 N^7+560 N^6+2755 N^5+3729 N^4-14194 N^3-61920 N^2-89140 N -46600\big)
        \nonumber\\&\ \ -128 \varepsilon^2 (N+2) \big(12 N^7+254 N^6+2249 N^5+10758 N^4+30173 N^3+50610 N^2\nonumber\\&\ \ +49122 N+22706\big)
        \nonumber\\&\ \ +256 \varepsilon (N+2)^2 (N+3) (N+4) \big(44 N^4+501 N^3+2044 N^2+3455 N+1976\big)
        \nonumber\\&\ \ -512 (N+1) (N+2)^3 (N+3)^2 (N+4) \big(6 N^2+47 N+95\big)                
\big)\nonumber,\\
\tilde{A}_1&(N,\varepsilon)=(N+2)\big(
        -22 \varepsilon^{11}
        -2 \varepsilon^{10}  (157 N+435)
        -\varepsilon^9 \big(1500 N^2+8611 N+11745\big)
        \nonumber\\&\ \ -\varepsilon^8 \big(2548 N^3+22936 N^2+63597 N+54229\big)
        \nonumber\\&\ \ +4 \varepsilon^7  \big(266 N^4+1857 N^3+6065 N^2+14351 N+15987\big)
        \nonumber\\&\ \ +8 \varepsilon^6 \big(994 N^5+12961 N^4+67246 N^3+174692 N^2+226821 N+116092\big)
        \nonumber\\&\ \ +16 \varepsilon^5 \big(336 N^6+5348 N^5+33569 N^4+104918 N^3+165290 N^2+108259 N+6100\big)
        \nonumber\\&\ \ -16 \varepsilon^4  \big(404 N^7+7578 N^6+61778 N^5+284762 N^4+802660 N^3+1382074 N^2\nonumber\\&\ \ +1340455 N+560287\big)
        \nonumber\\&\ \ -64 \varepsilon^3 \big(94 N^8+1823 N^7+14305 N^6+55870 N^5+96299 N^4-37256 N^3\nonumber\\&\ \ -447044 N^2-704959 N-379338\big)
        \nonumber\\&\ \ +128 \varepsilon^2 (N+3) \big(30 N^8+715 N^7+7667 N^6+48253 N^5+194086 N^4+507439 N^3\nonumber\\&\ \ +835393 N^2+785327 N+320382\big)
        \nonumber\\&\ \ -256 \varepsilon (N+2) (N+3)^2 \big(107 N^6+2070 N^5+16342 N^4+67226 N^3+151557 N^2\nonumber\\&\ \ +176932 N+83196\big)                
        \nonumber\\&\ \ +256 (N+2)^3 (N+3)^3 (N+4) \big(30 N^3+331 N^2+1193 N+1386\big)                
\big),\nonumber\\
\tilde{A}_2&(N,\varepsilon)=\big(
        12 \varepsilon^{12}
        +12 \varepsilon^{11} (17 N+45)
        +2 \varepsilon^{10} \big(620 N^2+3553 N+4795\big)
        \nonumber\\&\ \ +2 \varepsilon^9 \big(1504 N^3+14190 N^2+41901 N+38907\big)
        \nonumber\\&\ \ +4 \varepsilon^8 \big(172 N^4+4983 N^3+30942 N^2+69119 N+50850\big)
        \nonumber\\&\ \ -4 \varepsilon^7 \big(1996 N^5+24056 N^4+113313 N^3+269119 N^2+337198 N+185290\big)
        \nonumber\\&\ \ -16 \varepsilon^6 \big(450 N^6+8210 N^5+59749 N^4+227386 N^3+486841 N^2+563176 N+275664\big)
        \nonumber\\&\ \ +16 \varepsilon^5 \big(340 N^7+4314 N^6+19137 N^5+25532 N^4-55105 N^3-206516 N^2-191528 N\nonumber\\&\ \ -23458\big)
        \nonumber\\&\ \ +32 \varepsilon^4 \big(140 N^8+2940 N^7+26550 N^6+139926 N^5+493839 N^4+1240186 N^3\nonumber\\&\ \ +2161699 N^2+2304248 N+1100084\big)
        \nonumber\\&\ \ +64 \varepsilon^3 \big(4 N^9+506 N^8+8651 N^7+63510 N^6+236215 N^5+395334 N^4-105413 N^3\nonumber\\&\ \ -1551017 N^2-2362944 N-1217770\big)
        \nonumber\\&\ \ -128 \varepsilon^2 (N+3) \big(12 N^9+314 N^8+3782 N^7+29105 N^6+160727 N^5+640273 N^4\nonumber\\&\ \ +1750874 N^3+3052505 N^2+3017094 N+1276604\big)
        \nonumber\\&\ \ +256 \varepsilon (N+2) (N+3)^2 (N+4) \big(26 N^6+825 N^5+8967 N^4+46529 N^3+125411 N^2\nonumber\\&\ \ +168628 N+88652\big)                
        \nonumber\\&\ \ -512 (N+1) (N+2)^2 (N+3)^3 (N+4)^2 \big(6 N^3+98 N^2+459 N+655\big)                
\big),\nonumber\\
\tilde{A}_3&(N,\varepsilon)=\big(
        -64 \varepsilon^{12}
        -8 \varepsilon^{11} (113 N+298)
        -8 \varepsilon^{10} \big(519 N^2+2948 N+3896\big)
        \nonumber\\&\ \ -4 \varepsilon^9 \big(1444 N^3+13839 N^2+39746 N+34305\big)
        \nonumber\\&\ \ +4 \varepsilon^8 \big(1948 N^4+17868 N^3+63837 N^2+112966 N+84655\big)
        \nonumber\\&\ \ +16 \varepsilon^7 \big(1456 N^5+20460 N^4+112365 N^3+304963 N^2+412258 N+221769\big)
        \nonumber\\&\ \ -8 \varepsilon^6 \big(320 N^6+2050 N^5+4192 N^4+27408 N^3+174901 N^2+411759 N+324872\big)
        \nonumber\\&\ \ -16 \varepsilon^5 \big(1756 N^7+33154 N^6+265889 N^5+1186719 N^4+3218059 N^3+5349388 N^2\nonumber\\&\ \ +5071913 N+2113696\big)
        \nonumber\\&\ \ +32 \varepsilon^4 \big(188 N^8+4802 N^7+59527 N^6+439922 N^5+2025336 N^4+5813984 N^3\nonumber\\&\ \ +10076450 N^2+9621283 N+3878602\big)
        \nonumber\\&\ \ +64 \varepsilon^3 \big(140 N^9+2768 N^8+22500 N^7+99545 N^6+287700 N^5+723136 N^4\nonumber\\&\ \ +1854572 N^3+3714620 N^2+4272517 N+2031600\big)
        \nonumber\\&\ \ -128 \varepsilon^2 \big(24 N^{10}+830 N^9+14362 N^8+152630 N^7+1053620 N^6+4834279 N^5\nonumber\\&\ \ +14824351 N^4+29964399 N^3+38244797 N^2+27875896 N+8824032\big)
        \nonumber\\&\ \ +256 \varepsilon (N+2) (N+3) (N+4) \big(118 N^7+2639 N^6+24247 N^5+118311 N^4+329565 N^3\nonumber\\&\ \ +520306 N^2+426076 N+136854\big)                
        \nonumber\\&\ \ -512 (N+1) (N+2)^2 (N+3)^2 (N+4)^2 (N+5) \big(12 N^3+97 N^2+230 N+144\big)               
\big),\nonumber\\
\tilde{A}_4&(N,\varepsilon)=\big(
        64 \varepsilon^{12}
        +192 \varepsilon^{11} (5 N+14)
        +16 \varepsilon^{10} \big(297 N^2+1769 N+2451\big)
        \nonumber\\&\ \ +16 \varepsilon^9 \big(453 N^3+4462 N^2+13094 N+11244\big)
        \nonumber\\&\ \ -8 \varepsilon^8 \big(1084 N^4+11117 N^3+47258 N^2+103981 N+94650\big)
        \nonumber\\&\ \ -8 \varepsilon^7 \big(3304 N^5+51138 N^4+311957 N^3+948722 N^2+1440105 N+858544\big)
        \nonumber\\&\ \ +16 \varepsilon^6 \big(420 N^6+5507 N^5+36275 N^4+169650 N^3+536911 N^2+952507 N+694370\big)
        \nonumber\\&\ \ +16 \varepsilon^5 \big(1828 N^7+38868 N^6+353301 N^5+1801014 N^4+5604391 N^3+10664390 N^2\nonumber\\&\ \ +11433064 N+5260048\big)
        \nonumber\\&\ \ -32 \varepsilon^4 \big(316 N^8+8356 N^7+105800 N^6+802421 N^5+3836854 N^4+11588223 N^3\nonumber\\&\ \ +21401558 N^2+22066744 N+9745752\big)
        \nonumber\\&\ \ -64 \varepsilon^3 \big(116 N^9+2424 N^8+19923 N^7+82966 N^6+208191 N^5+530980 N^4+1847484 N^3\nonumber\\&\ \ +4687014 N^2+6120858 N+3111104\big)
        \nonumber\\&\ \ +128 \varepsilon^2 \big(24 N^{10}+826 N^9+14897 N^8+172000 N^7+1314686 N^6+6710299 N^5\nonumber\\&\ \ +22873183 N^4+51298261 N^3+72551278 N^2+58573022 N+20544948\big)
        \nonumber\\&\ \ -256 \varepsilon (N+2) (N+3) \big(106 N^8+3278 N^7+42903 N^6+310942 N^5+1366350 N^4\nonumber\\&\ \ +3729418 N^3+6173159 N^2+5657732 N+2191212\big)                
        \nonumber\\&\ \ +512 (N+1) (N+2)^2 (N+3)^2 (N+4) (N+5) (N+6) \big(12 N^3+121 N^2+396 N+431\big)                
\big),\nonumber\\
\tilde{A}_5&(N,\varepsilon)=(N+5)\big(
        -128 \varepsilon^{11} 
        -128 \varepsilon^{10} (11 N+26)
        -32 \varepsilon^9  \big(115 N^2+592 N+647\big)
        \nonumber\\&\ \ +32 \varepsilon^8  \big(63 N^3+430 N^2+1665 N+2384\big)
        \nonumber\\&\ \ +16 \varepsilon^7  \big(714 N^4+7881 N^3+33802 N^2+66225 N+47654\big)
        \nonumber\\&\ \ -16 \varepsilon^6  \big(234 N^5+2444 N^4+13989 N^3+50862 N^2+104083 N+87848\big)
        \nonumber\\&\ \ -16 \varepsilon^5  \big(580 N^6+10181 N^5+76586 N^4+319207 N^3+772120 N^2+1012046 N+547832\big)
        \nonumber\\&\ \ +16 \varepsilon^4  \big(244 N^7+5456 N^6+61605 N^5+401216 N^4+1536277 N^3+3408574 N^2\nonumber\\&\ \ +4066436 N+2026928\big)
        \nonumber\\&\ \ +64 \varepsilon^3  \big(26 N^8+357 N^7+583 N^6-11139 N^5-65193 N^4-120264 N^3+11864 N^2\nonumber\\&\ \ +272830 N+222624\big)
        \nonumber\\&\ \ -64 \varepsilon^2 (N+3) \big(12 N^8+298 N^7+4684 N^6+49024 N^5+306907 N^4+1122441 N^3\nonumber\\&\ \ +2350650 N^2+2607576 N+1185072\big)
        \nonumber\\&\ \ +256 \varepsilon (N+2) (N+3) \big(25 N^7+743 N^6+8856 N^5+55358 N^4+197497 N^3+404131 N^2\nonumber\\&\ \ +439902 N+196128\big)
        \nonumber\\&\ \ -256 (N+1) (N+2)^2 (N+3)^2 (N+4) (N+6) (N+7) \big(6 N^2+35 N+54\big)                
\big).
\end{align*}
%---------------------------------------------------------------------------------------------------------------
Finally, plugging
%---------------------------------------------------------------------------------------------------------------
$$\tilde{J}_5(N,\varepsilon)= \frac{\Gamma \left(\frac{\varepsilon}{2}+N+2\right)}{\Gamma \left(-\frac{\varepsilon}{2}\right)^2 \Gamma (N+1)}J_5(N,\ep)$$
into~\eqref{Equ:tileARec}, dividing through 
$\frac{\Gamma \left(\frac{\varepsilon}{2}+N+2\right)}{\Gamma \left(-\frac{\varepsilon}{2}\right)^2 \Gamma (N+1)}$ 
and clearing denominators yield the recurrence
%---------------------------------------------------------------------------------------------------------------
\begin{eqnarray}\label{eq:RecJ5a}
\sum_{k = 0}^5 {A}_k(N,\varepsilon) J_5(N+k,\varepsilon) = 0,
\label{Equ:ARec}
\end{eqnarray}
%---------------------------------------------------------------------------------------------------------------
for our input integral $J_5(N,\ep)$ with the polynomials $A_i(N,\ep)$ in $N$ and $\ep$, see~(\ref{Equ:tileARec}).

%---------------------------------------------------------------------------------------------------------------
\subsubsection*{Further speed-ups}
%---------------------------------------------------------------------------------------------------------------

\vspace*{1mm}
\noindent
We emphasize that the underlying systems to be solved can be rather large. This becomes even worse if one switches 
from~\eqref{mAZansatz} to the modified  Ansatz~\eqref{mAZansatz2}. In order to derive recurrences for heavy integrals 
such as~\eqref{Equ:J5Integral}, further essential improvements are built into \texttt{MultiIntegrate}. E.g., the underlying 
linear system of Ansatzes~\eqref{mAZintegrand} can be reduced substantially by homomorphic image testing. In this way one can 
hunt for exactly one solution (and not several), one can find out if, e.g., certain unknown coefficients are obsolete, and one 
can remove redundant constraints of the linear system. Further details on these improvements can be found 
in~\cite{Ablinger:PhDThesis}. E.g., for $J_5(1,1,1)$ the package \texttt{MultiIntegrate} produces for the order $L=5$ a linear 
system with 805 equations and 762 variables with entries from $\set K=\set Q(N,\ep)$ which requires in total  3.9 GB of memory. 
After two weeks of calculation we still failed to extract the solution from the system. However, by optimizing the linear system, it 
consists only of $339$ equations with $340$ unknowns, i.e., it possesses exactly one solution. In its reduced form it requires 
only $1.6~{\rm GB}$ of memory and we 
could 
derive the desired solution within $32755$ seconds. We obtain similar improvements  for the remaining integrals~\eqref{Equ:J5Integral} 
summarized in the following table.
%---------------------------------------------------------------------------------------------------------------
\begin{center}
\begin{tabular}{|c||c|c|c|c|c|c|c|c|}
\hline
&&\multicolumn{3}{c|}{original system}&\multicolumn{3}{c}{optimized system}&\\
integral & L & eqs & variables & size & eqs & variables & size & time\\
\hline
\hline
$J_5(1,1,1)$ & 5 & 805 & 762 & 3933 MB & 339 & 340 & 1603 MB &
32755 s\\
$J_5(1,1,2)$& 6 &1222 & 1351 & 6938 MB & 486 & 487 & 2327 MB & 473048s\\
$J_5(2,1,1)$& 6 & 1105 & 1183 & 6099 MB & 485 & 486 & 2355 MB & 507467s\\
$J_5(2,1,2)$& 6 & 1222 & 1351 & 6921 MB & 486 & 487 & 2321 MB & 398723s\\
$J_5(2,2,1)$& 6 & 1105 & 1183 & 6116 MB & 485 & 486 & 2366 MB & 669255s\\
$J_5(3,1,1)$& 6 & 1105 & 1183 & 6087 MB & 485 & 486 & 2353 MB & 390360s\\
\hline
\end{tabular}
\end{center}
%---------------------------------------------------------------------------------------------------------------

%---------------------------------------------------------------------------------------------------------------
\subsubsection{Finding Laurent series solutions of a linear recurrence}\label{Sec:RecSolver}
%---------------------------------------------------------------------------------------------------------------

\vspace*{1mm}
\noindent
In order to calculate the $\ep$-expansion
%---------------------------------------------------------------------------------------------------------------
\begin{equation}\label{Equ:J5Ansatz}
J_5(N,\ep)=F_{-3}(N)\ep^{-3}+F_{-2}(N)\ep^{-2}+\dots+F_{4}(N)\ep^{4}+\dots,
\end{equation}
%---------------------------------------------------------------------------------------------------------------
we will need besides the recurrence~\eqref{eq:RecJ5a} five initial values expanded up to $\ep^4$. From the sum 
representation~\eqref{Equ:J5ForInitialValues} of $J_5(1,1,1)$ the following initial values for $J_5(1,1,1)$ can be extracted using {\tt Sigma} and {\tt EvaluateMultiSums}:
%---------------------------------------------------------------------------------------------------------------
\begin{eqnarray}
&& J_5(2,\varepsilon)=
\frac{20}{27 \varepsilon^3}
-\frac{40}{27 \varepsilon^2}
+\frac{1}{\ep} \left(\frac{1393}{486}
+\frac{5 \zeta_2}{18}\right)
-\frac{9601}{1944}
-\frac{5 \zeta_2}{9}
+\frac{49 \zeta_3}{54}
\nonumber\\&&\phantom{J_5(1)}+\varepsilon \Bigl(
        \frac{565297}{69984}
        +\frac{1393 \zeta_2}{1296}
        +\frac{1151 \zeta_2^2}{1440}
        -\frac{137 \zeta_3}{54}
\Bigr)
\nonumber\\&&\phantom{J_5(1)}+\varepsilon^2 \Bigl(
        -\frac{1150003}{93312}
        -\frac{9601 \zeta_2}{5184}
        -\frac{1619 \zeta_2^2}{720}
        +\frac{17831 \zeta_3}{3888}
        +\frac{49 \zeta_2 \zeta_3}{144}
        +\frac{265 \zeta_5}{72}
\Bigr)
\nonumber\\&&\phantom{J_5(1)}+\varepsilon^3 \Bigl(
        \frac{184376401}{10077696}
        +\frac{565297 \zeta_2}{186624}
        +\frac{420979 \zeta_2^2}{103680}
        +\frac{449843 \zeta_2^3}{241920}
        -\frac{103607 \zeta_3}{15552}
\nonumber\\&&\phantom{J_5(1)}-\frac{137}{144} \zeta_2 \zeta_3
	-\frac{259 \zeta_3^2}{864}
        -\frac{191 \zeta_5}{18}
\Bigr)+O(\varepsilon^4),\nonumber\\
&&J_5(3,\varepsilon)=
\frac{1}{6 \varepsilon^3}
-\frac{11}{48 \varepsilon^2}
+\frac{1}{\ep} \left(
\frac{703}{3456}
+\frac{\zeta_2}{16}\right)
-\frac{9773}{9216}
-\frac{11 \zeta_2}{128}
+\frac{41 \zeta_3}{48}
\nonumber\\&&\phantom{J_5(1)}+\varepsilon \Bigl(
        \frac{2157295}{1990656}
        +\frac{703 \zeta_2}{9216}
        +\frac{979 \zeta_2^2}{1280}
        -\frac{931 \zeta_3}{384}
\Bigr)
\nonumber\\&&\phantom{J_5(1)}+\varepsilon^2 \Bigl(
        -\frac{60535183}{15925248}
        -\frac{9773 \zeta_2}{24576}
        -\frac{22289 \zeta_2^2}{10240}
        +\frac{137591 \zeta_3}{27648}
        +\frac{41 \zeta_2 \zeta_3}{128}
        +\frac{1201 \zeta_5}{320}
\Bigr)
\nonumber\\&&\phantom{J_5(1)}+\varepsilon^3 \Bigl(
        \frac{2116767175}{1146617856}
        +\frac{2157295 \zeta_2}{5308416}
        +\frac{3298669 \zeta_2^2}{737280}
        +\frac{406711 \zeta_2^3}{215040}
        -\frac{550229 \zeta_3}{73728}
\nonumber\\&&\phantom{J_5(1)}-\frac{931 \zeta_2 \zeta_3}
        {1024}
         -
        \frac{239 \zeta_3^2}{768}
        -\frac{27611 \zeta_5}{2560}
\Bigr)
+O(\varepsilon^4),\nonumber\\
&&J_5(4,\varepsilon)=
\frac{64}{225 \varepsilon^3}
-\frac{1748}{3375 \varepsilon^2}
+\frac{1}{\ep} \left(\frac{102181}{101250}
+\frac{8 \zeta_2}{75}\right)
-\frac{5738207}{2430000}
-\frac{437 \zeta_2}{2250}
+\frac{196 \zeta_3}{225}
\nonumber\\&&\phantom{J_5(1)}+\varepsilon \Bigl(
        \frac{1681164919}{486000000}
        +\frac{102181 \zeta_2}{270000}
        +\frac{583 \zeta_2^2}{750}
        -\frac{17059 \zeta_3}{6750}
\Bigr)
\nonumber\\&&\phantom{J_5(1)}+\varepsilon^2 \Bigl(
        -\frac{423112175849}{58320000000}
        -\frac{5738207 \zeta_2}{6480000}
        -\frac{407231 \zeta_2^2}{180000}
        +\frac{4288637 \zeta_3}
        {810000}
        +
        \frac{49 \zeta_2 \zeta_3}{150}
        +\frac{1412 \zeta_5}{375}
\Bigr)
\nonumber\\&&\phantom{J_5(1)}+\varepsilon^3 \Bigl(
        \frac{157023072517301}{20995200000000}
        +\frac{1681164919 \zeta_2}{1296000000}
        +\frac{102416383 \zeta_2^2}{21600000}
        +\frac{239203 \zeta_2^3}{126000}
\nonumber\\&&\phantom{J_5(1)}-\frac{155753563 \zeta_3}{19440000}
	-\frac{17059 \zeta_2 \zeta_3}{18000}
	-\frac{14 \zeta_3^2}{45}
        -\frac{499097 \zeta_5}{45000}
\Bigr)
+O(\varepsilon^4),\nonumber\\
&&J_5(5,\varepsilon)=
\frac{2}{27 \varepsilon^3}
-\frac{17}{162 \varepsilon^2}
+\frac{1}{\ep} \left(
\frac{7583}{97200}
+\frac{\zeta_2}{36}\right)
-\frac{1666837}{1296000}
-\frac{17 \zeta_2}{432}
+\frac{113 \zeta_3}{108}
\nonumber\\&&\phantom{J_5(1)}+\varepsilon \Bigl(
        \frac{187423951}{155520000}
        +\frac{7583 \zeta_2}
        {259200}
        +
        \frac{2707 \zeta_2^2}{2880}
        -\frac{19769 \zeta_3}{6480}
\Bigr)
\nonumber\\&&\phantom{J_5(1)}+\varepsilon^2 \Bigl(
        -\frac{6827006887}{1244160000}
        -\frac{1666837 \zeta_2}{3456000}
        -\frac{474031 \zeta_2^2}{172800}
        +\frac{5247499 \zeta_3}{777600}
        +\frac{113 \zeta_2 \zeta_3}{288}
        +\frac{3361 \zeta_5}{720}
\Bigr)
\nonumber\\&&\phantom{J_5(1)}+\varepsilon^3 \Bigl(
        \frac{40517946316703}{20155392000000}
        +\frac{187423951 \zeta_2}{414720000}
        +\frac{125902061 \zeta_2^2}{20736000}
        +\frac{227531 \zeta_2^3}{96768}
        -\frac{321933739 \zeta_3}{31104000}
\nonumber\\&&\phantom{J_5(1)}-\frac{19769 \zeta_2 \zeta_3}{17280}
        -\frac{671 \zeta_3^2}{1728}
        -\frac{118121 \zeta_5}{8640}
\Bigr)
+O(\varepsilon^4),\nonumber\\
&&J_5(6,\varepsilon)=
+\frac{22}{147 \varepsilon^3}
-\frac{535}{2058 \varepsilon^2}
+\frac{1}{\ep} \left(\frac{630043}{1234800}
+\frac{11 \zeta_2}{196}\right)
-\frac{1949958721}{871274880}
-\frac{535 \zeta_2}{5488}
+\frac{775 \zeta_3}{588}
\nonumber\\&&\phantom{J_5(1)}+\varepsilon \Bigl(
        \frac{1873033251331}{677658240000}
        +\frac{630043 \zeta_2}{3292800}
        +\frac{3709 \zeta_2^2}{3136}
        -\frac{321149 \zeta_3}{82320}
\Bigr)
\nonumber\\&&\phantom{J_5(1)}+\varepsilon^2 \Bigl(
        -\frac{181315912291471}{20910597120000}
        -\frac{1949958721 \zeta_2}{2323399680}
        -\frac{7694201 \zeta_2^2}{2195200}
        +\frac{86628433 \zeta_3}{9878400}
        +\frac{775 \zeta_2 \zeta_3}{1568}
\nonumber\\&&\phantom{J_5(1)}+\frac{22931 \zeta_5}{3920}
\Bigr)
\nonumber\\&&\phantom{J_5(1)}+\varepsilon^3 \Bigl(
        \frac{73004792897853520153}{12910202661888000000}
        +\frac{1873033251331 \zeta_2}{1807088640000}
        +\frac{2075932177 \zeta_2^2}{263424000}
        +\frac{7763069 \zeta_2^3}{2634240}
\nonumber\\&&\phantom{J_5(1)}-\frac{2376849605591 \zeta_3}{174254976000}
	-\frac{321149 \zeta_2 \zeta_3}{219520}
        -\frac{4573 \zeta_3^2}{9408}
        -\frac{1911379 \zeta_5}{109760}
\Bigr)
+O(\varepsilon^4).
\label{initvaluesJ}
\end{eqnarray}
%---------------------------------------------------------------------------------------------------------------
The calculations up to $\ep^3$ took 70000 seconds. For the $\ep^4$-coefficient (which we did not print here) seven further days 
have been spent. 

Eventually, given this information we can activate the recurrence solver worked out in~\cite{Blumlein:2010zv} and implemented in 
\texttt{Sigma}. In the following we will illustrate the basic calculation steps of this algorithm.
Inserting the Ansatz~\eqref{Equ:J5Ansatz} into~\eqref{eq:RecJ5a} yields
%----------------------------------------------------------------------------------------------------
\begin{equation}\label{Equ:EpRecAnsatz}
\begin{split}
A_0(N,\ep)&\Big[F_{-3}(N)\ep^{-3}+F_{-2}(N)\ep^{-2}+F_{-1}(N)\ep^{-1}+F_{0}(N)\ep^{0}+\dots\Big]+\\
 A_1(N,\ep)&\Big[F_{-3}(N+1)\ep^{-3}+F_{-2}(N+1)\ep^{-2}+F_{-1}(N+1)\ep^{-1}
+F_{0}(N+1)\ep^{0}+\dots\Big]\\
  +\dots+\\
 A_5(N,\ep)&\Big[F_{-3}(N+5)\ep^{-3}+F_{-2}(N+5)\ep^{-2}+F_{-1}(N+5)\ep^{-1}
+F_{0}(N+5)\ep^{0}+\dots\Big]=0.
\end{split}
 \end{equation}
%----------------------------------------------------------------------------------------------------
Since two Laurent series agree if they agree coefficient-wise, we obtain the following constraint for $F_{-3}(N)$ by coefficient 
comparison: 
%----------------------------------------------------------------------------------------------------
\begin{equation}
\sum_{k=0}^5 A_k(N,0) F_{-3}(N+k) = 0,
\end{equation}
%----------------------------------------------------------------------------------------------------
with
%----------------------------------------------------------------------------------------------------
\begin{align*}
A_0(N,0)&=-512 (N+1)^3 (N+2)^4 (N+3)^3 (N+4)^2 \big(
        6 N^2+47 N+95\big),\\
A_1(N,0)&=256 (N+1) (N+2)^4 (N+3)^4 (N+4)^2 \big(
        30 N^3+331 N^2+1193 N+1386\big),\\
A_2(N,0)&=-512 (N+1) (N+2)^2 (N+3)^5 (N+4)^3 \big(
        6 N^3+98 N^2+459 N+655\big),\\
A_3(N,0)&=-512 (N+1) (N+2)^2 (N+3)^3 (N+4)^4 (N+5) \big(
        12 N^3+97 N^2+230 N+144\big)\\
A_4(N,0)&=512 (N\!+\!1) (N\!+\!2)^2 (N\!+\!3)^3 (N\!+\!4)^2 (N\!+\!5)^2 (N\!+\!6) \big(
        12 N^3+121 N^2+396 N+431\big),\\
A_5(N,0)&=-256 (N+1) (N+2)^2 (N+3)^3 (N+4)^2 (N+5) (N+6)^2 (N+7) \big(
        6 N^2+35 N+54\big).
\end{align*}
%----------------------------------------------------------------------------------------------------
Hence, with the first initial values, i.e.,
%----------------------------------------------------------------------------------------------------
\begin{equation}\label{Equ:Fm3Init}
F_{-3}(2)=\frac{20}{27},\, F_{-3}(3)=\frac{1}{6},\, F_{-3}(4)=\frac{64}{225},\, F_{-3}(5)=\frac{2}{27},\, F_{-3}(2)=\frac{22}{147},
\end{equation}
%----------------------------------------------------------------------------------------------------
all integer points of $F_{-3}(N)$ are uniquely determined. Even better, we are now in the position to find an explicit representation 
using \texttt{Sigma}'s recurrence solver~\cite{Blumlein:2010zv}. 

%----------------------------------------------------------------------------------------------------
\begin{remark}
\rm
In general, given a recurrence with polynomial coefficients where the inhomogeneous part is given as an indefinite nested product-sum expression, \texttt{Sigma} finds all solutions that can be represented in terms of an indefinite nested product-sum expression~\cite{SIG2,RecSolver,Schneider:01,Schneider:05a}. In particular, the occurring sums and products (except $(-1)^N$) are algebraically independent among each other.
\end{remark}
%----------------------------------------------------------------------------------------------------

\noindent In our example, Sigma finds the solution set
%----------------------------------------------------------------------------------------------------
\begin{equation}
L_{-3}=\{c_1\,g_1+c_2\,g_2+c_3\,g_3+c_4\,g_4+c_5\,g_5|c_1,c_2,c_3,c_3,c_5\in\set R\},
\end{equation}
%----------------------------------------------------------------------------------------------------
where the expressions
%----------------------------------------------------------------------------------------------------
\begin{align*}
g_1&=\frac{1}{(N+1)^2 (N+2)},\quad
g_2=\frac{N}{(N+1)^2 (N+2)},\\ 
g_3&=-\frac{1}{4 (N+1) (N+2)}
        +\frac{(-1)^N}{4 (N+1) (N+2)},\\ 
g_4&=\frac{(-1)^N}{2 (N+1) (N+2)}
        +\frac{1}{(N+1)^2 (N+2)} \frac{1}{2} \big(
                -9
                +2^{N+3}
                -5 N
        \big),\\ 
g_5&=-\frac{3 (-1)^N}{64 (N+1) (N+2)}
        -\frac{2^N S_1\left({{\frac{1}{2}}}\right)}{(N+1)^2 (N+2)} (3)
        -\frac{3\big(
                55
                -7 2^{N+3}
                +27 N
        \big)}{64(N+1)^2 (N+2)} 
        +\frac{3 S_1}{2 (N+1)^2 (N+2)}
\end{align*}
%----------------------------------------------------------------------------------------------------
are linearly independent over $\set R$.
Since the solution set is completely determined, it follows that $F_{-3}(N)\in L_{-3}$.
Together with the initial values~\eqref{Equ:Fm3Init}
the $c_i$ are uniquely determined with $c_1=\frac{32}{3}$, $c_2=8$, $c_3=\frac{32}{3}$, $c_4=0$, $c_5=0$ and we 
obtain the triple pole term
%----------------------------------------------------------------------------------------------------
\begin{equation}
F_{-3}(N)=\frac{8 (-1)^N}{3 (N+1) (N+2)}
+\frac{8 (2 N+3)}{3 (N+1)^2 (N+2)}.
\end{equation}
%----------------------------------------------------------------------------------------------------
In order to obtain the next coefficient of the Laurent series in $\ep$, we insert this information 
into~\eqref{Equ:EpRecAnsatz}, which yields
%----------------------------------------------------------------------------------------------------
\begin{align*}
 A_0(N,\ep)&\Big[F_{-2}(N)\ep^{-2}+F_{-1}(N)\ep^{-1}+F_{0}(N)\ep^{0}+\dots\Big]+\\
 A_1(N,\ep)&\Big[F_{-2}(N+1)\ep^{-2}+F_{-1}(N+1)\ep^{-1}+F_{0}(N+1)\ep^{0}+\dots\Big]\\
  +\dots+\\
 A_5(N,\ep)&\Big[F_{-2}(N+5)\ep^{-2}+F_{-1}(N+5)\ep^{-1}+F_{0}(N+5)\ep^{0}+\dots\Big]= r(N,\ep),
 \end{align*}
%----------------------------------------------------------------------------------------------------
with
%----------------------------------------------------------------------------------------------------
\begin{align*}
r(N,\ep)=&-A_0(N,\ep)F_{-3}(N)\ep^{-3}-
 A_1(N,\ep) F_{-3}(N+1)\ep^{-3}-A_2(N,\ep) F_{-3}(N+2)\ep^{-3}\\
 &-A_3(N,\ep) F_{-3}(N+3)\ep^{-3}-A_4(N,\ep) F_{-3}(N+4)\ep^{-3}-A_5(N,\ep) F_{-3}(N+5)\ep^{-3}\\
&=h_{-2}(N)\ep^{-2}+h_{-1}(N)\ep^{-1}+h_0(N)\ep^0+\dots,
\end{align*}
%----------------------------------------------------------------------------------------------------
where the $h_{i}(N)$ are explicitly given. E.g., we have that
%----------------------------------------------------------------------------------------------------
\begin{eqnarray}
h_{-2}(N) &=&-\frac{R_1}{(N+5)(N+6)(N+7)} 
+\frac{(-1)^N (N+2) (N+3)^2 (N+4) R_2}{3(N+5) (N+6) (N+7)},  
\end{eqnarray}
where
\begin{eqnarray}
R_1 &=& 8192 (N+2) (N+3)^2 (N+4)(3 N^6+76 N^5+649 N^4+2372 N^3+2844 
N^2, \nonumber\\ &&
-3324 N-7612) \\
R_2 &=& 16384\big(
        54 N^9+1854 N^8+27561 N^7+232537 N^6+1225070 N^5+4169707 N^4
\nonumber\\ &&
+9140480 N^3+12388516 N^2+9356363 N+2966634\big)~.
\end{eqnarray}
%----------------------------------------------------------------------------------------------------
Now we repeat the above procedure: by coefficient comparison we obtain the following constraint for $F_{-2}(N)$:
%----------------------------------------------------------------------------------------------------
\begin{align}
& \sum_{k=0}^5 A_k(N,0) F_{-2}(N+k) = h_{-2}(N).
\end{align}
%----------------------------------------------------------------------------------------------------
\texttt{Sigma} provides the solution set
%----------------------------------------------------------------------------------------------------
\begin{equation}
L_{-2}=\{c_1\,g_1+c_2\,g_2+c_3\,g_3+c_4\,g_4+c_5\,g_5+p|c_1,c_2,c_3,c_3,c_5\in\set R\},
\end{equation}
%----------------------------------------------------------------------------------------------------
with the particular solution
%----------------------------------------------------------------------------------------------------
$$p=-\frac{8 (-1)^N \big(
        3 N^2+8 N+6\big)}{3 (N+1)^3 (N+2)^2}
+\frac{4 (3 N+2)}{3 (N+1)^3 (N+2)^2}.$$
Again the solution set is completely determined and it follows that $F_{-2}(N)\in L_{-2}$. Thus with the initial values from above the $c_i$ are uniquely determined and we obtain the double pole term
%----------------------------------------------------------------------------------------------------
\begin{equation}
F_{-2}(N)=-\frac{4(-1)^N\big(
        3 N^3+18 N^2+31 N+18\big)}{3(N+1)^3 (N+2)^2} 
-\frac{4\big(
        6 N^3+32 N^2+51 N+26\big)}{3(N+1)^3 (N+2)^2}.
\end{equation}
%----------------------------------------------------------------------------------------------------
In a completely analogous way we calculate 
%----------------------------------------------------------------------------------------------------
\begin{align*}
 F_{-1}(N)&=(-1)^N \left(
        \frac{2\big(
                9 N^5+81 N^4+295 N^3+533 N^2+500 N+204\big)}{3(N+1)^4 (N+2)^3} 
        +\frac{\zeta_2}{(N+1) (N+2)}
\right)\\
&+\frac{2\big(
        18 N^5+150 N^4+490 N^3+755 N^2+536 N+132\big)}{3(N+1)^4 (N+2)^3}  
+\frac{(2 N+3) \zeta_2}{(N+1)^2 (N+2)}\\
&+\left(
        -\frac{4}{(N+1)^2 (N+2)}
        +\frac{4 (-1)^N}{(N+1) (N+2)}
\right) S_2\\
&+\left(
        \frac{4 (-1)^N}{3 (N+1) (N+2)}
        -\frac{4 (N+9)}{3 (N+1)^2 (N+2)}
\right) S_{-2}
\end{align*}
%----------------------------------------------------------------------------------------------------
and the coefficients $F_0(N),\dots,F_4(N)$.
The constant term requires besides harmonic sums the generalized harmonic sums
$S_ {1,2}\left({{\frac{1}{2},-1},N}\right)$ and $S_{1,2}\left({{\frac{1}{2},1},N}\right)$.
Finally, the $\ep^4$ term consists of 819 (generalized) harmonic sums (which are algebraically independent among each other, see Remark~\ref{Equ:AlgebraicIndependence} below) where one of the most complicated ones is
%----------------------------------------------------------------------------------------------------
\begin{equation}\label{Equ:SSumsForJ}
S_{1,2,1,1,1,1}\left({{\tfrac{1}{2},-1,1,1,1,1},N}\right).
\end{equation}
%----------------------------------------------------------------------------------------------------

\noindent The illustrated calculation steps can be summarized with the following theorem; for a rigorous proof and further 
algorithmic improvements see Ref.~\cite{Blumlein:2010zv}.

\medskip

%----------------------------------------------------------------------------------------------------
\noindent\textbf{Theorem REC.} Suppose we are given a linear recurrence 
\begin{multline}
A_0(N,\ep)J(N,\ep)+A_1(N,\ep)J(N+1,\ep)+\dots+A_d(N,\ep)J(N+d,\ep)\\
=h_{l}(N)\ep^l+h_{l+1}(N)\ep^{l+1}+\dots+h_{r}(N)\ep^r+\dots
\end{multline}
of order $d$ where the $A_i(N,\ep)$ are polynomials in $N$ and $\ep$ and where the inhomogeneous part can be expanded in $\ep$ up to order $r$.
Consider a function which has a Laurent series expansion
\begin{equation}
J(N,\ep)=F_{l}(N)\ep^l+F_{l+1}(N)\ep^{l+1}+\dots
\end{equation}
and which is a solution of the given recurrence for all $N\geq n_0$ for some $n_0\in\set N$. Then together with the
$d$ initial values $F_{j}(n_0),\dots,F_j(n_0+d-1)$
with $1\leq j\leq r$, all values $F_l(n),\dots,F_r(n)$ with $n\geq n_0$ can be computed provided that the values 
$h_i(n)$ for all $i$ with $l\leq i\leq r$ and all integers $n$ with $n\geq n_0$ can be computed. In addition, if the $h_l(N),\dots,h_r(N)$ are given explicitly in terms of indefinite nested product-sum expressions, there is an algorithm which decides constructively if the $F_{l}(N),\dots,F_r(N)$ can be given in terms of indefinite nested product-sum expressions.
%----------------------------------------------------------------------------------------------------

\medskip

Using the above technology we obtained the $\ep$-expansions for the integrals
%----------------------------------------------------------------------------------------------------
\begin{equation}\label{JEqu:5List}
J_5(1,1,1), \, J_5(1,1,2), \, J_5(2,1,1), \, J_5(2,1,2), \, J_5(2,2,1), \, J_5(3,1,1)
\end{equation}
%----------------------------------------------------------------------------------------------------
up to the needed orders in $\ep$ of $4,3,4,4$, respectively.
Given the recurrence and initial values, the above calculations for $J_5(1,1,1)$ can be executed within 5992 seconds. 
The other integrals can be handled with similar timings. In particular, the result is of similar type, i.e., up to a small variation it is built by the same set of generalized harmonic sums. Concerning these excellent timings the following remarks are in place.

%----------------------------------------------------------------------------------------------------
\begin{remark}\label{Equ:AlgebraicIndependence}
\rm
\texttt{Sigma} is capable to perform these calculations in which about 1000 sums are represented in a difference ring; 
based on the difference field/ring theory~\cite{Schneider:10c} it is guaranteed that the arising sums are algebraically 
independent among each other. But, using in addition the \texttt{HarmonicSums} package, big parts of these complicated 
difference ring calculations can be outsourced. In particular, representing these sums in an algebraically independent 
basis can be performed rather efficiently. In this regard, complete tables of algebraic relations of (generalized) 
harmonic sums are calculated up to certain weights using the underlying quasi-shuffle algebra. For further details and 
the derived tables we refer to~\cite{Blumlein:2009ta,Blumlein:2009fz,Ablinger:2013cf}. However, if 
the weights of the sums are too large or
if one tackles generalized harmonic sums and binomial sums involving several letters, the tables grow dramatically in size. 
To overcome this situation, we used special algorithms of \texttt{HarmonicSums} which produce the needed relations from scratch. 
For the treatment of  (generalized) harmonic sums we refer to~\cite{Ablinger:2013cf,HARMONICSUMS,Ablinger:PhDThesis} and for binomial 
sums we refer to~\cite{BinomRelations}. 
Summarizing, if only (generalized) harmonic sums, cyclotomic sums or (inverse) binomial sums are involved (and not different 
classes of indefinite nested product-sum expressions), we activate these \texttt{HarmonicSums} features to support \texttt{Sigma}. 
This will be of particular importance for further calculations presented in  Subsection~\ref{Sec:CoupledSolver} below.
\end{remark}
%----------------------------------------------------------------------------------------------------
%%%%%%%%%%%%%%%%%%%%%%%%%%%%%%%%%%%%%%%%%%%%%%%%%%%%%%%%%%%%%%%%%%%%%%%%%%%%%%%%%%%%%%%%%%%%%%%%%%%%%%%%%%%%%%%%%%%%%%%%%%%%%%%%%%%%%%%%%%%%%%

%%%%%%%%%%%%%%%%%%%%%%%%%%%%%%%%%%%%%%%%%%%%%%%%%%%%%%%%%%%%%%%%%%%%%%%
\subsection{Solving coupled differential equations or difference equations}
\label{sec:DiffEq}
%%%%%%%%%%%%%%%%%%%%%%%%%%%%%%%%%%%%%%%%%%%%%%%%%%%%%%%%%%%%%%%%%%%%%%%

\vspace*{1mm}\noindent
In the following we will prepare the stage to tackle diagram~12, the hardest diagram from Figures~~\ref{samplediagrams} and 
\ref{samplediagrams2}. Similar steps apply for the calculation of the simpler diagrams.
Using \texttt{Reduze\!~2} we can express it in the form~\eqref{Equ:DExp} in terms of $r=92$ master 
integrals. For 60 of these master integrals the necessary $\ep$-expansions can be calculated with the 
techniques presented in the previous subsections. In what follows we will denote them by $B_1(N),\dots,B_{60}(N)$ and define
the generating function
%------------------------------------------------------------------------------------------------------------------------------
\begin{equation}
\label{BixDef}
\hat{B}_i(x)=\sum_{N=0}^{\infty}B_i(N)\,x^N.
\end{equation}
%------------------------------------------------------------------------------------------------------------------------------
However, for the 32 remaining master integrals the previous tools fail to calculate the $\ep$-expansion to the required order, either due to 
cumbersome 
sum/integral representations or due to limited time and space resources. These master integrals are denoted by $\IIntegral_1(N),\dots,\IIntegral_{32}(N)$ and
%------------------------------------------------------------------------------------------------------------------------------
\begin{equation}\label{IixDef}
\hat{\IIntegral}_i(x)=\sum_{N=0}^{\infty}\IIntegral_i(N)\,x^N.
\end{equation}
Summarizing, diagram~12 is decomposed in the form
\begin{equation}\label{Equ:D12MI}
\hat{D}_{12}(x)=c_1(x,\ep)\,\hat{B}_1(x)+\dots+c_{60}(x,\ep)\,\hat{B}_{60}(x)+c_{61}(x,\ep)\,\hat{\IIntegral}_1(x)+\dots+c_{92}(x,\ep)\,\hat{\IIntegral}_{32}(x).
\end{equation}
%------------------------------------------------------------------------------------------------------------------------------

In order to tackle the integrals $I_i$, we will combine our symbolic summation tools with the differential equation approach, which 
is a very efficient method for calculating Feynman integrals. The idea behind this method is to 
take derivatives of the integrals with respect to the variable $x$ that we introduced before in order to be able to use Laporta's reduction algorithm.
Doing this to a master integral usually leads to a linear combination of scalar integrals that can then be substituted by their corresponding expressions in terms
of master integrals. In this way, we obtain a hierarchically organized system of coupled differential equations for the master 
integrals. We remark that the hierarchical structure is induced by the sector decomposition of the differential equation method. In the following we aim to extract the desired $\ep$-expansions from these coupled systems by exploiting our available difference field/ring tools and uncoupling algorithms. 

First, we will elaborate this new Ansatz by a concrete example. Next, we will work out the complete algorithm with its input and output specification. In particular, we will exemplify the most complicated system that occurred so far within our calculations. Finally, we present the general tactic how the hierarchically organized coupled system in terms of the 
unknown master integrals $\IIntegral_i$ can be solved completely automatically with these tools.
We remark, that our methods {\it do not} request particular forms of the coefficient matrix. 

%----------------------------------------------------------------------------------------------------
\subsubsection{Solving coupled systems --  a first example}\label{Sec:CoupledSolver}

\vspace*{1mm}\noindent
We start with the master integrals $\hat{\IIntegral}_1(x),\hat{\IIntegral}_2(x),\hat{\IIntegral}_3(x)$ out of the 32 
unknown master integrals. These integrals are given explicitely by
%----------------------------------------------------------------------------------------------------
\begin{equation}
\hat{I}_1(x) = J_6(1,1;x), \quad
\hat{I}_2(x) = J_6(2,1;x), \quad
\hat{I}_3(x) = J_6(1,2;x),
\end{equation}
%----------------------------------------------------------------------------------------------------
where
%----------------------------------------------------------------------------------------------------
\begin{equation}
J_6(\nu_2,\nu_4;x) =
\int \frac{d^Dk_1}{(2 \pi)^D}  \frac{d^Dk_2}{(2 \pi)^D} \frac{d^Dk_3}{(2 \pi)^D} \,\,
\frac{1}{P_2^{\nu_2} P_4^{\nu_4} P_5 P_7 P_8 P_{10}} \ ,
\end{equation}
%----------------------------------------------------------------------------------------------------
and the propagators are those of family {\tt B3a} shown in Table~\ref{families1}. We will not show the explicit forms of the 
remaining $\hat{I}_i(x)$ integrals, since these forms are not used once we generate the differential equation systems.

Analyzing the coefficients $c_i(\ep,x)$ in~\eqref{Equ:D12MI} it turns out that we have to expand them up to the orders $2,3,3$, respectively, in order to have the full information to expand $D_{12}$ up to the constant term.
Using \texttt{Reduze 2} we obtain a coupled system 
%----------------------------------------------------------------------------------------------------
\begin{equation*}
D_x\left(\begin{matrix}
	  \hat{\IIntegral_1}(x)\\ \hat{\IIntegral}_2(x)\\ \hat{\IIntegral}_3(x)
	  \end{matrix}\right)
      = {\ds A
        \left(\begin{matrix}
	  \hat{\IIntegral}_1(x)\\
	  \hat{\IIntegral}_2(x)\\
	  \hat{\IIntegral}_3(x)
        \end{matrix}\right)}
        +\left(\begin{matrix} \hat{R}_1(x)\\\hat{R}_2(x)\\-\hat{R}_2(x)\end{matrix}\right)
\end{equation*}
%----------------------------------------------------------------------------------------------------
in terms of the unknown integrals $\hat{\IIntegral}_i(x)$ with the matrix
$$A=\left(
        \begin{matrix}
         \frac{\ds 1
        +\ep
        -x
        }{\ds (x-1) x}  & \frac{\ds -2}{\ds (x-1) x} & 0\\
         \frac{\ds -\ep (3 \ep+2) (x-2)}{\ds4 (x-1) x}  & \frac{\ds -2
        -5 \ep
        +x
        +3 \ep x
        }{\ds 2(x-1) x}& \frac{\ds (-2 \ep
        -x
        +\ep x
        )}{\ds 2(x-1) x}  \\
         \frac{\ds \ep (3 \ep+2)}{\ds 4 (x-1)}  & \frac{\ds 2
        +\ep
        -3 x
        -3 \ep x
        }{\ds 2(x-1) x} & -\frac{\ds \ep+1}{\ds 2 (x-1)}
        \end{matrix}\right)$$
%----------------------------------------------------------------------------------------------------
and
\begin{align*}
\hat{R}_1(x)=&\frac{\hat{B}_4}{(x-1) x},\\
\hat{R}_2(x)=&\frac{(\ep+2)^3}{16 (\ep+1) (x-1) x}\hat{B}_1(x)-\frac{(\ep+2) (3 \ep+4) \big(
        19 \ep^2+36 \ep+16\big) }{16\ep (5 \ep+6) (x-1) x} \hat{B}_2(x)\\
      &-\frac{(\ep+1)^2 (3 \ep+4)^2 }{2\ep (5 \ep+6) x} \hat{B}_3(x)-\frac{ -24
        -50 \ep
        -25 \ep^2
        +8 x
        +14 \ep x
        +6 \ep^2 x}{4(5 \ep+6) (x-1) x} \hat{B}_4(x)~.
\end{align*}
%----------------------------------------------------------------------------------------------------
Here $D_x$ stands for the derivative operator w.r.t.\ $x$.
Notice that the expressions $\hat{R}_i(x)$ depend on the integrals
$\hat{B}_1(x),\hat{B}_2(x),\hat{B}_3(x),\hat{B}_4(x)$ which can be handled with 
our symbolic summation and integration methods introduced earlier. These integrals $\hat{B}_i(x)$ will be also 
called ``base-case integrals'', since their solutions are available using other methods having been described before.
%----------------------------------------------------------------------------------------------------
\subsubsection*{Step 1: Transformation to a first-order coupled recurrence system}
%----------------------------------------------------------------------------------------------------

\vspace*{1mm}
\noindent
In a first step, we transform this system of differential equations to a system of recurrences. For instance, take the first equation of the system:
%----------------------------------------------------------------------------------------------------
\begin{equation}
(x-1)\,x\,D_x \hat{\IIntegral}_1(x)=-(-1
        -\ep
        +x)\hat{\IIntegral}_1(x) -2\hat{\IIntegral}_2(x)+\hat{B}_4(x).
\label{eq:DXEQ}
\end{equation}
%----------------------------------------------------------------------------------------------------
Inserting~\eqref{BixDef} and~\eqref{IixDef} into this equation and carrying out the derivative on the formal power series yields
%----------------------------------------------------------------------------------------------------
\begin{align*}
x(x-1)\sum_{N=0}^{\infty}N\,\IIntegral_1(N)x^{N-1}=-(-1-\ep+x)\sum_{N=0}^{\infty}\IIntegral_1(N)x^N
-2\sum_{N=0}^{\infty}\IIntegral_2(N)x^N+\sum_{N=0}^{\infty}B_4(N)x^N.
\end{align*}
%----------------------------------------------------------------------------------------------------
Finally, comparing the $N$th coefficient delivers the constraint
%----------------------------------------------------------------------------------------------------
\begin{equation}
-(N+\ep+1)\IIntegral_1(N)+N \IIntegral_1(N-1)+2\IIntegral_2(N)=r_1(N)
\label{eq:DNEQ}
\end{equation}
with
\begin{equation}\label{Equ:r1}
r_1(N)=B_1(N).
\end{equation}
%----------------------------------------------------------------------------------------------------
In a completely analogous way, we transform the remaining two differential equations to the recurrences
%----------------------------------------------------------------------------------------------------
\begin{alignat*}{2}
&\ep (3 \ep+2) \IIntegral_1(N-1)
        -2 \ep (3 \ep+2) \IIntegral_1(N)
        -2 (3 \ep+1) \IIntegral_2(N-1)
        +2 (5 \ep+2) \IIntegral_2(N)\\
        & -2 (1
        +\ep
        -2 N
        ) \IIntegral_3(N-1)
        +4 (\ep
        -N
        ) \IIntegral_3(N) &= r_2(N)\\
&-\ep (3 \ep+2) \IIntegral_1(N-1)
        +2 (1
        +3 \ep
        +2 N
        ) \IIntegral_2(N-1)
-2(2
        +\ep
        +2 N
        ) \IIntegral_2(N)
\\
        &
        +2 (\ep+1) \IIntegral_3(N-1) &= r_3(N) \, ,
\end{alignat*}
%----------------------------------------------------------------------------------------------------
with
%----------------------------------------------------------------------------------------------------
\begin{align}
r_2(N)=&
\frac{(\ep+2)^3 }{4 (\ep+1)}B_1(N)
-\frac{(\ep+2) (3 \ep+4) \big(
        19 \ep^2+36 \ep+16\big)}{4\ep (5 \ep+6)}B_2(N)\label{Equ:r2}\\
&-\frac{2 (\ep+1)^2 (3 \ep+4)^2}{\ep (5 \ep+6)} B_3(N-1)
+\frac{2 (\ep+1)^2 (3 \ep+4)^2}{\ep (5 \ep+6)} B_3(N)\nonumber\\
&-\frac{2 (\ep+1) (3 \ep+4) B_4(N-1)}{5 \ep+6}
-(-5 \ep-4) B_4(N),\nonumber\\
r_3(N)=&-r_2(N) \, .
\label{Equ:r3}
\end{align}
By construction, the right hand sides $r_i(N)$ are given as a linear combination of the $B_1(N),\dots,B_4(N)$ and their shifted versions. 
In general, any system of linear differential equations can be transformed in this way to such a system of linear recurrences. For 
further details on these holonomic closure properties we refer to~\cite{Kauers:13} and references therein.
Finally, we transform the three coupled recurrences to the form 
%----------------------------------------------------------------------------------------------------
{\small
\begin{equation}
\label{Equ:PreUncoupled}
\left(\begin{matrix}\IIntegral_1(N+1)\\ \IIntegral_2(N+1)\\ \IIntegral_3(N+1)
        \end{matrix}\right) = 
A\,\left(\begin{matrix}\IIntegral_1(N)\\ \IIntegral_2(N)\\ \IIntegral_3(N)
        \end{matrix}\right)+\left(\begin{matrix}v_1(N)\\v_2(N)\\v_3(N) \end{matrix}\right) 
\end{equation}
}
%----------------------------------------------------------------------------------------------------
with the matrix
$$A= \left(
        \begin{matrix}\frac{\ds (1-\ep+N) (4
        +3 \ep
        +2 N
        )}{\ds (2
        +\ep
        +N
        ) (4
        +\ep
        +2 N
        )}  & \frac{\ds 2 (3
        +3 \ep
        +2 N
        )}{\ds (2
        +\ep
        +N
        ) (4
        +\ep
        +2 N
        )} & \frac{\ds 2 (\ep+1)}{\ds (2
        +\ep
        +N
        ) (4
        +\ep
        +2 N
        )}  \\
         -\frac{\ds \ep (3 \ep+2)}{\ds 2 (4
        +\ep
        +2 N
        )} & \frac{\ds 3
        +3 \ep
        +2 N
        }{\ds 4
        +\ep
        +2 N
        } & \frac{\ds \ep+1}{\ds 4
        +\ep
        +2 N
        } \\
a(N)&b(N)&c(N) \\
        \end{matrix}\right)$$
%----------------------------------------------------------------------------------------------------
where
\begin{align*}          
	a(N)=&-\frac{\ep (3 \ep+2) \big(
                -2
                -2 \ep
                +\ep^2
                -3 N
                -2 \ep N
                -N^2
        \big)}{2(-1
        +\ep
        -N
        ) (2
        +\ep
        +N
        ) (4
        +\ep
        +2 N
        )},\\
        b(N)=& \frac{
                -2
                -3 \ep
                -\ep^2
                +3 \ep^3
                -3 N
                -5 \ep N
                -2 \ep^2 N
                -N^2
                -2 \ep N^2
        }{(-1
        +\ep
        -N
        ) (2
        +\ep
        +N
        ) (4
        +\ep
        +2 N
        )},\\
        c(N)=&\frac{-6
                -5 \ep
                -\ep^2
                +\ep^3
                -13 N
                -7 \ep N
                -2 \ep^2 N
                -9 N^2
                -2 \ep N^2
                -2 N^3}{(-1
        +\ep
        -N
        ) (2
        +\ep
        +N
        ) (4
        +\ep
        +2 N
        )}
\end{align*}
%----------------------------------------------------------------------------------------------------
        and
%----------------------------------------------------------------------------------------------------
\begin{eqnarray*}
v_1(n)&=& -\frac{\ds r_3(N+1)}{\ds (2
+\ep
+N
) (4
+\ep
+2 N
)} 
-\frac{\ds r_1(N+1)}{\ds 2
+\ep
+N
}\\
v_2(N)&=&-\frac{\ds r_3(N+1)}{\ds 2 (\ep
+2 N
+4
)}\\
v_3(N)&=&\frac{\ds -\ep (3 \ep+2) r_1(N+1)}{\ds 2(-1
+\ep
-N
) (2
+\ep
+N
)}
\\
&&+\frac{\ds r_2(N+1)}{\ds 4 (-1
+\ep
-N
)}+\frac{\ds 
        (-4
        -8 \ep
        +\ep^2
        -2 N
        -5 \ep N)r_3(N+1)}{\ds 4(1
-\ep
+N
) (2
+\ep
+N
) (4
+\ep
+2 N
)}.
\nonumber
\end{eqnarray*}
%----------------------------------------------------------------------------------------------------

\normalsize
\noindent This can be achieved, e.g., by Gaussian elimination. More precisely, one writes the system of equations in matrix form where the 
rows 
encode the equations and where the first three columns contain the contributions of $\IIntegral_1(N+1)$, 
$\IIntegral_2(N+1)$ and $\IIntegral_3(N+1)$, respectively, and where the fourth column contains the rest of the system, 
i.e., the contributions of $\IIntegral_1(N)$, $\IIntegral_2(N)$ and $\IIntegral_3(N)$ and all the $B$-integrals. 
Finally, one transforms this $3\times 4$ matrix by row operations over the field $\set Q(\ep,N)$ to a matrix where the first part 
of the matrix is the $3\times3$ identity matrix. This yields exactly the desired shape given in~\eqref{Equ:PreUncoupled}.

As well-known \cite{NOERLUND}, linear difference and differential equations are related by a Mellin(Laplace) transformation.
Therefore, difference equations of higher order correspond to differential equations of higher order. As has been outlined before in 
Eqs.~(\ref{eq:DXEQ},\ref{eq:DNEQ}), equivalently using formal power series representations also transforms the differential system in the
difference one. 
%----------------------------------------------------------------------------------------------------
\begin{remark}
\rm
An extra difficulty is that the derived recurrence system is usually not of first-order but of higher-order type. 
In this case, the system can always be transformed to a first-order one. E.g., if the sub-expression 
$$a_3(N) \IIntegral_1(N+3)+a_2(N) \IIntegral_1(N+2)+a_1(N) \IIntegral_1(N+1)+a_0(N) \IIntegral_1(N)+...=0$$ 
occurs, one introduces 
the auxiliary functions $h_1(N)=\IIntegral_1(N+1)$ and $h_2(N)=\IIntegral_1(N+2)$ 
and obtains 
the modified system 
%-----------------------------------------------------------------------------
\begin{align}
a_3(N) h_2(N+1)+a_2(N) h_2(N)+a_1(N) h_1(N)+a_0(N) \IIntegral_1(N)+ ... =&0,\\
h_1(N+1)-h_2(N) =& 0,\\
\IIntegral_1(N+1)-h_1(N) =&0.
\end{align}
%-----------------------------------------------------------------------------
Performing this transformation for all the higher order shifts will eventually yield a coupled first-order system.
\end{remark}
%----------------------------------------------------------------------------------------------------

%----------------------------------------------------------------------------------------------------
\subsubsection*{Step 2: Uncoupling to a scalar recurrence}
%----------------------------------------------------------------------------------------------------

\vspace*{1mm}
\noindent
Finally, given the system in the form~\eqref{Equ:PreUncoupled}, we can execute any 
uncoupling algorithm which is on the market. For our calculations we chose  Z\"urcher's algorithm~\cite{Zuercher:94} implemented in Gerhold's \texttt{OreSys} 
package~\cite{ORESYS}.
Executing this code (with some slight improvements) we get the following scalar linear recurrence
%----------------------------------------------------------------------------------------------------
\begin{equation}\label{Equ:ScalarRec}
\begin{split}
&-2 (N+1) (N+2) (2
+\ep
+N
) \IIntegral_1(N)\\
&\quad-(N+2) \big(
        -32
        -7 \ep
        +2 \ep^2
        -28 N
        -5 \ep N
        -6 N^2
\big)\IIntegral_1(N+1)\\
&\quad\quad-\big(120
        +3 \ep
        -14 \ep^2
        -\ep^3
        +136 N
        +13 \ep N
        -4 \ep^2 N
        +50 N^2
        +4 \ep N^2
        +6 N^3
\big) \IIntegral_1(N+2)
\\ 
&\quad\quad\quad+(2
-\ep
+N
) (4
+\ep
+N
) (8
+\ep
+2 N
) \IIntegral_1(N+3)
\\ &=
\big(
        -24
        -\ep
        +\ep^2
        -20 N
        -\ep N
        -4 N^2
\big) r_1(N+2)
+2 (N+2) (2
+\ep
+N
) r_1(N+1)
\\ &
\quad\quad-(-2
+\ep
-N
) (8
+\ep
+2 N
) r_1(N+3)
+\frac{1}{2} (\ep+1) r_2(N+2)
\\ 
&\quad\quad\quad+\frac{1}{2} (-3
+\ep
-2 N
) r_3(N+2)
+(2
-\ep
+N
) r_3(N+3)
\end{split}
\end{equation}
%----------------------------------------------------------------------------------------------------
in $\IIntegral_1(N)$, and we can express the remaining functions $\IIntegral_2(N)$ and 
$\IIntegral_3(N)$ in terms of $\IIntegral_1(N)$ and its shifted versions:
%----------------------------------------------------------------------------------------------------
\begin{equation}\label{Equ:I2Ev}
\begin{split}
\IIntegral_2&(N)=
\frac{
        -8
        +2 \ep
        +4 \ep^2
        -12 N
        -\ep N
        -4 N^2}{4 (1
+\ep
+N
)}\IIntegral_1(N)\\
&+\frac{
        28
        -6 \ep
        -10 \ep^2
        -\ep^3
        +54 N
        +5 \ep N
        -4 \ep^2 N
        +32 N^2
        +4 \ep N^2
        +6 N^3}{4 (N+1) (1
+\ep
+N
)}\IIntegral_1(N+1)\\
&+\frac{-1
+\ep
-N
) (3
+\ep
+N
) (6
+\ep
+2 N
}{4 (N+1) (1
+\ep
+N
)}\IIntegral_1(N+2)+\frac{
        8
        -\ep^2
        +12 N
        +\ep N
        +4 N^2
}{4 (N+1) (1
+\ep
+N
)}r_1(N+1)\\
&+\frac{(-1
+\ep
-N
) (6
+\ep
+2 N
)}{4 (N+1) (1
+\ep
+N
)}r_1(N+2)
-\frac{\ep+1}{8 (N+1) (1
+\ep
+N
)}r_2(N+1)\\
&+\frac{1
-\ep
+2 N
}{8 (N+1) (1
+\ep
+N
)}r_3(N+1)
-\frac{1
-\ep
+N
}{4 (N+1) (1
+\ep
+N
)}r_3(N+2),
\end{split}
\end{equation}
%----------------------------------------------------------------------------------------------------
%----------------------------------------------------------------------------------------------------
\begin{eqnarray}\label{Equ:I3Ev}
\IIntegral_3(N)&=&-\frac{
        -16
        -12 \ep
        +10 \ep^2
        +6 \ep^3
        -32 N
        -23 \ep N
        +\ep^2 N
        -20 N^2
        -8 \ep N^2
        -4 N^3
}{4 (\ep+1) (1
+\ep
+N
)}\IIntegral_1(N)\nonumber\\
&&+\frac{R_3(N)
}{4 (\ep+1) (N+1) (1
+\ep
+N
)}\IIntegral_1(N+1)\nonumber\\
&&-\frac{(-1
+\ep
-N
) (3
+\ep
+N
) (6
+\ep
+2 N
) (3
+3 \ep
+2 N
)}{4 (\ep+1) (N+1) (1
+\ep
+N
)}\IIntegral_1(N+2)\nonumber\\
&&+\frac{
        -16
        -14 \ep
        +5 \ep^2
        +3 \ep^3
        -32 N
        -23 \ep N
        +\ep^2 N
        -20 N^2
        -8 \ep N^2
        -4 N^3
}{4 (\ep+1) (N+1) (1
+\ep
+N
)}r_1(N+1)\nonumber\\
&&-\frac{(-1
+\ep
-N
) (6
+\ep
+2 N
) (3
+3 \ep
+2 N
)}{4 (\ep+1) (N+1) (1
+\ep
+N
)}r_1(N+2)
\nonumber\\ &&
+\frac{3
+3 \ep
+2 N
}{8 (N+1) (1
+\ep
+N
)}r_2(N+1)
+\frac{3 \ep+1}{8 (N+1) (1
+\ep
+N
)}r_3(N+1)
\nonumber\\ &&
-\frac{(-1
+\ep
-N
) (3
+3 \ep
+2 N
)}{4 (\ep+1) (N+1) (1
+\ep
+N
)}r_3(N+2),
\end{eqnarray}
with
\begin{align*}
R_3(N)=& 
        -68
        -38 \ep
        +62 \ep
        ^2
        +35 \ep^3
        +3 \ep^4
        -170 N
        -103 \ep N
        +39 \ep^2 N
        +16 \ep^3 N  
        -152 N^2\\ 
        &
        -74 \ep N^2
        +4 \ep^2 N^2
        -58 N^3 
        -16 \ep N^3
        -8 N^4.
\end{align*}
%----------------------------------------------------------------------------------------------------

%----------------------------------------------------------------------------------------------------
\begin{remark}
\rm
In general, we are given a first-order coupled system of linear difference equations (or a higher-order coupled system that we transform to a first-order coupled system) in the unknown functions $I_1(N),\dots,I_n(N)$. Then it can be uncoupled into scalar recurrences in some of the unknown functions, say $I_1(N),\dots,I_u(N)$  with $u\leq n$, i.e., each scalar recurrence depends exactly on one of the $I_i(N)$ with $1\leq i\leq u$. Finally, the remaining unknown functions $I_{u+1}(N),\dots,I_n(N)$ (which do not occur in the scalar recurrences) can be expressed as a linear combination of the $I_1(N),\dots,I_u(N)$ and their shifted versions. As a consequence, the system is completely determined by the first initial values of $I_1(N),\dots,I_u(N)$, i.e., the values of the other unknown functions are just consequences of them. In addition, an explicit representation of the $I_1(N),\dots,I_u(N)$ produces immediately closed form solutions of the remaining unknown functions, provided that the 
base-case integrals can be calculated. 
\end{remark}
%----------------------------------------------------------------------------------------------------

%----------------------------------------------------------------------------------------------------
\subsubsection*{Step 3: Analyzing the system}
%----------------------------------------------------------------------------------------------------

\vspace*{1mm}
\noindent
Given this uncoupled form, we will proceed as follows: Use the algorithm presented in Subsection~\ref{Sec:RecSolver} to determine the 
Laurent series representation of $\IIntegral_1(N)$. Then plug the derived $\ep$-expansion of $\IIntegral_1(N)$ and the 
$\ep$-expansions of the base-case integrals into~\eqref{Equ:I2Ev} and~\eqref{Equ:I3Ev} to derive the $\ep$-expansions of $\IIntegral_2(N)$ and $\IIntegral_3(N)$, respectively.

In order to carry out these steps, the following ingredients are needed.

\begin{enumerate}
\item For the $\ep$-expansions of $\IIntegral_2(N)$ and $\IIntegral_3(N)$ up to order $\ep^{3}$ we have to guarantee that the objects 
occurring in the right hand sides of~\eqref{Equ:I2Ev} and~\eqref{Equ:I3Ev} are expanded sufficiently high. 
Hence we extract the coefficients of $I_1(N)$ and their shifted versions in~\eqref{Equ:I2Ev} and~\eqref{Equ:I3Ev}, factorize these coefficients and read off the factor $\ep^o$ with $o\in\set Z$. In our situation the maximal value among all such factors is $o=0$. Hence an $\ep$-expansion of $\IIntegral_1(N)$ up to $\ep^3$ is sufficient to obtain the expansions of~\eqref{Equ:I2Ev} and~\eqref{Equ:I3Ev} up to $\ep^3$.\\
In addition, we plug~\eqref{Equ:r1}, \eqref{Equ:r2} and~\eqref{Equ:r3} into \eqref{Equ:I2Ev} and~\eqref{Equ:I3Ev}. Then applying the same procedure for each $B_i(N)$ with $1\leq i\leq 4$ tells us how high the occurring $B_i(N)$ have to be expanded. In this particular instance, we need for $B_1(N), B_2(N), B_3(N), B_4(N)$ the $\ep$-expansions up to the orders $3,4,4,3$, respectively. 
\item Finally, we want to activate our $\ep$-expansion solver of Subsection~\ref{Sec:RecSolver} summarized in Theorem~REC. 
\begin{enumerate}
\item The recurrence~\eqref{Equ:ScalarRec} has order $3$. Thus we need the $\ep$-expansion up to $\ep^3$ for three initial values, say, $\IIntegral_1(1)$, $\IIntegral_1(2)$, and $\IIntegral_1(3)$.
\item The right hand side of~\eqref{Equ:ScalarRec} needs to be expanded up to $\ep^3$. Thus we proceed as in item 1: We plug~\eqref{Equ:r1}, \eqref{Equ:r2} and~\eqref{Equ:r3} into the right hand side of ~\eqref{Equ:ScalarRec}, extract the coefficients of the $B_i(N)$, derive the necessary $\ep$-expansion orders of the $B_i(N)$ and update the global information about how high the $\ep$-expansions of the $B_i(N)$ must be carried out. In our case, the orders remain unchanged.
\end{enumerate}
\end{enumerate}

\noindent Summarizing, we have to expand $B_1(N),B_2(N),B_3(N),B_4(N)$ up to the orders $3,4,4,3$, respectively. Note that this is 
exactly the order that we also need in order to assemble diagram~12. In addition, we have to expand 
$\IIntegral_1,\IIntegral_2,\IIntegral_3$, all of them to order $\ep^3$. Note that in this case we have to expand $\IIntegral_1(N)$ one term higher than originally required to determine the expansion of diagram~12. In order to get these expansions, 
we need in total $3$ initial values of $\IIntegral_1(N)$ up to order $\ep^3$. 

%----------------------------------------------------------------------------------------------------
\begin{remark}\label{Equ:AnalyzeSystem1}
\rm
Similarly, we can also uncouple the system such that the scalar recurrence is given in $\IIntegral_2(N)$ or $\IIntegral_3(N)$. But in these cases the situation gets worse: In both cases, we must expand $B_1(N),B_2(N),B_3(N),B_4(N)$ up to the orders $4,5,5,4$, the expansions of $\IIntegral_1,\IIntegral_2,\IIntegral_3$ have to be carried out up to the orders $3,4,4$, respectively, and we would need 3 initial values of $\IIntegral_2(N)$ (or $\IIntegral_3(N)$) up to the order 4. 
Summarizing, choosing $\IIntegral_1(N)$ for the scalar recurrence is the optimal choice. We remark that the calculations of the Steps 1--3 in each case take in total 2.5 seconds. Hence finding the optimal choice (in checking all cases and choosing $\IIntegral_1(N)$), takes about 8 seconds.
\end{remark}
%----------------------------------------------------------------------------------------------------

%----------------------------------------------------------------------------------------------------
\subsubsection*{Step 4: The calculation of the base-case integrals and initial values}
%----------------------------------------------------------------------------------------------------

\vspace*{1mm}
\noindent
First, we calculate the $\ep$-expansions of the base-case integrals $B_i(N)$ up to the determined order or we check 
if the existing database contains these expansions already. Then we insert these $\ep$-expansions into the right hand side 
of~\eqref{Equ:ScalarRec} and calculate its $\ep$-expansion up to $\ep^3$. Here we print only the terms up to
$\ep^{-1}$:
%----------------------------------------------------------------------------------------------------
\begin{eqnarray}\label{Equ:ExpandedRHS}
&&\frac{1}{\ep^3}\frac{-4 (N+2)}{3 (N+3)}
 +\frac{1}{\ep^2}\Big[-\frac{2 (2 N+7) S_1}{3 (N+3)}
-\frac{2\big(
        4 N^4+35 N^3+101 N^2+105 N+25\big)}{3(N+1) (N+2) (N+3)^2}\Bigr]\nonumber\\
&&+\frac{1}{\ep}\Big[\frac{
        8 N^4+73 N^3+229 N^2+273 N+83}{3(N+1) (N+2) (N+3)^2}S_1
+\frac{(-4 N-17) S_1^2}{6 (N+3)}\nonumber\\
&&+\frac{-15 N^7-216 N^6-1275 N^5-3962 N^4-6937 N^3-6823 N^2-3573 N-851}{3(N+1)^2 (N+2)^2 (N+3)^3}\nonumber \\
&&+\frac{(-4 N-17) S_2(N)}{6 (N+3)}+\frac{(-N-2) \zeta_2}{2 (N+3)}\Big]+O(\ep^0).
\end{eqnarray}
%----------------------------------------------------------------------------------------------------
In addition, we calculate the necessary initial values up to the required order $3$:
%----------------------------------------------------------------------------------------------------
\begin{eqnarray}\label{Equ:InitialI}
\IIntegral_1(1) &=& \frac{5}{\ep^3} -\frac{163}{12 \ep^2} + \frac{1}{\ep}\left(\frac{1223}{48}
        +\frac{15 \zeta_2}{8}\right)
        -\frac{7975}{192}
        -\frac{163 \zeta_2}{32}
        +\frac{21 \zeta_3}{8}
         + \ep \Bigl(
                \frac{16285}{256}
                +\frac{1223 \zeta_2}{128}
\nonumber\\ &&
                +\frac{1437 \zeta_2^2}{640}
                -\frac{467 \zeta_3}{96}
        \Bigr)
        + \ep^2 \Bigl(
                -\frac{96989}{1024}
                +\Bigl(
                        -\frac{7975}{512}
                        +\frac{63 \zeta_3}{64}
                \Bigr) \zeta_2
                -\frac{10393 \zeta_2^2}{2560}
                +\frac{1975 \zeta_3}{384}
\nonumber\\ &&
                +\frac{291 \zeta_5}{32}
        \Bigr)
        + \ep^3 \Bigl(
                \frac{571037}{4096}
                +\Bigl(
                        \frac{48855}{2048}
                        -\frac{467 \zeta_3}{256}
                \Bigr) \zeta_2
                +\frac{8257 \zeta_2^2}{2048}
                +\frac{165587 \zeta_2^3}{35840}
                -\frac{3095 \zeta_3}{1536}
\nonumber\\ &&
                -\frac{91 \zeta_3^2}{128}
                -\frac{9283 \zeta_5}{640}
        \Bigr) 
\\
%----------------------------
  \IIntegral_1(2) &=&  \frac{130}{27 \ep^3} -\frac{695}{54 \ep^2} + \frac{1}{\ep} \left(\frac{46379}{1944}
        +\frac{65 \zeta_2}{36} \right)
        -\frac{99761}{2592}
        -\frac{695 \zeta_2}{144}
        +\frac{241 \zeta_3}{108}
        + \ep \Bigl(\frac{16419323}{279936}
\nonumber\\ &&
                +\frac{46379 \zeta_2}{5184}
                +\frac{5459 \zeta_2^2}
                {2880}
                -
                \frac{1591 \zeta_3}{432}
        \Bigr) 
        + \ep^2 \Bigl(
                -\frac{97523803}{1119744}
                +\Bigl(
                        -\frac{99761}{6912}
                        +\frac{241 \zeta_3}{288}
                \Bigr) \zeta_2
                -\frac{34709 \zeta_2^2}{11520}
\nonumber\\ &&
                +\frac{48427 \zeta_3}{15552}
                +\frac{1069 \zeta_5}{144}
        \Bigr)+ \ep^3 \Bigr(
                \frac{5167209971}{40310784}
                +\Bigl(
                        \frac{16419323}{746496}
                        -\frac{1591 \zeta_3}{1152}
                \Bigr) \zeta_2
                +\frac{930353 \zeta_2^2}{414720}
\nonumber\\ &&
                +\frac{1827767 \zeta_2^3}{483840}
                +\frac{14735 \zeta_3}{20736}
                -\frac{991 \zeta_3^2}{1728}
                -\frac{5515 \zeta_5}{576}
        \Bigr) 
\\
%--------------------------
   \IIntegral_1(3) &=&  \frac{169}{36 \ep^3} -\frac{395}{32 \ep^2} + \frac{1}{\ep}\left(\frac{470071}{20736}
        +\frac{169 \zeta_2}{96}\right)
        -\frac{6021247}{165888}
        -\frac{1185 \zeta_2}{256}
        +\frac{569 \zeta_3}{288}+ \ep \Bigl(
                \frac{658074919}{11943936}
\nonumber\\ &&
                +\frac{470071 \zeta_2}{55296}
                +\frac{12811 \zeta_2^2}{7680}
                -\frac{2257 \zeta_3}{768}
        \Bigr)
        +\ep^2 \Bigl(
                -\frac{7813200839}{95551488}
                +\Bigl(
                        -\frac{6021247}{442368}
                        +\frac{569 \zeta_3}{768}
                \Bigr) \zeta_2
\nonumber\\ &&
                -\frac{48243 \zeta_2^2}{20480}
                +\frac{320327 \zeta_3}
                {165888}
                +\frac{12169 \zeta_5}{1920}
        \Bigr)+ \ep^3 \Bigl(
                \frac{826541143999}{6879707136}
                +\Bigl(
                        \frac{658074919}{31850496}
                        -\frac{2257 \zeta_3}{2048}
                \Bigr) \zeta_2
\nonumber\\ &&
                +\frac{5337493 \zeta_2^2}{4423680}
                +\frac{4167439 \zeta_2^3}{1290240}
                +\frac{2946673 \zeta_3}{1327104}
                -\frac{2231 \zeta_3^2}{4608}
                -\frac{6669 \zeta_5}{1024}
        \Bigr).
\end{eqnarray}        
%----------------------------------------------------------------------------------------------------
To get these initial values, we exploited the $\alpha$-parameterization of the integral; for further details on this method we refer 
to~\cite{Ablinger:2014uka}. Note that in other situations we also used our summation tools, provided a reasonable sum representation 
has been derived, see for instance Section~\ref{Sec:RecSolver}.

%----------------------------------------------------------------------------------------------------
\subsubsection*{Step 5: The solution of the scalar recurrence}
%----------------------------------------------------------------------------------------------------

\vspace*{1mm}
\noindent
Finally, we can activate the algorithm in Subsection~\ref{Sec:RecSolver}: Given 
the recurrence~\eqref{Equ:ScalarRec} together with the right hand side~\eqref{Equ:ExpandedRHS} expanded up to $\ep^3$ and given the initial values~\eqref{Equ:InitialI} expanded up to $\ep^3$, we succeed in calculating the $\ep$-expansion of $\IIntegral_1(N)$ up to order 3. Here we print the solution up to $\ep^{-1}$: 
%----------------------------------------------------------------------------------------------------
\begin{align*}
 \IIntegral_1(N)= \frac{1}{\ep^3}&\Big[\frac{4 \big(
                3 N^2+6 N+4\big)}{3 (N+1)^2}
        +\frac{4 S_1}{3 (N+1)}
        \Big]\\
        +\frac{1}{\ep^2}&\Big[-\frac{2(20 N^3+58 N^2+57 N+22)}{3(N+1)^3}
        +\frac{2 (N+2) (2 N-1) S_1}{3 (N+1)^2}
        -\frac{S_1^2}{N+1}
        -\frac{S_2}{N+1}
        \Big]\\
       +\frac1{\ep}&\Big[\frac{89 N^4+344 N^3+495 N^2+317 N+84}{3(N+1)^4}
        +\left(
                \frac{3 N^2+6 N+4}{2 (N+1)^2}
                +\frac{S_1}{2 (N+1)}
        \right) \zeta_2\\
        &\quad+\left(
                \frac{-12 N^3-39 N^2-39 N-5}{3(N+1)^3} 
                +\frac{S_2}{6 (N+1)}
        \right) S_1
        +\frac{(3 N+4) S_1^2}{6 (N+1)^2}
       \\
        &\quad+\frac{S_1^3}{18 (N+1)}+\frac{(3 N+4) S_2}{6 (N+1)^2}
        +\frac{S_3}{9 (N+1)}
        +\frac{2 S_{2,1}}{N+1}
        \Big] +O(\ep^0)
\end{align*}
%----------------------------------------------------------------------------------------------------
For the coefficients up to $\ep^3$ in addition to $\zeta_3, \, \zeta_5$ the harmonic sums
%----------------------------------------------------------------------------------------------------
\begin{equation}\label{Equ:SSumsListForI1}
\begin{split}
 &S_1,S_2,S_3,S_4,S_5,S_6,S_7,S_{2,1},S_{3,1},S_{3,2}, S_{4,1},S_{4,2},\\
 &S_{4,3},S_{5,1},S_{5,2},S_{6,1},S_{2,1,1},
 S_{2,2,1},S_{3,1,1},S_{3,1,2},S_{3,2,1},S_{3,2,2},\\
 &S_{3,3,1},S_{4,1,1}, S_{4,1,2},S_{4,2,1},S_{5,1,1},S_{2,1,1,1},S_{2,2,1,1},S_{2,2,2,1},\\
 &S_{3,1,1,1},S_{3,1,1,2},S_{3,1,2,1},S_{3,2,1,1},S_{4,1,1,1},S_{2,1,1,1,1}, S_{2,1,2,1,1},\\
 &S_{2,2,1,1,1},S_{3,1,1,1,1},S_{2,1,1,1,1,1}
\end{split}
\end{equation}
%----------------------------------------------------------------------------------------------------
contribute.

%----------------------------------------------------------------------------------------------------
\subsubsection*{Step 6: Calculate the remaining integrals}
%----------------------------------------------------------------------------------------------------

\vspace*{1mm}
\noindent
Finally, we insert the expansions of the $B_i(N)$ and the solution of $\IIntegral_1(N)$ into the right hand sides of~\eqref{Equ:I2Ev} and~\eqref{Equ:I3Ev} and calculate the $\ep$-expansions up to $\ep^3$. 
The expansions up to the constant term are
%----------------------------------------------------------------------------------------------------
\begin{align*}
   \IIntegral_2(N)&=\frac{4}{3 \ep^3} -\frac{2}{\ep^2} 
   + \frac1{\ep}
     \left(
        \frac{5 N+7}{3 (N+1)}
        +\frac{2}{3} S_1
        -\frac{1}{3} S_1^2
        -\frac{1}{3} S_2
        +\frac{\zeta_2}{2}
        \right)\\ 
        +&\frac{N^2-12 N-15}{6 (N+1)^2}
        +\left(
                \frac{-7 N-8}{3 (N+1)}
                +\frac{(1-N) S_2}{2 (N+1)}
        \right) S_1
        +S_1^2
        +\frac{(1-N) S_1({N}
        )^3}{6 (N+1)}\\
       &+S_2
        +\frac{(1-N) S_3}{3 (N+1)}
        -\frac{3 \zeta_2}{4}
        +\frac{(5 N+17) \zeta_3}{6 (N+1)}+O(\ep),\\
   \IIntegral_3(N)&=     
         \frac{8}{3 \ep^3} +\frac1{\ep^2}\left(-\frac{4 \big(
                4 N^2+7 N+2\big)}{3 (N+1)^2}
        +\frac{4 (N+2) S_1}{3 (N+1)}
       \right)\\
       +\frac{1}{\ep} & \left(\frac{2(12 N^3+32 N^2+25 N+2)}{3(N+1)^3} 
        -\frac{2 \big(
                4 N^2+11 N+10\big) S_1}{3 (N+1)^2} \right. \\
        &\left. +\frac{(N-2) S_1^2}{3 (N+1)}
        +\frac{(N-2) S_2}{3 (N+1)}
        +\zeta_2\right)\\
        +&\frac{-32 N^4-116 N^3-148 N^2-69 N+2}{3 (N+1)^4} 
        +\left(
                \frac{-4 N^2-7 N-2}{2 (N+1)^2}
                +\frac{(N+2) S_1}{2 (N+1)}
        \right) \zeta_2\\
        &+\left(
                \frac{12 N^3+44 N^2+59 N+34}{3(N+1)^3} 
                +\frac{(N-4) S_2}{6 (N+1)}
        \right) S_1\\
        &+\frac{\big(
                -4 N^2-7 N-2\big) S_1^2}{6 (N+1)^2}
        +\frac{(N-4) S_1^3}{18 (N+1)}
        +\frac{\big(
                -4 N^2-7 N-2\big) S_2}{6 (N+1)^2}\\
       &+\frac{(N-4) S_3}{9 (N+1)}
        +\frac{2 (N+2) S_{2,1}}{N+1}
        +\frac{(-N-7) \zeta_3}{3 (N+1)}+O(\ep).
\end{align*}
%----------------------------------------------------------------------------------------------------
The expressions up to $\ep^3$ can be written in terms of the harmonic sums given in~\eqref{Equ:SSumsListForI1}. 
The total computation time to obtain these $\ep$-expansions was 229 seconds.

%----------------------------------------------------------------------------------------------------
\subsubsection{The full algorithm to solve coupled systems}
%----------------------------------------------------------------------------------------------------

\vspace*{1mm}
\noindent
In the following we will present the full algorithm in order to solve coupled difference equations in terms of indefinite nested product-sum expressions. Afterwards we elaborate the case of coupled differential equations.

Let $\set K$ be a field containing the rational numbers as sub-field (e.g., $\set K=\set Q$) and let $\set K(N,\ep)$ be a rational function field (i.e., the elements of $\set K(N,\ep)$ are built by numerator and denominator polynomials in $N$ and $\ep$). 

\subsubsection*{Solving coupled difference equations}

Consider the coupled system of difference equations
\begin{equation}\label{Equ:GenericCRS}
A_0 \left(\begin{matrix}I_1(N)\\ \vdots\\ I_n(N)\end{matrix}\right)
+A_1 \left(\begin{matrix}I_1(N+1)\\ \vdots\\ I_n(N+1)\end{matrix}\right)
\dots+A_l \left(\begin{matrix}I_1(N+l)\\ \vdots\\ I_n(N+l)\end{matrix}\right)=
\left(\begin{matrix}r_1(N)\\ \vdots\\ r_n(N)\end{matrix}\right)
\end{equation}
where the $n\times n$ matrices $A_i$ with entries from
$\set K(N,\ep)$ are given and where the functions $r_i(N)$ on the right hand side can be represented in terms of an $\ep$-expansion.

Then also the unknown functions $I_i(N)$ in~\eqref{Equ:GenericCRS} can be expanded in $\ep$ and we are 
interested in the calculation of the first coefficients in terms of indefinite nested product-sum expressions. In 
our approach this is possible under the following two algorithmic assumptions.\footnote{In the following the symbol
$\bot$ means `no solution'.}

\begin{description}
 \item[(A1)] There is the following decision procedure: given $\rho\in\set Z$, output the $\ep$-expansion of the 
$r_i(N)$ up to $\ep^{\rho}$ where the coefficients are given in terms of indefinite nested product-sums; or return $\bot$ if the coefficients cannot be expressed in this form.
\item[(A2)] There is the following algorithm: given $\lambda\in\set N$ and $\mu\in\set Z$, compute the $\ep$-expansion of the $I_i(\lambda)$ up to $\ep^{\mu}$. 
\end{description}
More precisely, we obtain the following result.

\medskip

\noindent\textbf{Theorem 1.} Suppose we are given the coupled system~\eqref{Equ:GenericCRS} with unknown functions $I_1(N),\dots,I_n(N)$ as stated above and suppose that we are given algorithms with the specifications given in (A1) and (A2). Then there exists the following decision procedure: Given $\rho_1,\dots,\rho_n\in\set Z$, output the coefficients of the $\ep$-expansion of the $I_i(N)$ up to $\ep^{\rho_i}$ for all $i$ ($1\leq i\leq n$) in terms of indefinite nested product-sum expression; or return $\bot$ if these coefficients cannot be expressed in terms of indefinite nested product-sum expressions.

\medskip

The proof follows immediately by the following procedure.

\medskip

\noindent\textbf{Algorithm} \texttt{SolveCoupledRECs}:
\begin{enumerate}
 \item Uncouple the system~\eqref{Equ:GenericCRS}. Say, we obtain $u$ scalar recurrences in terms of the $I_1(N),\dots,I_u(N)$ with orders $\nu_1,\dots,\nu_u$, respectively. Note: The right hand sides of the scalar recurrences can be written in terms of a linear combination of the $r_i(N)$ and their shifted versions. 
 In addition, we obtain relations that determine the remaining unknown integrals $I_{u+1}(N),\dots,I_n(N)$ in terms of a linear combination of the $I_1(N),\dots,I_u(N)$ and their shifted versions (see Step~2 of Subsection~\ref{Sec:CoupledSolver}). In particular, one can determine the needed orders of the $\ep$-expansions of the $I_1(N),\dots,I_u(N)$ in order to determine the coefficients of the $I_{u+1}(N),\dots,I_n(N)$ up to the orders $\rho_{u+1},\dots,\rho_{n}$, respectively (see Step~3 of Subsection~\ref{Sec:CoupledSolver}). 
 \item Activate Theorem REC in order to compute the required expansions for the functions $I_1(N),\dots,I_u(N)$. Here we need the following preparation steps (see Step~4 of Subsection~\ref{Sec:CoupledSolver}):
 \begin{enumerate}
 \item  Using (A1) expand the right hand sides of the scalar recurrences sufficiently high in terms of the indefinite nested product-sum expressions. If this fails, the $I_i(N)$ cannot be represented in terms of nested product-sum expressions. Hence return $\bot$.
 \item Using (A2) calculate for the $I_1(N),\dots,I_u(N)$ the first $\nu_1,\dots,\nu_u$ initial values, respectively, in terms of an $\ep$-expansion sufficiently high.
\end{enumerate}
Using this information compute the coefficients of the $\ep$-expansions of the $I_1(N),\dots,I_u(N)$ in terms of indefinite nested product-sums with the underlying algorithm of Theorem REC (see Step~5 of Subsection~\ref{Sec:CoupledSolver}). If the decision procedure fails, return $\bot$.
\item Otherwise, take the found $\ep$-expansions in terms of nested product-sums and derive the $\ep$-expansions of the $I_{u+1}(N),\dots,I_n(N)$. Since they are given by shifts of the coefficients of $I_1(N),\dots,I_u(N)$, the coefficients up to $\ep^{\mu_i}$ can be represented also by indefinite nested product-sum expressions (see Step~6 of Subsection~\ref{Sec:CoupledSolver}). Finally, return the derived coefficients of the desired expansions in terms of indefinite nested product-sum expressions.
\end{enumerate}

\subsubsection*{Solving coupled difference equations}

Consider the coupled system of differential equations
\begin{equation}\label{Equ:GenericCDS}
A_0 \left(\begin{matrix}\hat{I}_1(x)\\ \vdots\\ \hat{I}_n(x)\end{matrix}\right)
+A_1 D_x\left(\begin{matrix}\hat{I}_1(x)\\ \vdots\\ \hat{I}_n(x)\end{matrix}\right)
\dots+A_l D_x^l\left(\begin{matrix}\hat{I}_1(x)\\ \vdots\\ \hat{I}_n(x)\end{matrix}\right)=
\left(\begin{matrix}\hat{r}_1(x)\\ \vdots\\ \hat{r}_n(x)\end{matrix}\right)
\end{equation}
where the $n\times n$ matrices $A_i$ with entries from
$\set K(x,\ep)$ are given and where the functions on the right hand side can be represented as formal power series
$$\hat{r}_i(x)=\sum_{N=0}^{\infty}r_i(N)x^N$$ 
where the coefficients $r_i(N)$ can be represented in terms of an $\ep$-expansion.
Then also the functions $\hat{I}_i(x)$ in~\eqref{Equ:GenericCDS} have power series representations in $x$: 
$$\hat{I}_i(x)=\sum_{N=0}^{\infty}I_i(N)x^N$$
where the coefficients $I_i(N)$ can be expanded in $\ep$. As above, we are interested in the calculation of the first coefficients of the $\ep$-expansion of $I_i(N)$ in terms of indefinite nested product-sum expressions.  More precisely, we obtain the following result.

\medskip

\noindent\textbf{Theorem 2.} Suppose we are given the coupled system~\eqref{Equ:GenericCDS} with unknown functions $\hat{I}_1(x),\dots,\hat{I}_n(x)$ with the coefficients $I_1(N),\dots,I_n(N)$ as stated above, and suppose that we are given algorithms with the specifications given in (A1) and (A2). Then there exists the following decision procedure: Given $\rho_1,\dots,\rho_n\in\set Z$, output the coefficients of the $\ep$-expansion of the $I_i(N)$ up to $\ep^{\rho_i}$ in terms of indefinite nested product-sum expression; or return $\bot$ if these coefficients cannot be expressed in terms of indefinite nested product-sum expressions.

\medskip

The proof is given by the following Algorithm \texttt{SolveCoupledDEs}.

\medskip

\noindent\textbf{Algorithm} \texttt{SolveCoupledDEs}. 
\begin{enumerate}
\item Transform the coupled differential system~\eqref{Equ:GenericCDS} to a coupled difference system by holonomic closure properties (see Step~1 of Subsection~\ref{Sec:CoupledSolver}). This will lead to a coupled difference system of the form~\eqref{Equ:GenericCRS}.

\item Apply Algorithm \texttt{SolveCoupledRECs} (see Steps~2--6 of Subsection~\ref{Sec:CoupledSolver}) to the coupled difference system and return the corresponding output.

\end{enumerate}

%----------------------------------------------------------------------------------------------------
\subsubsection*{The most complicated coupled system}
%----------------------------------------------------------------------------------------------------

\vspace*{1mm}
\noindent
We conclude this subsection with the most complicated system that we have considered so far. It is a coupled system in terms of the unknown integrals $\IIntegral_4,\IIntegral_5,\IIntegral_6,\IIntegral_7$,
which have to be expanded up to the orders $4,3,2,3$, respectively.
Carrying out the Steps 1--4 of Subsection~\ref{Sec:CoupledSolver}
takes in total 28187~seconds. At this point we obtain a scalar linear recurrence for
%----------------------------------------------------------------------------------------------------
$\IIntegral_4(N)$ of order 5:
\begin{multline}\label{Equ:I4Rec}
a_0(N,\ep)\IIntegral_4(N)+a_1(N,\ep)\IIntegral_4(N+1)+\dots+a_5(N,\ep)\IIntegral_4(N+5)\\
=h_{-1}(N)\ep^{-1}+h_{0}(N)\ep^{0}+h_{1}(N)\ep^{1}+h_{2}(N)\ep^{2}+h_{3}(N)\ep^{3}+h_{4}(N)\ep^{4}+\dots \,\, .
\end{multline}
%----------------------------------------------------------------------------------------------------
Here the $a_i(N,\ep)$ are polynomials in $\ep$ and $N$, and the right hand side has been expanded up to order 
$\ep^4$ where the coefficients $h_i(N)$ are given 
in terms of 726 $S$-sums up to weight {\sf w $\leq$ 7}. Note that the coupled system depends on the base-case 
integrals~\eqref{JEqu:5List} 
handled in Section~\ref{sec:Zeilberger}, and the derived generalized harmonic sums within this calculation occur now in the $h_i(N)$. The recurrence~\eqref{Equ:I4Rec} requires 5 MB of memory. Together with the initial values of $\IIntegral_4(N)$, we are now ready to determine the coefficients of the $\ep$-expansion
%----------------------------------------------------------------------------------------------------
\begin{equation}\label{Equ:I4Expansion}
\IIntegral_4(N)=F_{-1}(N)\ep^{-1}+F_{0}(N)\ep^{0}+F_{1}(N)\ep^{1}+F_{2}(N)\ep^{2}+F_{3}(N)\ep^{3}+F_{4}(N)\ep^{4}+\dots~.
\end{equation}
%----------------------------------------------------------------------------------------------------
According to our algorithm from Section~\ref{Sec:RecSolver}, we obtain the following constraint
%----------------------------------------------------------------------------------------------------
$$a_0(N,0)F_{-1}(N)+a_1(N,0)F_{-1}(N+1)+\dots+a_5(N,0)F_{-1}(N+5)=h_{-1}(N)$$
%----------------------------------------------------------------------------------------------------
for the single-pole term $F_{-1}(N)$. Here the given ingredients are
%----------------------------------------------------------------------------------------------------
\begin{align*}
a_0(N,0)&=27 (N+1)^2 (N+3)^2 (N+4) (N+5),\\
a_1(N,0)&=-54 (N+2)^2 (N+3) (N+4) (N+5) (7 N+25),\\
a_2(N,0)&=27 (N+3)^3 (N+4) (N+5) (73 N+305),\\
a_3(N,0)&=-54 (N+3) (N+4)^3 (N+5) (86 N+417),\\
a_4(N,0)&=432 (N+3) (N+4) (N+5)^3 (11 N+62),\\
a_5(N,0)&=-864 (N+3) (N+4) (N+5) (N+6)^2 (2 N+13)
\end{align*}
%----------------------------------------------------------------------------------------------------
and
%----------------------------------------------------------------------------------------------------
\begin{align*}
h_{-1}(N)=&4 (N+6) (2 N+13) \\
&\times\Bigg[(-1)^N \Big(
                \frac{27p_1(N)}{2 (N+1)^2 (N+2)^2 (N+3) (N+4) (N+5) (N+6)^3 (2 N+13)}\\
               &\quad+\frac{1350 (N+3) (N+4) (N+5) (5 N+26) S_2}{(N+6) (2 N+13)}\\
&               \quad+\frac{2700 (N+3) (N+4) (N+5) (5 N+26) S_{-2}}{(N+6) (2 N+13)}        \Big)\\
&\quad-\frac{54p_2(N)}{(N+1)^2 (N+2)^2 (N+3) (N+4) (N+5) (N+6)^3 (2 N+13)}
\Bigg]
\end{align*}
%----------------------------------------------------------------------------------------------------
with
%----------------------------------------------------------------------------------------------------
\begin{align*}
p_1(N)=&2700 N^{11}+110540 N^{10}+2023863 N^9+21861837 N^8+154727026 N^7\\
&+753018778 N^6+2570551555 N^5+6154149425 N^4+10128965992 N^3\\
&+10925337340 N^2+6965691744 N+1997724960\\
p_2(N)=&125 N^{11}+5775 N^{10}+120248 N^9+1489952 N^8+12208573 N^7+69468823 N^6,\\
&+280119686 N^5+800347450 N^4+1586935256 N^3+2076154736 N^2\\
&+1607189760 N+554226912.
\end{align*}
%----------------------------------------------------------------------------------------------------
Using \texttt{Sigma} we obtain the solution set
%----------------------------------------------------------------------------------------------------
\begin{equation*}
L_{-1}=\{c_1\,g_1+c_2\,g_2+c_3\,g_3+c_4\,g_4+c_5\,g_5+p|c_1,c_2,c_3,c_3,c_5\in\set R\},
\end{equation*}
%----------------------------------------------------------------------------------------------------
with
%----------------------------------------------------------------------------------------------------
\begin{align*}         
 g_1=&\frac{1}{(N+1)^2},\quad g_2=\frac{2^{2-2 N}}{3 (N+1)^2},\quad g_3=\frac{2^{2-2 N} (3 N+1)}{9 (N+1)^2}, \\
 g_4=& -\frac{1}{(N+1) (2 N+1) (2 N+3) \binom{2 N}{N}} \frac{64 (N+2)}{9}, \\
 g_5=& \frac{8}{9 (N+1)^2 (2 N+3)}
        +\frac{8(N+2)}{9(N+1) (2 N+1) (2 N+3) \binom{2 N}{N}} 
        \sum_{i=1}^N \binom{2 i}{i},\\
 p=& -\frac{4 \big(
                19 N^2+50 N+37\big)}{3 (N+1)^4 (2 N+3)}
        -\frac{76(N+2)}{3(N+1) (2 N+1) (2 N+3) \binom{2 N}{N}}  
        \sum_{i=1}^N \binom{2 i}{i}\\
        &
        +(-1)^N \big(
                \frac{6}{(N+1)^4}
                +\frac{2 S_2}{(N+1)^2}
                +\frac{4 S_{-2}}{(N+1)^2}
        \big)-\frac{4 S_2}{(N+1)^2}
        -\frac{4 S_{-2}}{(N+1)^2}.
\end{align*}
%----------------------------------------------------------------------------------------------------
Since the solution space is completely determined, it follows that $F_{-1}(N)\in L_{-1}$. Together with the initial values
%----------------------------------------------------------------------------------------------------
$$F_{-1}(1)=-\frac{3}{8},F_{-1}(2)=-\frac{49}{162},F_{-1}(3)=-\frac{155}{1152},F_{-1}(4)=-\frac{5269}{45000},F_{-1}(5)=-\frac{1477}{21600}$$
%----------------------------------------------------------------------------------------------------
we end up at the representation
%----------------------------------------------------------------------------------------------------
\begin{equation}\label{Equ:I4epM1}
F_{-1}(N)=\big(
        (-1)^N-2
\big) \frac{2S_2}{(N+1)^2}
+\big(
        (-1)^N-1
\big) \frac{4S_{-2}}{(N+1)^2}
-\frac{8}{(N+1)^4}
+\frac{6 (-1)^N}{(N+1)^4}.
\end{equation}
%----------------------------------------------------------------------------------------------------
Repeating the algorithmic steps in Section~\ref{Sec:RecSolver}, we set up the linear recurrence 
%----------------------------------------------------------------------------------------------------
$$a_0(N,0)F_{0}(N)+a_1(N,0)F_{0}(N+1)+\dots+a_1(N,0)F_{0}(N+5)=h'(N)$$
%----------------------------------------------------------------------------------------------------
for $F_0(N)$ where the expression $h'(N)$ is given in terms of 
%----------------------------------------------------------------------------------------------------
\begin{equation}\label{Equ:CoupledHardep0}
\begin{split}
&S_{-3},S_{-2},S_1,S_2,S_3,S_{-2,1},S_{2,1},\\
&S_1\left({{\tfrac{1}{2}}}\right),S_3\left({{-\tfrac{1}{2}}}\right),S_3\left({{\tfrac{1}{2}}}\right),
S_{2,1}\left({{-1,\tfrac{1}{2}}}\right),S_{2,1}\left({{1,\tfrac{1}{2}}}\right).
\end{split}
\end{equation}
%----------------------------------------------------------------------------------------------------
Finally, we solve this recurrence with \texttt{Sigma}, and using the initial values of $F_0(N)$ we obtain the representation of 
$F_0(N)$ in terms of the (generalized) harmonic sums~\eqref{Equ:CoupledHardep0} and 19 (inverse) binomial sums, where one of them 
is 
%----------------------------------------------------------------------------------------------------
$$\sum_{i_1=1}^N (-2)^{i_1} i_1^2\binom{2 i_1}{i_1}
        \sum_{i_2=1}^{i_1} \frac{\displaystyle 2^{-i_2} 
        \sum_{i_3=1}^{i_2} \frac{(-1)^{i_3}}{i_3^2}}{i_2}.$$
%----------------------------------------------------------------------------------------------------
The calculation of the full expansion~\eqref{Equ:I4Expansion}         
up to $\ep^4$ took more than 300000 seconds. The final result has a size of 20MB  and is written in terms of 928 
(generalized) 
harmonic sums up to weight {\sf w=8} and 2598 binomial sums up to nesting depth 7. Due to the strong engine of 
\texttt{HarmonicSums}, we 
obtain a representation where the arising sums form an algebraic independent basis; for further details we refer to Remark~\ref{Equ:AlgebraicIndependence}.
Inside the binomial sums at most two binomials letters arise. More precisely, there are
517 binomial sums involving two binomial coefficients. 38 of these sums have nesting depth 6. A typical example is
%----------------------------------------------------------------------------------------------------
%$\sum_{i_1=1}^N \frac{
%\sum_{i_2=1}^{i_1} \frac{
%\sum_{i_3=1}^{i_2} \frac{
%\sum_{i_4=1}^{i_3} (-2)^{i_4} \binom{2 i_4}{i_4} \big(
%        \sum_{i_5=1}^{i_4} \frac{2^{-i_5} 
%        \sum_{i_6=1}^{i_5} \frac{(-1)^{i_6}}{i_6^2}}{i_5}\right) i_4^2}{\binom{2 i_3}{i_3} i_3^2}}{i_2}}{i_1}$
%----------------------------------------------------------------------------------------------------
$$\displaystyle\sum_{i_1=1}^N \frac{1}{i_1}
\sum_{i_2=1}^{i_1} \frac{1}{i_2}
\sum_{i_3=1}^{i_2} \frac1{\binom{2 i_3}{i_3} i_3^2}
\sum_{i_4=1}^{i_3} (-2)^{i_4} \binom{2 i_4}{i_4}i_4^2
        \sum_{i_5=1}^{i_4} \frac{2^{-i_5}}{i_5} 
        \sum_{i_6=1}^{i_5} \frac{(-1)^{i_6}}{i_6^2}.$$
%----------------------------------------------------------------------------------------------------
In the 63 binomial sums with nesting depth 7 only one binomial coefficient is involved, with a typical example for these sums given 
by
%----------------------------------------------------------------------------------------------------
%$\sum_{i_1=1}^N \frac{
%\sum_{i_2=1}^{i_1} \frac{
%\sum_{i_3=1}^{i_2} \frac{
%\sum_{i_4=1}^{i_3} \frac{
%\sum_{i_5=1}^{i_4} (-2)^{i_5} \binom{2 i_5}{i_5} \big(
%        \sum_{i_6=1}^{i_5} \frac{2^{-i_6} 
%        \sum_{i_7=1}^{i_6} \frac{1}{i_7^2}}{i_6}\right) i_5^2}{i_4}}{i_3}}{i_2}}{i_1}$
%----------------------------------------------------------------------------------------------------
$$\sum_{i_1=1}^N \frac{1}{i_1}
\sum_{i_2=1}^{i_1} \frac{1}{i_2}
\sum_{i_3=1}^{i_2} \frac{1}{i_3}
\sum_{i_4=1}^{i_3} \frac{1}{i_4}
\sum_{i_5=1}^{i_4} (-2)^{i_5} \binom{2 i_5}{i_5}i_5^2
        \sum_{i_6=1}^{i_5} \frac{2^{-i_6}}{i_6} 
        \sum_{i_7=1}^{i_6} \frac{1}{i_7^2}.$$
%----------------------------------------------------------------------------------------------------
Summarizing, we calculated the expansion of $\IIntegral_4(N)$ up to $\ep^4$ in terms of nested binomial sums which 
completes Step~5 of Subsection~\ref{Sec:CoupledSolver}. Following Step~6 of Subsection~\ref{Sec:CoupledSolver} we 
turn to the integrals $\IIntegral_5(N),\IIntegral_6(N), \IIntegral_7(N)$. They are expressed in terms of 
$\IIntegral_4(N)$ in its shifted versions and various base-case integrals (again in their shifted versions). Using 
the expansion of $\IIntegral_4(N)$ and the expansions of the base-case integrals, which we expanded sufficiently 
high in $\ep$, we obtain the $\ep$-expansions of $\IIntegral_5(N), \IIntegral_6(N), \IIntegral_7(N)$ to the desired orders $3,2,3$, respectively.

%----------------------------------------------------------------------------------------------------
\subsubsection{Solving hierarchically defined coupled systems}
%----------------------------------------------------------------------------------------------------

\vspace*{1mm}
\noindent
So far we have explained how one can solve the integrals $\IIntegral_1,\IIntegral_2,\IIntegral_3$ and 
$\IIntegral_4,\IIntegral_5,\IIntegral_6,\IIntegral_7$ of the 32 master integrals of diagram~12. In addition, $\IIntegral_8$ and 
$\IIntegral_9$ are already given in uncoupled form and can be solved directly. However, the coupled system 
$\IIntegral_{10},\IIntegral_{11}$ also depends on the integrals $\IIntegral_1,\IIntegral_2,\IIntegral_3$. By this intrinsic 
structure we are forced to calculate the first three integrals first before we can turn to the 10th and 11th integral. In 
general, the coupled systems of the $\IIntegral_1,\dots,\IIntegral_{32}$ 
are hierarchically structured as can be seen in Table~\ref{Table:D12Systems}. Here the second column contains the unknown functions of the system and the fourth column indicates that the corresponding system depends on functions of other systems that have to be treated first. 

%----------------------------------------------------------------------------------------------------
\begin{table}[ht]
\begin{center}
    \begin{tabular}{|c||c|c|c|c|c|}
        \hline
        \textbf{system}& unknown integrals & up to $\ep^i$ & $\begin{array}{l} \text{depends on}\\ \text{lower systems}\end{array}$& $\begin{array}{l} \text{scalar rec.}\\ \text{in $\IIntegral_i$ (order)}\end{array}$ &  used time\\
        \hline\hline
         1& $\IIntegral_1,\IIntegral_2,\IIntegral_3$ & $3,3,3$ & -- & $
                \begin{array}{cc}
                 \IIntegral_1 & 3 \\
                \end{array}$ &  229s \\
         \hline
         2& $\begin{array}[t]{c}\IIntegral_4,\IIntegral_5,
           \IIntegral_6,\IIntegral_7\end{array}$
	    & $4,3,2,3$ & -- &
	   $\begin{array}{cc}
                 \IIntegral_4 & 5 \\
                \end{array}$ & 125 h \\
         \hline
         3&$\IIntegral_8$ & 1 &  -- & 
		$\begin{array}{cc}
                 \IIntegral_8 & 1 \\
                \end{array}$ &12s \\
         \hline
         4&$\IIntegral_9$ & 0 &  -- &
                $\begin{array}{cc}
                 \IIntegral_9 & 1 \\
                \end{array}$ &8s \\
        \hline
        5& $\IIntegral_{10},\IIntegral_{11}$ & 2,1 & 1& 
	      $\begin{array}[t]{cc}
                 \IIntegral_{10} & 3 \\
                 \IIntegral_{11} & 1 \\
                \end{array}$ & 115s \\
        \hline
        6& $\IIntegral_{12},\IIntegral_{13},\IIntegral_{14}$ & 3,1,2 & -- &
	      $\begin{array}{cc}
                 \IIntegral_{13} & 1 \\
                \end{array}$ & 68s \\
        \hline
        7& $\IIntegral_{15},\IIntegral_{16},\IIntegral_{17}$ & 2,1,1 & -- & 
	      $\begin{array}{cc}
                 \IIntegral_{15} & 5 \\
                \end{array}$ & 530s \\
        \hline
        8& $\IIntegral_{18}$ & 3 & 2& 
	      $\begin{array}{cc}
                 \IIntegral_{18} & 2 \\
                \end{array}$ & 19.6h \\
        \hline
        9& $\IIntegral_{19}$ & 3 & 2& 
	      $\begin{array}{cc}
                 \IIntegral_{19} & 1 \\
                \end{array}$ & 13.3h \\
        \hline
        10& $\IIntegral_{20},\IIntegral_{21}$ & 1,-1 & 1,2& 
	    $\begin{array}{cc}
                 \IIntegral_{20} & 5 \\
                \end{array}$ & 3754s \\
        \hline
        11& $\IIntegral_{22},\IIntegral_{23}$ & 1,0 & 2,6& 
	    $\begin{array}{cc}
                 \IIntegral_{22} & 5 \\
                \end{array}$ & 22.5h \\
        \hline
        12& $\IIntegral_{24}$ & 1 & -- & 
	    $\begin{array}{cc}
                 \IIntegral_{24} & 1 \\
                \end{array}$& 17.5s \\
        \hline
        13& $\IIntegral_{25}$ & 0 & 1,5& 
	    $\begin{array}{cc}
                 \IIntegral_{25} & 2 \\
                \end{array}$ & 6s \\
        \hline
        14& $\IIntegral_{26}$ & 1 & 1& 
	      $\begin{array}{cc}
                 \IIntegral_{26} & 3 \\
                \end{array}$ & 169s \\
        \hline
        15& $\IIntegral_{27},\IIntegral_{28}$ & 1,0 & 1,5,6,7& 
	    $\begin{array}[t]{cc}
                 \IIntegral_{27} & 5 \\
                 \IIntegral_{28} & 1 \\
                \end{array}$ & 1852s \\
        \hline
        16& $\IIntegral_{29}$ & 1 & 1,2,6,10,11& 
	    $\begin{array}{cc}
                 \IIntegral_{29} & 2 \\
                \end{array}$ & 1708s \\
        \hline
        17& $\IIntegral_{30}$ & 0 & 1,5,6,7,15 &
	    $\begin{array}{cc}
                 \IIntegral_{30} & 1 \\
                \end{array}$ &41s \\
        \hline
        18& $\IIntegral_{31}$ & 1 & $\begin{array}[t]{l}1,2,5,6,7,\\10,11,15\end{array}$ &
	    $\begin{array}{cc}
                 \IIntegral_{31} & 2 \\
                \end{array}$ &2816s \\
        \hline
        19& $\IIntegral_{32}$ & 1 & $\begin{array}[t]{l}1,5,6,\\7,14,15\end{array}$& 
	    $\begin{array}{cc}
                 \IIntegral_{32} & 3 \\
                \end{array}$ & 953s \\
        \hline
        \end{tabular}\\
 \end{center}
\caption{\sf \small The hierarchically structured system of coupled differential equations for diagram 12.}
\label{Table:D12Systems}
\end{table}
%----------------------------------------------------------------------------------------------------

In order to attack, e.g., system 19, one needs first the $\ep$-expansions of the integrals of the systems 1,5,6,7,14,15. In 
general, we have to solve the systems 
with a bottom-up strategy, namely, we have to process them, e.g., in the order $1\to2\to3\to\dots\to19$.
However, in order to obtain the $\ep$-expansion of $\IIntegral_{19}$, the orders of the $\ep$-expansions of the integrals in the systems 1,5,6,7,14,15 might change.
Hence, we first have to analyze the hierarchical system with a top-down strategy. More precisely, we analyze the systems in the order 
$19\to18\to\dots\to1$ and update step by step the necessary expansion orders of the $\IIntegral_i$ and $B_i$ accordingly.

In the following we will give some more details on our strategy executing it on the system of diagram~12. The general method can be 
easily derived from this concrete presentation.

%----------------------------------------------------------------------------------------------------
\subsubsection*{Step A: Analysis of the diagram}
%----------------------------------------------------------------------------------------------------

\vspace*{1mm}
\noindent
First we analyze all master integrals in~\eqref{Equ:D12MI} and determine how high the $\ep$-expansions are needed (for the integrals 
$\IIntegral_i$ and the base-case integrals $B_i$) in order to extract the constant term. E.g., for the integrals 
$\IIntegral_1(N),\IIntegral_2(N),\IIntegral_3(N)$ we need expansions up to the orders $2,3,3$ and for the integrals 
$B_1(N),B_2(N),B_3(N),B_4(N)$ up to the orders $3,4,4,3$.

%----------------------------------------------------------------------------------------------------
\subsubsection*{Step B: Analysis of the hierarchical system}
%----------------------------------------------------------------------------------------------------

\vspace*{1mm}
\noindent
Now we analyze the systems top-down as follows.

%----------------------------------------------------------------------------------------------------
\begin{enumerate}
\item Set $S:=19$.

\item If $S=0$, {\tt Stop}.\\ 
Otherwise let $\IIntegral_l,\IIntegral_{l+1},\dots,\IIntegral_r$ be the unknown functions of the system $S$ (see the second column in 
Table~\ref{Table:D12Systems}).
\item Perform steps 1--3 of Subsection~\ref{Sec:CoupledSolver} for the system $S$ and extract the following information:
\begin{enumerate}
 \item Update the required expansion orders of the base-case integrals $B_i$ and integrals $\IIntegral_i$ occurring in the systems $<S$.
 \item Update the necessary expansion orders of $\IIntegral_l,\IIntegral_{l+1},\dots,\IIntegral_r$.
 \item Save the recurrence order of the scalar recurrence in $\IIntegral_i$.
\end{enumerate}
\item Uncouple the system to scalar recurrences in terms of the other integrals $\IIntegral_{i+1},\dots,\IIntegral_r$ and perform the 
analysis for each case. Given all this information, choose the optimal\footnote{Here one can consider different variations, like minimizing the 
total sum of all expansion orders of the base-case integrals or minimizing the total sum of expansion orders of the initial values.} version and store the selected data (compare Remark~\ref{Equ:AnalyzeSystem1}, where we analyze system~1).

\item Set $S:=S-1$ and go to Step 2.
\end{enumerate}
%----------------------------------------------------------------------------------------------------

Within the package \texttt{SumProduction} we implemented a new function call which needs as input the hierarchical system of coupled 
differential equations and outputs the necessary expansion orders of the base-case integrals, the necessary number of initial values of the unknown functions arising in the scalar recurrences (plus the necessary expansion order of the initial values), plus some extra information which is convenient for further processing. E.g. in Table~\ref{Table:D12Systems} some of this data is presented: the necessary order to which the unknown integrals have to be expanded (column 3) and the recurrence order of the scalar recurrence (column 5).
Executing this function call and analyzing the system as described above took in total 1302 seconds.

%----------------------------------------------------------------------------------------------------
\subsubsection*{Step C: Calculation of the base-case integrals and initial values}
%----------------------------------------------------------------------------------------------------
	
\vspace*{1mm}\noindent
Next, we check our existing database that contains the initial values of master integrals and $\ep$-expansions of already calculated base-case integrals. If necessary, update the database by new calculations using the technologies presented in the previous Subsections~\ref{sec:Sigma-MIs}--\ref{sec:Zeilberger}.

%----------------------------------------------------------------------------------------------------
\subsubsection*{Step D: Calculation of the $\boldsymbol \ep$-expansions of the master integrals}
%----------------------------------------------------------------------------------------------------

\vspace*{1mm}\noindent
Given a completed database, we can start to solve the full hierarchically structured coupled system. Here we provide again a function call within the package \texttt{SumProduction} which takes as input the data of the analyzed system and the databases of the initial values and the $\ep$-expansions of the base-case integrals.
The function call loops through all the systems (starting with System 1) and carries out steps 5 and 6 of 
Subsection~\ref{Sec:CoupledSolver}.
When one system is completed, it turns to the next system providing the previously calculated integrals as new base-case integrals. 
The calculation times of all the systems are presented in the last column of Table~\ref{Table:D12Systems}. 
Solving all 32 integrals took in total 184.4 hours (=7.68 days).

%---------------------------------------------------------------------------------------------------------------------------
\section{From the master integrals to the final result}\label{sec:GetMoment}
%---------------------------------------------------------------------------------------------------------------------------

\vspace*{1mm}
\noindent
Suppose we are given the representation~\eqref{Equ:DExp} of a diagram in terms of master integrals $\hat{M}_i(x)$, 
and suppose that we succeeded in calculating the Laurent series expansions~\eqref{Equ:epExpansionMI} sufficiently high 
where the $M_{i,j}(N)$ are indefinite nested product-sum expressions.
In some instances these symbolic representations hold only from a certain point on, say $N\geq l_i$, and %that 
for $N=0,\dots,l_i-1$ special values have to be calculated.
In a nutshell, we can write the power series representation of the master integrals in the form
%---------------------------------------------------------------------------------------------------------------------------
\begin{equation}\label{Equ:MIExp}
\hat{M}_i(x)=\ep^{-3}\big[p_{i,-3}(x)+\sum_{N=l_i}^{\infty} M_{i,-3}(N)x^N\big]
+\ep^{-2}\big[p_{i,-2}(x)+\sum_{N=l_i}^{\infty} M_{i,-2}(N)x^N\big]+...,
\end{equation}
%---------------------------------------------------------------------------------------------------------------------------
where the $p_{i,j}(x)$ are polynomials in $x$.
Since the $c_i(x,\ep)$ in~\eqref{Equ:DExp} are rational functions in $x$ and $\ep$, we can easily calculate their Laurent series 
expansions
%---------------------------------------------------------------------------------------------------------------------------
\begin{equation}\label{Equ:CExp}
c_i(x,\ep)=c_{i,\mu_i}(x)\ep^{\mu_i}+c_{i,\mu_i+1}(x)\ep^{\mu_i+1}...,\quad\mu_i\in\set Z,
\end{equation}
%---------------------------------------------------------------------------------------------------------------------------
with rational functions $c_{i,j}(x)$ in $x$.
Finally, we plug the $\ep$-expansions~\eqref{Equ:MIExp} and~\eqref{Equ:CExp}  into~\eqref{Equ:DExp}, collect the terms w.r.t.\ $\ep^i$ with $i\in\set Z$ and obtain~\eqref{equ:DxExp} with
%---------------------------------------------------------------------------------------------------------------------------
\begin{equation}\label{Equ:FXRep}
\hat{F}_k(x)=d_{k,0}(x)+d_{k,1}(x)\sum_{N=l_1}^{\infty}h_{k,1}(N)x^N+d_{k,2}(x)\sum_{N=l_2}^{\infty}h_{k,2}(N)x^N+\dots
+d_{k,s}(x)\sum_{N=l_s}^{\infty}h_{k,s}(N)x^N,
\end{equation}
%---------------------------------------------------------------------------------------------------------------------------
where the $d_{k,1}(x)$ are rational functions in $x$ and the $h_{k,l}(N)$ are indefinite nested product-sum expressions in $N$.
Note that the $\hat{F}_k(x)$ can be written as power series in $x$, i.e.,
%---------------------------------------------------------------------------------------------------------------------------
$$\hat{F}_k(x)=\sum_{N=0}^{\infty}F_k(N)x^N,$$
%---------------------------------------------------------------------------------------------------------------------------
where the $F_k(N)$ are precisely the coefficients of the desired $\ep$-expansion given in~\eqref{Equ:epExpansion}.

In the remaining part of this section we will focus on the challenges to extract the coefficients $F_k(N)$ in terms of symbolic sums 
from the given input expressions~\eqref{Equ:FXRep}. As it will turn out, we can choose an appropriate $\delta\in\set N$ such that for $N\geq\delta$ these $F_i(N)$ can be represented in terms of indefinite nested product-sum expressions in $N$. The underlying method is implemented in \texttt{SumProduction} and utilizes various features of \texttt{Sigma}, \texttt{EvaluateMultiSums} and 
\texttt{HarmonicSums}.

\medskip

We will illustrate our general method by tackling diagram~2 from Figure~\ref{samplediagrams}. It can be expressed in the 
form~\eqref{Equ:DExp} with $r=43$ master integrals. Now we proceed as described above: we calculate the 
$\ep$-expansions~\eqref{Equ:MIExp} up to the needed order where the coefficients $M_{i,j}(N)$ are given in terms of indefinite nested 
product-sum expressions, and we expand the coefficients $c_i(x,\ep)$ in $\ep$. Then we insert all the $\ep$-expansions 
%%into~\eqref{samplediagrams} 
and get the coefficients~\eqref{Equ:FXRep}. Finally, we are in the position to calculate the coefficients $F_{-3}(N)$, $F_{-2}(N)$, $F_{-1}(N)$ and $F_{0}(N)$ within 66~seconds, 104~seconds, 221~seconds and 1309~seconds, respectively. In the following we will focus on the problem of extracting the constant term $F_0(N)$ from $\hat{F}_0(x)$. 

%---------------------------------------------------------------------------------------------------------------------------
\subsubsection*{Step 1: Crunching sums}
%---------------------------------------------------------------------------------------------------------------------------

\vspace*{1mm}\noindent
Using the functionality of the package \texttt{SumProduction} described in~\cite{EMSSP} we crunch the expression~\eqref{Equ:FXRep} 
to an expression $A(x)$ which consists of a sum of sub-expressions, each of the form
%---------------------------------------------------------------------------------------------------------------------------
\begin{equation}\label{Equ:MasterSums}
\hat{H}(x)=\hat{e}(x)\sum_{N=o}^{\infty}s_1(N)s_2(N)\dots s_u(N)r(N)\,x^N \, .
\end{equation}
%---------------------------------------------------------------------------------------------------------------------------
Here $\hat{e}(x)$ is a rational function in $x$, $o\in\set N$, and $r(N)$ is a rational function in $N$. In particular, each $s_i(N)$ is a hypergeometric expression or an indefinite nested sum defined over 
hypergeometric expressions. Note that $\hat{e}(x)$ might have the factor $x^{\nu_0}$ in the denominator, but this is compensated with $o-\nu_0\geq0$. In this article, the $s_i(N)$ stand for $(2^N)^j$ or $\binom{2N}{N}^j$ with $j\in\set Z$ or stand for (generalized) harmonic sums or (inverse) binomial sums. 

\medskip

%---------------------------------------------------------------------------------------------------------------------------
\noindent\textit{Example.} Now let us turn to diagram~2. The expression $\hat{F}_0(x)$ uses 4.2MB of memory and consists of $s=193$ 
formal power series. Here the summands are given in terms of 1080 generalized harmonic sums~\eqref{Equ:SSumsIntro} with $x_i\in\{2,-2,1/2,-1/2\}$ and weight $\leq6$. Using \texttt{SumProduction} we can crunch $\hat{F}_0(x)$ within 800s to the expression $A(x)$ of size 0.7MB which can be written in terms of 150 sums of the form~\eqref{Equ:MasterSums}. In this condensed form, the summands consist of 47 generalized harmonic sums up to weight 6. For the calculation of the algebraic relations between the (generalized) harmonic sums the package \texttt{HarmonicSums} was used.
%---------------------------------------------------------------------------------------------------------------------------

%---------------------------------------------------------------------------------------------------------------------------
\subsubsection*{The extraction of the \boldmath $N$th coefficient}
%---------------------------------------------------------------------------------------------------------------------------

\vspace*{1mm}\noindent
In principle, we are now ready to calculate for each term~\eqref{Equ:MasterSums} in $A(x)$ the coefficient $H(N)$ of the underlying 
Laurent series expansion 
%---------------------------------------------------------------------------------------------------------------------------
\begin{equation}\label{Equ:ExpMasterSums}
\hat{H}(x)=\sum_{N=\mu'}^{\infty}H(N)x^N.
\end{equation}
%---------------------------------------------------------------------------------------------------------------------------
Then combining all these $H(N)$ delivers the coefficient $F_i(N)$.

In a first step, we can calculate the Laurent series expansion 
%---------------------------------------------------------------------------------------------------------------------------
\begin{equation}\label{Equ:eExp}
\hat{e}(x)=\sum_{N=\mu}^{\infty}e(N)x^i
\end{equation}
%---------------------------------------------------------------------------------------------------------------------------
with $\mu\in\set Z$ as follows. Write $\hat{e}(x)=\frac{a(x)}{b(x)}$ where $a(x)$ and $b(x)$ are polynomials in $x$ and consider the complete factorization 
%---------------------------------------------------------------------------------------------------------------------------
\begin{equation}\label{Equ:RootsOfB}
b(x)=c\,x^{\nu_0}(x-\rho_1)^{\nu_1}(x-\rho_2)^{\nu_2}\dots(x-\rho_v)^{\nu_v}
\end{equation}
%---------------------------------------------------------------------------------------------------------------------------
with $c\in\set R\setminus\{0\}$ and $\rho_i\in\set C\setminus\{0\}$ where $\nu_i\in\set N$ counts the multiplicity of the roots 
$\rho_i$ ($\rho_0=0$). Then, as worked out in~\cite[Thm. 4
1.1]{Stanley:97}, we can calculate the 
expansion
%---------------------------------------------------------------------------------------------------------------------------
$$\frac{1}{b(x)}=\frac{1}{x^{\nu_0}}\sum_{N=0}^{\infty}\beta(N)x^N$$
%---------------------------------------------------------------------------------------------------------------------------
with
%---------------------------------------------------------------------------------------------------------------------------
$$\beta(N)=p_1(N)\,\rho_1^N+p_2(N)\,\rho_2^N+\dots+p_v(N)\,\rho_v^N,$$
%---------------------------------------------------------------------------------------------------------------------------
where the $p_i(N)$ are polynomials in $N$ with degree at most $\nu_i-1$. Now we perform the Cauchy product on $\hat{e}(x)=a(x)\,\frac{1}{x^{\nu_0}}\sum_{N=0}^{\infty}\beta(N)x^N$, and it follows that the coefficient $e(N)$ of the expansion~\eqref{Equ:eExp} can be written again as a linear combination of the $\rho_i^N$ with polynomial coefficients in $N$.
Finally, we obtain
%---------------------------------------------------------------------------------------------------------------------------
$$H(N)=\sum_{k=\mu}^{N+\nu_0}s_1(k)s_2(k)\dots s_u(k)r(k)e(N-k)$$
%---------------------------------------------------------------------------------------------------------------------------
by applying once more the Cauchy product.
Since $e(N-k)$ is given as a linear combination of the $\rho_i^{N-k}=\rho_i^{N}\rho_i^{-k}$ where the coefficients are polynomials in $N$ and $k$,
we can pull out all expressions that depend on $N$. 

Summarizing, we can write $H(N)$ as an indefinite nested product-sum expression in $N$. In particular, the summands of the arising sums are built by the objects 
$s_1(k)s_2(k)\dots s_u(k)r(k)$ given in~\eqref{Equ:MasterSums} and the roots $\rho_i^k$  from~\eqref{Equ:RootsOfB}. 

\medskip

%---------------------------------------------------------------------------------------------------------------------------
\noindent\textit{Example.} For diagram 2 we are faced with expressions of the form
%---------------------------------------------------------------------------------------------------------------------------
$$\frac{1}{(1-x) x}
\sum_{N=2}^{\infty } \frac{x^{N} S_{2,1,1}(N)}{(1+N)^2 (2+N)}=\sum_{N=1}^{\infty}f(N)x^N$$
%---------------------------------------------------------------------------------------------------------------------------
and we calculate the $N$th coefficient
%---------------------------------------------------------------------------------------------------------------------------
\begin{equation}\label{Equ:D2TypicalSum}
\begin{split}
f(N)=&\sum_{k=2}^N \frac{S_{2,1,1}\left({k}\right)}{\big(
        1+k\big)^2 \big(
        2+k\big)}
+\frac{S_{2,1,1}(N)}{(N+2)^2 (N+3)}
+\frac{S_1^2(N)}{2(N+1)^2 (N+2)^2 (N+3)}\\
&+\frac{S_2(N)}{2(N+1)^2 (N+2)^2 (N+3)}
+\frac{S_1(N)}{(N+1)^3 (N+2)^2 (N+3)}
+\frac{1}{(N+1)^4 (N+2)^2 (N+3)}.
\end{split}
\end{equation}
%---------------------------------------------------------------------------------------------------------------------------
Besides such objects, also expressions of the form
%---------------------------------------------------------------------------------------------------------------------------
$$\frac{1}{x^2-2 x-1}
\sum_{N=2}^{\infty } \frac{x^{N} S_{3,1}(N)}{(1+N)^2 (2+N)}$$
%---------------------------------------------------------------------------------------------------------------------------
arise, where the denominator polynomial has the factorization
%---------------------------------------------------------------------------------------------------------------------------
$$x^2-2x-1=(x-(1+\sqrt{2}))(x-(1-\sqrt{2})).$$
%---------------------------------------------------------------------------------------------------------------------------
Thus, as predicted above, the powers $(1+\sqrt{2})^N$ and $(1-\sqrt{2})^N$ occur in the expansion:
%---------------------------------------------------------------------------------------------------------------------------
$$\frac1{x^2-2x-1} = 
\frac{1}{2 \sqrt{2}}
\sum_{j=0}^{\infty}
x^j \left[(1-\sqrt{2})^{-j-1} - (1+\sqrt{2})^{-j-1} \right] 
$$
%---------------------------------------------------------------------------------------------------------------------------
Hence, by the Cauchy product we end up at the $x$-expansion
%---------------------------------------------------------------------------------------------------------------------------
\begin{eqnarray}\label{Equ:AlienSums}
\frac{1}{x^2-2 x-1}
\sum_{k=2}^{\infty } 
\frac{x^k S_{3,1}(k)}{(1+k)^2 (2+k)}
&=& 
\frac{1}{2 \sqrt{2}}
\sum_{j=2}^{\infty} x^j
\sum_{k=2}^j \left[(1-\sqrt{2})^{-j+k-1} - (1+\sqrt{2})^{-j+k-1} \right] 
\nonumber\\ && \times
\frac{S_{3,1}(k)}{(1+k)^2 (2+k)}.
\end{eqnarray} 
%---------------------------------------------------------------------------------------------------------------------------
In a nutshell, we produce (generalized) harmonic sums with the arguments $(1-\sqrt{2})$ and $(1+\sqrt{2})$, which actually 
are not expected to occur in the expressions of diagram~2. Hence carrying out this naive approach produces a large expression with a 
huge number of alien sums. In addition, it is extremely cumbersome to eliminate these alien sums in the available computer algebra systems, i.e., discovering their algebraic relations when such algebraic numbers are involved.

In order to overcome these difficulties, we continue with the following preparation step.

%---------------------------------------------------------------------------------------------------------------------------
\subsubsection*{Step 2: Elimination of bad denominators}
%---------------------------------------------------------------------------------------------------------------------------

\vspace*{1mm}
\noindent
After crunching the expression~\eqref{Equ:FXRep} to $A(x)$, where the sub-expressions are of the form~\eqref{Equ:MasterSums}, we 
extract all those sub-expressions whose denominators of $\hat{e}(x)$ do not factor linearly over $\set Q$. Hereafter, we denote 
this bad part by $B(x)$. Next, we truncate the formal power series in $B(x)$ and denote this new expression by $T(x)$. More precisely, $T(x)$ consists of  sums of the form
%---------------------------------------------------------------------------------------------------------------------------
\begin{equation}\label{Equ:TruncatedVersion}
\hat{e}(x)\sum_{N=o}^{a}s_1(N)s_2(N)\dots s_u(N)r(N)\,x^N,
\end{equation}
%---------------------------------------------------------------------------------------------------------------------------
which are now turned to indefinite nested product-sum expressions w.r.t.\ a new variable $a$.
Now we use \texttt{Sigma} to calculate all algebraic relations among these indefinite nested sums. If these expressions can be 
written in terms of (generalized) harmonic sums or (inverse) binomial sums, we exploit the corresponding features of 
\texttt{HarmonicSums}. This will lead to a modified expression $B(x)$ where the variable $a$ occurs not only in the upper bounds of the sum, but also in rational expressions outside of the indefinite nested sums. Therefore we calculate the limit\footnote{In order to perform this limit, we utilize the asymptotic expansions of the occurring sums that can be computed with 
\texttt{HarmonicSums}.} $a\to\infty$ and denote this new expression by $B'(x)$. Note that the obtained expression $B'(x)$ equals $B(x)$ if one considers their formal power series representations.  

We remark that we computed in some sense generalized contiguous relations for sums where $x^N$ occurs only in the out-most summation. In particular, after applying these relations to the expression $B(x)$, we obtain the alternative expression $B'(x)$ where again $x^N$ occurs only in the out-most summation. In addition, also the modified expression $B'(x)$ can be written as a linear combination of formal power series with rational coefficients in $x$. Hence we can update the expression $A(x)$ by replacing the sub-expression $B(x)$ with $B'(x)$ and obtain a new expression $A'(x)$ that represents~\eqref{Equ:FXRep} written again in terms of sub-expressions of the form~\eqref{Equ:MasterSums}.

With this manipulation of $A(x)$, all the bad denominators cancel\footnote{This phenomenon occurred not only in the ladder graph 
project, but also in~\cite{Ablinger:2014uka,Ablinger:2014vwa,Ablinger:2014nga}.}. 
More precisely, in the sub-expressions~\eqref{Equ:MasterSums} the denominators of $\hat{c}(x)$ factor linearly over $\set Q$.
Finally, we crunch once more the expression $A'(x)$ with the techniques described in Step~1.

\medskip

%---------------------------------------------------------------------------------------------------------------------------
\noindent\textit{Example.} The calculations of Step~2 for diagram~2 in total took 200~seconds. In the end, we obtain an expression 
with 142 sums of the form~\eqref{Equ:MasterSums} where the denominators of the $\hat{e}(x)$ factor linearly over $\set Q$.
%---------------------------------------------------------------------------------------------------------------------------

%---------------------------------------------------------------------------------------------------------------------------
\begin{remark}
\rm
In principle, one could calculate the algebraic relations of the truncated versions of all arising formal power series~\eqref{Equ:TruncatedVersion} in the expression $A(x)$. However, in order to reduce the calculation time, we restrict ourselves to handle only those sub-expressions with bad denominators collected in $B(x)$.\\
We emphasize that this recipe can be applied also to other situations. E.g., for calculations outside the ladder graph project, we 
struggled with denominators~\eqref{Equ:RootsOfB} which have, e.g., the factor $x-3$. Thus generalized harmonic 
sums~\eqref{Equ:SSumsIntro} occur with the additional letters $x_i=3,1/3,-3,-1/3$ (and the interactions with the letters
already arising). If these extra letters are not expected, they cause again a huge amount of alien sums. In such 
situations, we can apply exactly the same tactic as above: collect those sub-expressions with factors $x-3$ in the expression 
$B(x)$ and apply  the elimination strategy outlined above. Then we observed again the same phenomenon: in all our calculations 
the unexpected 
denominator factors vanish. Thus we avoided completely the problem of treating exotic generalized harmonic sums. \end{remark}
%---------------------------------------------------------------------------------------------------------------------------

%%%%%%%%%%%%%%%%%%%%%%%%%%%%%%%%%%%%%%%%%%%%%%%%%%%%%%%%%%%%%%%%%%%%%%%%%%%%%%%%%%%%%%%%%%%%%%%%%%%%%%%%%%%%%%%%%%%%%%%%%%%%%%%%%%%%%%%%%
\begin{table}[H]
\begin{center}
\begin{tabular}{|l|r|}
\hline
Diagram  &  $\#$ Master Integrals \\
\hline
$D_1$         &  28 \\
$D_2$         &  43\\
$D_3$         &  43 \\
$D_4$         &  72 \\
$D_5$         &  16\\
$D_6$         &  13 \\
$D_7$         &  13 \\
$D_8$         &   8\\
$D_9$         &  19 \\
$D_{10}$      &  46 \\
$D_{11}$      &  39 \\
$D_{12a}$     &  61 \\
$D_{12b}$     &  80 \\
$D_{12}$      &  92 \\

\hline
\end{tabular}
\end{center}
\caption{\sf \small The number of master integrals contributing to different diagrams.}
\label{nummas1}
\end{table}

%---------------------------------------------------------------------------------------------------------------------------
\subsubsection*{Step 3: Calculation of the \boldmath $\bf N$th coefficient of the series expansion}
%---------------------------------------------------------------------------------------------------------------------------

\vspace*{1mm}
\noindent
Now we determine the $N$th coefficient for each sub-expression in $A'(x)$ analytically and end up 
at the coefficient $F_i(N)$ in~\eqref{Equ:epExpansion} given in terms of indefinite nested product-sums expressions. This calculation is done with the function call \texttt{GetMoment} in \texttt{HarmonicSums} which can tackle also more general situations needed, e.g., for the treatment of binomial sums~\cite{Ablinger:2014bra}.

\medskip

%---------------------------------------------------------------------------------------------------------------------------
\noindent\textit{Example.} For diagram~2 we calculated the $N$th coefficient for the 142 sums of the form~\eqref{Equ:MasterSums} 
and 10 sum-free terms. This took in total 130~seconds. Since we succeeded in eliminating all bad denominators, we avoided alien sums like~\eqref{Equ:AlienSums}. In total we obtain an expression of size 2.6MB with 853 indefinite nested sums.
%---------------------------------------------------------------------------------------------------------------------------

%---------------------------------------------------------------------------------------------------------------------------
\subsubsection*{Step 4: Representation of the indefinite nested product-sum expressions in terms of special functions and 
elimination of algebraic dependencies}
%---------------------------------------------------------------------------------------------------------------------------

\vspace*{1mm}\noindent
Finally, we crunch the expression (as described in Step 1) and reduce the indefinite nested sums to sums of the form~\eqref{Equ:MasterSums} with $x=1$.
Next, we transform the derived sums in terms of (generalized, cyclotomic) harmonic sums or (inverse) binomial sums whenever this is 
possible. Here we used the package \texttt{HarmonicSums}.
Finally, we eliminate all algebraic relations among the derived sums. If the sums are given in terms of (generalized, cyclotomic) 
harmonic 
sums or (inverse) binomial sums, we activate \texttt{HarmonicSums}, otherwise we use the more general (but also more time consuming) setting of \texttt{Sigma}.
\medskip

%---------------------------------------------------------------------------------------------------------------------------
\noindent\textit{Example.} For diagram~2 we reduced the expression produced in Step~3 from 2.6MB to 0.2MB within 15~seconds.
In particular, the number of indefinite nested sums (such as~\eqref{Equ:D2TypicalSum}) has been reduced from 853 to 26. 
Afterwards, the sums were transformed to generalized harmonic sums and all algebraic relations between the different terms 
have been eliminated, which took 20~seconds. 
%---------------------------------------------------------------------------------------------------------------------------

\medskip

\noindent\textit{Example.} Similarly, we calculated the $\ep$-expansion for diagram~12b, the crossed box part. Here we 
start
with~\eqref{Equ:D12MI}, take all the calculated $\ep$-expansions of the master integrals in terms of indefinite nested 
product-sum
expressions, and obtain the expressions~\eqref{Equ:FXRep}. Finally, we are in the position to calculate the coefficients 
$F_{-3}(N)$, $F_{-2}(N)$, $F_{-1}(N)$ and
$F_{0}(N)$ within 669~seconds, 2046~seconds, 11622~seconds and 141451~seconds (about 39 hours), respectively.\\
\noindent We give some further information for the constant term. The generation of $\hat{F}_0(x)$ takes 1021~seconds. The 
size of
$\hat{F}_0(x)$ is 52 MB; it contains 468 formal power series where the summands are built by 655 (generalized) harmonic 
sums and
1948 (inverse) binomial sums. In Step 1 we crunch this expression to 21MB. In total we obtain 3312 formal power series 
where the
summands are given in terms of 380 (generalized) harmonic sums and 1551 (inverse) binomial sums. This reduction took 19396 
seconds. In
this particular instance, no bad denominators in $x$ occur and we skip Step 2. In Step 3 we calculate the $N$th 
coefficient of the
3312 sum expressions in terms of indefinite nested sums (plus some sum-free expressions). This took in total 
37872~seconds. The expression size is 117MB. It
contains 21319 indefinite nested sums (with $x=1$) where the summands are built in terms of 353 generalized harmonic sums 
and 1551 (inverse) binomial sums.
Finally, we crunch these expressions further in Step 4 within 58219~seconds. We obtain an expression of size 21MB with 
3292
indefinite nested sums and 3273 terms free of sums. Finally, we transform these sums to (generalized) harmonic sums and 
(inverse)
binomial sums and calculate all algebraic relations among them. This took in total 24943~seconds. The final result 
amounts to 0.8 MB
in size and is given in terms of 44 (generalized) harmonic sums and 62 (inverse) binomial sums.

\medskip

The final calculation of all ladder and $V$-diagrams proceeds the way having been outlined so far. In Table~\ref{nummas1} we 
summarize the number of contributing master integrals. The results for diagrams 1-3 and 5-11 are presented in the appendix. They obey 
representations in terms of harmonic and generalized harmonic sums. To perform their analytic continuation to complex values of 
$N$, needed in the physical applications, requires to know the asymptotic expansion of these expressions in the analyticity
region for $|N| \rightarrow \infty$ analytically, cf.~Ref.~\cite{Blumlein:2009ta}.

%------------------------------------------------------------------------------------------------------------------------------
\section{The Ladder Diagrams}
\label{sec:LDIA}
%------------------------------------------------------------------------------------------------------------------------------

\vspace{1mm}
\noindent
In the following we present the results for the ladder diagrams. A more involved example is diagram~4 containing
an operator insertion on a triple $qqg$-vertex. The techniques being described in the previous sections lead to the
following expression of the physical graph.
%---------------------------------------------------------------------------------------------------------------------------
\begin{eqnarray}
%%%texparser:LHS:testD4phys%%%
D_{4} &=&
\textcolor{blue}{\left(\frac{C_A}{2}-C_F\right)^2 T_F} \Biggl\{
        \frac{256}{\varepsilon^3} \frac{\left(1-S_1\right)}{3 (N+1) (N+2)}
%        \frac{1}{\varepsilon^3} \Biggl\{
%                \frac{256}{3 (N+1) (N+2)}
%                -\frac{256}{3 (N+1) (N+2)} S_1
%        \Biggr\}
        +\frac{16}{\varepsilon^2} \Biggl[
                \frac{2 \big(7 N^2+36 N+38\big)}{3 (N+1)^2 (N+2)^2}
\nonumber\\ &&
                +\frac{(N-1) (4 N+9)}{3 N (N+1) (N+2)} S_2
                +\frac{5 N+17}{3 N (N+1) (N+2)} S_1^2
                -\frac{P_1 S_1}{3 N^2 (N+1)^2 (N+2)^2}
        \Biggr]
\nonumber\\ &&
        +\frac{4}{\varepsilon} \Biggl[
                -\frac{23 N^2-57 N+198}{9 N (N+1) (N+2) (N+3)} S_1^3
                -\frac{2 \big(10 N^2+13 N+40\big)}{3 N (N+1) (N+2)} S_{2,1}
\nonumber\\ &&
                +\frac{2 \big(30 N^3+289 N^2+357 N+432\big)}{9 N (N+1) (N+2) (N+3)} S_3
                -\frac{4 P_3}{3 (N+1)^3 (N+2)^3 (N+3)}
\nonumber\\ &&
                +\frac{P_4 S_1^2-P_5 S_2}{3 N^2 (N+1)^2 (N+2)^2 (N+3)}
%                +\frac{4 P_4 S_1^2}{3 N^2 (N+1)^2 (N+2)^2 (N+3)}
%                -\frac{4 P_5 S_2}{3 N^2 (N+1)^2 (N+2)^2 (N+3)}
                +\frac{8 (1-S_1)}{(N+1) (N+2)} \zeta_2
%                +\biggl(
%                        \frac{32}{(N+1) (N+2)}
%                        -\frac{32 S_1}{(N+1) (N+2)}
%                \biggr) \zeta_2
\nonumber\\ &&
                -\biggl(
                        \frac{8 N^3+47 N^2-113 N+30}{3 N (N+1) (N+2) (N+3)} S_2
                        +\frac{2 P_{13}}{3 N^3 (N+1)^3 (N+2)^3 (N+3)}
                \biggr) S_1
        \Biggr]
\nonumber\\ &&
        +\frac{\big(-37 N^2+38 N-345\big)}{9 N (N+1) (N+2) (N+3)} S_1^4
        +\frac{4 \big(26 N^3+401 N^2+2969 N+3408\big)}{3 N (N+1) (N+2) (N+3)} S_{3,1}
\nonumber\\ &&
        -\frac{2 \big(86 N^3+558 N^2+2723 N+2661\big)}{3 N (N+1) (N+2) (N+3)} S_4
        -\frac{4 P_7 S_{2,1}}{3 N^2 (N+1)^2 (N+2)^2 (N+3)}
\nonumber\\ &&
        +\frac{\big(204 N^3+885 N^2+616 N-537\big)}{3 N (N+1) (N+2) (N+3)} S_2^2
        +\frac{4 P_8 S_1^3}{9 N^2 (N+1)^2 (N+2)^2 (N+3)^2}
\nonumber\\ &&
        -\frac{4 \big(94 N^3+505 N^2+1513 N+1236\big)}{3 N (N+1) (N+2) (N+3)} S_{2,1,1}
        -\frac{4 P_{11} S_3}{9 N^2 (N+1)^2 (N+2)^2 (N+3)^2}
\nonumber\\ &&
        +\frac{8 P_{12}}{3 (N+1)^4 (N+2)^4 (N+3)^2}
        -\frac{2 P_{14} S_2}{3 N^3 (N+1)^3 (N+2)^3 (N+3)^2}
\nonumber\\ &&
        +\biggl(
                \frac{8 \big(3 N^3-193 N^2-2221 N-2829\big)}{9 N (N+1) (N+2) (N+3)} S_3
                +\frac{4 P_{10} S_2}{3 N^2 (N+1)^2 (N+2)^2 (N+3)^2}
\nonumber\\ &&
                +\frac{4 \big(52 N^3+235 N^2+661 N+408\big) S_{2,1}}{3 N (N+1) (N+2) (N+3)}
                +\frac{4 P_{16}}{3 N^4 (N+1)^4 (N+2)^4 (N+3)^2}
        \biggr) S_1
\nonumber\\ &&
        -\biggl(
                \frac{2 \big(4 N^3-41 N^2-300 N-207\big)}{3 N (N+1) (N+2) (N+3)} S_2
                +\frac{2 P_{15}}{3 N^3 (N+1)^3 (N+2)^3 (N+3)^2}
        \biggr) S_1^2
\nonumber\\ &&
        +\biggl(
                \frac{4 \big(7 N^2+36 N+38\big)}{(N+1)^2 (N+2)^2}
                +\frac{2 (N-1) (4 N+9)}{N (N+1) (N+2)} S_2
                +\frac{2 (5 N+17)}{N (N+1) (N+2)} S_1^2
\nonumber\\ &&
                -\frac{2 P_1 S_1}{N^2 (N+1)^2 (N+2)^2}
        \biggr) \zeta_2
        +\biggl(
                -\frac{16 \big(21 N^3+64 N^2-147 N-306\big)}{3 N (N+1) (N+2) (N+3)} S_1
\nonumber\\ &&
                +\frac{32 P_9}{3 N^2 (N+1)^2 (N+2)^2 (N+3)^2}
        \biggr) \zeta_3
        +(-1)^N \frac{8 P_6 \left(S_{-3}+2 S_{-2,1}+2 \zeta_3\right)}{N^2 (N+1)^2 (N+2)^2 (N+3)^2} %\bigl[
%                S_{-3}
%                +2 S_{-2,1}
%                +2 \zeta_3
%        \bigr]
\nonumber\\ &&
        +\frac{32 (N^3+5 N^2-2 N-12)}{N (N+1) (N+2) (N+3)} \Biggl[
                \biggl(
                        S_{2,1}\left({{1,\frac{1}{2}}}\right)
                        -S_3\left({{\frac{1}{2}}}\right)
                \biggr) S_1({{2}})
                -S_2 S_{1,1}\left({{2,\frac{1}{2}}}\right)
\nonumber\\ &&
                +S_{3,1}\left({{\frac{1}{2},2}}\right)
                -S_{2,1,1}\left({{1,\frac{1}{2},2}}\right)
                -\frac{1}{2} S_{1,1,1,1}\left({{2,\frac{1}{2},1,1}}\right)
                +S_1({{2}}) \zeta_3
        \Biggr]
\nonumber\\ &&
        +\frac{2^N P_2}{N (N+1)^2 (N+2)^2 (N+3)^2} \Biggl[
                8 S_1^2 S_1\left({{\frac{1}{2}}}\right)
                +40 S_2 S_1\left({{\frac{1}{2}}}\right)
                +32 S_3\left({{\frac{1}{2}}}\right)
\nonumber\\ &&
                -16 S_1 S_{1,1}\left({{1,\frac{1}{2}}}\right)
                +16 S_{2,1}\left({{\frac{1}{2},1}}\right)
                -48 S_{2,1}\left({{1,\frac{1}{2}}}\right)
                +16 S_{1,1,1}\left({{1,1,\frac{1}{2}}}\right)
\nonumber\\ &&
                -32 \zeta_3
        \Biggr]
\Biggr\}.
\end{eqnarray}
%---------------------------------------------------------------------------------------------------------------------------
The polynomials $P_i$ read
%---------------------------------------------------------------------------------------------------------------------------
\begin{eqnarray}
P_1    &=& 3 N^4+61 N^3+134 N^2+80 N+16 \\
P_2    &=& N^5-N^4-27 N^3-147 N^2-290 N-144 \\
P_3    &=& 22 N^5+336 N^4+1780 N^3+4431 N^2+5299 N+2496 \\
P_4    &=& 25 N^5-16 N^4-154 N^3-435 N^2-704 N-168 \\
P_5    &=& 69 N^5+384 N^4-258 N^3-2455 N^2-2048 N-168 \\
P_6    &=& N^6+18 N^5+100 N^4+256 N^3+319 N^2+210 N+72 \\
P_7    &=& 10 N^6+173 N^5+922 N^4+1196 N^3+295 N^2+1040 N+600 \\
P_8    &=& 111 N^6+415 N^5-13 N^4-1561 N^3-2018 N^2+150 N+1332 \\
P_9    &=& 6 N^7+8 N^6-300 N^5-1526 N^4-2739 N^3-1590 N^2+423 N+270 \\
P_{10} &=& 16 N^7+169 N^6+43 N^5-2819 N^4-7717 N^3-5562 N^2+2250 N+2052 \\
P_{11} &=& 156 N^7-33 N^6-12269 N^5-60280 N^4-109729 N^3-82685 N^2-24312 N
\nonumber\\ &&
-4176 \\
P_{12} &=& 33 N^8+929 N^7+9058 N^6+45257 N^5+132016 N^4+235379 N^3+255487 N^2
\nonumber\\ &&
+157641 N+43128 \\
P_{13} &=& 94 N^8+532 N^7+607 N^6-2110 N^5-6389 N^4-6022 N^3-2064 N^2-464 N
\nonumber\\ &&
-96 \\
P_{14} &=& 52 N^9-320 N^8-2503 N^7+7573 N^6+82014 N^5+218117 N^4+255071 N^3
\nonumber\\ &&
+134820 N^2+33912 N+7344 \\
P_{15} &=& 220 N^9+1504 N^8+2695 N^7-5585 N^6-31902 N^5-54489 N^4-41095 N^3
\nonumber\\ &&
-13260 N^2-2664 N+432 \\
P_{16} &=& 368 N^{12}+4462 N^{11}+20941 N^{10}+38782 N^9-27919 N^8-266895 N^7
\nonumber\\ &&
-523028 N^6-498037 N^5-240514 N^4-53824 N^3-4064 N^2+1536 N
\nonumber\\ &&
+576.
\end{eqnarray}
%---------------------------------------------------------------------------------------------------------------------------

Here and in the following we have checked the result for general values of $N$ with the results obtained earlier in 
\cite{Bierenbaum:2009mv} for a set of fixed moments calculated using the package {\tt MATAD3.0} \cite{Steinhauser:2000ry}, 
which has also been used for the calculation of initial values of some of the master integrals and in cross checks. 

In the result for diagram $D_4$ generalized harmonic sums contribute, which are partly weighted by powers $\propto 2^N$ and one has to 
check explicitely whether the asymptotic expansion is regular\footnote{In Ref.~\cite{Ablinger:2014yaa} we found one massive scalar 
3-loop diagram, where this has not been the case.}. In case of the appearing sums {\tt HarmonicSums} provides algorithms to derive 
the asymptotic expansion $N \rightarrow \infty$ analytically. It is given by 
%---------------------------------------------------------------------------------------------------------------------------
\begin{eqnarray}
%%%texparser:LHS:testD4physAsyExp%%%
D_{4}^{\rm asy} &=&
\textcolor{blue}{\left(\frac{C_A}{2}-C_F\right)^2 T_F}
\biggl(
        \ln^4(\bar{N}) \biggl[
                -\frac{37}{9 N^2}
                +\frac{260}{9 N^3}
                -\frac{1498}{9 N^4}
                +\frac{6350}{9 N^5}
                -\frac{23182}{9 N^6}
                +\frac{78230}{9 N^7}
\nonumber \\ &&
                -\frac{252478}{9 N^8}
                +\frac{793430}{9 N^9}
                -\frac{2452702}{9 N^{10}}
        \biggr]
        +\ln^3(\bar{N}) \biggl[
                \frac{148}{3 N^2}
                -\frac{3742}{9 N^3}
                +\frac{56233}{27 N^4}
                -\frac{237188}{27 N^5}
\nonumber \\ &&
                +\frac{3030901}{90 N^6}
                -\frac{364316}{3 N^7}
                +\frac{1191908966}{2835 N^8}
                -\frac{802611251}{567 N^9}
                +\frac{7571739197}{1620 N^{10}}
        \biggr]
\nonumber \\ &&
        +\ln^2(\bar{N}) \biggl[
                -\frac{476}{3 N^2}
                +\frac{3916}{3 N^3}
                -\frac{37963}{6 N^4}
                +\frac{487889}{18 N^5}
                -\frac{115619381}{1080 N^6}
                +\frac{214156793}{540 N^7}
\nonumber \\ &&
                -\frac{10531100887}{7560 N^8}
                +\frac{1953123301}{420 N^9}
                -\frac{6574421070137}{453600 N^{10}}
                +\zeta_2 \biggl(
                        -\frac{8}{3 N}
                        +\frac{160}{3 N^2}
                        -\frac{80}{3 N^3}
\nonumber \\ &&
                        -\frac{512}{3 N^4}
                        +\frac{2992}{3 N^5}
                        -\frac{11840}{3 N^6}
                        +\frac{41200}{3 N^7}
                        -\frac{134912}{3 N^8}
                        +\frac{427312}{3 N^9}
                        -\frac{1327040}{3 N^{10}}
                \biggr)
        \biggr]
\nonumber \\ &&
        +\ln(\bar{N}) \biggl[
                \frac{1568}{3 N^2}
                -\frac{8678}{3 N^3}
                +\frac{107683}{9 N^4}
                -\frac{288043}{6 N^5}
                +\frac{17224157}{90 N^6}
                -\frac{794327263}{1080 N^7}
\nonumber \\ &&
                +\frac{12708918931159}{4762800 N^8}
                -\frac{21264661886953}{2381400 N^9}
                +\frac{9733380027047}{381024 N^{10}}
                +\zeta_3 \biggl(
                        \frac{88}{3 N}
                        -\frac{560}{9 N^2}
\nonumber \\ &&
                        +\frac{5608}{9 N^3}
                        -\frac{27224}{9 N^4}
                        +\frac{105016}{9 N^5}
                        -\frac{364280}{9 N^6}
                        +\frac{1193848}{9 N^7}
                        -\frac{3786104}{9 N^8}
                        +\frac{11769976}{9 N^9}
\nonumber \\ &&
                        -\frac{36135800}{9 N^{10}}
                \biggr)
                +\zeta_2 \biggl(
                        \frac{64}{3 N}
                        -\frac{118}{3 N^2}
                        -\frac{7598}{9 N^3}
                        +\frac{44842}{9 N^4}
                        -\frac{304364}{15 N^5}
                        +\frac{222694}{3 N^6}
\nonumber \\ &&
                        -\frac{48984386}{189 N^7}
                        +\frac{832600922}{945 N^8}
                        -\frac{2779325083}{945 N^9}
                        +\frac{1830552406}{189 N^{10}}
                \biggr)
        \biggr]
        +\zeta_4 \biggl(
                -\frac{148}{N}
                +\frac{1481}{6 N^2}
\nonumber \\ &&
                -\frac{6998}{3 N^3}
                +\frac{29773}{3 N^4}
                -\frac{106103}{3 N^5}
                +\frac{351103}{3 N^6}
                -\frac{1118123}{3 N^7}
                +\frac{3483223}{3 N^8}
                -\frac{10706603}{3 N^9}
\nonumber \\ &&
                +\frac{32632903}{3 N^{10}}
        \biggr)
        +\zeta_3 \biggl(
                -\frac{32}{N}
                -\frac{128}{3 N^2}
                +\frac{2762}{3 N^3}
                -\frac{87808}{27 N^4}
                +\frac{1288903}{135 N^5}
                -\frac{249484}{9 N^6}
\nonumber \\ &&
                +\frac{79258867}{945 N^7}
                -\frac{107614439}{405 N^8}
                +\frac{4928363147}{5670 N^9}
                -\frac{1638402274}{567 N^{10}}
        \biggr)
        +\zeta_2 \biggl(
                -\frac{256}{3 N^2}
                +\frac{8477}{9 N^3}
\nonumber \\ &&
                -\frac{105305}{18 N^4}
                +\frac{3102209}{135 N^5}
                -\frac{40907239}{540 N^6}
                +\frac{168345925}{756 N^7}
                -\frac{704036903}{1260 N^8}
                +\frac{91944867169}{113400 N^9}
\nonumber \\ &&
                +\frac{18035424397}{4536 N^{10}}
        \biggr)
        +\frac{632}{3 N^2}
        +\frac{3767}{3 N^3}
        -\frac{1025027}{162 N^4}
        +\frac{6991955}{324 N^5}
        -\frac{132245691}{2000 N^6}
\nonumber \\ &&
        +\frac{4030291117}{20250 N^7}
        -\frac{6815103143399}{12348000 N^8}
        +\frac{297957468833173}{333396000 N^9}
        +\frac{6493831316846623}{1143072000 N^{10}}
\biggr) \nonumber\\ &&
+ O\left(\frac{1}{N^{11}} \ln^4(\bar{N})\right)
\end{eqnarray}
%---------------------------------------------------------------------------------------------------------------------------

and turns out to be regular. In case of the usual harmonic sums the structure of the asymptotic expansion is well-known, 
cf.~\cite{Blumlein:2009ta,Blumlein:2009fz}.

The explicit expressions for the diagrams 1-3 and 5-9 are given in the Appendix. In case generalized harmonic sums contribute,
we also provide the asymptotic representation. We now turn to the discussion of the result for diagrams 12a and 12b.
%------------------------------------------------------------------------------------------------------------------------------
\section{The \boldmath $V$-Diagrams}
\label{sec:VDIA}
%------------------------------------------------------------------------------------------------------------------------------

\vspace{1mm}
\noindent
Diagram~12, Figure~2, consist of two pieces, diagrams~12a and b, the former of which stems from a ladder and the latter from
a crossed box topology prior to the contraction of one massive line leading to the local operator insertion, cf.~Figures~3 and 4.
Diagram~12a is therefore expected to have a simpler sum representation than diagram 12b and the corresponding  calculation
turns out to be also much simpler. The corresponding scalar topologies, which are much easier to compute, have been dealt with
in Ref.~\cite{Ablinger:2014yaa} before. 

Diagram~12a depends on 61 master integrals, cf.~Table~4. It is given by
%-------------------------------------------------------------------------
%>>D12a:
%-----------------------------------------------------------------------------------------------------------------------------------
\begin{eqnarray}
%% Calculated as D5physLadder (D5a)
%% New convention: D_{12,a}
%%%texparser:LHS:testD5physLadder%%%
D_{12,a} &=&
        \textcolor{blue}{\left(\frac{C_A}{2}-C_F\right)^2 T_F} \Biggl\{
        \frac{1}{\varepsilon^3} \Biggl[
                \frac{192 S_2 - 192 S_1^2 - 256}{3 (N+1) (N+2)}
                +\frac{128 (3 N+1) S_1}{3 (N+1)^2 (N+2)}
        \Biggr]
\nonumber \\ &&
        +\frac{1}{\varepsilon^2} \Biggl[
                -\frac{32 \big(5 N^3+53 N^2+99 N+44\big) S_1^2}{3 N (N+1)^2 (N+2)^2}
                -\frac{32 \big(5 N^3+7 N^2-5 N-4\big) S_2}{N (N+1)^2 (N+2)^2}
\nonumber \\ &&
                -\frac{64 \big(9 N^2+37 N+32\big)}{3 (N+1)^2 (N+2)^2}
                +\frac{64 S_1^3}{3 (N+1) (N+2)}
                -\frac{64 (4 N+11) S_{2,1}}{3 (N+1) (N+2)}
\nonumber \\ &&
                +\frac{64 (8 N+15) S_3}{3 (N+1) (N+2)}
                +\biggl(
                        \frac{64 P_{17}}{3 N (N+1)^3 (N+2)^2}
                        +\frac{64 S_2}{(N+1) (N+2)}
                \biggr) S_1
        \Biggr]
\nonumber \\ &&
        +\frac{1}{\varepsilon} \Biggl[
                \frac{32 P_{20}}{3 (N+1)^3 (N+2)^3}
                +\frac{20 S_1^4}{3 (N+1) (N+2)}
                +\frac{32 \big(26 N^2+37 N-12\big) S_{2,1,1}}{3 N (N+1) (N+2)}
\nonumber \\ &&
                +\frac{32 P_{22} S_{2,1}}{3 N^2 (N+1)^2 (N+2)^2}
                -\frac{4 \big(72 N^2+93 N-16\big) S_2^2}{3 N (N+1) (N+2)}
                -\frac{16 P_{23} S_2}{3 N (N+1)^3 (N+2)^3}
\nonumber \\ &&
                -\frac{32 \big(15 N^3+34 N^2-19 N-42\big) S_1^3}{9 N (N+1)^2 (N+2)^2}
                -\frac{32 \big(74 N^2+115 N-20\big) S_{3,1}}{3 N (N+1) (N+2)}
\nonumber \\ &&
                +\biggl(
                        \frac{16 P_{36}}{3 N^3 (N+1)^3 (N+2)^3}
                        +\frac{8 S_2}{(N+1) (N+2)}
                \biggr) S_1^2
                +\frac{8 \big(124 N^2+225 N-24\big) S_4}{3 N (N+1) (N+2)}
\nonumber \\ &&
                -\frac{32 P_{25} S_3}{9 N^2 (N+1)^2 (N+2)^2}
                +\biggl(
                        \frac{32 P_{19} S_2}{3 N^2 (N+1)^2 (N+2)^2}
                        +\frac{32 \big(28 N^2+41 N-8\big) S_3}{3 N (N+1) (N+2)}
\nonumber \\ &&
                        -\frac{32 \big(14 N^2+33 N-4\big) S_{2,1}}{3 N (N+1) (N+2)}
                        -\frac{32 P_{34}}{3 N (N+1)^4 (N+2)^3}
                \biggr) S_1
                +\biggl(
                        -\frac{32}{(N+1) (N+2)}
\nonumber \\ &&
                        -\frac{24 S_1^2}{(N+1) (N+2)}
                        +\frac{16 (3 N+1) S_1}{(N+1)^2 (N+2)}
                        +\frac{24 S_2}{(N+1) (N+2)}
                \biggr) \zeta_2
\nonumber \\ &&
                +(-1)^N \Biggl[
                        \frac{640 S_{-3}}{(N+1)^2}
                        -\frac{768 S_{-2,1}}{(N+1)^2}
                \Biggr]
        \Biggr]
        +\frac{16 P_{32}}{3 (N+1)^4 (N+2)^4}
        +\frac{8 S_1^5}{5 (N+1) (N+2)}
\nonumber \\ &&
        +\frac{128 (6 N+11) S_{-2} S_{-2,1}}{(N+1) (N+2)}
        -\frac{128 (6 N+11) S_{-2,1,-2}}{(N+1) (N+2)}
        +\frac{64 (19 N+22) S_{4,1}}{3 N (N+1) (N+2)}
\nonumber \\ &&
        -\frac{64 \big(105 N^2+126 N-40\big) S_5}{15 N (N+1) (N+2)}
        -\frac{16 \big(116 N^2+289 N+168\big) S_{2,2,1}}{3 N (N+1) (N+2)}
\nonumber \\ &&
        -\frac{16 \big(148 N^2+197 N-72\big) S_{2,1,1,1}}{3 N (N+1) (N+2)}
        +\frac{16 \big(190 N^2+289 N+28\big) S_{2,3}}{3 N (N+1) (N+2)}
\nonumber \\ &&
        +\frac{16 \big(354 N^2+441 N-284\big) S_{3,1,1}}{3 N (N+1) (N+2)}
        -\frac{2 \big(77 N^3+107 N^2-255 N-292\big) S_1^4}{9 N (N+1)^2 (N+2)^2}
\nonumber \\ &&
        +\frac{2 P_{27} S_2^2 + 16 P_{28} S_{3,1} - 16 P_{26} S_{2,1,1} - 4 P_{30} S_4}{3 N^2 (N+1)^2 (N+2)^2}
        +\frac{16 (P_{40} S_3 - 3 P_{39} S_{2,1})}{9 N^3 (N+1)^3 (N+2)^3}
\nonumber \\ &&
        +\biggl(
                \frac{8 P_{37} S_2}{3 N^3 (N+1)^3 (N+2)^3}
                +\frac{16 P_{41}}{3 N (N+1)^5 (N+2)^4}
                -\frac{16 \big(35 N^2+44 N-10\big) S_2^2}{3 N (N+1) (N+2)}
\nonumber \\ &&
                +\frac{16 \big(112 N^2+129 N-32\big) S_{2,1,1}}{3 N (N+1) (N+2)}
                -\frac{16 \big(294 N^2+443 N-84\big) S_{3,1}}{3 N (N+1) (N+2)}
\nonumber \\ &&
                +\frac{16 (3 P_{24} S_{2,1}-P_{29} S_3)}{9 N^2 (N+1)^2 (N+2)^2}
                +\frac{8 \big(238 N^2+371 N-68\big) S_4}{3 N (N+1) (N+2)}
        \biggr) S_1
        +\biggl(
                \frac{32 S_2}{3 (N+1) (N+2)}
\nonumber \\ &&
                +\frac{8 P_{38}}{9 N^3 (N+1)^3 (N+2)^3}
        \biggr) S_1^3
        -\frac{256 (N-1) S_{-2,-3}}{N (N+1)}
        +\biggl(
                \frac{8 P_{43}}{3 N^4 (N+1)^4 (N+2)^4}
\nonumber \\ &&
                -\frac{32 \big(104 N^2+153 N-36\big) S_3}{3 N (N+1) (N+2)}
                +\frac{8 \big(316 N^2+605 N+88\big) S_{2,1}}{3 N (N+1) (N+2)}
        \biggr) S_2
\nonumber \\ &&
        +\biggl(
                \frac{16 \big(56 N^2+87 N-16\big) S_3}{3 N (N+1) (N+2)}
                -\frac{8 P_{42}}{3 N^4 (N+1)^4 (N+2)^4}
                -\frac{4 P_{18} S_2}{3 N^2 (N+1)^2 (N+2)^2}
\nonumber \\ &&
                -\frac{8 \big(56 N^2+57 N-16\big) S_{2,1}}{3 N (N+1) (N+2)}
        \biggr) S_1^2
        +\biggl[
                -\frac{8 \big(9 N^2+37 N+32\big)}{(N+1)^2 (N+2)^2}
                +\frac{8 S_1^3}{(N+1) (N+2)}
\nonumber \\ &&
                -\frac{12 \big(5 N^3+7 N^2-5 N-4\big) S_2}{N (N+1)^2 (N+2)^2}
                -\frac{4 \big(5 N^3+53 N^2+99 N+44\big) S_1^2}{N (N+1)^2 (N+2)^2}
\nonumber \\ &&
                +\biggl(
                        \frac{8 P_{17}}{N (N+1)^3 (N+2)^2}
                        +\frac{24 S_2}{(N+1) (N+2)}
                \biggr) S_1
                +\frac{8 (8 N+15) S_3}{(N+1) (N+2)}
\nonumber \\ &&
                -\frac{8 (4 N+11) S_{2,1}}{(N+1) (N+2)}
        \biggr] \zeta_2
        +\biggl[
                \frac{32 P_{21}}{3 N^2 (N+1)^3 (N+2)^2}
                -\frac{72 S_1^2}{(N+1) (N+2)}
\nonumber \\ &&
                +\frac{64 \big(3 N^2+8 N+6\big) S_{-2}}{N (N+1) (N+2)}
                +\frac{8 \big(16 N^2+57 N+32\big) S_2}{N (N+1) (N+2)}
                +\frac{16 (27 N+17) S_1}{3 (N+1)^2 (N+2)}
        \biggr] \zeta_3
\nonumber \\ &&
        +(-1)^N \Biggl[
                \frac{32 (2 N+3)}{(N+1) (N+2)} \big(
                        2 S_{-2} S_{2,1}
                        -2 S_{2,1,-2}
                        -4 S_{-2,2,1}
                        -5 S_{-3,1,1}
                        +6 S_2 S_{-2,1}
\nonumber \\ &&
                        +6 S_{-2,1,1,1}
                        -2 S_{-5}
                        -S_{2,-3}
                        +3 S_{-4,1}
                \big)
                +\biggl(
                        -\frac{32 P_{35}}{N^3 (N+1)^3 (N+2)^2}
                        -\frac{576 S_1}{(N+1)^2}
\nonumber \\ &&
                        -\frac{32 (2 N+3) S_2}{(N+1) (N+2)}
                \biggr) S_{-3}
                -\frac{2176 S_{-2,1,1}}{(N+1)^2}
                +\frac{896 S_1 S_{-2,1}}{(N+1)^2}
                +\frac{64 P_{33} S_{-2,1}}{N^3 (N+1)^3 (N+2)^2}
\nonumber \\ &&
                -\frac{64 (2 N+3) S_{-2,3}}{(N+1) (N+2)}
                +\frac{1024 S_{-2,2}}{(N+1)^2}
                -\frac{64 S_{-4}}{(N+1)^2}
                +\frac{960 S_{-3,1}}{(N+1)^2}
                +\biggl(
                        \frac{64 (2 N+3) S_2}{(N+1) (N+2)}
\nonumber \\ &&
                        +\frac{128 S_1}{(N+1)^2}
                        +\frac{64 P_{31}}{N^3 (N+1)^3 (N+2)^2}
                        +\frac{128 (2 N+3) S_{-2}}{(N+1) (N+2)}
                \biggr) \zeta_3
        \Biggr]
\Biggr\}~.
\end{eqnarray}
%-----------------------------------------------------------------------------------------------------------------------------------
It is represented by nested harmonic sums only. Here the polynomials $P_i$ read
%-----------------------------------------------------------------------------------------------------------------------------------
\begin{eqnarray}
P_{17}    &=& 7 N^4+53 N^3+93 N^2+53 N+16 \\
P_{18}    &=& 9 N^4-25 N^3-131 N^2-100 N-32 \\
P_{19}    &=& 13 N^4+92 N^3+149 N^2+66 N+16 \\
P_{20}    &=& 17 N^4+129 N^3+351 N^2+403 N+168 \\
P_{21}    &=& N^5-56 N^4-193 N^3-232 N^2-132 N-48 \\
P_{22}    &=& 4 N^5+39 N^4+79 N^3+31 N^2-32 N-16 \\
P_{23}    &=& 12 N^5+167 N^4+564 N^3+807 N^2+595 N+240 \\
P_{24}    &=& 16 N^5-31 N^4-315 N^3-467 N^2-248 N-80 \\
P_{25}    &=& 24 N^5+105 N^4+194 N^3+139 N^2-48 N-24 \\
P_{26} &=& 32 N^5+3 N^4-259 N^3-283 N^2-40 N-32 \\
P_{27} &=& 80 N^5+151 N^4-63 N^3-77 N^2+164 N+96 \\
P_{28} &=& 84 N^5+167 N^4-189 N^3-453 N^2-224 N-120 \\
P_{29} &=& 96 N^5+101 N^4-577 N^3-831 N^2-268 N-216 \\
P_{30} &=& 136 N^5+467 N^4+417 N^3-153 N^2-500 N-208 \\
P_{31} &=& 2 N^6+2 N^5-14 N^4-43 N^3-47 N^2-20 N-4 \\
P_{32} &=& 7 N^6+16 N^5-139 N^4-705 N^3-1288 N^2-1079 N-352 \\
P_{33} &=& 20 N^6+104 N^5+214 N^4+197 N^3+103 N^2+76 N+20 \\
P_{34} &=& 22 N^6+203 N^5+632 N^4+800 N^3+280 N^2-95 N+24 \\
P_{35} &=& 28 N^6+168 N^5+394 N^4+443 N^3+297 N^2+180 N+44 \\
P_{36} &=& 26 N^7+221 N^6+624 N^5+821 N^4+555 N^3+120 N^2-112 N-32 \\
P_{37} &=& 30 N^7-639 N^6-3480 N^5-5561 N^4-2370 N^3+1232 N^2+1024 N+320 \\
P_{38} &=& 94 N^7+457 N^6+652 N^5+131 N^4-466 N^3-760 N^2-624 N-96 \\
P_{39} &=& 4 N^8+16 N^7-206 N^6-1314 N^5-3390 N^4-4971 N^3-4164 N^2
        \nonumber \\ &&
        -1752 N-336 \\
P_{40} &=& 24 N^8+157 N^7+487 N^6-14 N^5-5221 N^4-14311 N^3-15376 N^2
        \nonumber \\ &&
        -7056 N-1440 \\
P_{41} &=& 30 N^8+378 N^7+1635 N^6+2867 N^5-112 N^4-8176 N^3-11681 N^2
        \nonumber \\ &&
        -5423 N-32 \\
P_{42} &=& 63 N^{10}+786 N^9+3556 N^8+8205 N^7+10280 N^6+4952 N^5-4737 N^4
        \nonumber \\ &&
        -9296 N^3-6048 N^2-1728 N-192 \\
P_{43} &=& 83 N^{10}+1156 N^9+5266 N^8+11679 N^7+15630 N^6+16718 N^5+16891 N^4
        \nonumber \\ &&
        +13040 N^3+6240 N^2+1728 N+192.
\end{eqnarray}
%-----------------------------------------------------------------------------------------------------------------------------------

%%\input{D5phys-ladder-asyexp.tex}

\noindent
The diagram can be fully expressed by nested harmonic sums up to weight {\sf w = 5}. Its asymptotic representation 
is therefore regular, growing logarithmically at most with the highest contributing power of $S_1$, 
cf.~\cite{Blumlein:2009ta}.

More involved structures are found for diagram $D_{12 b}$. It is given by

%>>D12b:
\begin{eqnarray}
%% Calculated as D5physCBox (D5b)
%% New convention: D_{12,b}
%%%texparser:LHS:testD5physCBox%%%
D_{12,b} &=&
\textcolor{blue}{T_F \left(\frac{C_A}{2}-C_F\right) \left(C_A-C_F\right)}
\Biggl\{
        \frac{1}{\varepsilon^3} \Biggl[
                -\frac{128 \big(N^2+N+1\big)}{3 N (N+1)^2 (N+2)}
                +\frac{128 (N+3)}{3 (N+1)^2 (N+2)} S_1
\nonumber \\ &&
                -\frac{64}{3 (N+1) (N+2)} \bigl[3 S_2+S_1^2+4 S_{-2}\bigr]
                +(-1)^N \frac{128}{3 N (N+1)^2 (N+2)}
        \Biggr]
\nonumber \\ &&
        +\frac{1}{\varepsilon^2} \Biggl[
                -\frac{64 P_{45}}{3 N (N+1)^3 (N+2)^2}
                -\frac{32 (2 N+1) \big(4 N^3+10 N^2+17 N+20\big)}{3 N (N+1)^2 (N+2)^2} S_2
\nonumber \\ &&
                -\frac{32 \big(2 N^3+20 N^2+35 N+12\big)}{3 N (N+1)^2 (N+2)^2} S_1^2
                +(-1)^N \frac{64 \big(4 N^3+16 N^2+28 N+21\big)}{3 N (N+1)^3 (N+2)^2}
\nonumber \\ &&
                +\biggl(
                        \frac{64 P_{46}}{3 N (N+1)^3 (N+2)^2}
                        -\frac{96}{(N+1) (N+2)} S_2
                \biggr) S_1
                +\frac{32}{3 (N+1) (N+2)} S_1^3
\nonumber \\ &&
                -\frac{256 (2 N+5)}{3 (N+1) (N+2)} S_3
                +\biggl(
                        -\frac{128 P_{44}}{3 N (N+1)^2 (N+2)^2}
                        -\frac{512}{3 (N+1) (N+2)} S_1
                \biggr) S_{-2}
\nonumber \\ &&
                +\frac{256 (N+4)}{3 (N+1) (N+2)} S_{2,1}
                +\frac{512}{3 (N+1) (N+2)} S_{-2,1}
        \Biggl]
        +\frac{1}{\varepsilon} \Biggl[
                -\frac{32 P_{71}}{3 N (N+1)^4 (N+2)^3}
\nonumber \\ &&
                +\frac{128 \big(N^2-5 N+2\big)}{3 N (N+1) (N+2)} S_{-2,1,1}
                +\frac{16 \big(10 N^3+62 N^2+111 N+60\big)}{9 N (N+1)^2 (N+2)^2} S_1^3
\nonumber \\ &&
                -\frac{128 \big(N^2-N+2\big)}{3 N (N+1) (N+2)} S_{-2,2}
                +\frac{128 \big(N^2+4 N+2\big)}{3 N (N+1) (N+2)} S_{-4}
                -\frac{64 \big(3 N^2+N+6\big)}{3 N (N+1) (N+2)} S_{-3,1}
\nonumber \\ &&
                +\frac{128 \big(19 N^2+37 N-4\big)}{3 N (N+1) (N+2)} S_{3,1}
                +\frac{64 P_{51} S_{2,1}}{3 N^2 (N+1)^2 (N+2)^2}
                -\frac{32 P_{61} S_3}{9 N^2 (N+1)^2 (N+2)^2}
\nonumber \\ &&
                +\frac{16 P_{72}}{3 N (N+1)^3 (N+2)^3} S_2
                +(-1)^N \biggl(
                        -\frac{64 \big(4 N^3+18 N^2+29 N+16\big)}{N (N+1)^3 (N+2)^3} S_2
\nonumber \\ &&
                        -\frac{32 P_{55}}{3 N (N+1)^4 (N+2)^3}
                        -\frac{128 \big(13 N^3+59 N^2+95 N+52\big)}{3 N (N+1)^3 (N+2)^3} S_{-2}
                        -\frac{640}{(N+1)^2} S_{-3}
\nonumber \\ &&
                        +\frac{768}{(N+1)^2} S_{-2,1}
                        +\frac{16 \zeta_2}{N (N+1)^2 (N+2)}
                \biggr)
                -\frac{8 \big(108 N^2+127 N-56\big)}{3 N (N+1) (N+2)} S_4
\nonumber \\ &&
                +\frac{128 \big(7 N^3+4 N^2+14 N+4\big)}{3 N^2 (N+1) (N+2)} S_{-2,1}
                +\biggl(
                        -\frac{32 \big(78 N^2+187 N-36\big)}{9 N (N+1) (N+2)} S_3
\nonumber \\ &&
                        +\frac{32 P_{69}}{3 N (N+1)^4 (N+2)^3}
                        -\frac{128 \big(N^2-5 N+2\big)}{3 N (N+1) (N+2)} S_{-2,1}
                        +\frac{64 \big(7 N^2+19 N-2\big)}{3 N (N+1) (N+2)} S_{2,1}
\nonumber \\ &&
                        -\frac{16 P_{59}}{3 N^2 (N+1)^2 (N+2)^2} S_2
                \biggr) S_1
                -\frac{28 S_1^4}{9 (N+1) (N+2)}
                +\biggl(
                        -\frac{16 P_{57}}{3 N (N+1)^3 (N+2)^3}
\nonumber \\ &&
                        +\frac{8 \big(4 N^2-19 N+8\big)}{3 N (N+1) (N+2)} S_2
                \biggr) S_1^2
                +\frac{4 (80 N+233)}{3 (N+1) (N+2)} S_2^2
                +\frac{512}{3 (N+1) (N+2)} S_{-2}^2
\nonumber \\ &&
                +\biggl(
                        \frac{64 P_{70}}{3 N (N+1)^3 (N+2)^3}
                        +\frac{64 \big(N^2-5 N+2\big)}{3 N (N+1) (N+2)} S_1^2
                        +\frac{64 \big(N^2+9 N+2\big)}{3 N (N+1) (N+2)} S_2
\nonumber \\ &&
                        -\frac{128 \big(7 N^3+4 N^2+14 N+4\big)}{3 N^2 (N+1) (N+2)} S_1
                \biggr) S_{-2}
                +\biggl(
                        \frac{64 \big(3 N^2+N+6\big)}{3 N (N+1) (N+2)} S_1
\nonumber \\ &&
                        -\frac{64 P_{50}}{3 N^2 (N+1)^2 (N+2)^2}
                \biggr) S_{-3}
                -\frac{512 (2 N+5)}{3 (N+1) (N+2)} S_{2,1,1}
                +\biggl(
                        \frac{16 (N+3)}{(N+1)^2 (N+2)} S_1
\nonumber \\ &&
                        -\frac{16 \big(N^2+N+1\big)}{N (N+1)^2 (N+2)}
                        -\frac{8 \bigl[S_1^2+3 S_2+4 S_{-2}\bigr]}{(N+1) (N+2)}
                \biggr) \zeta_2
        \Biggr]
\nonumber \\ &&
        -\frac{P_{49}}{N^3 (N+1) (N+2) (2 N+1) (2 N+3) \binom{2 N}{N}} \Biggl[
                16 \left(\sum_{i_1=1}^N (-2)^{i_1}
                        \binom{2 i_1}{i_1}\right) \zeta_3
\nonumber \\ &&
                +16 \sum_{i_1=1}^N (-2)^{i_1}
                        \binom{2 i_1}{i_1} S_{1,2}\left({{\frac{1}{2},1},i_1}\right)
                +48 \sum_{i_1=1}^N (-2)^{i_1}
                        \binom{2 i_1}{i_1} S_{1,2}\left({{\frac{1}{2},-1},i_1}\right)
        \Biggr]
\nonumber \\ &&
        +\frac{N^2+4 N+2}{N (N+1) (N+2)} \Biggl[
                -192 \sum_{i_1=1}^N (-2)^{i_1} \binom{2 i_1}{i_1}
                        \left(\sum_{i_2=1}^{i_1} \frac{1}{\binom{2 i_2}{i_2} i_2^2}\right)
                        S_{1,2}\left({{\frac{1}{2},1},i_1}\right)
\nonumber \\ &&
                -576 \sum_{i_1=1}^N (-2)^{i_1} \binom{2 i_1}{i_1}
                        \left(\sum_{i_2=1}^{i_1} \frac{1}{\binom{2 i_2}{i_2} i_2^2}\right)
                        S_{1,2}\left({{\frac{1}{2},-1},i_1}\right)
                -32 \sum_{i_1=1}^N \frac{\sum_{i_2=1}^{i_1} \frac{(-1)^{i_2} \binom{2 i_2}{i_2}}{i_2^3}}{\binom{2 i_1}{i_1} \big(1+i_1\big)}
\nonumber \\ &&
                +\Biggl(
                        192 \sum_{i_1=1}^N (-2)^{i_1} \binom{2 i_1}{i_1}
                                S_{1,2}\left({{\frac{1}{2},1},i_1}\right)
                        +576 \sum_{i_1=1}^N (-2)^{i_1} \binom{2 i_1}{i_1}
                                S_{1,2}\left({{\frac{1}{2},-1},i_1}\right)
                \Biggr)
\nonumber \\ &&
                \times \sum_{i_1=1}^N \frac{1}{\binom{2 i_1}{i_1} i_1^2}
                -64 \sum_{i_1=1}^N \frac{\sum_{i_2=1}^{i_1} \binom{2 i_2}{i_2} \frac{S_1\left({i_2}\right)}{i_2^2}}{\binom{2 i_1}{i_1} \big(1+i_1\big)}
                +64 \sum_{i_1=1}^N \frac{\sum_{i_2=1}^{i_1} \frac{(-1)^{i_2} \binom{2 i_2}{i_2} S_2\left({i_2}\right)}{i_2}}{\binom{2 i_1}{i_1} \big(1+i_1\big)}
\nonumber \\ &&
                +96 \sum_{i_1=1}^N \frac{\sum_{i_2=1}^{i_1} \frac{\binom{2 i_2}{i_2} S_2\left({i_2}\right)}{i_2}}{\binom{2 i_1}{i_1} \big(1+i_1\big)}
                +96 \sum_{i_1=1}^N \frac{\sum_{i_2=1}^{i_1} \frac{\binom{2 i_2}{i_2} S_{-2}\left({i_2}\right)}{i_2}}{\binom{2 i_1}{i_1} \big(1+i_1\big)}
                +96 \sum_{i_1=1}^N \frac{\sum_{i_2=1}^{i_1} \frac{\binom{2 i_2}{i_2} S_{1,1}\left({i_2}\right)}{i_2}}{\binom{2 i_1}{i_1} \big(1+i_1\big)}
\nonumber \\ &&
                +192 \sum_{i_1=1}^N \frac{\sum_{i_2=1}^{i_1} \frac{(-1)^{i_2} \binom{2 i_2}{i_2} S_{-2}\left({i_2}\right)}{i_2}}{\binom{2 i_1}{i_1} \big(1+i_1\big)}
                -64 S_{2,1,2}\left({{-2,\frac{1}{2},1}}\right)
                -192 S_{2,1,2}\left({{-2,\frac{1}{2},-1}}\right)
\nonumber \\ &&
                +\Biggl(
                        192 \sum_{i_1=1}^N \frac{\sum_{i_2=1}^{i_1} (-2)^{i_2} \binom{2 i_2}{i_2}}{\binom{2 i_1}{i_1} i_1^2}
                        -256 S_2({{-2}})
                \Biggr) \zeta_3
        \Biggr]
\nonumber \\ &&
        +\frac{\big(3 N^2+16\big)}{N (N+1) (N+2)} \Biggl[
                \frac{64}{3} \sum_{i_1=1}^N \frac{\sum_{i_2=1}^{i_1} \frac{(-1)^{i_2} \binom{2 i_2}{i_2}}{i_2^3}}{\binom{2 i_1}{i_1} \big(1+2 i_1\big)}
                -128 \sum_{i_1=1}^N \frac{\sum_{i_2=1}^{i_1} \frac{(-1)^{i_2} \binom{2 i_2}{i_2} S_{-2}\left({i_2}\right)}{i_2}}{\binom{2 i_1}{i_1} \big(1+2 i_1\big)}
\nonumber \\ &&
                +\frac{128}{3} \sum_{i_1=1}^N \frac{\sum_{i_2=1}^{i_1} \frac{\binom{2 i_2}{i_2} S_1\left({i_2}\right)}{i_2^2}}{\binom{2 i_1}{i_1} \big(1+2 i_1\big)}
                -64 \sum_{i_1=1}^N \frac{\sum_{i_2=1}^{i_1} \frac{\binom{2 i_2}{i_2} S_2\left({i_2}\right)}{i_2}}{\binom{2 i_1}{i_1} \big(1+2 i_1\big)}
                -64 \sum_{i_1=1}^N \frac{\sum_{i_2=1}^{i_1} \frac{\binom{2 i_2}{i_2} S_{-2}\left({i_2}\right)}{i_2}}{\binom{2 i_1}{i_1} \big(1+2 i_1\big)}
\nonumber \\ &&
                -\frac{128}{3} \sum_{i_1=1}^N \frac{\sum_{i_2=1}^{i_1} \frac{(-1)^{i_2} \binom{2 i_2}{i_2} S_2\left({i_2}\right)}{i_2}}{\binom{2 i_1}{i_1} \big(1+2 i_1\big)}
                -64 \sum_{i_1=1}^N \frac{\sum_{i_2=1}^{i_1} \frac{\binom{2 i_2}{i_2} S_{1,1}\left({i_2}\right)}{i_2}}{\binom{2 i_1}{i_1} \big(1+2 i_1\big)}
        \Biggr]
\nonumber \\ &&
        +\frac{6 N-5}{N (N+1) (N+2)} \Biggl[
                \Biggr(
                        -256 \sum_{i_1=1}^N (-2)^{i_1} \binom{2 i_1}{i_1} S_{1,2}\left({{\frac{1}{2},1},i_1}\right)
\nonumber \\ &&
                        -768 \sum_{i_1=1}^N (-2)^{i_1} \binom{2 i_1}{i_1} S_{1,2}\left({{\frac{1}{2},-1},i_1}\right)
                \Biggr) \sum_{i_1=1}^N \frac{1}{\binom{2 i_1}{i_1} i_1}
\nonumber \\ &&
                +768 \sum_{i_1=1}^N (-2)^{i_1} \binom{2 i_1}{i_1} \left(\sum_{i_2=1}^{i_1} \frac{1}{\binom{2 i_2}{i_2} i_2}\right) S_{1,2}\left({{\frac{1}{2},-1},i_1}\right)
\nonumber \\ &&
                +256 \sum_{i_1=1}^N (-2)^{i_1} \binom{2 i_1}{i_1} \left(\sum_{i_2=1}^{i_1} \frac{1}{\binom{2 i_2}{i_2} i_2}\right) S_{1,2}\left({{\frac{1}{2},1},i_1}\right)
\nonumber \\ &&
                +\Biggl(
                        256 S_1({{-2}})
                        -256 \sum_{i_1=1}^N \frac{\sum_{i_2=1}^{i_1} (-2)^{i_2} \binom{2 i_2}{i_2}}{\binom{2 i_1}{i_1} i_1}
                \Biggr) \zeta_3
        \Biggr]
\nonumber \\ &&
        +\frac{P_{64}}{N^3 (N+1)^2 (N+2) (2 N+1) (2 N+3) \binom{2 N}{N}} \Biggl[
                \frac{32}{3} \sum_{i_1=1}^N \frac{(-1)^{i_1} \binom{2 i_1}{i_1} S_2\left({i_1}\right)}{i_1}
\nonumber \\ &&
                +32 \sum_{i_1=1}^N \frac{(-1)^{i_1} \binom{2 i_1}{i_1} S_{-2}\left({i_1}\right)}{i_1}
                -\frac{16}{3} \sum_{i_1=1}^N \frac{(-1)^{i_1} \binom{2 i_1}{i_1}}{i_1^3}
                -\frac{32}{3} \sum_{i_1=1}^N \frac{\binom{2 i_1}{i_1} S_1\left({i_1}\right)}{i_1^2}
\nonumber \\ &&
                +16 \sum_{i_1=1}^N \frac{\binom{2 i_1}{i_1} S_2\left({i_1}\right)}{i_1}
                +16 \sum_{i_1=1}^N \frac{\binom{2 i_1}{i_1} S_{-2}\left({i_1}\right)}{i_1}
                +16 \sum_{i_1=1}^N \frac{\binom{2 i_1}{i_1} S_{1,1}\left({i_1}\right)}{i_1}
        \Biggr]
\nonumber \\ &&
        +\frac{64 \big(5 N^2-N+10\big)}{3 N (N+1) (N+2)} S_{-2,2,1}
        +\frac{2 \big(34 N^3+34 N^2-119 N-108\big)}{9 N (N+1)^2 (N+2)^2} S_1^4
\nonumber \\ &&
        -\frac{64 \big(5 N^2+19 N+10\big)}{3 N (N+1) (N+2)} S_{-4,1}
        -\frac{64 \big(25 N^3+4 N^2+58 N+20\big)}{3 N^2 (N+1) (N+2)} S_{-2,1,1}
\nonumber \\ &&
        +\frac{32 \big(5 N^2+57 N+10\big)}{3 N (N+1) (N+2)} S_{2,-3}
        -\frac{64 \big(7 N^2-11 N+14\big)}{3 N (N+1) (N+2)} S_{-2,1,1,1}
\nonumber \\ &&
        +\frac{64 \big(9 N^2-7 N+18\big)}{3 N (N+1) (N+2)} S_{2,1,-2}
        +\frac{32 \big(19 N^3-20 N^2+62 N+28\big)}{3 N^2 (N+1) (N+2)} S_{-3,1}
\nonumber \\ &&
        +\frac{64 \big(9 N^2+4 N+18\big)}{3 N (N+1) (N+2)} S_{-2,3}
        +\frac{64 \big(11 N^3-4 N^2+30 N+12\big)}{3 N^2 (N+1) (N+2)} S_{-2,2}
\nonumber \\ &&
        +\frac{32 \big(13 N^2+7 N+26\big)}{3 N (N+1) (N+2)} S_{-3,1,1}
        -\frac{64 \big(13 N^2+30 N+26\big)}{3 N (N+1) (N+2)} S_{-5}
\nonumber \\ &&
        +\frac{128 \big(15 N^2+10 N-6\big)}{3 N (N+1) (N+2)} S_{-2,1,-2}
        -\frac{32 \big(15 N^2+412 N+530\big)}{15 N (N+1) (N+2)} S_5
\nonumber \\ &&
        +\frac{64 \big(16 N^2+43 N+16\big)}{3 N (N+1) (N+2)} S_{2,2,1}
        +\frac{128 \big(21 N^2+58 N+18\big)}{3 N (N+1) (N+2)} S_{-2,-3}
\nonumber \\ &&
        +\frac{32 \big(97 N^2+167 N+10\big)}{3 N (N+1) (N+2)} S_{2,1,1,1}
        -\frac{32 \big(214 N^2+335 N-68\big)}{3 N (N+1) (N+2)} S_{3,1,1}
\nonumber \\ &&
        +\frac{32 \big(27 N^2+14 N+10\big)}{3 N (N+1) (N+2)} S_{4,1}
        -\frac{32 \big(65 N^2+50 N-46\big)}{3 N (N+1) (N+2)} S_{2,3}
        -\frac{2 S_1^5}{5 (N+1) (N+2)}
\nonumber \\ &&
        +\frac{16 P_{84}}{9 N^3 (N+1)^3 (N+2)^3} S_3
        +\frac{1}{3 N^2 (N+1)^2 (N+2)^2} \bigl[
                -32 P_{54} S_{2,1,1}
                -64 P_{60} S_{-2,1}
\nonumber \\ &&
                -32 P_{53} S_{3,1}
                +2 P_{66} S_2^2
                +4 P_{67} S_4
        \bigr]
        -\frac{32 P_{80} S_{2,1}}{3 N^3 (N+1)^3 (N+2)^3}
\nonumber \\ &&
        +\biggl(
                -\frac{8 P_{76}}{9 N^3 (N+1)^3 (N+2)^3}
                +\frac{4 \big(28 N^2-45 N+56\big)}{9 N (N+1) (N+2)} S_2
        \biggr) S_1^3
\nonumber \\ &&
        +\biggl(
                \frac{32 (11 N-2) (13 N+25)}{3 N (N+1) (N+2)} S_{3,1}
                -\frac{64 \big(5 N^2-N+10\big)}{3 N (N+1) (N+2)} S_{-2,2}
\nonumber \\ &&
                +\frac{64 \big(7 N^2-11 N+14\big)}{3 N (N+1) (N+2)} S_{-2,1,1}
                -\frac{32 \big(13 N^2+7 N+26\big)}{3 N (N+1) (N+2)} S_{-3,1}
\nonumber \\ &&
                -\frac{64 \big(28 N^2+55 N-8\big)}{3 N (N+1) (N+2)} S_{2,1,1}
                +\frac{2 \big(288 N^2+759 N-64\big)}{3 N (N+1) (N+2)} S_2^2
\nonumber \\ &&
                -\frac{4 \big(412 N^2+461 N-264\big)}{3 N (N+1) (N+2)} S_4
                +\frac{64 \big(25 N^3+4 N^2+58 N+20\big)}{3 N^2 (N+1) (N+2)} S_{-2,1}
\nonumber \\ &&
                +\frac{64 P_{47} S_{2,1}}{3 N^2 (N+1)^2 (N+2)}
                -\frac{16 P_{65} S_3}{9 N^2 (N+1)^2 (N+2)^2}
                +\frac{8 P_{82} S_2}{3 N^3 (N+1)^3 (N+2)^3}
\nonumber \\ &&
                -\frac{16 P_{85}}{3 N (N+1)^5 (N+2)^4 (2 N+3)}
        \biggr) S_1
        +\biggl(
                \frac{8 P_{83}}{3 N (N+1)^4 (N+2)^4 (2 N+3)}
\nonumber \\ &&
                +\frac{128 \big(3 N^2+8 N-2\big)}{3 N (N+1) (N+2)} S_{2,1}
                -\frac{8 \big(94 N^2+213 N-68\big)}{3 N (N+1) (N+2)} S_3
                -\frac{4 P_{63} S_2}{3 N^2 (N+1)^2 (N+2)^2}
\nonumber \\ &&
                -\frac{32 \big(7 N^2-11 N+14\big)}{3 N (N+1) (N+2)} S_{-2,1}
        \biggr) S_1^2
        +\frac{16 P_{87}}{3 N (N+1)^5 (N+2)^4 (2 N+3)}
\nonumber \\ &&
        +\biggl(
                \frac{32 \big(7 N^2-47 N+14\big)}{3 N (N+1) (N+2)} S_{-2,1}
                -\frac{64 \big(35 N^2+75 N+2\big)}{3 N (N+1) (N+2)} S_{2,1}
\nonumber \\ &&
                +\frac{8 \big(1054 N^2+1245 N-820\big)}{9 N (N+1) (N+2)} S_3
                -\frac{8 P_{88}}{3 N (N+1)^4 (N+2)^4 (2 N+3)}
        \biggr) S_2
\nonumber \\ &&
        +\biggl(
                \frac{32 P_{79}}{3 N^3 (N+1)^3 (N+2)^3}
                -\frac{128 \big(3 N^2+7 N+6\big)}{N (N+1) (N+2)} S_{-2}
                +\frac{16 \big(13 N^2+7 N+26\big)}{3 N (N+1) (N+2)} S_1^2
\nonumber \\ &&
                -\frac{16 \big(37 N^2+67 N+74\big)}{3 N (N+1) (N+2)} S_2
                -\frac{32 \big(19 N^3-20 N^2+62 N+28\big)}{3 N^2 (N+1) (N+2)} S_1
        \biggr) S_{-3}
\nonumber \\ &&
        +\biggl(
                \frac{64 P_{52}}{3 N^2 (N+1)^2 (N+2)^2}
                +\frac{64 \big(5 N^2+19 N+10\big)}{3 N (N+1) (N+2)} S_1
        \biggr) S_{-4}
\nonumber \\ &&
        +\biggl(
                -\frac{64 \big(5 N^2-7 N+10\big)}{3 N (N+1) (N+2)} S_{2,1}
                -\frac{256 \big(5 N^2+14 N+10\big)}{9 N (N+1) (N+2)} S_3
\nonumber \\ &&
                +\frac{32 \big(7 N^2-11 N+14\big)}{9 N (N+1) (N+2)} S_1^3
                -\frac{128 \big(15 N^2+32 N-6\big)}{3 N (N+1) (N+2)} S_{-2,1}
\nonumber \\ &&
                -\frac{32 \big(25 N^3+4 N^2+58 N+20\big)}{3 N^2 (N+1) (N+2)} S_1^2
                +\frac{32 P_{58}}{3 N^2 (N+1)^2 (N+2)^2} S_2
\nonumber \\ &&
                +\biggl(
                        \frac{64 P_{60}}{3 N^2 (N+1)^2 (N+2)^2}
                        +\frac{32 \big(N^2+11 N+2\big)}{N (N+1) (N+2)} S_2
                \biggr) S_1
\nonumber \\ &&
                -\frac{32 P_{86}}{3 N (N+1)^4 (N+2)^4 (2 N+3)}
        \biggr) S_{-2}
        +\biggl(
                \frac{64 P_{48}}{3 N (N+1)^2 (N+2)^2}
\nonumber \\ &&
                +\frac{1024}{3 (N+1) (N+2)} S_1
        \biggr) S_{-2}^2
        +\Biggl[
                -\frac{4 (2 N+1) \big(4 N^3+10 N^2+17 N+20\big)}{N (N+1)^2 (N+2)^2} S_2
\nonumber \\ &&
                -\frac{8 P_{45}}{N (N+1)^3 (N+2)^2}
                -\frac{4 \big(2 N^3+20 N^2+35 N+12\big)}{N (N+1)^2 (N+2)^2} S_1^2
                +\biggl(
                        -\frac{36 S_2}{(N+1) (N+2)}
\nonumber \\ &&
                        +\frac{8 P_{46}}{N (N+1)^3 (N+2)^2}
                \biggr) S_1
                +\frac{4}{(N+1) (N+2)} S_1^3
                -\frac{32 (2 N+5)}{(N+1) (N+2)} S_3
\nonumber \\ &&
                +\biggl(
                        -\frac{16 P_{44}}{N (N+1)^2 (N+2)^2}
                        -\frac{64}{(N+1) (N+2)} S_1
                \biggr) S_{-2}
                +\frac{32 (N+4)}{(N+1) (N+2)} S_{2,1}
\nonumber \\ &&
                +\frac{64}{(N+1) (N+2)} S_{-2,1}
        \Biggr] \zeta_2
        +\Biggl[
                -\frac{136}{(N+1) (N+2)} S_2
                -\frac{16 P_{56}}{3 N^3 (N+1)^3 (N+2)}
\nonumber \\ &&
                +\frac{32 \big(6 N^2+19 N+12\big)}{3 N (N+1) (N+2)} S_{-2}
                +\frac{16 (17 N+27)}{3 (N+1)^2 (N+2)} S_1
                -\frac{136}{3 (N+1) (N+2)} S_1^2
        \Biggr] \zeta_3
\nonumber \\ &&
        +(-1)^N \Biggl[
                -\frac{64 \big(4 N^3+18 N^2+29 N+16\big)}{N (N+1)^3 (N+2)^3} S_{2,1}
                +\frac{32 P_{68}}{3 N^3 (N+1)^3 (N+2)^3} S_3
\nonumber \\ &&
                -\frac{64 P_{75}}{N^3 (N+1)^3 (N+2)^3} S_{-2,1}
                +\frac{16 P_{81}}{3 N (N+1)^5 (N+2)^4 (2 N+3)}
                +\biggl(
                        -\frac{896 S_{-2,1}}{(N+1)^2}
\nonumber \\ &&
                        +\frac{64 \big(4 N^3+18 N^2+29 N+16\big)}{N (N+1)^3 (N+2)^3} S_2
                \biggr) S_1
                +\biggl(
                        -\frac{32 P_{74}}{N (N+1)^4 (N+2)^4 (2 N+3)}
\nonumber \\ &&
                        -\frac{192 (2 N+3)}{(N+1) (N+2)} S_{-2,1}
                \biggr) S_2
                +\biggl(
                        \frac{128 \big(4 N^3+18 N^2+29 N+16\big)}{N (N+1)^3 (N+2)^3} S_1
\nonumber \\ &&
                        -\frac{32 P_{77}}{3 N (N+1)^4 (N+2)^4 (2 N+3)}
                        -\frac{64 (2 N+3)}{(N+1) (N+2)} S_{2,1}
                \biggr) S_{-2}
\nonumber \\ &&
                +\biggl(
                        \frac{32 P_{78}}{3 N^3 (N+1)^3 (N+2)^3}
                        +\frac{576}{(N+1)^2} S_1
                        +\frac{32 (2 N+3)}{(N+1) (N+2)} S_2
                \biggr) S_{-3}
\nonumber \\ &&
                +\frac{2 N+3}{(N+1) (N+2)} \bigl[
                        64 S_{-5}
                        +32 S_{2,-3}
                        +64 S_{-2,3}
                        -96 S_{-4,1}
                        +64 S_{2,1,-2}
\nonumber \\ &&
                        +128 S_{-2,2,1}
                        +160 S_{-3,1,1}
                        -192 S_{-2,1,1,1}
                \bigr]
                +\frac{64}{(N+1)^2} \bigl[
                        S_{-4}
                        -16 S_{-2,2}
                        -15 S_{-3,1}
\nonumber \\ &&
                        +34 S_{-2,1,1} 
                \bigr]
                +\frac{8 \big(4 N^3+16 N^2+28 N+21\big)}{N (N+1)^3 (N+2)^2} \zeta_2
                +\biggl(
                        -\frac{16 P_{73}}{3 N^3 (N+1)^3 (N+2)^2}
\nonumber \\ &&
                        -\frac{128}{(N+1)^2} S_1
                        -\frac{64 (2 N+3)}{(N+1) (N+2)} S_2
                        -\frac{128 (2 N+3)}{(N+1) (N+2)} S_{-2}
                \biggr) \zeta_3
\nonumber \\ &&
                + \frac{64 P_{62} 2^N}{3 N^3 (N+1)^2 (N+2)^2 (2 N+3)} \biggl(
                        S_{1,2}\left({{\frac{1}{2},1}}\right)
                        +3 S_{1,2}\left({{\frac{1}{2},-1}}\right)
                        +\zeta_3
                \biggr)
        \Biggr]
\Biggr\},
\end{eqnarray}
%--------------------------------------------------------------------------------------------------------------------
with the polynomials
%--------------------------------------------------------------------------------------------------------------------
\begin{eqnarray}
P_{44}    &=& 4 N^4+16 N^3+34 N^2+41 N+16 \\
P_{45}    &=& 7 N^4+37 N^3+63 N^2+49 N+21 \\
P_{46}    &=& 8 N^4+37 N^3+69 N^2+52 N+2 \\
P_{47}    &=& 10 N^4+23 N^3+48 N^2+41 N+10 \\
P_{48}    &=& 16 N^4+67 N^3+97 N^2+71 N+28 \\
P_{49}    &=& 187 N^4+284 N^3+78 N^2+114 N+36 \\
P_{50}    &=& 5 N^5+7 N^4-20 N^3-12 N^2+24 N+8 \\
P_{51}    &=& 6 N^5+16 N^4+20 N^3+43 N^2+52 N+16 \\
P_{52}    &=& 9 N^5+57 N^4+88 N^3-3 N^2-64 N-24 \\
P_{53} &=& 10 N^5+4 N^4-293 N^3-766 N^2-564 N-128 \\
P_{54} &=& 12 N^5+98 N^4+343 N^3+521 N^2+296 N+56 \\
P_{55} &=& 16 N^5+96 N^4+234 N^3+329 N^2+327 N+177 \\
P_{56} &=& 17 N^5+38 N^4+66 N^3+105 N^2+60 N+24 \\
P_{57} &=& 19 N^5+111 N^4+253 N^3+295 N^2+203 N+76 \\
P_{58} &=& 27 N^5+117 N^4+172 N^3+116 N^2+32 N-8 \\
P_{59} &=& 28 N^5+106 N^4+206 N^3+269 N^2+148 N+32 \\
P_{60} &=& 29 N^5+62 N^4-88 N^3-220 N^2-44 N+48 \\
P_{61} &=& 42 N^5+113 N^4+70 N^3+234 N^2+408 N+96 \\
P_{62} &=& 77 N^5+348 N^4+532 N^3+390 N^2+192 N+72 \\
P_{63} &=& 100 N^5+322 N^4+542 N^3+911 N^2+716 N+160 \\
P_{64} &=& 103 N^5+377 N^4+514 N^3+348 N^2+150 N+36 \\
P_{65} &=& 138 N^5+584 N^4+1592 N^3+2753 N^2+1860 N+432 \\
P_{66} &=& 248 N^5+1362 N^4+2730 N^3+2505 N^2+948 N+64 \\
P_{67} &=& 296 N^5+1466 N^4+2014 N^3-67 N^2-1516 N-576 \\
P_{68} &=& 6 N^6+110 N^5+448 N^4+787 N^3+668 N^2+264 N+48 \\
P_{69} &=& 12 N^6+144 N^5+563 N^4+911 N^3+560 N^2+72 N+42 \\
P_{70} &=& 16 N^6+94 N^5+202 N^4+175 N^3+13 N^2-65 N-24 \\
P_{71} &=& 31 N^6+212 N^5+551 N^4+653 N^3+259 N^2-176 N-177 \\
P_{72} &=& 32 N^6+195 N^5+461 N^4+491 N^3+229 N^2+131 N+132 \\
P_{73} &=& 48 N^6+264 N^5+595 N^4+621 N^3+314 N^2+144 N+48 \\
P_{74} &=& 8 N^7+65 N^6+190 N^5+128 N^4-529 N^3-1346 N^2-1223 N-400 \\
P_{75} &=& 22 N^7+167 N^6+522 N^5+839 N^4+736 N^3+408 N^2+200 N+48 \\
P_{76} &=& 55 N^7+535 N^6+1873 N^5+3245 N^4+3365 N^3+2516 N^2+1344 N+384 \\
P_{77} &=& 64 N^7+555 N^6+1859 N^5+2402 N^4-1266 N^3-7244 N^2-7614 N-2664 \\
P_{78} &=& 78 N^7+603 N^6+1934 N^5+3247 N^4+3148 N^3+2120 N^2+1128 N+240 \\
P_{79} &=& 23 N^8+71 N^7-296 N^6-1490 N^5-1942 N^4-342 N^3+976 N^2+624 N+96 \\
P_{80} &=& 30 N^8+167 N^7+269 N^6-14 N^5-170 N^4+601 N^3+1208 N^2+696 N+144 \\
P_{81} &=& 112 N^8+1112 N^7+4488 N^6+9374 N^5+10942 N^4+8442 N^3+7578 N^2
        \nonumber \\ &&
        +7338 N+3243 \\
P_{82} &=& 116 N^8+565 N^7+593 N^6-1037 N^5-2717 N^4-2385 N^3-1108 N^2
        \nonumber \\ &&
        -288 N-192 \\
P_{83} &=& 142 N^8+1555 N^7+7455 N^6+20242 N^5+33670 N^4+34794 N^3+21892 N^2
        \nonumber \\ &&
        +8089 N+1572 \\
P_{84} &=& 246 N^8+1256 N^7+635 N^6-5665 N^5-8519 N^4+2968 N^3+12124 N^2
        \nonumber \\ &&
        +7008 N+1344 \\
P_{85} &=& 60 N^9+1074 N^8+7144 N^7+24222 N^6+45713 N^5+46281 N^4+18168 N^3
        \nonumber \\ &&
        -5654 N^2-4918 N+438 \\
P_{86} &=& 112 N^9+1100 N^8+4378 N^7+8577 N^6+6819 N^5-3954 N^4-13288 N^3
        \nonumber \\ &&
        -11414 N^2-4303 N-600 \\
P_{87} &=& 190 N^9+2147 N^8+10535 N^7+29503 N^6+52015 N^5+59287 N^4+41233 N^3
        \nonumber \\ &&
        +12273 N^2-3909 N-3243 \\
P_{88} &=& 224 N^9+2310 N^8+10351 N^7+25395 N^6+34556 N^5+21274 N^4-1556 N^3
        \nonumber \\ &&
        -4934 N^2+3961 N+3516.
\end{eqnarray}

%------------------------------------------------------------------------------------------------------------------------------
%------------------------------------------------------------------------------------------------------------------------------
The representation of diagram~12b was chosen such that the basis of sums which was used in the scalar case~\cite{Ablinger:2014yaa}
has been referred to first. It accidentally turns out that no other binomial sums are needed in the physical case and we
can use the relations derived in Ref.~\cite{Ablinger:2014bra} not needing any further extension here. 

The derivation of the asymptotic expansion for diagram~12b request to use the integral representations and new special 
numbers having been derived in  \cite{Ablinger:2014bra}. One obtains
\begin{eqnarray}
%% Old notation used in the calculation: D5physCBox
%% New notation: D_12,b
%%%texparser:LHS:testD5physCBoxAsyExp%%%
D_{12,b}^{\rm asy} &=&
\textcolor{blue}{\left(\frac{C_A}{2}-C_F\right) \left(C_A-C_F\right) T_F} \Biggl\{
        \ln(\bar{N})^5 \biggl[
                -\frac{2}{5 N^2}
                +\frac{6}{5 N^3}
                -\frac{14}{5 N^4}
                +\frac{6}{N^5}
                -\frac{62}{5 N^6}
\nonumber \\ &&
                +\frac{126}{5 N^7}
                -\frac{254}{5 N^8}
                +\frac{102}{N^9}
                -\frac{1022}{5 N^{10}}
        \biggr]
        +\ln(\bar{N})^4 \biggl[
                \frac{68}{9 N^2}
                -\frac{349}{9 N^3}
                +\frac{631}{6 N^4}
                -\frac{4375}{18 N^5}
\nonumber \\ &&
                +\frac{94787}{180 N^6}
                -\frac{66407}{60 N^7}
                +\frac{2886593}{1260 N^8}
                -\frac{1185505}{252 N^9}
                +\frac{3462851}{360 N^{10}}
        \biggr]
        +\ln(\bar{N})^3 \biggl[
                \zeta_2 \biggl(
                        \frac{32}{9 N^2}
\nonumber \\ &&
                        -\frac{32}{3 N^3}
                        +\frac{224}{9 N^4}
                        -\frac{160}{3 N^5}
                        +\frac{992}{9 N^6}
                        -\frac{224}{N^7}
                        +\frac{4064}{9 N^8}
                        -\frac{2720}{3 N^9}
                        +\frac{16352}{9 N^{10}}
                \biggr)
                -\frac{568}{9 N^2}
                -\frac{308}{9 N^3}
\nonumber \\ &&
                +\frac{1925}{9 N^4}
                -\frac{7282}{9 N^5}
                +\frac{429499}{180 N^6}
                -\frac{3454373}{540 N^7}
                +\frac{45579011}{2835 N^8}
                -\frac{217525589}{5670 N^9}
                +\frac{2230906393}{25200 N^{10}}
        \biggr]
\nonumber \\ &&
        +\ln(\bar{N})^2 \biggl[
                \zeta_3 \biggl(
                        -\frac{32}{N^2}
                        +\frac{96}{N^3}
                        -\frac{224}{N^4}
                        +\frac{480}{N^5}
                        -\frac{992}{N^6}
                        +\frac{2016}{N^7}
                        -\frac{4064}{N^8}
                        +\frac{8160}{N^9}
\nonumber \\ &&
                        -\frac{16352}{N^{10}}
                \biggr)
                +\zeta_2 \biggl(
                        -\frac{16}{N^2}
                        -\frac{184}{3 N^3}
                        +\frac{2824}{9 N^4}
                        -\frac{2848}{3 N^5}
                        +\frac{12076}{5 N^6}
                        -\frac{28348}{5 N^7}
                        +\frac{12055388}{945 N^8}
\nonumber \\ &&
                        -\frac{1765420}{63 N^9}
                        +\frac{8190218}{135 N^{10}}
                \biggr)
                +\frac{376}{3 N^2}
                -\frac{648}{N^3}
                +\frac{22174}{9 N^4}
                -\frac{373883}{54 N^5}
                +\frac{49074289}{2700 N^6}
\nonumber \\ &&
                -\frac{79605269}{1800 N^7}
                +\frac{57961374941}{529200 N^8}
                -\frac{5050028843}{19600 N^9}
                +\frac{10982047297}{16800 N^{10}}
        \biggr]
        +\ln(\bar{N}) \biggl[
                \zeta_4 \biggl(
                        \frac{59}{N^2}
\nonumber \\ &&
                        -\frac{177}{N^3}
                        +\frac{413}{N^4}
                        -\frac{885}{N^5}
                        +\frac{1829}{N^6}
                        -\frac{3717}{N^7}
                        +\frac{7493}{N^8}
                        -\frac{15045}{N^9}
                        +\frac{30149}{N^{10}}
                \biggr)
                +\zeta_3 \biggl(
                        \frac{1168}{9 N^2}
\nonumber \\ &&
                        -\frac{5264}{9 N^3}
                        +\frac{5008}{3 N^4}
                        -\frac{36736}{9 N^5}
                        +\frac{418456}{45 N^6}
                        -\frac{306376}{15 N^7}
                        +\frac{13849384}{315 N^8}
                        -\frac{5892536}{63 N^9}
\nonumber \\ &&
                        +\frac{8889164}{45 N^{10}}
                \biggr)
                +\zeta_2 \biggl(
                        \frac{152}{N^2}
                        -\frac{24}{N^3}
                        -\frac{7136}{9 N^4}
                        +\frac{22748}{9 N^5}
                        -\frac{757274}{135 N^6}
                        +\frac{93404}{9 N^7}
\nonumber \\ &&
                        -\frac{17233624}{945 N^8}
                        +\frac{33655894}{945 N^9}
                        -\frac{2246721079}{28350 N^{10}}
                \biggr)
                -\frac{672}{N^2}
                +\frac{2636}{3 N^3}
                +\frac{58048}{81 N^4}
                -\frac{172960}{27 N^5}
\nonumber \\ &&
                +\frac{561942949}{27000 N^6}
                -\frac{4643682457}{81000 N^7}
                +\frac{20970432670037}{166698000 N^8}
                -\frac{6200063134883}{20837250 N^9}
\nonumber \\ &&
                +\frac{123812203727083}{266716800 N^{10}}
        \biggr]
        +C_1 \biggl(
                -\frac{2}{N}
                +\frac{14}{N^2}
                -\frac{166}{3 N^3}
                +\frac{138}{N^4}
                -\frac{910}{3 N^5}
                +\frac{634}{N^6}
                -\frac{3886}{3 N^7}
\nonumber \\ &&
                +\frac{2618}{N^8}
                -\frac{15790}{3 N^9}
                +\frac{10554}{N^{10}}
        \biggr)
        +C_2 \frac{\sqrt{N}}{2^{2 N}} \biggl(
                -\frac{128}{N^3}
                +\frac{10448}{75 N^4}
                -\frac{47827}{675 N^5}
                -\frac{81793}{216 N^6}
\nonumber \\ &&
                +\frac{112032149}{57600 N^7}
                -\frac{84661250029}{12441600 N^8}
                +\frac{282488802823}{13271040 N^9}
                -\frac{98593205185337}{1592524800 N^{10}}
        \biggr)
\nonumber \\ &&
        +C_3 \biggl(
                -\frac{2}{N^2}
                +\frac{23}{3 N^3}
                -\frac{19}{N^4}
                +\frac{125}{3 N^5}
                -\frac{87}{N^6}
                +\frac{533}{3 N^7}
                -\frac{359}{N^8}
                +\frac{2165}{3 N^9}
                -\frac{1447}{N^{10}}
        \biggr)
\nonumber \\ &&
        +C_4 \biggl(
                -\frac{2}{N}
                -\frac{2}{N^2}
                +\frac{6}{N^3}
                -\frac{14}{N^4}
                +\frac{30}{N^5}
                -\frac{62}{N^6}
                +\frac{126}{N^7}
                -\frac{254}{N^8}
                +\frac{510}{N^9}
                -\frac{1022}{N^{10}}
        \biggr)
\nonumber \\ &&
        +\ln\left(\sqrt{5}-1\right) \zeta_2 \biggl(
                -\frac{1536}{5 N}
                +\frac{13312}{5 N^2}
                -\frac{156928}{15 N^3}
                +\frac{130304}{5 N^4}
                -\frac{171776}{3 N^5}
                +\frac{598272}{5 N^6}
\nonumber \\ &&
                -\frac{3666688}{15 N^7}
                +\frac{2470144}{5 N^8}
                -\frac{2979584}{3 N^9}
                +\frac{9957632}{5 N^{10}}
        \biggr)
        +\zeta_5 \biggl(
                -\frac{1034}{3 N}
                -\frac{3878}{5 N^2}
                +\frac{11634}{5 N^3}
\nonumber \\ &&
                -\frac{27146}{5 N^4}
                +\frac{11634}{N^5}
                -\frac{120218}{5 N^6}
                +\frac{244314}{5 N^7}
                -\frac{492506}{5 N^8}
                +\frac{197778}{N^9}
                -\frac{1981658}{5 N^{10}}
        \biggr)
\nonumber \\ &&
        +\zeta_4 \biggl(
                \frac{10502}{9 N^2}
                -\frac{290539}{54 N^3}
                +\frac{495337}{36 N^4}
                -\frac{3328457}{108 N^5}
                +\frac{2618423}{40 N^6}
                -\frac{146561723}{1080 N^7}
\nonumber \\ &&
                +\frac{233550391}{840 N^8}
                -\frac{856245007}{1512 N^9}
                +\frac{827962831}{720 N^{10}}
        \biggr)
        +\zeta_2 \zeta_3 \biggl(
                \frac{29036}{45 N}
                +\frac{14116}{45 N^2}
                -\frac{14116}{15 N^3}
\nonumber \\ &&
                +\frac{98812}{45 N^4}
                -\frac{14116}{3 N^5}
                +\frac{437596}{45 N^6}
                -\frac{98812}{5 N^7}
                +\frac{1792732}{45 N^8}
                -\frac{239972}{3 N^9}
                +\frac{7213276}{45 N^{10}}
        \biggr)
\nonumber \\ &&
        +\ln(2) \zeta_2 \biggl(
                \frac{2176}{5 N}
                -\frac{17792}{5 N^2}
                +\frac{70016}{5 N^3}
                -\frac{174464}{5 N^4}
                +\frac{76672}{N^5}
                -\frac{801152}{5 N^6}
                +\frac{1636736}{5 N^7}
\nonumber \\ &&
                -\frac{3307904}{5 N^8}
                +\frac{1330048}{N^9}
                -\frac{13334912}{5 N^{10}}
        \biggr)
        +\zeta_3 \biggl(
                -\frac{3328}{15 N}
                +\frac{78128}{45 N^2}
                -\frac{325864}{45 N^3}
\nonumber \\ &&
                +\frac{799276}{45 N^4}
                -\frac{364124}{9 N^5}
                +\frac{1331092}{15 N^6}
                -\frac{5163116}{27 N^7}
                +\frac{1156829872}{2835 N^8}
                -\frac{490844206}{567 N^9}
\nonumber \\ &&
                +\frac{5764973003}{3150 N^{10}}
        \biggr)
        +(-1)^N \zeta_3 \biggl(
                \frac{80}{N^6}
                -\frac{400}{N^7}
                +\frac{1120}{N^8}
                -\frac{2400}{N^9}
                +\frac{5040}{N^{10}}
        \biggr)
        +\zeta_2 \biggl(
                -\frac{160}{3 N^2}
\nonumber \\ &&
                -\frac{1616}{3 N^3}
                +\frac{17822}{3 N^4}
                -\frac{618754}{27 N^5}
                +\frac{3154276}{45 N^6}
                -\frac{28040113}{150 N^7}
                +\frac{284086748789}{595350 N^8}
\nonumber \\ &&
                -\frac{229916539121}{198450 N^9}
                +\frac{854848983379}{297675 N^{10}}
        \biggr)
        +(-1)^N \zeta_2 \biggl(
                \frac{40}{3 N^6}
                -\frac{2968}{3 N^7}
                +\frac{38504}{3 N^8}
\nonumber \\ &&
                -\frac{128560}{N^9}
                +\frac{3754024}{3 N^{10}}
        \biggr)
        +(-1)^N \biggl(
                -\frac{32}{N^5}
                -\frac{112}{N^6}
                +\frac{2304}{N^7}
                -\frac{222448}{9 N^8}
                +\frac{51413552}{225 N^9}
\nonumber \\ &&
                -\frac{476716016}{225 N^{10}}
        \biggr)
        +\frac{656}{3 N^2}
        -\frac{8614}{3 N^3}
        +\frac{2690786}{243 N^4}
        -\frac{30575645}{972 N^5}
        +\frac{48619290341}{607500 N^6}
\nonumber \\ &&
        -\frac{479063546623}{2430000 N^7}
        +\frac{1856312420341649}{3889620000 N^8}
        -\frac{3315935066084731}{2917215000 N^9}
\nonumber \\ &&
        +\frac{579849854424507841}{210039480000 N^{10}}
\Biggr\} + O\left(\frac{\ln^5(\bar{N})}{N^{11}}\right)~.
\end{eqnarray}

%------------------------------------------------------------------------------------------------------------------------------
Here we define $\bar{N} = N \exp(\gamma_E)$.
The constants $C_i$ can be expressed in terms of iterated integrals over alphabets including 
root-valued letters \cite{Ablinger:2014bra} at $x = 0$ and 1 and Mellin transforms of iterated 
integrals at $N=0$, beyond the multiple zeta values \cite{Blumlein:2009cf} and infinite generalized harmonic and cyclotomic sums 
\cite{Ablinger:2011te,Ablinger:2013cf}.
They are given by
%------------------------------------------------------------------------------------------------------------------------------
\begin{eqnarray}
C_1 &=&  \frac{96}{\pi} \Mvec\left[\frac{\HA_{\sf w_2, w_2, w_1}(x)}{x-\frac{1}{4}}\right](0)
        +\frac{48}{\pi} \Mvec\left[\frac{\HA_{\sf w_2, w_2, w_1}(x)}{x+\frac{1}{4}}\right](0)
        +\frac{32}{\pi} \HA_{\sf r_4, 0, w_1,1}(0)
\nonumber\\ &&
        -64 \text{Li}_3\left(\frac{1}{2} \left(\sqrt{5}-1\right)\right)
-\frac{64 \pi ^3}{9 \sqrt{3}}
+64 \ln^3(2)
-192 \ln^2(2) \ln \left(
        \sqrt{5}-1\right)
\nonumber\\ &&
+192 \ln(2) \ln^2\left(
        \sqrt{5}-1\right)
-64 \ln^3\left(
        \sqrt{5}-1\right)
+\frac{32 \pi }{3 \sqrt{3}}
\psi ^{(1)}\left(
        \frac{1}{3}\right)\nonumber\\ 
&\simeq&
101.3243556488913992
\\
%------------------------------------------------------------------------------------------------------------------------------
C_2 &=&
        -\frac{225}{32 \sqrt{\pi}} 
        \Mvec\left[\frac{\sqrt{x} \HA_{\sf w_5, w_2, w_1}(x)}
        {\big(x-\frac{1}{4}\big) \sqrt{x+2}}\right](0)
%--        
        -\frac{75}{32 \sqrt{\pi}} 
        \Mvec\left[\frac{\sqrt{x} \HA_{\sf w_6, w_1, w_1}(x)}{\sqrt{2-x} \big(
                x+\frac{1}{4}\big)}\right](0)
\nonumber\\ &&
%--
        -\frac{75 \zeta_2}{64 \sqrt{\pi}} 
        \Mvec\left[
        \frac{\sqrt{x} \HA_{\sf w_6}(x)}
             {\sqrt{2-x} \big(x+\frac{1}{4}\big)}
        \right](0) 
+\frac{75}{16 \sqrt{\pi }}
 \Mvec\left[
        \frac{\HA_{\sf w_2, w_2, w_1}(x)}{x-\frac{1}{4}}\right](0)
\nonumber\\ &&
%--
+\frac{75}{32 \sqrt{\pi }}
 \Mvec\left[\frac{\HA_{\sf w_2, w_2, w_1}(x)}{x+\frac{1}{4}}\right](0)
+\frac{25}{16 \sqrt{\pi }} \HA_{\sf r_4, 0, w_1,1}(0)
-\frac{25 \pi ^{3/2} \zeta_2}{12 \sqrt{3}}
+\frac{65 \sqrt{\pi } \zeta_3}{12}
\nonumber\\ &&
-\frac{25}{8} \ln^3(\sqrt{5}-1) \sqrt{\pi }
+\frac{75}{8} \ln^2(\sqrt{5}-1) \ln(2) \sqrt{\pi }
+\frac{145}{16} \ln(\sqrt{5}-1) \sqrt{\pi } \zeta_2
\nonumber\\ &&
-\frac{75}{8} \ln(\sqrt{5}-1) \ln^2(2) \sqrt{\pi }
+\frac{25 \ln^3(2) \sqrt{\pi }}{8}
-\frac{195}{16} \ln(2) \sqrt{\pi } \zeta_2
+\frac{25\pi ^{3/2}}{48 \sqrt{3}}  \psi^{(1)}\left(\frac{1}{3}\right)
\nonumber\\ &&
-\frac{25 \sqrt{\pi }}{8} \Li_3\left(\frac{1}{2}\left(\sqrt{5}-1\right)\right)
\nonumber\\ 
&<& 10^{-53}~,
\end{eqnarray}
%------------------------------------------------------------------------------------------------------------------------------
suggesting $C_2$ being zero. The other two constants are
%------------------------------------------------------------------------------------------------------------------------------
\begin{eqnarray}
C_3 &=& 
-128 \pi ^2 \Mvec\left[
        \frac{\sqrt{x} \HA_{\sf w_{13}}(x)}{\sqrt{8-x} (x+1)}\right](0)
-\frac{64 \pi ^2}{\sqrt{3}} \Mvec\left[
        \frac{\sqrt{x} \HA_{\sf w_{25}}(x)}{\sqrt{8-x} (x+1)}\right](0)
+\frac{512 \HA_{\sf r_4, 0, w_1, 1}(0)}{3 \sqrt{3}}
\nonumber\\ &&
-\frac{3328 \pi ^3 \ln (2)}{15 \sqrt{3}}
+\frac{1024 \pi  \ln^3 (2)}{3 \sqrt{3}}
+\frac{7424 \pi ^3 \ln \left(
        -1+\sqrt{5}\right)}{45 \sqrt{3}}
-\frac{1024 \pi  \ln^2 (2) \ln \left(
        -1+\sqrt{5}\right)}{\sqrt{3}}
\nonumber\\ &&
+\frac{1024 \pi  \ln (2) \ln^2 \left(
        -1+\sqrt{5}\right)}{\sqrt{3}}
-\frac{1024 \pi  \ln^3 \left(
        -1+\sqrt{5}\right)}{3 \sqrt{3}}
+\frac{320}{27} \pi ^2 \psi ^{(1)}\left(
        \frac{1}{3}\right)
\nonumber\\ &&
-\frac{1024 \pi}{3 \sqrt{3}}  \text{Li}_3\left(
        \frac{1}{2} \left(
                -1+\sqrt{5}\right)\right)
+\frac{26624 \pi  \zeta_3}{45 \sqrt{3}}
-\frac{32}{3} \pi ^2 \Mvec\left[
        \frac{x \HA_{\sf w_{14}}(x)}{\sqrt{x+\frac{1}{4}} (x+1)}\right](0)
\nonumber\\ &&
-1536 \Mvec\left[
        \frac{\sqrt{x} \HA_{\sf w_{13}, 1, 0}(x)}{\sqrt{8-x} (x+1)}\right](0)
-64 \Mvec\left[
        \frac{x \HA_{\sf w_{14}, 0, 0}(x)}{\sqrt{x+
        \frac{1}{4}} (x+1)}\right](0)
\nonumber\\ &&
-128 \Mvec\left[
        \frac{x \HA_{\sf w_{14}, 1, 0}(x)}{\sqrt{x+\frac{1}{4}} (x+1)}\right](0)
+192 \Mvec\left[
        \frac{x \HA_{\sf w_{14}, -1, 0}(x)}{\sqrt{x+\frac{1}{4}} (x+1)}\right](0)
\nonumber\\ &&
-4608 \Mvec\left[
        \frac{\sqrt{x} \HA_{\sf w_{18}, -1, 0}(x)}{(x-1) \sqrt{x+8}}\right](0)
+\frac{512}{\sqrt{3}} \Mvec\left[
        \frac{\HA_{\sf w_2, w_2, w_1}(x)}{x-\frac{1}{4}}\right](0)
\nonumber\\ &&
+\frac{256}{\sqrt{3}} \Mvec\left[
        \frac{\HA_{\sf w_2, w_2, w_1}(x)}{x+\frac{1}{4}}\right](0)
-768 \sqrt{3} \Mvec\left[
        \frac{\sqrt{x} \HA_{\sf w_{21}, w_{20}, w_{19}}(x)}{(x-1) \sqrt{x+8}}\right](0)
\nonumber\\ &&
-2304 \Mvec\left[
        \frac{\sqrt{x} \HA_{\sf w_{21}, w_{23}, 0}(x)}{(x-1) \sqrt{x+8}}\right](0)
-256 \sqrt{3} \Mvec\left[
        \frac{\sqrt{x} \HA_{\sf w_{25}, w_{19}, w_{19}}(x)}{\sqrt{8-x}(x+1)}\right](0)
\nonumber\\ &&
-768 \Mvec\left[
        \frac{\sqrt{x} \HA_{\sf w_{25}, w_{26}, 0}(x)}{\sqrt{8-x} (x+1)}\right](0)
+128 \Mvec\left[
        \frac{x \HA_{\sf w_8, 0, 1}(x)}{(x-1) \sqrt{x-\frac{1}{4}}}\right](0)
\nonumber\\ &&
-192 \Mvec\left[
        \frac{x \HA_{\sf w_8, 1, 0}(x)}{(x-1) \sqrt{x-\frac{1}{4}}}\right](0)
+192 \Mvec\left[
        \frac{x \HA_{\sf w_8, 1, 1}(x)}{(x-1) \sqrt{x-\frac{1}{4}}}\right](0)
\nonumber\\
&&
+384 \Mvec\left[
        \frac{x \HA_{\sf w_8 -1, 0}(x)}{(x-1) \sqrt{x-\frac{1}{4}}}\right](0)
\nonumber\\
&\simeq&
-3.90077878750796324
\\
%------------------------------------------------------------------------------------------------------------------------------
C_4 &=& 
\frac{64 \pi ^5}{81 \sqrt{3}}
+8 \pi ^2 \Mvec\left[
        \frac{\sqrt{x} \HA_{\sf w_{12}, 0}(x)}{\sqrt{8-x} (x+1)}\right](0)
+8 \pi ^2 \Mvec\left[
        \frac{\sqrt{x} \HA_{\sf w_{12}, 2}(x)}{\sqrt{8-x} (x+1)}\right](0)
\nonumber\\ &&
+8 \pi ^2 \Mvec\left[
        \frac{\sqrt{x} \HA_{\sf w_{25}, w_{19}}(x)}{\sqrt{8-x} (x+1)}\right](0)
-32 \HHH_{-1,0,2,1,0}(1)
-96 \HHH_{0,-2,-1,0,1}(1)
-96 \HHH_{0,-2,-1,1,0}(1)
\nonumber\\ &&
-96 \HHH_{0,-2,1,-1,0}(1)
-96 \HHH_{0,1,-2,-1,0}(1)
+2 \pi ^2 \HA_{\sf 1, w_8, w_8}(0)
- \frac{4}{3} \pi ^2 \HA_{\sf -1, w_{14}, w_{14}}(0)
\nonumber\\ &&
- \frac{32}{9} \pi   \HA_{\sf r_4, 0, w_1, 1}(0)
+ \frac{187}{45} \pi ^4 \ln (2)
+ \frac{32}{9} \pi ^2 \ln^3(2)
-4 \ln^5(2)
-12 \pi ^2 \ln^2(2) \ln (3)
\nonumber\\ &&
+4 \pi ^2 \ln (2) \ln^2 (3)
-\frac{464}{135} \pi ^4 \ln \left(
        \sqrt{5}-1\right)
+\frac{64}{3} \pi ^2 \ln (2)^2 \ln \left(
        \sqrt{5}-1\right)
\nonumber\\ &&
-\frac{64}{3} \pi ^2 \ln (2) \ln^2 \left(
        \sqrt{5}-1\right)
+\frac{64}{9} \pi ^2 \ln^3 \left(
        \sqrt{5}-1\right)
-\frac{32 \pi ^3} {27 \sqrt{3}}
\psi ^{(1)}\left(
        \frac{1}{3}\right)
\nonumber\\ &&
-8 \pi ^2 \ln (2) \text{Li}_2\left(
        -
        \frac{1}{2}\right)
-8 \pi ^2 \ln (3) \text{Li}_2\left(
        -\frac{1}{2}\right)
-32 \pi ^2 \text{Li}_3\left(
        -\frac{1}{2}\right)
-4 \pi ^2 \text{Li}_3\left(
        \frac{3}{4}\right)
\nonumber\\ &&
+\frac{64}{9} \pi ^2 \text{Li}_3\left(
        \frac{1}{2} \left(
                \sqrt{5}-1\right)\right)
-96 \ln (2) \text{Li}_4\left(
        \frac{1}{2}\right)
-\frac{32}{3} \pi  \Mvec\left[
        \frac{\HA_{\sf w_2, w_2, w_1}(x)}{x-\frac{1}{4}}\right](0)
\nonumber\\ &&
-\frac{16}{3} \pi  \Mvec\left[
        \frac{\HA_{\sf w_2, w_2, w_1}(x)}{x+\frac{1}{4}}\right](0)
+96 \Mvec\left[
        \frac{\sqrt{x} \HA_{\sf w_{12}, 0, 1, 0}(x)}{\sqrt{8-x} 
(x+1)}\right](0)
+96 \Mvec\left[
        \frac{\sqrt{x} \HA_{\sf w_{12}, 2, 1, 0}(x)}{\sqrt{8-x} 
(x+1)}\right](0)
\nonumber\\ &&
+8 \Mvec\left[
        \frac{x \HA_{\sf w_{14}, w_{14}, 0, 0} (x)}{x+1}\right](0)
+16 \Mvec\left[
        \frac{x \HA_{\sf w_{14}, w_{14}, 1, 0}(x)}{x+1}\right](0)
-24 \Mvec\left[
        \frac{x \HA_{\sf w_{14}, w_{14}, -1, 0}(x)}{x+1}\right](0)
\nonumber\\ &&
+288 \Mvec\left[
        \frac{\sqrt{x} \HA_{\sf w_{17}, 0, -1, 0}(x)}{(x-1) \sqrt{x+8}}\right](0)
-288 \Mvec\left[
        \frac{\sqrt{x} \HA_{\sf w_{17}, -2, -1, 0}(x)}{(x-1) \sqrt{x+8}}\right](0)
\nonumber\\ &&
+288 \Mvec\left[
        \frac{\sqrt{x} \HA_{\sf w_{21}, w_{20}, 0, 0}(x)}{(x-1) \sqrt{x+8}}\right](0)
-288 \Mvec\left[\frac{\sqrt{x} \HA_{\sf w_{21}, w_{20}, -1, 0}(x)}{(x-1) \sqrt{x+8}}\right](0)
\nonumber\\ &&
+288 \Mvec\left[
        \frac{\sqrt{x} \HA_{\sf w_{21}, w_{20}, w_{19}, w_{19}}(x)}{(x-1) \sqrt{x+8}}\right](0)
+96 \Mvec\left[\frac{\sqrt{x} \HA_{\sf w_{25}, w_{19}, 0, 0}(x)}{\sqrt{8-x} (x+1)}\right](0)
\nonumber\\ &&
+96 \Mvec\left[
        \frac{\sqrt{x} \HA_{\sf w_{25}, w_{19}, 1, 0}(x)}{\sqrt{8-x} (x+1)}\right](0)
+96 \Mvec\left[
        \frac{\sqrt{x} \HA_{\sf w_{25}, w_{19}, w_{19}, w_{19}}(x)}{\sqrt{8-x} (x+1)}\right](0)
\nonumber\\ &&
-16 \Mvec\left[
        \frac{x \HA_{\sf w_8, w_8, 0, 1}(x)}{x-1}\right](0)
+24 \Mvec\left[
        \frac{x \HA_{\sf w_8, w_8, 1, 0}(x)}{x-1}\right](0)
\nonumber\\ &&
-24 \Mvec\left[
        \frac{x \HA_{\sf w_8, w_8, 1, 1}(x)}{x-1}\right](0)
-48 \Mvec\left[
        \frac{x \HA_{\sf w_8, w_8, -1, 0}(x)}{x-1}\right](0)
\nonumber\\
&\simeq&
278.9253246705036914.
\end{eqnarray}
%------------------------------------------------------------------------------------------------------------------------------
Here letters ${\sf w_i}$ and ${\sf r_4}$ and the (generalized) harmonic polylogarithms $\HA$ are defined in 
Ref.~\cite{Ablinger:2014bra}. The latter are iterated integrals over the support $[x,1]$.
The asymptotic expansion of diagram $D_{12,b}$ is regular unlike the case for the corresponding scalar diagram, which contains
divergent terms $\propto 2^N$, cf.~\cite{Ablinger:2014yaa}.

The results for the $V$-diagrams 10 and 11 are given in Appendix~\ref{results}. 
%------------------------------------------------------------------------------------------------------------------------------
\section{Conclusions}
\label{sec:Conc}
%------------------------------------------------------------------------------------------------------------------------------

\vspace{1mm}
\noindent
We presented different computer algebra methods to calculate the massive 3-loop Feynman diagrams of the ladder and 
$V$-topologies containing local operator insertions. These techniques are widely automated. Beyond the by now well-known 
summation methods tackling the $\ep$-expansion of (generalized) hypergeometric functions and their extensions to other 
higher transcendental functions, we described the use of representations obtained by Mellin-Barnes integrals, the method 
of differential equations, and multi-integration using the multivariate Almkvist-Zeilberger algorithm. These symbolic 
integration methods map to systems of recursions in the Mellin variable $N$, which can be uncoupled, obtaining a  single 
difference equation, also depending on the dimensional parameter $\ep$. The solution of this recursion is found applying the 
summation technologies in $\Pi\Sigma^*$-fields and $R\Pi\Sigma^*$-rings. In parallel the $\ep$-expansion is performed. The 
corresponding algorithms are implemented in the packages {\tt Sigma} \cite{SIG1,SIG2}, {\tt EvaluateMultiSum}, and 
{\tt SumProduction} \cite{EMSSP}. Here, mutual use is made of the package {\tt HarmonicSums} 
\cite{HARMONICSUMS,Ablinger:PhDThesis,Ablinger:2011te,Ablinger:2013cf,Ablinger:2014bra}. The multivariate Almkvist-Zeilberger 
algorithm is implemented in the package {\tt MultiIntegrate} \cite{Ablinger:PhDThesis}.

All these symbolic summation and integration methods apply to the master integrals, since the Feynman integrals have been 
reduced before using the integration-by-parts method implemented in the package {\tt Reduze~2} \cite{Reduze2}. In the calculation 
the $\ep$-expansion is performed automatically, needing no specific basis representation for the master integrals or some sub-system 
thereof. The complexity of the final results grows with the power of $\ep$, requesting the solution of larger systems, which is 
generally more time consuming. Usually, the structure of the result is not more involved but only appears at higher weight.
The results of the calculation can be represented in terms of harmonic sums, generalized harmonic sums and nested generalized
harmonic sums weighted by (inverse) binomial sums. For all diagrams regular asymptotic series in $N$ 
are obtained in the analyticity region. In case of the binomially weighted sums new special constants appear beyond those
related to the infinite nested harmonic, generalized harmonic, and (generalized) cyclotomic sums. 
We would like to mention that the present algorithms allow to to decouple and solve all single-scale systems of differential 
or difference equations, which can be solved in $\Pi\Sigma^*$-fields and $R\Pi\Sigma^*$-rings, in any given basis in automated 
form and perform the $\ep$-expansion.

In all cases the asymptotic representation of the Feynman diagrams could be derived for $|N| \rightarrow \infty,
|\arg(N)| < \pi$ analytically. Furthermore, the diagrams obey recursions, such that the step in the argument $N 
\rightarrow N-1$ can be performed algebraically within the analyticity region. In this way, the complete analytic continuation 
for each diagram is in principle available at any accuracy, being determined by the (numerical) accuracy reached for the asymptotic 
representation.

%%%%%%%%%%%%%%%%%%%%%%%%%%%%%%%%%%%%%%%%%%%%%%%%%%%%%%%%%%%%%%%%%%%%%%%%%%%%%%%%%%%%%%%%%%%%%%%%%%%%%%%%%%%%%%%%%%%%%%%%%%%%%%%%%%%%%%%%%
%%%%%%%%%%%%%%%%%%%%%%%%%%%%%%%%%%%%%%%%%%%%%%%%%%%%%%%%%%%%%%%%%%%%%%%%%%%%%%%%%%%%%%%%%%%%%%%%%%%%%%%%%%%%%%%%%%%%%%%%%%%%%%%%%%%%%%%%%%%%%%%%%%%%%%%%%%%%%%%%%%%
\appendix
%%%%%%%%%%%%%%%%%%%%%%%%%%%%%%%%%%%%%%%%%%%%%%%%%%%%%%%%%%%%%%%%%%%%%%%%%%%%%%%%%%%%%%%%%%%%%%%%%%%%%%%%%%%%%%%%%%%%%%%%%%%%%%%%%%%%%%%%%%%%%%%%%%%%%%%%%%%%%%%%%%%
%------------------------------------------------------------------------------------------------------------------------------
\section{Results for Diagrams 1-3 and 5-11}
\label{results}
%------------------------------------------------------------------------------------------------------------------------------

\vspace{1mm}
\noindent
In this appendix we present the results obtained for diagrams 1-3 and 5-11 of 
Figures~\ref{samplediagrams} and \ref{samplediagrams2}. 
In case generalized harmonic sums contribute, we also calculate the first 10 terms of the asymptotic expansion and show that 
its behaviour is regular.

%\iffalse
Diagram~1 is given by
%-----------------------------------------------------------------------------------------------------------------
\begin{eqnarray}
%%%texparser:LHS:testD1phys%%%
D_{1} &=&
\textcolor{blue}{\left(\frac{C_A}{2}-C_F\right)^2 T_F} \Biggl\{
        -\frac{1}{\varepsilon^3} \frac{32 (N+3)}{3 (N+1)^2}
        +\frac{1}{\varepsilon^2} \Biggl[
                \frac{16 \big(2 N^3+9 N^2+15 N+6\big)}{3 N (N+1)^2 (N+2)} S_1
\nonumber \\ &&
                +\frac{16 P_{89}}{3 N^2 (N+1)^3 (N+2)}
        \Biggr]
        +\frac{1}{\varepsilon} \Biggl[
                -\frac{4 (N+3)}{(N+1)^2} \zeta_2
                +\frac{4 P_{90} S_2}{3 N (N+1)^2 (N+2) (N+3)}
\nonumber \\ &&
                -\frac{4 \big(19 N^3+64 N^2+149 N+96\big)}{3 N (N+1)^2 (N+2) (N+3)} S_1^2
                -\frac{8 P_{95} S_1}{3 N^2 (N+1)^3 (N+2) (N+3)}
\nonumber \\ &&
                -\frac{8 P_{97}}{3 N^3 (N+1)^4 (N+2)^2 (N+3)}
        \Biggr]
        +\frac{4 S_1^4}{N (N+1) (N+2)}
        +\frac{8 \big(4 N^2+N-21\big) S_4}{N (N+1) (N+2)}
\nonumber \\ &&
        -\frac{16 \big(4 N^2+N-22\big) S_{2,1,1}}{N (N+1) (N+2)}
        -\frac{4 \big(8 N^2+2 N-39\big) S_2^2}{N (N+1) (N+2)}
        +\frac{16 \big(12 N^2+3 N-62\big) S_{3,1}}{N (N+1) (N+2)}
\nonumber \\ &&
        +\frac{4 P_{96} S_3 - 2 P_{91} S_1^3 - 72 P_{94} S_{2,1}}{9 N^2 (N+1)^2 (N+2) (N+3)}
%        -\frac{2 P_{91} S_1^3}{9 N^2 (N+1)^2 (N+2) (N+3)}
%        -\frac{8 P_{94} S_{2,1}}{N^2 (N+1)^2 (N+2) (N+3)}
%        +\frac{4 P_{96} S_3}{9 N^2 (N+1)^2 (N+2) (N+3)}
        -\frac{2 P_{99} S_2}{3 (N-1) N^3 (N+1)^3 (N+2)^2 (N+3)^2}
\nonumber \\ &&
        -\frac{4 P_{101}}{3 N^4 (N+1)^5 (N+2)^3 (N+3)^2}
        +\biggl(
                \frac{4 P_{100}}{3 (N-1) N^3 (N+1)^4 (N+2)^2 (N+3)^2}
\nonumber \\ &&
                -\frac{64 S_{2,1} + 32 \big(4 N^2+N-21\big) S_3}{N (N+1) (N+2)}
%                -\frac{64 S_{2,1}}{N (N+1) (N+2)}
%                -\frac{32 \big(4 N^2+N-21\big) S_3}{N (N+1) (N+2)}
                -\frac{2 P_{92} S_2}{3 N^2 (N+1)^2 (N+2) (N+3)}
        \biggr) S_1
\nonumber \\ &&
        +\biggl(
                \frac{40 S_2}{N (N+1) (N+2)}
                +\frac{2 P_{98}}{3 (N-1) N^3 (N+1)^3 (N+2)^2 (N+3)^2}
        \biggr) S_1^2
\nonumber \\ &&
        +\biggl(
                \frac{2 P_{89}}{N^2 (N+1)^3 (N+2)}
                +\frac{2 \big(2 N^3+9 N^2+15 N+6\big) S_1}{N (N+1)^2 (N+2)}
        \biggr) \zeta_2
\nonumber \\ &&
        +\biggl(
                -\frac{4 P_{93}}{3 N^2 (N+1)^2 (N+2)}
                +\frac{32 (N-2) (4 N+9)}{N (N+1) (N+2)}  S_1
        \biggr) \zeta_3
\Biggr\},
\end{eqnarray}
%-----------------------------------------------------------------------------------------------------------------
with the polynomials
%-----------------------------------------------------------------------------------------------------------------
\begin{eqnarray}
P_{89}    &=& 6 N^4+16 N^3+21 N^2+11 N+2 \\
P_{90}    &=& 32 N^4+157 N^3+304 N^2+47 N-84 \\
P_{91}    &=& 4 N^5-69 N^4-24 N^3+453 N^2+3204 N+1944 \\
P_{92}    &=& 20 N^5-29 N^4+160 N^3+869 N^2+5316 N+3240 \\
P_{93}    &=& 48 N^5-19 N^4-899 N^3-1134 N^2+720 N+432 \\
P_{94}    &=& 2 N^6+11 N^5-81 N^3-140 N^2-360 N-216 \\
P_{95}    &=& 8 N^6+46 N^5+111 N^4-8 N^3-205 N^2-224 N-96 \\
P_{96}    &=& 216 N^6+902 N^5-2121 N^4-12390 N^3-10629 N^2-1170 N-1944 \\
P_{97}    &=& 39 N^8+260 N^7+781 N^6+1203 N^5+1007 N^4+691 N^3+511 N^2
\nonumber\\ &&
+248 N+60 \\
P_{98} &=& 8 N^{10}-104 N^9-1001 N^8-2252 N^7+6164 N^6+39130 N^5+61001 N^4
\nonumber \\ &&
+20882 N^3-24108 N^2-22824 N-7776 \\
P_{99} &=& 40 N^{10}+820 N^9+4561 N^8+8988 N^7-11128 N^6-75274 N^5-88213 N^4
\nonumber \\ &&
+7026 N^3+53172 N^2+23112 N+7776 \\
P_{100} &=& 4 N^{11}+121 N^{10}+770 N^9+2942 N^8+5148 N^7-6299 N^6-37910 N^5
\nonumber \\ &&
-57772 N^4-39492 N^3-8392 N^2+2352 N+288 \\
P_{101} &=& 12 N^{13}+180 N^{12}+902 N^{11}+2247 N^{10}+4439 N^9+13622 N^8+40302 N^7
\nonumber \\ &&
+66192 N^6+44874 N^5-1519 N^4-6897 N^3+7138 N^2+5460 N+1224~.
\end{eqnarray}
%-----------------------------------------------------------------------------------------------------------------

Diagram~2 also contains generalized harmonic sums, unlike diagram~1. 
The numerator weights include $\{1,2,\tfrac{1}{2}\}$. It is given by
%------------------------------------------------------------------------------------------------------------------------
\begin{eqnarray}
%%%texparser:LHS:testD2phys%%%
D_{2} &=&
\textcolor{blue}{\left(\frac{C_A}{2}-C_F\right)^2 T_F} \Biggl\{
        -\frac{1}{\varepsilon^3} \frac{16 \big(N^2+5 N-2\big)}{3 N (N+1)^2}
        +\frac{1}{\varepsilon^2} \Biggl[
                \frac{4 P_{103}}{3 N^2 (N+1)^3 (N+2)}
\nonumber \\ &&
                +\frac{8 \big(2 N^3+7 N^2-3 N-12\big)}{3 N^2 (N+1)^2} S_1
                +\frac{32 S_2}{N (N+1)}
        \Biggr]
        +\frac{1}{\varepsilon} \Biggl[
                -\frac{2 \big(N^2+5 N-2\big)}{N (N+1)^2} \zeta_2
\nonumber \\ &&
                +\frac{2 P_{104} S_2-2 P_{102} S_1^2}{3 N^2 (N+1)^2 (N+2) (N+3)}
%                -\frac{2 P_{102} S_1^2}{3 N^2 (N+1)^2 (N+2) (N+3)}
%                +\frac{2 P_{104} S_2}{3 N^2 (N+1)^2 (N+2) (N+3)}
                +\frac{P_{114}}{3 N^3 (N+1)^4 (N+2)^2 (N+3)}
                -\frac{48 S_3}{N (N+1)}
\nonumber \\ &&
                +\biggl(
                        -\frac{2 P_{107}}{3 N^3 (N+1)^3 (N+2) (N+3)}
                        -\frac{16 S_2}{N (N+1)}
                \biggr) S_1
                +\frac{16 S_{2,1}}{N (N+1)}
        \Biggr]
\nonumber \\ &&
        +\frac{16 (3 N+8) S_{2,1,1}-48 (N+4) S_{3,1}}{(N+1) (N+2) (N+3)}
%        -\frac{48 (N+4) S_{3,1}}{(N+1) (N+2) (N+3)}
%        +\frac{16 (3 N+8) S_{2,1,1}}{(N+1) (N+2) (N+3)}
        +\frac{4 \big(2 N^2+5 N-6\big)}{3 N (N+1) (N+2) (N+3)} S_1^4
\nonumber \\ &&
        +\frac{4 P_{106} S_{2,1}}{3 N^2 (N+1)^2 (N+2)^2 (N+3)^2}
        +\frac{P_{109} S_1^3+2 P_{116} S_3}{9 N^2 (N+1)^2 (N+2)^2 (N+3)^2 (N+4)}
%        +\frac{P_{109} S_1^3}{9 N^2 (N+1)^2 (N+2)^2 (N+3)^2 (N+4)}
%        +\frac{2 P_{116} S_3}{9 N^2 (N+1)^2 (N+2)^2 (N+3)^2 (N+4)}
\nonumber \\ &&
        -\frac{4 \big(3 N^2-13 N-54\big)}{N (N+1) (N+2) (N+3)} S_4
        +\frac{P_{117} S_2}{6 N^3 (N+1)^3 (N+2)^2 (N+3)^2 (N+4)}
\nonumber \\ &&
        -\frac{4 \big(7 N^2-2 N-48\big)}{N (N+1) (N+2) (N+3)} S_2^2
        +\frac{P_{120}}{12 N^4 (N+1)^5 (N+2)^3 (N+3)^2 (N+4)}
\nonumber \\ &&
        +\biggl(
                \frac{P_{108} S_2}{3 N^2 (N+1)^2 (N+2)^2 (N+3)^2 (N+4)}
                -\frac{8 (N-30) S_3}{3 N (N+2) (N+3)}
\nonumber \\ &&
                -\frac{8 \big(5 N^2+15 N-6\big) S_{2,1}}{N (N+1) (N+2) (N+3)}
                +\frac{P_{119}}{6 N^4 (N+1)^4 (N+2)^2 (N+3)^2 (N+4)}
        \biggr) S_1
\nonumber \\ &&
        +\biggl(
                \frac{8 \big(2 N^2+7 N-6\big) S_2}{N (N+1) (N+2) (N+3)}
                +\frac{P_{118}}{6 N^3 (N+1)^3 (N+2)^2 (N+3)^2 (N+4)}
        \biggr) S_1^2
\nonumber \\ &&
        +\biggl(
                \frac{P_{103}}{2 N^2 (N+1)^3 (N+2)}
                +\frac{\big(2 N^3+7 N^2-3 N-12\big) S_1}{N^2 (N+1)^2}
                +\frac{12 S_2}{N (N+1)}
        \biggr) \zeta_2
\nonumber \\ &&
        +\biggl(
                \frac{48 (3 N-2) S_1}{N (N+1) (N+2)} 
                -\frac{2 P_{113}}{3 N (N+1)^2 (N+2)^2 (N+3)^2 (N+4)}
        \biggr) \zeta_3
\nonumber \\ &&
        +(-1)^N \frac{8 P_{115}}{N^2 (N+1)^2 (N+2)^2 (N+3)^2 (N+4)} \bigl(
                S_{-3}
                +2 S_{-2,1}
                +2 \zeta_3
        \bigr)
\nonumber \\ &&
        +\frac{16 \big(5 N^2+11 N-6\big)}{N (N+1) (N+2) (N+3)} \Biggl[
                \frac{1}{N^2} \left(
                S_1\left({{\frac{1}{2}}}\right) S_1({{2}})
                -S_{1,1}\left({{\frac{1}{2},2}}\right)
                -S_{1,1}\left({{2,\frac{1}{2}}}\right)
                \right)
%                \frac{16}{N^2} S_1\left({{\frac{1}{2}}}\right) S_1({{2}})
%                -\frac{16}{N^2} S_{1,1}\left({{\frac{1}{2},2}}\right)
%                -\frac{16}{N^2} 
\nonumber \\ &&
                +\biggl(
                        S_3\left({{\frac{1}{2}}}\right)
                        -S_{2,1}\left({{1,\frac{1}{2}}}\right)
                \biggr) S_1({{2}})
                +S_2 S_{1,1}\left({{2,\frac{1}{2}}}\right)
                +S_{2,1,1}\left({{1,\frac{1}{2},2}}\right)
\nonumber \\ &&
                +\frac{1}{2} S_{1,1,1,1}\left({{2,\frac{1}{2},1,1}}\right)
                -S_{3,1}\left({{\frac{1}{2},2}}\right)
                -S_1({{2}}) \zeta_3
        \Biggr]
\nonumber \\ &&
        +\frac{2^N}{N (N+1)^2 (N+2)^2 (N+3)^2 (N+4)} \Biggl[
                \frac{8 P_{111}}{N} S_{2,1}\left({{1,\frac{1}{2}}}\right)
                -\frac{4 P_{112}}{N} S_2 S_1\left({{\frac{1}{2}}}\right)
\nonumber \\ &&
%                \frac{32 P_{105}}{N (N+1)^2 (N+2)^2 (N+3)^2 (N+4)} \zeta_3
%                -\frac{32 P_{105}}{N (N+1)^2 (N+2)^2 (N+3)^2 (N+4)} S_3\left({{\frac{1}{2}}}\right)
                +\frac{P_{110}}{N} \biggl(
                        -4 S_1^2 S_1\left({{\frac{1}{2}}}\right)
                        +8 S_1 S_{1,1}\left({{1,\frac{1}{2}}}\right)
                        -8 S_{2,1}\left({{\frac{1}{2},1}}\right)
                        -8 S_{1,1,1}\left({{1,1,\frac{1}{2}}}\right)
                \biggr)
\nonumber \\ &&
                +32 P_{105} \left(
                  \zeta_3-  S_3\left({{\frac{1}{2}}}\right) 
                \right)  
        \Biggr]
                -2^N \frac{8 \big(5 N^2+11 N-6\big)}{N^2 (N+1) (N+2) (N+3)} S_{1,1,1}\left({{\frac{1}{2},1,1}}\right)
\Biggr\},
\end{eqnarray}
%------------------------------------------------------------------------------------------------------------------------
with the polynomials
%------------------------------------------------------------------------------------------------------------------------
\begin{eqnarray}
P_{102}    &=& 21 N^4+62 N^3+127 N^2+54 N-144 \\
P_{103}    &=& 5 N^5+6 N^4+3 N^3+68 N^2+66 N-4 \\
P_{104}    &=& 28 N^5+155 N^4+390 N^3+129 N^2-294 N+144 \\
P_{105}    &=& N^7+5 N^6-16 N^5-17 N^4+411 N^3+956 N^2+524 N+48 \\
P_{106}    &=& 7 N^7+64 N^6+54 N^5-1192 N^4-4825 N^3-7224 N^2-4644 N-1296 \\
P_{107}    &=& 14 N^7+29 N^6-55 N^5-299 N^4-151 N^3-90 N^2-552 N-144 \\
P_{108}    &=& -20 N^8-153 N^7+1397 N^6+18797 N^5+75531 N^4+133444 N^3+101052 N^2
        \nonumber \\ &&
        +26064 N+3456 \\
P_{109}    &=& -4 N^8-217 N^7-1755 N^6-4067 N^5+4219 N^4+28932 N^3+35964 N^2
        \nonumber \\ &&
        +14544 N+3456 \\
P_{110}    &=& 2 N^8+10 N^7-37 N^6-95 N^5+543 N^4+1337 N^3+588 N^2+132 N+144 \\
P_{111} &=& 6 N^8+30 N^7-101 N^6-163 N^5+2187 N^4+5161 N^3+2684 N^2
        \nonumber \\ &&
+324 N+144 \\
P_{112} &=& 10 N^8+50 N^7-165 N^6-231 N^5+3831 N^4+8985 N^3+4780 N^2
        \nonumber \\ &&
+516 N+144 \\
P_{113} &=& 24 N^8+437 N^7+3791 N^6+19493 N^5+60209 N^4+102338 N^3+76868 N^2
        \nonumber \\ &&
        +11784 N+864 \\
P_{114} &=& -55 N^9-782 N^8-3418 N^7-6010 N^6-2529 N^5+2572 N^4-2062 N^3
        \nonumber \\ &&
        -5788 N^2-1176 N-144 \\
P_{115} &=& N^9+8 N^8-11 N^7-297 N^6-1179 N^5-2127 N^4-2071 N^3-1444 N^2
        \nonumber \\ &&
        -972 N-432 \\
P_{116} &=& 72 N^9+1166 N^8+7469 N^7+25731 N^6+56935 N^5+81259 N^4+43836 N^3
        \nonumber \\ &&
        -30564 N^2-20016 N+3456 \\
P_{117} &=& -136 N^{10}-2605 N^9-19654 N^8-74404 N^7-164938 N^6-202123 N^5
        \nonumber \\ &&
        -12264 N^4+263340 N^3+226992 N^2+60192 N+10368 \\
P_{118} &=& 28 N^{10}+859 N^9+7650 N^8+29956 N^7+51090 N^6+26005 N^5+3632 N^4
        \nonumber \\ &&
        +44412 N^3+54096 N^2+20448 N+3456 \\
P_{119} &=& 34 N^{12}-827 N^{11}-14079 N^{10}-74642 N^9-174832 N^8-177615 N^7-74407 N^6
        \nonumber \\ &&
        -124724 N^5-294188 N^4-256848 N^3-61248 N^2+33408 N+13824 \\
P_{120} &=& 485 N^{14}+15646 N^{13}+181445 N^{12}+1091656 N^{11}+3917869 N^{10}+9005374 N^9
        \nonumber \\ &&
        +13823259 N^8+14547908 N^7+10710534 N^6+5705512 N^5+2282792 N^4
        \nonumber \\ &&
        +507024 N^3-136576 N^2-95424 N-16128. 
\end{eqnarray}
%------------------------------------------------------------------------------------------------------------------------

Since powers $2^N$ emerge as factors in the present diagram at $O(\ep^0)$, one has to check 
whether the asymptotic expansion is regular.
For the constant term in $\ep$ one obtains 
%---------------------------------------------------------------------------------------------------------------------
\begin{eqnarray}
%%%texparser:LHS:testD2physAsyExp%%%
D_{2}^{\rm asy} &=&
\textcolor{blue}{\left(\frac{C_A}{2}-C_F\right)^2 T_F}
\biggl\{
        \ln^4(\bar{N}) \biggl[
                \frac{8}{3 N^2}
                -\frac{28}{3 N^3}
                +\frac{56}{3 N^4}
                -\frac{76}{3 N^5}
                +\frac{8}{3 N^6}
                +\frac{452}{3 N^7}
                -\frac{2344}{3 N^8}
\nonumber \\ &&
                +\frac{9044}{3 N^9}
                -\frac{31192}{3 N^{10}}
        \biggr]
        +\ln^3(\bar{N}) \biggl[
                -\frac{4}{9 N}
                -\frac{17}{N^2}
                +\frac{1165}{9 N^3}
                -\frac{4393}{9 N^4}
                +\frac{14257}{9 N^5}
                -\frac{227921}{45 N^6}
\nonumber \\ &&
                +\frac{758911}{45 N^7}
                -\frac{56214877}{945 N^8}
                +\frac{29853521}{135 N^9}
                -\frac{115308037}{135 N^{10}}
        \biggr]
        +\ln^2(\bar{N}) \biggl[
                \zeta_2 \biggl(
                        \frac{16}{N^2}
                        -\frac{40}{N^3}
\nonumber \\ &&
                        +\frac{16}{N^4}
                        +\frac{248}{N^5}
                        -\frac{1424}{N^6}
                        +\frac{5720}{N^7}
                        -\frac{20144}{N^8}
                        +\frac{66488}{N^9}
                        -\frac{211664}{N^{10}}
                \biggr)
                +\frac{14}{3 N}
                +\frac{427}{6 N^2}
                -\frac{4400}{9 N^3}
\nonumber \\ &&
                +\frac{25819}{12 N^4}
                -\frac{1423207}{180 N^5}
                +\frac{1126753}{40 N^6}
                -\frac{159213697}{1512 N^7}
                +\frac{1143011533}{2520 N^8}
                -\frac{2794326547}{1080 N^9}
\nonumber \\ &&
                +\frac{1617415434889}{75600 N^{10}}
        \biggr]
        +\ln(\bar{N}) \biggl[
                \zeta_3 \biggl(
                        \frac{184}{3 N^2}
                        -\frac{584}{3 N^3}
                        +\frac{1144}{3 N^4}
                        -\frac{1544}{3 N^5}
                        +\frac{184}{3 N^6}
                        +\frac{9016}{3 N^7}
\nonumber \\ &&
                        -\frac{46856}{3 N^8}
                        +\frac{180856}{3 N^9}
                        -\frac{623816}{3 N^{10}}
                \biggr)
                +\zeta_2 \biggl(
                        -\frac{14}{3 N}
                        +\frac{176}{3 N^2}
                        +\frac{860}{3 N^3}
                        -\frac{5012}{3 N^4}
                        +\frac{5696}{N^5}
\nonumber \\ &&
                        -\frac{232096}{15 N^6}
                        +\frac{33610}{N^7}
                        -\frac{13379896}{315 N^8}
                        -\frac{33628838}{315 N^9}
                        +\frac{385560506}{315 N^{10}}
                \biggr)
                +\frac{17}{3 N}
                -\frac{1379}{6 N^2}
                +\frac{10961}{9 N^3}
\nonumber \\ &&
                -\frac{43724}{9 N^4}
                +\frac{2483923}{135 N^5}
                -\frac{17564599}{240 N^6}
                +\frac{3512318441}{10800 N^7}
                -\frac{99300815819}{56700 N^8}
                +\frac{19967049648257}{1587600 N^9}
\nonumber \\ &&
                -\frac{110976048453281}{907200 N^{10}}
        \biggr]
        +\zeta_4 \biggl(
                -\frac{98}{N^2}
                +\frac{668}{N^3}
                -\frac{1898}{N^4}
                +\frac{4628}{N^5}
                -\frac{10898}{N^6}
                +\frac{25868}{N^7}
                -\frac{63098}{N^8}
\nonumber \\ &&
                +\frac{159428}{N^9}
                -\frac{417698}{N^{10}}
        \biggr)
        +\zeta_3 \biggl(
                -\frac{122}{9 N}
                -\frac{1216}{3 N^2}
                +\frac{13520}{9 N^3}
                -\frac{51074}{9 N^4}
                +\frac{184154}{9 N^5}
                -\frac{3387067}{45 N^6}
\nonumber \\ &&
                +\frac{12885377}{45 N^7}
                -\frac{1057710317}{945 N^8}
                +\frac{4204295167}{945 N^9}
                -\frac{33712183831}{1890 N^{10}}
        \biggr)
        +\zeta_2 \biggl(
                -\frac{121}{6 N}
                -\frac{62}{3 N^2}
\nonumber \\ &&
                -\frac{1871}{18 N^3}
                +\frac{17038}{9 N^4}
                -\frac{1981967}{180 N^5}
                +\frac{737924}{15 N^6}
                -\frac{11255201}{54 N^7}
                +\frac{398476655}{378 N^8}
                -\frac{8606975273}{1080 N^9}
\nonumber \\ &&
                +\frac{231958440413}{2700 N^{10}}
        \biggr)
        +\frac{485}{12 N}
        +\frac{1853}{4 N^2}
        -\frac{86443}{36 N^3}
        +\frac{997441}{108 N^4}
        -\frac{4347068}{135 N^5}
        +\frac{2572017299}{21600 N^6}
\nonumber \\ &&
        -\frac{1770870297697}{3402000 N^7}
        +\frac{4417562190127}{1512000 N^8}
        -\frac{14791421053852637}{666792000 N^9}
\nonumber \\ &&
        +\frac{149148442921884047}{666792000 N^{10}}
\biggr\} + O\left(\frac{1}{N^{11}} \ln^4(\bar{N})\right)
\end{eqnarray}
%---------------------------------------------------------------------------------------------------------------------
a regular representation. The pole terms contain harmonic sums only.

For diagram~3 structurally similar results are obtained as for diagram~2. It is given by
%---------------------------------------------------------------------------------------------------------------------------
\begin{eqnarray}
%%%texparser:LHS:testD3phys%%%
D_{3} &=&
\big(1+(-1)^N\big) \textcolor{blue}{\left(\frac{C_A}{2}-C_F\right)^2 T_F} \Biggl\{
        \frac{1}{\varepsilon^3} \frac{64}{3 (N+1) (N+2)}
        +\frac{1}{\varepsilon^2} \Biggl[
                \frac{32 (N+3) (3 N+4)}{3 (N+1)^2 (N+2)^2}
\nonumber \\ &&
                -\frac{8 \big(5 N^2+11 N-6\big) S_1}{3 N (N+1)^2 (N+2)}
                -\frac{32 S_2}{N (N+1)}
        \Biggr]
        +\frac{1}{\varepsilon} \Biggl[
                -\frac{2 \big(23 N^2+255 N-18\big) S_2}{3 N (N+1) (N+2) (N+3)}
\nonumber \\ &&
                +\frac{2 \big(3 N^3-2 N^2+9 N-2\big) S_1^2}{N (N+1)^2 (N+2) (N+3)}
                +\frac{2 P_{123}}{3 (N+1)^3 (N+2)^3 (N+3)}
                +\frac{48 S_3-16 S_{2,1}}{N (N+1)}
%                +\frac{48 S_3}{N (N+1)}
%                -\frac{16 S_{2,1}}{N (N+1)}
\nonumber \\ &&
                +\biggl(
                        -\frac{2 P_{122}}{3 N (N+1)^3 (N+2)^2 (N+3)}
                        +\frac{16 S_2}{N (N+1)}
                \biggr) S_1
                +\frac{8}{(N+1) (N+2)} \zeta_2
        \Biggr]
\nonumber \\ &&
        +\frac{12 \big(N^2-3 N-14\big)}{N (N+1) (N+2) (N+3)} S_4
        -\frac{4 \big(N^2+2 N-6\big)}{3 N (N+1) (N+2) (N+3)} S_1^4
\nonumber \\ &&
        -\frac{48 \big(N^2+3 N+1\big)}{N (N+1) (N+2) (N+3)} S_{2,1,1}
        +\frac{16 \big(N^2+5 N-3\big)}{N (N+1) (N+2) (N+3)} S_{3,1}
\nonumber \\ &&
        +\frac{P_{134}}{3 (N+1)^4 (N+2)^4 (N+3)^2 (N+4)}
        -\frac{4 P_{121} S_{2,1}}{3 N (N+1)^2 (N+2)^2 (N+3)^2}
\nonumber \\ &&
        +\frac{72 P_{126} \left(S_{-3}+2 S_{-2,1}\right)+P_{127} S_1^3-2 P_{129} S_3}{9 N (N+1)^2 (N+2)^2 (N+3)^2 (N+4)}
%        +\frac{8 P_{126} S_{-3}}{N (N+1)^2 (N+2)^2 (N+3)^2 (N+4)}
%        +\frac{16 P_{126} S_{-2,1}}{N (N+1)^2 (N+2)^2 (N+3)^2 (N+4)}
%        +\frac{P_{127} S_1^3}{9 N (N+1)^2 (N+2)^2 (N+3)^2 (N+4)}
%        -\frac{2 P_{129} S_3}{9 N (N+1)^2 (N+2)^2 (N+3)^2 (N+4)}
        +\frac{4 \big(8 N^2+7 N-30\big)}{N (N+1) (N+2) (N+3)} S_2^2
\nonumber \\ &&
        +\biggl(
                \frac{24 \big(N^2+3 N-2\big)}{N (N+1) (N+2) (N+3)} S_{2,1}
                +\frac{P_{124} S_2}{N (N+1)^2 (N+2)^2 (N+3)^2 (N+4)}
\nonumber \\ &&
                +\frac{8 \big(11 N^2+13 N+6\big)}{3 N (N+1) (N+2) (N+3)} S_3
                +\frac{P_{135}}{6 N (N+1)^4 (N+2)^3 (N+3)^2 (N+4)}
        \biggr) S_1
\nonumber \\ &&
        +\biggl(
                \frac{72 S_2}{N (N+1) (N+2) (N+3)}
                +\frac{P_{130}}{6 N (N+1)^3 (N+2)^2 (N+3)^2 (N+4)}
        \biggr) S_1^2
\nonumber \\ &&
        +\frac{P_{136} S_2}{6 (N-1)^2 N^2 (N+1)^3 (N+2)^2 (N+3)^2 (N+4)}
        +\biggl(
                \frac{4 (N+3) (3 N+4)}{(N+1)^2 (N+2)^2}
\nonumber \\ &&
                +\frac{-5 N^2-11 N+6}{N (N+1)^2 (N+2)} S_1
                -\frac{12 S_2}{N (N+1)}
        \biggr) \zeta_2
        +\biggl(
                -\frac{16 (7 N-4)}{N (N+1) (N+2)} S_1
\nonumber \\ &&
                +\frac{8 P_{128}}{3 N (N+1)^2 (N+2)^2 (N+3)^2 (N+4)}
        \biggr) \zeta_3
        +\frac{5 N+13}{(N+1) (N+2) (N+3)} 
\nonumber \\ &&
\times
\Biggl[
                \frac{16 \big(2 N^2-2 N+1\big)}{(N-1)^2 N^2} \biggl(
                        -S_1\left({{\frac{1}{2}}}\right) S_1({{2}}) 
                        +S_{1,1}\left({{\frac{1}{2},2}}\right)
                        +S_{1,1}\left({{2,\frac{1}{2}}}\right)
                \biggr)
\nonumber \\ &&
                +\biggl(
                        -16 S_3\left({{\frac{1}{2}}}\right)
                        +16 S_{2,1}\left({{1,\frac{1}{2}}}\right)
                \biggr) S_1({{2}})
                -16 S_2 S_{1,1}\left({{2,\frac{1}{2}}}\right)
                +16 S_{3,1}\left({{\frac{1}{2},2}}\right)
\nonumber \\ &&
                -16 S_{2,1,1}\left({{1,\frac{1}{2},2}}\right)
                -8 S_{1,1,1,1}\left({{2,\frac{1}{2},1,1}}\right)
                +16 S_1({{2}}) \zeta_3
        \Biggr]
\nonumber \\ &&
        +2^N \Biggl[
                \frac{P_{131}}{(N-1) N (N+1)^2 (N+2)^2 (N+3)^2 (N+4)} \biggl(
                        4 S_{1,1,1}\left({{1,1,\frac{1}{2}}}\right)
\nonumber \\ &&
                        +2 S_1^2 S_1\left({{\frac{1}{2}}}\right)
                        -4 S_1 S_{1,1}\left({{1,\frac{1}{2}}}\right)
                        +4 S_{2,1}\left({{\frac{1}{2},1}}\right)
                \biggr)
\nonumber \\ &&
                +\frac{4 (3 N-2) (5 N+13)}{(N-1) N (N+1) (N+2) (N+3)} S_{1,1,1}\left({{\frac{1}{2},1,1}}\right)
\nonumber \\ &&
                +\frac{32 P_{125}}{N (N+1)^2 (N+2)^2 (N+3)^2 (N+4)} \left(S_3\left({{\frac{1}{2}}}\right)-\zeta_3\right)
\nonumber \\ &&
                -\frac{4 P_{132}}{(N-1) N (N+1)^2 (N+2)^2 (N+3)^2 (N+4)} S_{2,1}\left({{1,\frac{1}{2}}}\right)
\nonumber \\ &&
                +\frac{2 P_{133}}{(N-1) N (N+1)^2 (N+2)^2 (N+3)^2 (N+4)} S_2 S_1\left({{\frac{1}{2}}}\right)
%                -\frac{32 P_{125}}{N (N+1)^2 (N+2)^2 (N+3)^2 (N+4)} \zeta_3
        \Biggr]
\Biggr\},
\end{eqnarray}
%---------------------------------------------------------------------------------------------------------------------------
with the polynomials
%---------------------------------------------------------------------------------------------------------------------------
\begin{eqnarray}
P_{121}    &=& 5 N^5-27 N^4-689 N^3-2925 N^2-4524 N-2304 \\
P_{122}    &=& 41 N^5+107 N^4-293 N^3-1615 N^2-2108 N-708 \\
P_{123}    &=& 79 N^5+255 N^4-1019 N^3-6147 N^2-10140 N-5604 \\
P_{124}    &=& -25 N^6-607 N^5-5059 N^4-19141 N^3-34876 N^2-28660 N-7968 \\
P_{125}    &=& N^6-8 N^5+55 N^4+664 N^3+1612 N^2+1396 N+432 \\
P_{126}    &=& N^6+N^5-26 N^4-73 N^3-11 N^2+84 N-12 \\
P_{127}    &=& 85 N^6+691 N^5+1031 N^4-5311 N^3-19668 N^2-20556 N-5184 \\
P_{128}    &=& 107 N^6+1271 N^5+6355 N^4+16369 N^3+22170 N^2+15600 N+5184 \\
P_{129}    &=& 167 N^6+3233 N^5+23953 N^4+78067 N^3+110424 N^2+56844 N+5616 \\
P_{130} &=& -347 N^7-3270 N^6-10312 N^5-2618 N^4+44967 N^3+72368 N^2
\nonumber \\ &&
+28140 N-2880 \\
P_{131} &=& 4 N^7-51 N^6+73 N^5+1647 N^4+2287 N^3-1764 N^2-3252 N-1104 \\
P_{132} &=& 12 N^7-123 N^6+577 N^5+6519 N^4+9871 N^3-3492 N^2-10964 N-4560 \\
P_{133} &=& 20 N^7-195 N^6+1081 N^5+11391 N^4+17455 N^3-5220 N^2-18676 N
\nonumber\\ &&
-8016 \\
P_{134} &=& -1461 N^9-24515 N^8-173198 N^7-665974 N^6-1494817 N^5-1892023 N^4
\nonumber \\ &&
-993988 N^3+501936 N^2+966024 N+390240 \\
P_{135} &=& 1169 N^9+16377 N^8+90672 N^7+234070 N^6+192009 N^5-435259 N^4
\nonumber \\ &&
-1371474 N^3-1622932 N^2-938136 N-213984 \\
P_{136} &=& 181 N^{10}+5072 N^9+28113 N^8+58856 N^7-26013 N^6-318280 N^5
\nonumber \\ &&
-306085 N^4
+151152 N^3+127548 N^2-105312 N-29952.
\end{eqnarray}
%---------------------------------------------------------------------------------------------------------------------------

The asymptotic expansion of diagram~3 reads
%--------------------------------------------------------------------------------------------------
\begin{eqnarray}
%%%texparser:LHS:testD3physAsyExp%%%
D_{3} &=&
\big(1+(-1)^N\big) \textcolor{blue}{\left(\frac{C_A}{2}-C_F\right)^2 T_F} \Biggl\{
        \ln(\bar{N})^4 \biggl[
                -\frac{4}{3 N^2}
                +\frac{16}{3 N^3}
                -\frac{28}{3 N^4}
                +\frac{16}{3 N^5}
                +\frac{116}{3 N^6}
\nonumber \\ &&
                -\frac{704}{3 N^7}
                +\frac{2852}{3 N^8}
                -\frac{10064}{3 N^9}
                +\frac{33236}{3 N^{10}}
        \biggr]
        +\ln(\bar{N})^3 \biggl[
                \frac{85}{9 N^2}
                -\frac{77}{N^3}
                +\frac{2825}{9 N^4}
                -\frac{10141}{9 N^5}
\nonumber \\ &&
                +\frac{185483}{45 N^6}
                -\frac{79433}{5 N^7}
                +\frac{60380711}{945 N^8}
                -\frac{248609357}{945 N^9}
                +\frac{147509366}{135 N^{10}}
        \biggr]
        +\ln(\bar{N})^2 \biggl[
                \zeta_2 \biggl(
                        \frac{72}{N^4}
\nonumber \\ &&
                        -\frac{432}{N^5}
                        +\frac{1800}{N^6}
                        -\frac{6480}{N^7}
                        +\frac{21672}{N^8}
                        -\frac{69552}{N^9}
                        +\frac{217800}{N^{10}}
                \biggr)
                -\frac{347}{6 N^2}
                +\frac{1273}{3 N^3}
                -\frac{69181}{36 N^4}
\nonumber \\ &&
                +\frac{27665}{4 N^5}
                -\frac{1567309}{72 N^6}
                +\frac{20509637}{360 N^7}
                -\frac{514689451}{7560 N^8}
                -\frac{359860531}{504 N^9}
                +\frac{764716299253}{75600 N^{10}}
        \biggr]
\nonumber \\ &&
        +\ln(\bar{N}) \biggl[
                \zeta_3 \biggl(
                        -\frac{104}{3 N^2}
                        +\frac{344}{3 N^3}
                        -\frac{584}{3 N^4}
                        +\frac{344}{3 N^5}
                        +\frac{2296}{3 N^6}
                        -\frac{14056}{3 N^7}
                        +\frac{57016}{3 N^8}
                        -\frac{201256}{3 N^9}
\nonumber \\ &&
                        +\frac{664696}{3 N^{10}}
                \biggr)
                +\zeta_2 \biggl(
                        -\frac{30}{N^2}
                        -\frac{198}{N^3}
                        +\frac{898}{N^4}
                        -\frac{2190}{N^5}
                        +\frac{2038}{N^6}
                        +\frac{14826}{N^7}
                        -\frac{644044}{5 N^8}
\nonumber \\ &&
                        +\frac{3562734}{5 N^9}
                        -\frac{23710824}{7 N^{10}}
                \biggr)
                +\frac{1169}{6 N^2}
                -\frac{3628}{3 N^3}
                +\frac{43766}{9 N^4}
                -\frac{210899}{12 N^5}
                +\frac{121102043}{2160 N^6}
\nonumber \\ &&
                -\frac{157661837}{1200 N^7}
                -\frac{2783060641}{28350 N^8}
                +\frac{1969090159787}{396900 N^9}
                -\frac{4511474336429}{78400 N^{10}}
        \biggr]
        +\zeta_4 \biggl(
                \frac{68}{N^2}
\nonumber \\ &&
                -\frac{518}{N^3}
                +\frac{1508}{N^4}
                -\frac{3758}{N^5}
                +\frac{9068}{N^6}
                -\frac{22118}{N^7}
                +\frac{55508}{N^8}
                -\frac{144158}{N^9}
                +\frac{387068}{N^{10}}
        \biggr)
\nonumber \\ &&
        +\zeta_3 \biggl(
                \frac{1970}{9 N^2}
                -\frac{846}{N^3}
                +\frac{38536}{9 N^4}
                -\frac{175460}{9 N^5}
                +\frac{3783667}{45 N^6}
                -\frac{1772353}{5 N^7}
                +\frac{1395767797}{945 N^8}
\nonumber \\ &&
                -\frac{5769027547}{945 N^9}
                +\frac{47377155641}{1890 N^{10}}
        \biggr)
        +\zeta_2 \biggl(
                \frac{253}{6 N^2}
                +\frac{1619}{6 N^3}
                -\frac{7021}{3 N^4}
                +\frac{25097}{3 N^5}
                -\frac{129287}{12 N^6}
\nonumber \\ &&
                -\frac{2176661}{20 N^7}
                +\frac{37152809}{28 N^8}
                -\frac{4537942783}{420 N^9}
                +\frac{206833895747}{2520 N^{10}}
        \biggr)
        -\frac{487}{N^2}
        +\frac{10497}{4 N^3}
\nonumber \\ &&
        -\frac{359903}{36 N^4}
        +\frac{1809355}{54 N^5}
        -\frac{451092851}{4320 N^6}
        +\frac{136197543}{500 N^7}
        -\frac{354166478429}{1944000 N^8}
\nonumber \\ &&
        -\frac{458350390393967}{74088000 N^9}
        +\frac{28558035712221709}{333396000 N^{10}}
\Biggr\} + O\left(\frac{1}{N^{11}} \ln^4(\bar{N})\right).
\end{eqnarray}
%--------------------------------------------------------------------------------------------------

%>>D5:

One obtains for diagram~5
%----------------------------------------------------------------------------------------------------------------------------------
\begin{eqnarray}
%% Present Notation: D5
%%%texparser:LHS:testD11phys%%%
%D_{11} &=&
D_5 &=&
\textcolor{blue}{C_A^2 T_F} \Biggl\{
        \frac{1}{\varepsilon^3} \Biggl[
                -\frac{16 P_{146}}{3 N^3 (N+1)^2 (N+2)}
                -\frac{32 P_{141} S_1}{3 N^2 (N+1)^2 (N+2)}
                +\frac{32 \left(S_2-S_1^2\right)}{3 N} 
%                -\frac{32 S_1^2}{3 N}
%                +\frac{32 S_2}{3 N}
        \Biggr]
\nonumber \\ &&
        +\frac{1}{\varepsilon^2} \Biggl[
                \frac{8 P_{158}}{3 N^4 (N+1)^3 (N+2)^2}
                -\frac{8 \big(2 N^4+7 N^3-7 N-4\big) S_1^2}{3 N^2 (N+1)^2 (N+2)}
                -\frac{16 S_1^3}{3 N}
                +\frac{32 S_3}{3 N}
\nonumber \\ &&
                -\frac{8 P_{142} S_2}{N^2 (N+1)^2 (N+2)}
                +\biggl(
                        \frac{8 P_{153}}{3 N^3 (N+1)^3 (N+2)^2}
                        -\frac{16 S_2}{3 N}
                \biggr) S_1
        \Biggr]
\nonumber \\ &&
        +\frac{1}{\varepsilon} \Biggl[
                -\frac{4 P_{163}}{3 N^5 (N+1)^4 (N+2)^3}
                +\frac{8 P_{148} S_3-4 P_{139} S_1^3-36 P_{145} S_{2,1}}{9 N^2 (N+1)^2 (N+2)}
%                -\frac{4 P_{139} S_1^3}{9 N^2 (N+1)^2 (N+2)}
%                -\frac{ 4 P_{145} S_{2,1}}{N^2 (N+1)^2 (N+2)}
%                +\frac{8 P_{148} S_3}{9 N^2 (N+1)^2 (N+2)}
                +\frac{26 S_2^2}{3 N}
                +\frac{52 S_4}{3 N}
\nonumber \\ &&
                +\frac{2 P_{156} S_2}{N^3 (N+1)^3 (N+2)^2}
                +\biggl(
                        -\frac{4 P_{160}}{3 N^4 (N+1)^4 (N+2)^3}
                        -\frac{16 S_3}{9 N}
                        -\frac{16 S_{2,1}}{N}
\nonumber \\ &&
                        +\frac{4 \big(N^2-3\big) \big(2 N^2+N-4\big) S_2}{N^2 (N+1)^2 (N+2)}
                \biggr) S_1
                +\biggl(
                        \frac{2 P_{154}}{3 N^3 (N+1)^3 (N+2)^2}
                        -\frac{20 S_2}{3 N}
                \biggr) S_1^2
\nonumber \\ &&
                -\frac{14 S_1^4}{9 N}
                +\biggl(
                        -\frac{2 P_{146}}{N^3 (N+1)^2 (N+2)}
                        -\frac{4 P_{141} S_1}{N^2 (N+1)^2 (N+2)}
                        -\frac{4 S_1^2}{N}
                        +\frac{4 S_2}{N}
                \biggr) \zeta_2
        \Biggr]
\nonumber \\ &&
        +\frac{2 P_{165}}{3 N^6 (N+1)^5 (N+2)^4}
        +\frac{P_{138} S_1^4+9 P_{137} S_2^2-144 P_{143} S_{3,1}-36 P_{144} S_{2,1,1}+18 P_{150} S_4}{18 N^2 (N+1)^2 (N+2)}
%        +\frac{P_{137} S_2^2}{2 N^2 (N+1)^2 (N+2)}
%        +\frac{P_{138} S_1^4}{18 N^2 (N+1)^2 (N+2)}
%        -\frac{8 P_{143} S_{3,1}}{N^2 (N+1)^2 (N+2)}
%        -\frac{2 P_{144} S_{2,1,1}}{N^2 (N+1)^2 (N+2)}
%        +\frac{P_{150} S_4}{N^2 (N+1)^2 (N+2)}
\nonumber \\ &&
        +\frac{18 P_{157} S_{2,1}-2 P_{159} S_3}{9 N^3 (N+1)^3 (N+2)^2}
%        +\frac{2 P_{157} S_{2,1}}{N^3 (N+1)^3 (N+2)^2}
%        -\frac{2 P_{159} S_3}{9 N^3 (N+1)^3 (N+2)^2}
        +\biggl(
                -\frac{8 \big(N^3-3 N-4\big) S_{2,1}}{N^2 (N+1)^2 (N+2)}
                +\frac{4 P_{149} S_3}{9 N^2 (N+1)^2 (N+2)}
\nonumber \\ &&
                +\frac{P_{152} S_2}{N^3 (N+1)^3 (N+2)^2}
                +\frac{2 P_{164}}{3 (N-1) N^5 (N+1)^5 (N+2)^4}
                -\frac{19 S_2^2}{3 N}
                -\frac{14 S_4}{3 N}
\nonumber \\ &&
                -\frac{16 S_{3,1}}{N}
                +\frac{8 S_{2,1,1}}{N}
        \biggr) S_1
        +\biggl(
                \frac{P_{162}}{3 (N-1) N^4 (N+1)^4 (N+2)^3}
                -\frac{4 S_3}{N}
                -\frac{12 S_{2,1}}{N}
\nonumber \\ &&
                +\frac{P_{140} S_2}{N^2 (N+1)^2 (N+2)}
        \biggr) S_1^2
        +\biggl(
                \frac{P_{155}}{9 N^3 (N+1)^3 (N+2)^2}
                -\frac{26 S_2}{9 N}
        \biggr) S_1^3
        -\frac{S_1^5}{3 N}
\nonumber \\ &&
        +\biggl(
                \frac{P_{161}}{(N-1) N^4 (N+1)^4 (N+2)^3} 
                +\frac{236 S_3}{9 N}
                +\frac{4 S_{2,1}}{N}
        \biggr) S_2
        +\frac{32 S_5}{N}
        +\frac{8 S_{4,1}}{N}
\nonumber \\ &&
        -\frac{16 S_{2,2,1}}{N}
        -\frac{16 S_{3,1,1}}{N}
        +\biggl[
                \frac{P_{158}}{N^4 (N+1)^3 (N+2)^2}
                -\frac{3 P_{142} S_2}{N^2 (N+1)^2 (N+2)}
                +\frac{4 S_3}{N} 
\nonumber \\ &&
                -\frac{\big(2 N^4+7 N^3-7 N-4\big) S_1^2}{N^2 (N+1)^2 (N+2)}
                +\biggl(
                        \frac{P_{153}}{N^3 (N+1)^3 (N+2)^2}
                        -\frac{2 S_2}{N}
                \biggr) S_1
                -\frac{2 S_1^3}{N}
        \biggr] \zeta_2
\nonumber \\ &&
        +\biggl(
                \frac{2 P_{151}}{3 N^3 (N+1)^2 (N+2)}
                -\frac{4 P_{147} S_1}{3 N^2 (N+1)^2 (N+2)}
                +\frac{4 S_1^2}{3 N}
                -\frac{4 S_2}{3 N}
        \biggr) \zeta_3
\Biggr\},
\end{eqnarray}
%----------------------------------------------------------------------------------------------------------------------------------
where
%----------------------------------------------------------------------------------------------------------------------------------
\begin{eqnarray}
P_{137}    &=& -6 N^4-103 N^3-290 N^2-139 N+108 \\
P_{138}    &=& -2 N^4+11 N^3+90 N^2+79 N+4 \\
P_{139}    &=& 2 N^4+N^3-30 N^2-31 N-4 \\
P_{140}    &=& 2 N^4+5 N^3+10 N^2+13 N+12 \\
P_{141}    &=& 2 N^4+10 N^3+15 N^2+5 N-4 \\
P_{142}    &=& 2 N^4+11 N^3+20 N^2+9 N-4 \\
P_{143}    &=& 2 N^5+N^4-9 N^3-17 N^2-15 N-12 \\
P_{144}    &=& 2 N^5+7 N^4-9 N^3-62 N^2-36 N+24 \\
P_{145}    &=& 2 N^5+7 N^4+11 N^3-2 N^2-12 N-8 \\
P_{146} &=& 2 N^5+10 N^4+19 N^3+20 N^2+20 N+8 \\
P_{147} &=& 6 N^5+7 N^4-37 N^3-111 N^2-71 N+4 \\
P_{148} &=& 18 N^5-5 N^4-205 N^3-438 N^2-260 N-8 \\
P_{149} &=& 18 N^5+13 N^4-148 N^3-378 N^2-221 N+28 \\
P_{150} &=& 32 N^5-14 N^4-367 N^3-742 N^2-439 N-36 \\
P_{151} &=& 6 N^6+68 N^5+124 N^4-197 N^3-442 N^2-160 N+8 \\
P_{152} &=& -8 N^7-35 N^6-39 N^5+192 N^4+476 N^3+384 N^2+184 N+48 \\
P_{153} &=& 8 N^7+71 N^6+215 N^5+342 N^4+286 N^3+104 N^2+32 N+16 \\
P_{154} &=& 8 N^7+107 N^6+419 N^5+1020 N^4+1516 N^3+1088 N^2+296 N+16 \\
P_{155} &=& 8 N^7+179 N^6+827 N^5+2376 N^4+3976 N^3+3056 N^2+824 N+16 \\
P_{156} &=& 16 N^7+105 N^6+265 N^5+318 N^4+102 N^3-128 N^2-96 N-16 \\
P_{157} &=& 3 N^8+2 N^7+13 N^6+159 N^5+503 N^4+698 N^3+448 N^2+136 N+16 \\
P_{158} &=& 8 N^8+71 N^7+244 N^6+385 N^5+326 N^4+100 N^3-140 N^2-160 N-48 \\
P_{159} &=& 72 N^8-56 N^7-1361 N^6-3485 N^5-3627 N^4-355 N^3+2656 N^2+2188 N
\nonumber \\ &&
+464 \\
P_{160} &=& 22 N^{10}+226 N^9+778 N^8+1363 N^7+1337 N^6+344 N^5-952 N^4-1276 N^3
\nonumber \\ &&
-744 N^2-208 N-32 \\
P_{161} &=& -42 N^{11}-362 N^{10}-1098 N^9-1237 N^8+980 N^7+4385 N^6+4096 N^5
\nonumber \\ &&
-130 N^4-2272 N^3-912 N^2+320 N+160 \\
P_{162} &=& -22 N^{11}-324 N^{10}-1188 N^9-2469 N^8-3046 N^7+1167 N^6+8880 N^5
\nonumber \\ &&
+8778 N^4+1952 N^3-1328 N^2-704 N-32 \\
P_{163} &=& 30 N^{11}+287 N^{10}+1292 N^9+3123 N^8+4363 N^7+3616 N^6+1436 N^5
\nonumber \\ &&
+152 N^4+1004 N^3+1736 N^2+1040 N+224 \\
P_{164} &=& 35 N^{14}+454 N^{13}+1614 N^{12}+2010 N^{11}-1156 N^{10}-9046 N^9-20721 N^8
\nonumber \\ &&
-27766 N^7-18672 N^6-1244 N^5+5180 N^4+896 N^3-1056 N^2-448 N
\nonumber \\ &&
-64 \\
P_{165} &=& 67 N^{14}+801 N^{13}+5001 N^{12}+18294 N^{11}+41329 N^{10}+59947 N^9+56854 N^8
\nonumber \\ &&
+36348 N^7+18704 N^6+7500 N^5-6492 N^4-16704 N^3-14240 N^2-5824 N
\nonumber \\ &&
-960.
\end{eqnarray}
%----------------------------------------------------------------------------------------------------------------------------------

Diagram~6 has the representation
%--------------------------------------------------------------------------------------------------------------------------------
\begin{eqnarray}
%%%texparser:LHS:testD6phys%%%
D_{6} &=&
\textcolor{blue}{C_A^2 T_F} \Biggl\{
        \frac{1}{\varepsilon^3} \Biggl[
                \frac{P_{176}}{3 N^3 (N+1)^3}
                -\frac{2 \big(2 N^3+N^2+2 N+6\big) S_1}{3 N^2 (N+1)^2}
                -\frac{4 S_1^2}{3 N}
                -\frac{2 (5 N+2) S_2}{3 N (N+1)}
        \Biggr]
\nonumber \\ &&
        +\frac{1}{\varepsilon^2} \Biggl[
                \frac{P_{185}}{36 N^4 (N+1)^4 (N+2)}
                +\frac{3 P_{166} S_2+P_{167} S_1^2}{6 N^2 (N+1)^2 (N+2)}
%                +\frac{P_{166} S_2}{2 N^2 (N+1)^2 (N+2)}
%                +\frac{P_{167} S_1^2}{6 N^2 (N+1)^2 (N+2)}
                -\frac{(13 N+4)}{3 N (N+1)} S_3
\nonumber \\ &&
                +\biggl(
                        \frac{P_{178}}{18 N^3 (N+1)^3 (N+2)}
                        -\frac{2 (2 N+1) S_2}{N (N+1)}
                \biggr) S_1
                -\frac{2 S_1^3}{3 N}
                +\frac{2 S_{2,1}}{N+1}
        \Biggr]
\nonumber \\ &&
        +\frac{1}{\varepsilon} \Biggl[
                -\frac{47 N^3+107 N^2+76 N+28}{6 N (N+1)^2 (N+2)} S_{2,1}
                +\frac{P_{168} S_1^3+P_{173} S_3}{36 N^2 (N+1)^2 (N+2)}
%                +\frac{P_{168} S_1^3}{36 N^2 (N+1)^2 (N+2)}
%                +\frac{P_{173} S_3}{36 N^2 (N+1)^2 (N+2)}
\nonumber \\ &&
                +\frac{P_{183} S_2}{24 N^3 (N+1)^3 (N+2)^2 (N+3)}
                +\frac{P_{189}}{432 N^5 (N+1)^5 (N+2)^2 (N+3)}
\nonumber \\ &&
                +\biggl(
                        \frac{P_{171} S_2}{12 N^2 (N+1)^2 (N+2)} 
%                        \frac{P_{171}}{12}
                        +\frac{P_{186}}{108 N^4 (N+1)^4 (N+2)^2 (N+3)}
                        -\frac{(41 N+14) S_3}{9 N (N+1)}
\nonumber \\ &&
                        -\frac{2 (N+2) S_{2,1}}{N (N+1)}
                \biggr) S_1
                +\biggl(
                        \frac{P_{184}}{24 N^3 (N+1)^3 (N+2)^2 (N+3)}
                        -\frac{(13 N+7) S_2}{6 N (N+1)}
                \biggr) S_1^2
\nonumber \\ &&
                -\frac{7}{36 N} S_1^4
                -\frac{49 N+31}{12 N (N+1)} S_2^2
                -\frac{43 N+19}{6 N (N+1)} S_4
                -\frac{4}{N} S_{3,1}
                +\frac{9 N+8}{N (N+1)} S_{2,1,1}
\nonumber \\ &&
                +\biggl(
                        \frac{P_{176}}{8 N^3 (N+1)^3}
                        -\frac{2 N^3+N^2+2 N+6}{4 N^2 (N+1)^2} S_1
                        -\frac{S_1^2}{2 N}
                        -\frac{(5 N+2) S_2}{4 N (N+1)}
                \biggr) \zeta_2
        \Biggr]
\nonumber \\ &&
        +\frac{6 N^2+35 N+30}{2 N (N+1) (N+2)} S_{3,1}
        -\frac{191 N^3+827 N^2+1084 N+460}{12 N (N+1)^2 (N+2)} S_{2,1,1}
\nonumber \\ &&
        +\frac{P_{170} S_1^4+18 P_{172} S_4+3 P_{175} S_2^2}{288 N^2 (N+1)^2 (N+2)}
%        +\frac{P_{170} S_1^4}{288 N^2 (N+1)^2 (N+2)}
%        +\frac{P_{172} S_4}{16 N^2 (N+1)^2 (N+2)}
%        +\frac{P_{175} S_2^2}{96 N^2 (N+1)^2 (N+2)}
        +\frac{P_{191}}{5184 N^6 (N+1)^6 (N+2)^3 (N+3)}
\nonumber \\ &&
        +\frac{P_{179} S_{2,1}}{6 N^2 (N+1)^3 (N+2)^2}
        +\frac{P_{181} S_3}{216 N^3 (N+1)^3 (N+2)^2 (N+3)}
\nonumber \\ &&
        +\biggl(
                \frac{7 N^3+130 N^2+233 N+98}{6 N (N+1)^2 (N+2)} S_{2,1}
                +\frac{P_{174} S_3}{36 N^2 (N+1)^2 (N+2)}
\nonumber \\ &&
                +\frac{P_{180} S_2}{144 N^3 (N+1)^3 (N+2)^2 (N+3)}
                +\frac{P_{190}}{1296 N^5 (N+1)^5 (N+2)^3 (N+3)}
\nonumber \\ &&
                -\frac{41 N+29}{8 N (N+1)} S_2^2
                -\frac{33 N+17}{4 N (N+1)} S_4
                -\frac{6}{N} S_{3,1}
                +\frac{7 N+6}{N (N+1)} S_{2,1,1}
        \biggr) S_1
\nonumber \\ &&
        +\biggl(
                \frac{P_{169} S_2}{16 N^2 (N+1)^2 (N+2)}
                +\frac{P_{187}}{144 N^4 (N+1)^4 (N+2)^3 (N+3)}
                -\frac{14 N+5}{6 N (N+1)} S_3
\nonumber \\ &&
                -\frac{(2 N+3) S_{2,1}}{N (N+1)}
        \biggr) S_1^2
        +\biggl(
                \frac{P_{182}}{432 N^3 (N+1)^3 (N+2)^2 (N+3)}
                -\frac{(9 N+5) S_2}{12 N (N+1)}
        \biggr) S_1^3
\nonumber \\ &&
        +\biggl(
                \frac{P_{188}}{144 N^4 (N+1)^4 (N+2)^3 (N+3)}
                -\frac{(81 N+41) S_3}{6 N (N+1)}
                +\frac{(5 N+3) S_{2,1}}{N (N+1)}
        \biggr) S_2
\nonumber \\ &&
        -\frac{S_1^5}{24 N}
        -\frac{(61 N+28) S_5}{4 N (N+1)}
        +\frac{3 (3 N+2) S_{2,3}}{N (N+1)}
        -\frac{2 (2 N+3) S_{2,2,1}}{N (N+1)}
        +\frac{3 (3 N+2) S_{3,1,1}}{N (N+1)}
\nonumber \\ &&
        -\frac{5 S_{2,1,1,1}}{2 (N+1)}
        +\biggl(
                \frac{P_{185}}{96 N^4 (N+1)^4 (N+2)}
                +\frac{3 P_{166} S_2+P_{167} S_1^2}{16 N^2 (N+1)^2 (N+2)}
%                +\frac{3 P_{166} S_2}{16 N^2 (N+1)^2 (N+2)}
%                +\frac{P_{167} S_1^2}{16 N^2 (N+1)^2 (N+2)}
                +\frac{3 S_{2,1}}{4 (N+1)}
\nonumber \\ &&
                +\biggl(
                        \frac{P_{178}}{48 N^3 (N+1)^3 (N+2)}
                        -\frac{3 (2 N+1) S_2}{4 N (N+1)}
                \biggr) S_1
                -\frac{S_1^3}{4 N}
                -\frac{(13 N+4) S_3}{8 N (N+1)}
        \biggr) \zeta_2
\nonumber \\ &&
        +\biggl(
                \frac{7 P_{177}}{24 N^3 (N+1)^3}
                +\frac{7 \big(2 N^3+N^2+2 N+6\big) S_1}{12 N^2 (N+1)^2}
                +\frac{7 S_1^2}{6 N}
                +\frac{7 (5 N+2) S_2}{12 N (N+1)}
        \biggr) \zeta_3
\Biggr\},
\nonumber\\
\end{eqnarray}
%--------------------------------------------------------------------------------------------------------------------------------
with the polynomials
%--------------------------------------------------------------------------------------------------------------------------------
\begin{eqnarray}
P_{166}    &=& 7 N^4+16 N^3+14 N^2+6 N-4 \\
P_{167}    &=& 10 N^4+55 N^3+80 N^2+26 N-12 \\
P_{168}    &=& 34 N^4+175 N^3+248 N^2+98 N-12 \\
P_{169}    &=& 58 N^4+185 N^3+204 N^2+54 N-52 \\
P_{170}    &=& 82 N^4+415 N^3+584 N^2+242 N-12 \\
P_{171}    &=& 104 N^4+329 N^3+352 N^2+34 N-156 \\
P_{172}    &=& 110 N^4+265 N^3+252 N^2+158 N-4 \\
P_{173}    &=& 167 N^4+287 N^3+154 N^2+124 N-24 \\
P_{174}    &=& 397 N^4+1108 N^3+1052 N^2-46 N-660 \\
P_{175} &=& 454 N^4+1651 N^3+2048 N^2+770 N-300 \\
P_{176} &=& -24 N^5-102 N^4-109 N^3-37 N^2-26 N-8 \\
P_{177} &=& 24 N^5+102 N^4+109 N^3+37 N^2+26 N+8 \\
P_{178} &=& -100 N^6-640 N^5-917 N^4+145 N^3+804 N^2+84 N-144 \\
P_{179} &=& 24 N^6-13 N^5-36 N^4+383 N^3+528 N^2+20 N-96 \\
P_{180} &=& -3052 N^8-27144 N^7-96193 N^6-163416 N^5-107509 N^4+35598 N^3
\nonumber \\ &&
+64140 N^2-1080 N-11232 \\
P_{181} &=& -1708 N^8-7944 N^7-3265 N^6+46764 N^5+114443 N^4+81690 N^3
\nonumber \\ &&
-44580 N^2-76680 N-16416 \\
P_{182} &=& -1276 N^8-12048 N^7-41389 N^6-57456 N^5-19585 N^4+14406 N^3
\nonumber \\ &&
+2076 N^2-5400 N-864 \\
P_{183} &=& -196 N^8-1288 N^7-2823 N^6-504 N^5+6869 N^4+7690 N^3-2092 N^2
\nonumber \\ &&
-5928 N-1440 \\
P_{184} &=& -164 N^8-1592 N^7-5647 N^6-8224 N^5-3059 N^4+2706 N^3+1364 N^2
\nonumber \\ &&
-648 N-288 \\
P_{185} &=& 979 N^8+7700 N^7+17466 N^6+16152 N^5+6481 N^4+934 N^3-1052 N^2
\nonumber \\ &&
-984 N-288 \\
P_{186} &=& 2489 N^{10}+31430 N^9+153854 N^8+371873 N^7+475283 N^6+324887 N^5
\nonumber \\ &&
+138288 N^4+71784 N^3+32544 N^2-1296 N-5184 \\
P_{187} &=& 2614 N^{11}+34275 N^{10}+185602 N^9+534893 N^8+901060 N^7+943574 N^6
\nonumber \\ &&
+681546 N^5+405044 N^4+202896 N^3+60432 N^2+864 N-3456 \\
P_{188} &=& 2736 N^{11}+28443 N^{10}+114862 N^9+171605 N^8-172622 N^7-973618 N^6
\nonumber \\ &&
-1319786 N^5-734692 N^4-178032 N^3-121872 N^2-97632 N-24192 \\
P_{189} &=& -27379 N^{12}-371222 N^{11}-1975289 N^{10}-5408128 N^9-8444761 N^8
\nonumber \\ &&
-7851970 N^7-4357311 N^6-1390736 N^5-368252 N^4-320280 N^3
\nonumber \\ &&
-251712 N^2-117792 N-24192 \\
P_{190} &=& -73433 N^{13}-1131224 N^{12}-7251945 N^{11}-25365270 N^{10}-53573073 N^9
\nonumber \\ &&
-71363328 N^8-60837685 N^7-33229966 N^6-11812452 N^5-2949336 N^4
\nonumber \\ &&
-463104 N^3-95040 N^2-279936 N-124416 \\
P_{191} &=& 649543 N^{15}+10473731 N^{14}+70598329 N^{13}+263274563 N^{12}
\nonumber \\ &&
+605379059 N^{11}
+901109445 N^{10}+880841027 N^9+557674945 N^8
\nonumber \\ &&
+221287154 N^7
+53363620 N^6-295496 N^5-18745008 N^4-18856224 N^3
\nonumber \\ &&
-11057472 N^2
-3867264 N-622080.
\end{eqnarray}
%--------------------------------------------------------------------------------------------------------------------------------

For diagram~7 one obtains
%--------------------------------------------------------------------------------------------------------------------
\begin{eqnarray}
%%%texparser:LHS:testD7phys%%%
D_{7} &=&
\big(1+(-1)^N\big) \textcolor{blue}{C_A^2 T_F} \Biggl\{
        \frac{1}{\varepsilon^3} \Biggl[
                \frac{2 P_{192}}{3 (N+1)^3 (N+2)^3}
                +\frac{2 (2 N-7) S_2+2 (2 N+3) S_1^2}{3 N (N+1) (N+2)}
%                +\frac{2 (2 N-7) S_2}{3 N (N+1) (N+2)}
%                +\frac{2 (2 N+3) S_1^2}{3 N (N+1) (N+2)}
\nonumber \\ &&
                -\frac{2 \big(N^3+12 N^2+39 N+26\big)}{3 N (N+1)^2 (N+2)^2} S_1
        \Biggr]
        +\frac{1}{\varepsilon^2} \Biggl[
                 \frac{20 S_{2,1}+4 (N-6) S_3+(2 N+3) S_1^3}{3 N (N+1) (N+2)}
%                 \frac{20 S_{2,1}}{3 N (N+1) (N+2)}
%                +\frac{4 (N-6) S_3}{3 N (N+1) (N+2)}
%                +\frac{(2 N+3) S_1^3}{3 N (N+1) (N+2)}
\nonumber \\ &&
                +\frac{P_{196}}{36 (N+1)^4 (N+2)^4}
                +\biggl(
                        \frac{P_{193}}{18 N (N+1)^3 (N+2)^3}
                        +\frac{(6 N-11) S_2}{3 N (N+1) (N+2)}
                \biggr) S_1
\nonumber \\ &&
                -\frac{\big(22 N^3+111 N^2+189 N+94\big) S_1^2}{6 N (N+1)^2 (N+2)^2}
                +\frac{\big(48 N^3+223 N^2+243 N+34\big) S_2}{6 N (N+1)^2 (N+2)^2}
        \Biggr]
\nonumber \\ &&
        +\frac{1}{\varepsilon} \Biggl[
                \frac{7 (2 N+3) S_1^4}{72 N (N+1) (N+2)}
                +\frac{2 (2 N+3) S_{3,1}}{N (N+1) (N+2)}
                -\frac{2 (12 N+13) S_{2,1,1}}{3 N (N+1) (N+2)}
\nonumber \\ &&
                +\frac{2 (38 N-103) S_4+(62 N-27) S_2^2}{24 N (N+1) (N+2)}
%                +\frac{(38 N-103) S_4}{12 N (N+1) (N+2)} 
%                +\frac{(62 N-27) S_2^2}{24 N (N+1) (N+2)}
                -\frac{64 N^3+309 N^2+489 N+230}{36 N (N+1)^2 (N+2)^2} S_1^3
\nonumber \\ &&
                -\frac{29 N^3+149 N^2+204 N+58}{3 N (N+1)^2 (N+2)^2} S_{2,1}
                +\frac{P_{198} S_2}{24 N (N+1)^3 (N+2)^3 (N+3)}
\nonumber \\ &&
                +\frac{251 N^3+1194 N^2+1455 N+346}{18 N (N+1)^2 (N+2)^2} S_3
                +\frac{P_{204}}{432 (N+1)^5 (N+2)^5 (N+3)}
\nonumber \\ &&
                +\biggl(
                        \frac{6 (6 N+19) S_{2,1}+(14 N-69) S_3}{9 N (N+1) (N+2)}
%                        \frac{2 (6 N+19) S_{2,1}}{3 N (N+1) (N+2)}
%                        +\frac{(14 N-69) S_3}{9 N (N+1) (N+2)}
                        +\frac{40 N^3+143 N^2-141 N-310}{12 N (N+1)^2 (N+2)^2} S_2
\nonumber \\ &&
                        +\frac{P_{201}}{108 N (N+1)^4 (N+2)^4 (N+3)}
                \biggr) S_1
                +\biggl(
                        \frac{P_{199}}{24 N (N+1)^3 (N+2)^3 (N+3)}
\nonumber \\ &&
                        +\frac{(14 N-19) S_2}{12 N (N+1) (N+2)}
                \biggr) S_1^2
                +\biggl(
                        \frac{P_{192}}{4 (N+1)^3 (N+2)^3}
                        +\frac{(2 N-7) S_2}{4 N (N+1) (N+2)} 
\nonumber \\ &&
                        +\frac{2 N+3}{4 N (N+1) (N+2)} S_1^2
                        -\frac{N^3+12 N^2+39 N+26}{4 N (N+1)^2 (N+2)^2} S_1
                \biggr) \zeta_2
        \Biggr]
\nonumber \\ &&
        -\frac{25 S_{2,1,1,1}}{3 N (N+1) (N+2)}
        +\frac{(2 N+3) S_1^5}{48 N (N+1) (N+2)}
        +\frac{(18 N+47) S_{2,2,1}}{3 N (N+1) (N+2)}
\nonumber \\ &&
        -\frac{148 N^3+705 N^2+1089 N+502}{288 N (N+1)^2 (N+2)^2} S_1^4
        -\frac{19 N^3+93 N^2+146 N+66}{2 N (N+1)^2 (N+2)^2} S_{3,1}
\nonumber \\ &&
        +\frac{97 N^3+445 N^2+696 N+350}{6 N (N+1)^2 (N+2)^2} S_{2,1,1}
        +\frac{164 N^3+639 N^2-441 N-1222}{96 N (N+1)^2 (N+2)^2} S_2^2
\nonumber \\ &&
        +\frac{744 N^3+3523 N^2+4071 N+754}{48 N (N+1)^2 (N+2)^2} S_4
        +\frac{P_{195} S_3}{216 N (N+1)^3 (N+2)^3 (N+3)}
\nonumber \\ &&
        +\frac{P_{206}}{5184 (N+1)^6 (N+2)^6 (N+3)}
        +\frac{(1-6 N) (S_{2,3}+S_{3,1,1})+(7 N-17) S_5}{N (N+1) (N+2)}
%        +\frac{(1-6 N) S_{2,3}}{N (N+1) (N+2)}
%        +\frac{(1-6 N) S_{3,1,1}}{N (N+1) (N+2)}
%        +\frac{(7 N-17) S_5}{N (N+1) (N+2)}
\nonumber \\ &&
        +\frac{P_{194} S_{2,1}}{12 N (N+1)^3 (N+2)^3}
        +\biggl(
                -\frac{(18 N+17) S_{2,1,1}}{3 N (N+1) (N+2)}
                +\frac{3 (2 N+3) S_{3,1}}{N (N+1) (N+2)}
\nonumber \\ &&
                +\frac{58 N+7}{16 N (N+1) (N+2)} S_2^2
                -\frac{127 N^3+613 N^2+870 N+326}{6 N (N+1)^2 (N+2)^2} S_{2,1}
\nonumber \\ &&
                +\frac{102 N-167}{24 N (N+1) (N+2)} S_4
                +\frac{320 N^3+1329 N^2+477 N-862}{36 N (N+1)^2 (N+2)^2} S_3
\nonumber \\ &&
                +\frac{P_{197} S_2}{144 N (N+1)^3 (N+2)^3 (N+3)}
                +\frac{P_{205}}{1296 N (N+1)^5 (N+2)^5 (N+3)}
        \biggr) S_1
\nonumber \\ &&
        +\biggl(
                \frac{5 (2 N-9)}{12 N (N+1) (N+2)} S_3
                +\frac{96 N^3+415 N^2+123 N-326}{48 N (N+1)^2 (N+2)^2} S_2
\nonumber \\ &&
                +\frac{18 N+47}{6 N (N+1) (N+2)} S_{2,1}
                +\frac{P_{202}}{144 N (N+1)^4 (N+2)^4 (N+3)}
        \biggr) S_1^2
\nonumber \\ &&
        +\biggl(
                \frac{5 (6 N-7)}{72 N (N+1) (N+2)} S_2
                +\frac{P_{200}}{432 N (N+1)^3 (N+2)^3 (N+3)}
        \biggr) S_1^3
\nonumber \\ &&
        +\biggl(
                \frac{(13-18 N) S_{2,1}}{6 N (N+1) (N+2)} 
                +\frac{P_{203}}{144 N (N+1)^4 (N+2)^4 (N+3)}
\nonumber \\ &&
                +\frac{(246 N-431) S_3}{36 N (N+1) (N+2)}
        \biggr) S_2
        +\biggl[
                \frac{P_{196}}{96 (N+1)^4 (N+2)^4}
                +\frac{5 S_{2,1}}{2 N (N+1) (N+2)}
\nonumber \\ &&
                +\frac{(N-6) S_3}{2 N (N+1) (N+2)}
                +\frac{(2 N+3) S_1^3}{8 N (N+1) (N+2)}
                -\frac{22 N^3+111 N^2+189 N+94}{16 N (N+1)^2 (N+2)^2} S_1^2
\nonumber \\ &&
                +\frac{48 N^3+223 N^2+243 N+34}{16 N (N+1)^2 (N+2)^2} S_2
%                +\frac{S_2}{N (N+1)^2 (N+2)^2} 
%                \frac{1}{16} \big(48 N^3+223 N^2+243 N+34\big)
                +\biggl(
                        \frac{P_{193}}{48 N (N+1)^3 (N+2)^3}
\nonumber \\ &&
                        +\frac{(6 N-11) S_2}{8 N (N+1) (N+2)}
                \biggr) S_1
        \biggr] \zeta_2
        +\biggl(
                -\frac{7 (2 N-7) S_2+7 (2 N+3) S_1^2}{12 N (N+1) (N+2)}
%                -\frac{7 (2 N-7) S_2}{12 N (N+1) (N+2)}
%                -\frac{7 (2 N+3) S_1^2}{12 N (N+1) (N+2)}
\nonumber \\ &&
                -\frac{7 P_{192}}{12 (N+1)^3 (N+2)^3}
                +\frac{7 \big(N^3+12 N^2+39 N+26\big) S_1}{12 N (N+1)^2 (N+2)^2}
        \biggr) \zeta_3
\Biggr\},
\end{eqnarray}
%--------------------------------------------------------------------------------------------------------------------
with
%--------------------------------------------------------------------------------------------------------------------
\begin{eqnarray}
P_{192}    &=& 7 N^4+88 N^3+334 N^2+491 N+250 \\
P_{193}    &=& 171 N^5+1337 N^4+3615 N^3+3575 N^2+426 N-664 \\
P_{194}    &=& 506 N^5+2767 N^4+4726 N^3+1821 N^2-1728 N-848 \\
P_{195}    &=& -13113 N^6-103096 N^5-298140 N^4-389110 N^3-248535 N^2-126382 N
\nonumber \\ &&
-69672 \\
P_{196}    &=& -2135 N^6-19091 N^5-67473 N^4-117737 N^3-101868 N^2
\nonumber\\ &&
-35360 N-280 \\
P_{197}    &=& -2013 N^6-9142 N^5+9594 N^4+109640 N^3+194103 N^2+116882 N
\nonumber\\ &&
+13416 \\
P_{198}    &=& -859 N^6-6482 N^5-17634 N^4-20856 N^3-12039 N^2-7850 N-5352 \\
P_{199}    &=& 261 N^6+2642 N^5+9962 N^4+17492 N^3+14157 N^2+4262 N+312 \\
P_{200}    &=& 2007 N^6+20078 N^5+74406 N^4+128588 N^3+105111 N^2+37130 N
\nonumber\\ &&
+6792 \\
P_{201} &=& -6318 N^8-83128 N^7-449157 N^6-1296762 N^5-2180700 N^4
\nonumber \\ &&
-2194500 N^3
-1324701 N^2-476078 N-88656 \\
P_{202} &=& -5377 N^8-73676 N^7-412043 N^6-1241556 N^5-2223893 N^4
\nonumber \\ &&
-2452398 N^3
-1652131 N^2-629902 N-101136 \\
P_{203} &=& 6059 N^8+51228 N^7+186109 N^6+452880 N^5+971491 N^4+1657638 N^3
\nonumber \\ &&
+1739921 N^2+898194 N+148464 \\
P_{204} &=& 75827 N^9+1104173 N^8+6907786 N^7+24379674 N^6+53621139 N^5
\nonumber \\ &&
+76740405 N^4+72626392 N^3+45288244 N^2+17789144 N+3530256 \\
P_{205} &=& 204975 N^{10}+3198139 N^9+21500052 N^8+82221572 N^7+198478617 N^6
\nonumber \\ &&
+316334823 N^5+336363930 N^4+232599230 N^3+95471202 N^2
\nonumber \\ &&
+17353444 N-289536 \\
P_{206} &=& -1784735 N^{11}-30708686 N^{10}-234329495 N^9-1048250278 N^8
\nonumber \\ &&
-3057729413 N^7-6108443586 N^6-8512411861 N^5-8217450082 N^4
\nonumber \\ &&
-5276614576 N^3-2021347784 N^2-327324352 N+16534752~. 
\end{eqnarray}
%--------------------------------------------------------------------------------------------------------------------

Diagram~8 is given by
%-----------------------------------------------------------------------------------------------------------------------------
\begin{eqnarray}
%%%texparser:LHS:testD8phys%%%
D_{8} &=&
\big(1+(-1)^N\big) \textcolor{blue}{C_A^2 T_F} \Biggl\{
        \frac{1}{\varepsilon^3} \Biggl[
                \frac{P_{207}}{3 N^3 (N+1)^3 (N+2)}
                +\frac{4 (2 N+3) S_1}{3 N^2 (N+1)^2}
        \Biggr]
\nonumber\\ &&
        +\frac{1}{\varepsilon^2} \Biggl[
                \frac{P_{213}}{18 N^4 (N+1)^4 (N+2)^3}
                +\frac{P_{211} S_1}{6 N^3 (N+1)^3 (N+2)^2}
                +\frac{(2 N+3) (S_1^2+S_2)}{3 N^2 (N+1)^2}
%                +\frac{(2 N+3) S_1^2}{3 N^2 (N+1)^2}
%                +\frac{(2 N+3) S_2}{3 N^2 (N+1)^2}
        \Biggr]
\nonumber\\ &&
        +\frac{1}{\varepsilon} \Biggl[
                \frac{P_{217}}{108 N^5 (N+1)^5 (N+2)^4}
                +\frac{P_{210} S_2+P_{211} S_1^2}{24 N^3 (N+1)^3 (N+2)^2}
%                +\frac{P_{210} S_2}{24 N^3 (N+1)^3 (N+2)^2}
%                +\frac{P_{211} S_1^2}{24 N^3 (N+1)^3 (N+2)^2}
                +\frac{(2 N+3) S_1^3}{18 N^2 (N+1)^2}
\nonumber\\ &&
                +\frac{2 N+3}{9 N^2 (N+1)^2} S_3
                +\biggl(
                        \frac{P_{214}}{36 N^4 (N+1)^4 (N+2)^3}
                        +\frac{13 (2 N+3)}{6 N^2 (N+1)^2} S_2
                \biggr) S_1
\nonumber\\ &&
                +\biggl(
                        \frac{P_{207}}{8 N^3 (N+1)^3 (N+2)}
                        +\frac{(2 N+3) S_1}{2 N^2 (N+1)^2}
                \biggr) \zeta_2
        \Biggr]
        +\frac{P_{218}}{648 N^6 (N+1)^6 (N+2)^5}
\nonumber\\ &&
        +\frac{2 P_{209} S_3+P_{211} S_1^3}{144 N^3 (N+1)^3 (N+2)^2}
%        +\frac{P_{209} S_3}{72 N^3 (N+1)^3 (N+2)^2}
%        +\frac{P_{211} S_1^3}{144 N^3 (N+1)^3 (N+2)^2}
        +\frac{P_{215} S_2}{144 N^4 (N+1)^4 (N+2)^3}
        +\frac{(2 N+3) S_1^4}{144 N^2 (N+1)^2}
\nonumber\\ &&
        +\biggl(
                \frac{P_{216}}{216 N^5 (N+1)^5 (N+2)^4}
                -\frac{13 P_{212} S_2}{48 N^3 (N+1)^3 (N+2)^2}
                +\frac{55 (2 N+3) S_3}{18 N^2 (N+1)^2}
        \biggr) S_1
\nonumber\\ &&
        +\biggl(
                \frac{P_{214}}{144 N^4 (N+1)^4 (N+2)^3}
                +\frac{13 (2 N+3) S_2}{24 N^2 (N+1)^2}
        \biggr) S_1^2
        +\frac{25 (2 N+3)}{48 N^2 (N+1)^2} S_2^2
\nonumber\\ &&
        +\frac{(2 N+3) S_4}{24 N^2 (N+1)^2}
        +\biggl(
                \frac{P_{213}}{48 N^4 (N+1)^4 (N+2)^3}
                +\frac{P_{211} S_1}{16 N^3 (N+1)^3 (N+2)^2}
\nonumber\\ &&
                +\frac{(2 N+3) (S_1^2+S_2)}{8 N^2 (N+1)^2}
%                +\frac{(2 N+3) S_1^2}{8 N^2 (N+1)^2}
%                +\frac{(2 N+3) S_2}{8 N^2 (N+1)^2}
        \biggr) \zeta_2
        +\biggl(
                \frac{7 P_{208}}{24 N^3 (N+1)^3 (N+2)}
%\nonumber\\ &&
                -\frac{7 (2 N+3) S_1}{6 N^2 (N+1)^2}
        \biggr) \zeta_3
\Biggr\},
\end{eqnarray}
%-----------------------------------------------------------------------------------------------------------------------------
where
%-----------------------------------------------------------------------------------------------------------------------------
\begin{eqnarray}
P_{207}    &=& -25 N^5-71 N^4-38 N^3+16 N^2+36 N+16 \\
P_{208}    &=& 25 N^5+71 N^4+38 N^3-16 N^2-36 N-16 \\
P_{209}    &=& -1369 N^6-6674 N^5-10133 N^4-3760 N^3+3528 N^2+4968 N+1824 \\
P_{210}    &=& -319 N^6-1592 N^5-2573 N^4-1240 N^3+672 N^2+1272 N+480 \\
P_{211}    &=& -19 N^6-140 N^5-413 N^4-520 N^3-144 N^2+216 N+96 \\
P_{212}    &=& 19 N^6+140 N^5+413 N^4+520 N^3+144 N^2-216 N-96 \\
P_{213}    &=& 349 N^9+3161 N^8+11044 N^7+18446 N^6+13690 N^5+836 N^4-3180 N^3
\nonumber\\ &&
+1056 N^2+2064 N+576 \\
P_{214}    &=& 349 N^9+3786 N^8+15467 N^7+29320 N^6+24332 N^5+3274 N^4-3524 N^3
\nonumber\\ &&
+3856 N^2+4320 N+1152 \\
P_{215}    &=& 4537 N^9+41718 N^8+147995 N^7+250672 N^6+188612 N^5+13306 N^4
\nonumber\\ &&
-41684 N^3+16528 N^2+29088 N+8064 \\
P_{216} &=& -2195 N^{12}-33171 N^{11}-186452 N^{10}-525287 N^9-817419 N^8-724548 N^7
\nonumber\\ &&
-404350 N^6-220482 N^5-90292 N^4+80480 N^3+130944 N^2+72576 N
\nonumber\\ &&
+13824 \\
P_{217} &=& -2195 N^{12}-28822 N^{11}-148626 N^{10}-395793 N^9-580980 N^8-443616 N^7
\nonumber\\ &&
-143782 N^6-58400 N^5-98892 N^4-24696 N^3+53136 N^2+39456 N
\nonumber\\ &&
+8064 \\
P_{218} &=& 13417 N^{15}+223085 N^{14}+1504820 N^{13}+5629246 N^{12}+13213194 N^{11}
\nonumber\\ &&
+20729910 N^{10}+22606046 N^9+17321518 N^8+8062942 N^7-588172 N^6
\nonumber\\ &&
-4173660 N^5-2213136 N^4+745632 N^3+1437696 N^2+654912 N
\nonumber\\ &&
+103680.
\end{eqnarray}
%-----------------------------------------------------------------------------------------------------------------------------

The result for diagram~9 reads
%------------------------------------------------------------------------------------------------------------------------------
\begin{eqnarray}
%%%texparser:LHS:testD9phys%%%
D_{9} &=&
\textcolor{blue}{C_A^2 T_F} \Biggl\{
        \frac{1}{\varepsilon^3} \Biggl[
                \frac{P_{246}}{3 N^3 (N+1)^3 (N+2)^2}
                -\frac{2 (2 N+1) \big(N^3+5 N^2-2 N-12\big)}{3 N^2 (N+1)^2 (N+2)} S_1
\nonumber\\ &&
                -\frac{14 N^2+21 N+22}{3 N (N+1) (N+2)} S_2
                -\frac{4 N^2+3 N+2}{3 N (N+1) (N+2)} S_1^2
                +\frac{16 N S_{-2}}{3 (N+1) (N+2)}
\nonumber\\ &&
                +(-1)^N \Biggl(
                        \frac{2 P_{230}}{3 N^3 (N+1)^3 (N+2)^2}
                        -\frac{2 \big(7 N^2+4 N+12\big)}{3 N^2 (N+1)^2 (N+2)} S_1
                \Biggr)
        \Biggr]
\nonumber\\ &&
        +\frac{1}{\varepsilon^2} \Biggl[
                -\frac{\big(4 N^2+3 N+2\big) S_1^3}{6 N (N+1) (N+2)}
                +\frac{2 \big(11 N^2+27 N+42\big) S_{2,1}}{3 N (N+1) (N+2)}
                +\frac{20 N S_{-3}}{3 (N+1) (N+2)}
\nonumber\\ &&
                +\frac{\big(21 N^2-6 N-40\big) S_3}{3 N (N+1) (N+2)}
                +\frac{P_{228} S_1^2}{6 N^2 (N+1)^2 (N+2)^2}
                +\frac{P_{233} S_2}{6 N^2 (N+1)^2 (N+2)^2}
\nonumber\\ &&
                +\frac{P_{260}}{18 (N-1) N^4 (N+1)^4 (N+2)^3}
                +\biggl(
                        \frac{P_{248}}{9 (N-1) N^3 (N+1)^3 (N+2)^2}
\nonumber\\ &&
                        -\frac{\big(16 N^2+53 N+94\big) S_2}{6 N (N+1) (N+2)}
                \biggr) S_1
                -\frac{8 N S_{-2,1}}{(N+1) (N+2)}
                +\biggl(
                        \frac{2 P_{222}}{3 N^2 (N+1)^2 (N+2)}
\nonumber\\ &&
                        +\frac{8 \big(4 N^2-N-6\big) S_1}{3 N (N+1) (N+2)}
                \biggr) S_{-2}
                +(-1)^N \Biggl(
                        \frac{P_{232} S_1}{6 (N-1) N^2 (N+1)^3 (N+2)^2}
\nonumber\\ &&
                        -\frac{\big(7 N^2+4 N+12\big) (S_1^2+S_2)}{6 N^2 (N+1)^2 (N+2)}
%                        +\frac{\big(-7 N^2-4 N-12\big) S_1^2}{6 N^2 (N+1)^2 (N+2)}
%                        +\frac{\big(-7 N^2-4 N-12\big) S_2}{6 N^2 (N+1)^2 (N+2)}
                        +\frac{P_{254}}{18 (N-1) N^4 (N+1)^4 (N+2)^3}
                \Biggr)
        \Biggr]
\nonumber\\ &&
        +\frac{1}{\varepsilon} \Biggl[
                -\frac{\big(436 N^2+261 N+14\big) S_2^2}{48 N (N+1) (N+2)}
                +\frac{12 N S_{-2,1,1}}{(N+1) (N+2)}
                -\frac{7 \big(4 N^2+3 N+2\big) S_1^4}{144 N (N+1) (N+2)} 
\nonumber\\ &&
                -\frac{\big(19 N^2+111 N+198\big) S_{2,1,1}}{3 N (N+1) (N+2)}
                +\frac{P_{253} S_2}{36 (N-1) N^3 (N+1)^3 (N+2)^3}
\nonumber\\ &&
                -\frac{2 \big(17 N^2-37 N-107\big) S_{3,1}}{3 N (N+1) (N+2)}
                +\frac{P_{264}}{108 (N-1) N^5 (N+1)^5 (N+2)^4}
\nonumber\\ &&
                +\frac{\big(468 N^2+299 N+58\big) S_4}{24 N (N+1) (N+2)}
                +\frac{P_{221} S_{-2,1}}{N^2 (N+1)^2 (N+2)}
                +\frac{P_{226} S_{2,1}}{6 N^2 (N+1)^2 (N+2)^2}
\nonumber\\ &&
                +\frac{P_{229} S_1^3+P_{238} S_3}{36 N^2 (N+1)^2 (N+2)^2}
%                +\frac{P_{229} S_1^3}{36 N^2 (N+1)^2 (N+2)^2}
%                +\frac{P_{238} S_3}{36 N^2 (N+1)^2 (N+2)^2}
                +\biggl(
                        \frac{2 \big(3 N^2+38 N+81\big) S_{2,1}}{3 N (N+1) (N+2)}
                        -\frac{4 \big(4 N^2-N-6\big) S_{-2,1}}{N (N+1) (N+2)}
\nonumber\\ &&
                        +\frac{314 N^2-159 N-962}{18 N (N+1) (N+2)} S_3
                        +\frac{P_{259}}{108 (N-1) N^4 (N+1)^4 (N+2)^3}
\nonumber\\ &&
                        +\frac{P_{234} S_2}{12 N^2 (N+1)^2 (N+2)^2}
                \biggr) S_1
                +\biggl(
                        \frac{P_{255}}{12 (N-1) N^3 (N+1)^3 (N+2)^3}
\nonumber\\ &&
                        -\frac{\big(20 N^2+233 N+526\big) S_2}{24 N (N+1) (N+2)}
                \biggr) S_1^2
                +\biggl(
                        \frac{P_{245}}{9 N^3 (N+1)^3 (N+2)^2}
\nonumber\\ &&
                        +\frac{2 \big(16 N^2-7 N-42\big) S_1^2}{3 N (N+1) (N+2)}
                        +\frac{2 \big(16 N^2-N-6\big) S_2}{3 N (N+1) (N+2)}
                        +\frac{P_{223} S_1}{3 N^2 (N+1)^2 (N+2)}
                \biggr) S_{-2}
\nonumber\\ &&
                +\biggl(
                        \frac{5 P_{222}}{6 N^2 (N+1)^2 (N+2)}
                        +\frac{10 \big(4 N^2-N-6\big) S_1}{3 N (N+1) (N+2)}
                \biggr) S_{-3}
                +\frac{22 N S_{-4}}{3 (N+1) (N+2)}
\nonumber\\ &&
                -\frac{8 N S_{-2,2}}{(N+1) (N+2)}
                -\frac{10 N S_{-3,1}}{(N+1) (N+2)}
                +\biggl(
                        -\frac{\big(14 N^2+21 N+22\big) S_2}{8 N (N+1) (N+2)}
\nonumber\\ &&
                        +\frac{P_{246}}{8 N^3 (N+1)^3 (N+2)^2}
                        +\frac{\big(-4 N^2-3 N-2\big) S_1^2}{8 N (N+1) (N+2)}
                        +\frac{2 N S_{-2}}{(N+1) (N+2)}
\nonumber\\ &&
                        -\frac{(2 N+1) \big(N^3+5 N^2-2 N-12\big) S_1}{4 N^2 (N+1)^2 (N+2)}
                \biggr) \zeta_2
                +(-1)^N \Biggl(
                        -\frac{\big(7 N^2+4 N+12\big) S_1^3}{36 N^2 (N+1)^2 (N+2)}
\nonumber\\ &&
                        -\frac{\big(7 N^2+4 N+12\big) S_3}{18 N^2 (N+1)^2 (N+2)}
                        +\frac{N P_{232} S_1^2+P_{241} S_2}{24 (N-1) N^3 (N+1)^3 (N+2)^2}
%                        +\frac{P_{232} S_1^2}{24 (N-1) N^2 (N+1)^3 (N+2)^2}
%                        +\frac{P_{241} S_2}{24 (N-1) N^3 (N+1)^3 (N+2)^2}
\nonumber\\ &&
                        +\frac{P_{262}}{108 (N-1) N^5 (N+1)^5 (N+2)^4}
                        +\biggl(
                                \frac{P_{243}}{36 (N-1) N^2 (N+1)^4 (N+2)^3}
\nonumber\\ &&
                                -\frac{13 \big(7 N^2+4 N+12\big) S_2}{12 N^2 (N+1)^2 (N+2)}
                        \biggr) S_1
                        +\biggl(
                                -\frac{7 N^2+4 N+12}{4 N^2 (N+1)^2 (N+2)} S_1
\nonumber\\ &&
                                +\frac{P_{230}}{4 N^3 (N+1)^3 (N+2)^2}
                        \biggr) \zeta_2
                \Biggr)
        \Biggr]
        +\frac{11 N^2+219 N+474}{6 N (N+1) (N+2)} S_{2,1,1,1}
\nonumber\\ &&
        +\frac{N}{(N+1) (N+2)} \left(12 S_{2,1,-2}+12 S_{-2,2,1}+15 S_{-3,1,1}-18 S_{-2,1,1,1}\right)
%        +\frac{12 N S_{2,1,-2}}{(N+1) (N+2)}
%        +\frac{12 N S_{-2,2,1}}{(N+1) (N+2)}
%        +\frac{15 N S_{-3,1,1}}{(N+1) (N+2)}
%        -\frac{18 N S_{-2,1,1,1}}{(N+1) (N+2)}
\nonumber\\ &&
        -\frac{\big(61 N^2-11 N-130\big) S_{4,1}}{3 N (N+1) (N+2)}
        -\frac{\big(4 N^2+3 N+2\big) S_1^5}{96 N (N+1) (N+2)}
        -\frac{5 P_{222} S_{-3,1}}{4 N^2 (N+1)^2 (N+2)}
\nonumber\\ &&
        +\frac{16 N^2+152 N+263}{3 N (N+1) (N+2)} S_{2,2,1}
        +\frac{41 N^2-183 N-493}{3 N (N+1) (N+2)} S_{3,1,1}
\nonumber\\ &&
        +\frac{59 N^2-123 N-361}{3 N (N+1) (N+2)} S_{2,3}
        +\frac{167 N^2+470 N+1012}{12 N (N+1) (N+2)} S_5
        +\frac{2 P_{221} S_{-2,2}+3 P_{222} S_{-2,1,1}}{2 N^2 (N+1)^2 (N+2)}
%        +\frac{3 P_{222} S_{-2,1,1}}{2 N^2 (N+1)^2 (N+2)}
\nonumber\\ &&
        +\frac{48 P_{227} S_{3,1}+P_{231} S_1^4+24 P_{235} S_{2,1,1}+6 P_{239} S_4+3 P_{240} S_2^2}{288 N^2 (N+1)^2 (N+2)^2}
%        +\frac{P_{227} S_{3,1}}{6 N^2 (N+1)^2 (N+2)^2}
%        +\frac{P_{231} S_1^4}{288 N^2 (N+1)^2 (N+2)^2}
%        +\frac{P_{235} S_{2,1,1}}{12 N^2 (N+1)^2 (N+2)^2}
%        +\frac{P_{239} S_4}{48 N^2 (N+1)^2 (N+2)^2}
%        +\frac{P_{240} S_2^2}{96 N^2 (N+1)^2 (N+2)^2}
        +\frac{P_{247} S_{-2,1}}{6 N^3 (N+1)^3 (N+2)^2}
\nonumber\\ &&
        +\frac{P_{251} S_3+3 (N-1) P_{249} S_{2,1}}{108 (N-1) N^3 (N+1)^3 (N+2)^3}
%        +\frac{P_{249} S_{2,1}}{36 N^3 (N+1)^3 (N+2)^3}
%        +\frac{P_{251} S_3}{108 (N-1) N^3 (N+1)^3 (N+2)^3}
        +\frac{P_{267}}{648 (N-1) N^6 (N+1)^6 (N+2)^5}
\nonumber\\ &&
        +\biggl[
                \frac{\big(-388 N^2-151 N+262\big) S_2^2}{32 N (N+1) (N+2)}
                +\frac{\big(-94 N^2+122 N+475\big) S_{3,1}}{3 N (N+1) (N+2)}
\nonumber\\ &&
                +\frac{4 N^2-N-6}{N (N+1) (N+2)} \left(6 S_{-2,1,1}-4 S_{-2,2}-5 S_{-3,1}\right)
%                -\frac{4 \big(4 N^2-N-6\big) S_{-2,2}}{N (N+1) (N+2)}
%                -\frac{5 \big(4 N^2-N-6\big) S_{-3,1}}{N (N+1) (N+2)}
%                +\frac{6 \big(4 N^2-N-6\big) S_{-2,1,1}}{N (N+1) (N+2)}
                +\frac{\big(9 N^2-124 N-303\big) S_{2,1,1}}{3 N (N+1) (N+2)}
\nonumber\\ &&
                +\frac{\big(1700 N^2+179 N-2598\big) S_4}{48 N (N+1) (N+2)}
                +\frac{P_{236} S_3+3 P_{225} S_{2,1}+18 (N+2) P_{220} S_{-2,1}}{36 N^2 (N+1)^2 (N+2)^2}
%                +\frac{P_{220} S_{-2,1}}{2 N^2 (N+1)^2 (N+2)}
%                +\frac{P_{225} S_{2,1}}{12 N^2 (N+1)^2 (N+2)^2}
%                +\frac{P_{236} S_3}{36 N^2 (N+1)^2 (N+2)^2}
\nonumber\\ &&
                +\frac{P_{252} S_2}{72 (N-1) N^3 (N+1)^3 (N+2)^3}
                +\frac{P_{265}}{648 (N-1) N^5 (N+1)^5 (N+2)^4}
        \biggr] S_1
\nonumber\\ &&
        +\biggl[
                \frac{\big(-26 N^2+105 N+309\big) S_{2,1}}{6 N (N+1) (N+2)}
                +\frac{\big(392 N^2-417 N-1890\big) S_3}{24 N (N+1) (N+2)}
\nonumber\\ &&
                +\frac{P_{237} S_2}{48 N^2 (N+1)^2 (N+2)^2}
                +\frac{P_{261}}{144 (N-1) N^4 (N+1)^4 (N+2)^4}
\nonumber\\ &&
                +\frac{\big(-16 N^2+7 N+42\big) S_{-2,1}}{N (N+1) (N+2)}
        \biggr] S_1^2
        +\biggl[
                \frac{\big(164 N^2-737 N-2254\big) S_2}{144 N (N+1) (N+2)}
\nonumber\\ &&
                +\frac{P_{256}}{216 (N-1) N^3 (N+1)^3 (N+2)^3}
        \biggr] S_1^3
        +\biggl[
                \frac{\big(-2068 N^2-371 N+1946\big) S_3}{72 N (N+1) (N+2)}
\nonumber\\ &&
                +\frac{\big(88 N^2-47 N-177\big) S_{2,1}}{6 N (N+1) (N+2)}
                +\frac{P_{263}}{432 (N-1) N^4 (N+1)^4 (N+2)^4}
\nonumber\\ &&
                +\frac{\big(-4 N^2+N+6\big) S_{-2,1}}{N (N+1) (N+2)}
        \biggr] S_2
        +\biggl[
                \frac{P_{219} S_1^2+P_{224} S_2}{12 N^2 (N+1)^2 (N+2)}
%                \frac{P_{219} S_1^2}{12 N^2 (N+1)^2 (N+2)}
%                +\frac{P_{224} S_2}{12 N^2 (N+1)^2 (N+2)}
                -\frac{12 N S_{2,1}}{(N+1) (N+2)}
\nonumber\\ &&
                +\frac{P_{258}}{54 N^4 (N+1)^4 (N+2)^3}
                +\frac{\big(64 N^2-37 N-222\big) S_1^3}{9 N (N+1) (N+2)}
                +\frac{2 \big(64 N^2-N-6\big) S_3}{9 N (N+1) (N+2)}
\nonumber\\ &&
                +\biggl(
                        \frac{P_{244}}{18 N^3 (N+1)^3 (N+2)^2}
                        +\frac{\big(64 N^2-19 N-114\big) S_2}{3 N (N+1) (N+2)}
                \biggr) S_1
        \biggr] S_{-2}
\nonumber\\ &&
        +\biggl[
                -\frac{5 P_{247}}{36 N^3 (N+1)^3 (N+2)^2}
                +\frac{5 \big(4 N^2-N-6\big) S_2}{6 N (N+1) (N+2)}
                +\frac{5 \big(16 N^2-7 N-42\big) S_1^2}{6 N (N+1) (N+2)}
\nonumber\\ &&
                +\frac{5 P_{223} S_1}{12 N^2 (N+1)^2 (N+2)}
        \biggr] S_{-3}
        +N \frac{12 S_{-2,3}-7 S_{-5}-6 S_{2,-3}-33 S_{-4,1}}{3 (N+1) (N+2)}
%        -\frac{7 N S_{-5}}{3 (N+1) (N+2)}
%        -\frac{2 N S_{2,-3}}{(N+1) (N+2)}
%        +\frac{4 N S_{-2,3}}{(N+1) (N+2)}
%        -\frac{11 N S_{-4,1}}{(N+1) (N+2)}
\nonumber\\ &&
        +\biggl(
                \frac{11 P_{222}}{12 N^2 (N+1)^2 (N+2)}
                +\frac{11 \big(4 N^2-N-6\big)}{3 N (N+1) (N+2)} S_1
        \biggr) S_{-4}
\nonumber\\ &&
        +\biggl[
                -\frac{\big(4 N^2+3 N+2\big) S_1^3}{16 N (N+1) (N+2)}
                +\frac{\big(11 N^2+27 N+42\big) S_{2,1}}{4 N (N+1) (N+2)}
                +\frac{\big(21 N^2-6 N-40\big) S_3}{8 N (N+1) (N+2)}
\nonumber\\ &&
                +\frac{P_{228} S_1^2+P_{233} S_2}{16 N^2 (N+1)^2 (N+2)^2}
%                +\frac{P_{228} S_1^2}{16 N^2 (N+1)^2 (N+2)^2}
%                +\frac{P_{233} S_2}{16 N^2 (N+1)^2 (N+2)^2}
                +\frac{P_{260}}{48 (N-1) N^4 (N+1)^4 (N+2)^3}
\nonumber\\ &&
                +\biggl(
                        -\frac{16 N^2+53 N+94}{16 N (N+1) (N+2)} S_2
                        +\frac{P_{248}}{24 (N-1) N^3 (N+1)^3 (N+2)^2}
                \biggr) S_1
\nonumber\\ &&
                +\biggl(
                        \frac{P_{222}}{4 N^2 (N+1)^2 (N+2)}
                        +\frac{\big(4 N^2-N-6\big) S_1}{N (N+1) (N+2)} 
                \biggr) S_{-2}
                +\frac{N \left(5 S_{-3}-6 S_{-2,1}\right)}{2 (N+1) (N+2)}
%                +\frac{5 N S_{-3}}{2 (N+1) (N+2)}
%                -\frac{3 N S_{-2,1}}{(N+1) (N+2)}
        \biggr] \zeta_2
\nonumber\\ &&
        +\biggl[
                \frac{7 \big(4 N^2+3 N+2\big) S_1^2}{24 N (N+1) (N+2)}
                +\frac{7 \big(14 N^2+21 N+22\big) S_2}{24 N (N+1) (N+2)}
                -\frac{7 P_{246}}{24 N^3 (N+1)^3 (N+2)^2}
\nonumber\\ &&
                +\frac{7 (2 N+1) \big(N^3+5 N^2-2 N-12\big)}{12 N^2 (N+1)^2 (N+2)} S_1
                -\frac{14 N S_{-2}}{3 (N+1) (N+2)}
        \biggr] \zeta_3
\nonumber\\ &&
        +(-1)^N \Biggl[
                \frac{P_{232} S_1^3}{144 (N-1) N^2 (N+1)^3 (N+2)^2}
                +\frac{P_{242} S_3}{72 (N-1) N^3 (N+1)^3 (N+2)^2}
\nonumber\\ &&
                -\frac{\left(7 N^2+4 N+12\right) \left(75 S_2^2+S_1^4+6 S_4\right)}{288 N^2 (N+1)^2 (N+2)} 
%                \frac{\big(-7 N^2-4 N-12\big) S_1^4}{288 N^2 (N+1)^2 (N+2)} 
%                +\frac{\big(-7 N^2-4 N-12\big) S_4}{48 N^2 (N+1)^2 (N+2)}
%                -\frac{25 \big(7 N^2+4 N+12\big) S_2^2}{96 N^2 (N+1)^2 (N+2)}
                +\frac{P_{250} S_2}{144 (N-1) N^4 (N+1)^4 (N+2)^3}
\nonumber\\ &&
                +\frac{P_{266}}{648 (N-1) N^6 (N+1)^6 (N+2)^5}
                +\biggl(
                        -\frac{55 \big(7 N^2+4 N+12\big)}{36 N^2 (N+1)^2 (N+2)} S_3
\nonumber\\ &&
                        +\frac{13 P_{232} S_2}{48 (N-1) N^2 (N+1)^3 (N+2)^2}
                        +\frac{P_{257}}{216 (N-1) N^2 (N+1)^5 (N+2)^4}
                \biggr) S_1
\nonumber\\ &&
                +\biggl(
                        -\frac{13 \big(7 N^2+4 N+12\big) S_2}{48 N^2 (N+1)^2 (N+2)}
                        +\frac{P_{243}}{144 (N-1) N^2 (N+1)^4 (N+2)^3}
                \biggr) S_1^2
\nonumber\\ &&
                +\biggl(
                        -\frac{\big(7 N^2+4 N+12\big)}{16 N^2 (N+1)^2 (N+2)} \left(S_1^2+S_2\right)
%                        \frac{\big(-7 N^2-4 N-12\big) S_1^2}{16 N^2 (N+1)^2 (N+2)}
%                        +\frac{\big(-7 N^2-4 N-12\big) S_2}{16 N^2 (N+1)^2 (N+2)}
                        +\frac{P_{232} S_1}{16 (N-1) N^2 (N+1)^3 (N+2)^2}
\nonumber\\ &&
                        +\frac{P_{254}}{48 (N-1) N^4 (N+1)^4 (N+2)^3}
                \biggr) \zeta_2
                +\biggl(
                        -\frac{7 P_{230}}{12 N^3 (N+1)^3 (N+2)^2}
\nonumber\\ &&
                        +\frac{7 \big(7 N^2+4 N+12\big)}{12 N^2 (N+1)^2 (N+2)} S_1
                \biggr) \zeta_3
        \Biggr]
\Biggr\},
\end{eqnarray}
%------------------------------------------------------------------------------------------------------------------------------
with
%------------------------------------------------------------------------------------------------------------------------------
\begin{eqnarray}
P_{219}    &=& -15 N^4+1586 N^3+2417 N^2+44 N-144 \\
P_{220}    &=& -3 N^4-326 N^3-419 N^2+76 N+48 \\
P_{221}    &=& -3 N^4-58 N^3-43 N^2+48 N+16 \\
P_{222}    &=& 3 N^4+58 N^3+43 N^2-48 N-16 \\
P_{223}    &=& 3 N^4+326 N^3+419 N^2-76 N-48 \\
P_{224}    &=& 39 N^4+1022 N^3+935 N^2-652 N-240 \\
P_{225}    &=& -199 N^5-2824 N^4-8641 N^3-7930 N^2-960 N+160 \\
P_{226}    &=& -110 N^5-897 N^4-2363 N^3-2078 N^2-296 N+32 \\
P_{227}    &=& -52 N^5-1917 N^4-6358 N^3-5840 N^2-680 N+96 \\
P_{228} &=& 2 N^5-25 N^4-55 N^3-16 N^2-8 \\
P_{229} &=& 10 N^5-45 N^4-119 N^3-96 N^2-128 N-72 \\
P_{230} &=& 11 N^5+28 N^4+34 N^3-6 N^2-56 N-16 \\
P_{231} &=& 26 N^5-85 N^4-247 N^3-256 N^2-384 N-200 \\
P_{232} &=& 69 N^5+90 N^4+161 N^3-24 N^2-544 N-136 \\
P_{233} &=& 81 N^5+476 N^4+1095 N^3+960 N^2+192 N-8 \\
P_{234} &=& 138 N^5+1161 N^4+3207 N^3+3084 N^2+832 N+152 \\
P_{235} &=& 148 N^5+2667 N^4+8551 N^3+8434 N^2+1744 N+32 \\
P_{236} &=& 281 N^5+10466 N^4+35243 N^3+32672 N^2+4200 N-56 \\
P_{237} &=& 312 N^5+4447 N^4+14127 N^3+12884 N^2+736 N-616 \\
P_{238} &=& 509 N^5+6027 N^4+16730 N^3+13260 N^2-112 N-720 \\
P_{239} &=& 622 N^5+9239 N^4+25261 N^3+17376 N^2-3088 N-1608 \\
P_{240} &=& 986 N^5+1343 N^4-619 N^3+1928 N^2+4320 N+760 \\
P_{241} &=& 333 N^6+498 N^5+305 N^4-984 N^3-1744 N^2+824 N+384 \\
P_{242} &=& 1257 N^6+1926 N^5+809 N^4-4344 N^3-5944 N^2+4184 N+1728 \\
P_{243} &=& -655 N^7-2737 N^6-6746 N^5-5147 N^4+10039 N^3+14110 N^2+2144 N
\nonumber\\ &&
+1664 \\
P_{244} &=& -41 N^7-3022 N^6-7001 N^5-316 N^4+4960 N^3-664 N^2-720 N-288 \\
P_{245} &=& -35 N^7-630 N^6-1123 N^5+736 N^4+1704 N^3+56 N^2-240 N-96 \\
P_{246} &=& 27 N^7+142 N^6+239 N^5+42 N^4-172 N^3-12 N^2+112 N+32 \\
P_{247} &=& 35 N^7+630 N^6+1123 N^5-736 N^4-1704 N^3-56 N^2+240 N+96 \\
P_{248} &=& 91 N^8+399 N^7-154 N^6-2226 N^5-1975 N^4+1041 N^3+1612 N^2+540 N
\nonumber\\ &&
+96 \\
P_{249} &=& 1315 N^8+12596 N^7+38983 N^6+40952 N^5-7580 N^4-34384 N^3-15696 N^2
\nonumber\\ &&
-5184 N-960 \\
P_{250} &=& -4591 N^9-20653 N^8-40250 N^7+3673 N^6+114127 N^5+81214 N^4
\nonumber\\ &&
-32752 N^3-12064 N^2+4608 N+2304 \\
P_{251} &=& -1982 N^9-30658 N^8-96466 N^7-60989 N^6+100196 N^5+107737 N^4
\nonumber\\ &&
+16630 N^3-700 N^2-28200 N-7296 \\
P_{252} &=& -872 N^9-10544 N^8-35123 N^7-39055 N^6-8753 N^5+3821 N^4+20130 N^3
\nonumber\\ &&
+35908 N^2+10296 N+1728 \\
P_{253} &=& -506 N^9-4494 N^8-13995 N^7-16365 N^6-3 N^5+18045 N^4+18554 N^3
\nonumber\\ &&
+4500 N^2-5736 N-1728 \\
P_{254} &=& -328 N^9-1493 N^8-2792 N^7+735 N^6+8674 N^5+5592 N^4-2908 N^3
\nonumber\\ &&
-1144 N^2+384 N+192 \\
P_{255} &=& 48 N^9+344 N^8+251 N^7-2097 N^6-4443 N^5-1377 N^4+3046 N^3+2684 N^2
\nonumber\\ &&
+776 N+192 \\
P_{256} &=& 250 N^9+1934 N^8+971 N^7-13805 N^6-27133 N^5-6575 N^4+20026 N^3
\nonumber\\ &&
+16628 N^2+4632 N+1344 \\
P_{257} &=& 4373 N^9+29306 N^8+96356 N^7+132842 N^6-14982 N^5-189180 N^4
\nonumber\\ &&
-178397 N^3-234014 N^2-226688 N-20960 \\
P_{258} &=& 295 N^{10}+5367 N^9+19123 N^8+29641 N^7+38806 N^6+49828 N^5+22980 N^4
\nonumber\\ &&
-9200 N^3-4416 N^2-2304 N-576 \\
P_{259} &=& -2800 N^{11}-19299 N^{10}-33325 N^9+30156 N^8+134148 N^7+92847 N^6
\nonumber\\ &&
-44207 N^5-79356 N^4-39380 N^3-5376 N^2+3456 N+1152 \\
P_{260} &=& -448 N^{11}-2849 N^{10}-5377 N^9+1150 N^8+12659 N^7+7001 N^6-6900 N^5
\nonumber\\ &&
-3080 N^4+4044 N^3+1288 N^2-384 N-192 \\
P_{261} &=& -1712 N^{12}-16317 N^{11}-47823 N^{10}-18734 N^9+151866 N^8+262309 N^7
\nonumber\\ &&
+72397 N^6-166982 N^5-161188 N^4-48600 N^3+2432 N^2
\nonumber\\ &&
+8064 N+2304 \\
P_{262} &=& 2276 N^{12}+15325 N^{11}+41891 N^{10}+22268 N^9-96177 N^8-122331 N^7
\nonumber\\ &&
+33640 N^6+48500 N^5-41816 N^4-14216 N^3+1776 N^2+4032 N+1152 \\
P_{263} &=& 5654 N^{12}+70953 N^{11}+348127 N^{10}+916098 N^9+1385076 N^8+847587 N^7
\nonumber\\ &&
-814529 N^6-1661106 N^5-566908 N^4+339000 N^3+70080 N^2-58752 N
\nonumber\\ &&
-20736 \\
P_{264} &=& 5495 N^{14}+50689 N^{13}+178557 N^{12}+268278 N^{11}+52269 N^{10}-333410 N^9
\nonumber\\ &&
-378425 N^8-130675 N^7+15234 N^6+73308 N^5+78768 N^4+22328 N^3
\nonumber\\ &&
-912 N^2-4032 N-1152 \\
P_{265} &=& 33626 N^{14}+327594 N^{13}+1183095 N^{12}+1765284 N^{11}-64157 N^{10}
\nonumber\\ &&
-4020690 N^9-5736551 N^8-2479656 N^7+2272011 N^6+3318528 N^5
\nonumber\\ &&
+1330292 N^4+168672 N^3+59904 N^2+31104 N+6912 \\
P_{266} &=& -14920 N^{15}-144143 N^{14}-620094 N^{13}-1362644 N^{12}-1506716 N^{11}
\nonumber\\ &&
-403094 N^{10}+2274640 N^9+6259607 N^8+6641532 N^7+1039004 N^6
\nonumber\\ &&
-2282692 N^5-901888 N^4-112512 N^3+50400 N^2+34560 N+6912 \\
P_{267} &=& -59572 N^{17}-710609 N^{16}-3480671 N^{15}-8595070 N^{14}-9308799 N^{13}
\nonumber\\ &&
+3900280 N^{12}+23518594 N^{11}+24867904 N^{10}+2652572 N^9-16072559 N^8
\nonumber\\ &&
-12326550 N^7-79156 N^6+3370116 N^5+1199200 N^4+168960 N^3
\nonumber\\ &&
-45216 N^2-34560 N-6912~. 
\end{eqnarray}
%------------------------------------------------------------------------------------------------------------------------------

Diagram~10 is given by
%-------------------------------------------------------------------------------------------------------------------------------
\begin{eqnarray}
%%%texparser:LHS:testD10phys%%%
D_{10} &=&
\textcolor{blue}{C_A^2 T_F} \Biggl\{
        \frac{1}{\varepsilon^3} \Biggl[
                -\frac{8 (2 N+5) S_{-2}}{3 (N+1) (N+2)}
                -\frac{2 \big(2 N^2+9 N-4\big) S_2}{3 N (N+1) (N+2)}
                -\frac{2 \big(4 N^2+N+8\big) S_1^2}{3 N (N+1) (N+2)}
\nonumber \\ &&
                +\frac{4 P_{274} S_1}{3 N^2 (N+1) (N+2)^2}
                -\frac{4 P_{292}}{3 N (N+1)^3 (N+2)^3}
                +\frac{(-1)^N}{3 N} \Biggl(
                        \frac{4 P_{276}}{(N+1)^3 (N+2)^3}
\nonumber \\ &&
                        -\frac{4 (N-10) S_2}{(N+1) (N+2)}
                        -\frac{8 (N-10) S_{-2}}{(N+1) (N+2)}
                        -\frac{4 \big(N^3-10 N^2-20 N-12\big) S_1}{N (N+1)^2 (N+2)^2}
                \Biggr)
        \Biggr]
\nonumber \\ &&
        +\frac{1}{\varepsilon^2} \Biggl[
                \frac{2 P_{326}}{3 N (N+1)^4 (N+2)^4}
                -\frac{\big(4 N^2+N+8\big) S_1^3}{3 N (N+1) (N+2)}
                +\frac{2 \big(20 N^2+43 N+12\big) S_{2,1}}{3 N (N+1) (N+2)}
\nonumber \\ &&
                +\frac{4 \big(4 N^3-14 N^2-41 N-40\big)}{3 N (N+1) (N+2)} S_{-3}
                +\frac{8 \big(4 N^3+30 N^2+37 N+24\big)}{3 N (N+1) (N+2)} S_{-2,1}
\nonumber \\ &&
                +\frac{2 \big(16 N^3+13 N^2-57 N-26\big)}{3 N (N+1) (N+2)} S_3
                +\frac{P_{287} S_2+P_{293} S_1^2}{3 N^2 (N+1)^2 (N+2)^2}
%                +\frac{P_{287} S_2}{3 N^2 (N+1)^2 (N+2)^2}
%                +\frac{P_{293} S_1^2}{3 N^2 (N+1)^2 (N+2)^2}
\nonumber \\ &&
                +\biggl(
                        -\frac{2 P_{336}}{3 N^3 (N+1)^3 (N+2)^3}
                        -\frac{24 N^2+57 N+8}{3 N (N+1) (N+2)} S_2
                \biggr) S_1
\nonumber \\ &&
                +\biggl(
                        -\frac{4 P_{296}}{3 N^2 (N+1)^2 (N+2)^2}
                        -\frac{4 \big(20 N^2+43 N+20\big)}{3 N (N+1) (N+2)} S_1
                \biggr) S_{-2}
\nonumber \\ &&
                +(-1)^N \Biggl(
                        \frac{P_{271} S_2}{3 N^2 (N+1)^2 (N+2)^2}
                        -\frac{2 (N+10) S_3}{3 N (N+1) (N+2)}
                        -\frac{4 (3 N-20) S_{-3}}{3 N (N+1) (N+2)}
\nonumber \\ &&
                        -\frac{4 (3 N-10) S_{2,1}}{3 N (N+1) (N+2)}
                        +\frac{-N^3+10 N^2+20 N+12}{3 N^2 (N+1)^2 (N+2)^2} S_1^2
                        -\frac{2 P_{315}}{3 N (N+1)^4 (N+2)^4}
\nonumber \\ &&
                        -\frac{8 S_{-2,1}}{3 (N+1) (N+2)}
                        +\biggl(
                                -\frac{2 P_{324}}{3 N^3 (N+1)^3 (N+2)^3}
                                +\frac{4 S_2}{3 (N+1) (N+2)}
                        \biggr) S_1
\nonumber \\ &&
                        +\biggl(
                                -\frac{8 \big(3 N^3+48 N^2+95 N+48\big)}{3 N (N+1)^2 (N+2)^2}
                                +\frac{8 S_1}{3 (N+1) (N+2)}
                        \biggr) S_{-2}
                \Biggr)
        \Biggr]
\nonumber \\ &&
        +\frac{1}{\varepsilon} \Biggl[
                -\frac{(3 N+4) \big(32 N^2-8 N-53\big)}{3 N (N+1) (N+2)} S_{3,1}
                +\frac{P_{342}}{3 N (N+1)^5 (N+2)^5}
\nonumber \\ &&
                -\frac{82 N^2+149 N+48}{3 N (N+1) (N+2)} S_{2,1,1}
                +\frac{16 \big(2 N^3+9 N^2+6 N+3\big)}{3 N (N+1) (N+2)} S_{-2}^2
\nonumber \\ &&
                -\frac{7 \big(4 N^2+N+8\big)}{72 N (N+1) (N+2)} S_1^4
                +\frac{4 \big(2 N^3+68 N^2+109 N+98\big)}{3 N (N+1) (N+2)} S_{-3,1}
\nonumber \\ &&
                +\frac{4 \big(8 N^3+80 N^2+103 N+84\big)}{3 N (N+1) (N+2)} S_{-2,2}
                +\frac{2 \big(12 N^3-40 N^2-131 N-136\big)}{3 N (N+1) (N+2)} S_{-4}
\nonumber \\ &&
                -\frac{8 \big(14 N^3+92 N^2+97 N+70\big)}{3 N (N+1) (N+2)} S_{-2,1,1}
                +\frac{64 N^3+388 N^2+329 N+136}{24 N (N+1) (N+2)} S_2^2
\nonumber \\ &&
                +\frac{P_{339} S_2}{6 N^3 (N+1)^3 (N+2)^3}
                +\frac{P_{295} S_1^3+6 P_{300} S_{2,1}+2 P_{306} S_3-12 P_{314} S_{-2,1}}{18 N^2 (N+1)^2 (N+2)^2}
%                +\frac{P_{295} S_1^3}{18 N^2 (N+1)^2 (N+2)^2}
%                +\frac{P_{300} S_{2,1}}{3 N^2 (N+1)^2 (N+2)^2}
%                +\frac{P_{306} S_3}{9 N^2 (N+1)^2 (N+2)^2}
%                -\frac{2 P_{314} S_{-2,1}}{3 N^2 (N+1)^2 (N+2)^2}
\nonumber \\ &&
                +\frac{384 N^3+536 N^2-781 N-968}{12 N (N+1) (N+2)} S_4
                +\biggl(
                        \frac{2 \big(96 N^3+10 N^2-359 N-250\big)}{9 N (N+1) (N+2)} S_3
\nonumber \\ &&
                        +\frac{4 \big(16 N^3+131 N^2+162 N+110\big)}{3 N (N+1) (N+2)} S_{-2,1}
                        +\frac{88 N^2+187 N+52}{3 N (N+1) (N+2)} S_{2,1}
\nonumber \\ &&
                        +\frac{P_{286} S_2}{6 N^2 (N+1)^2 (N+2)^2}
                        +\frac{P_{343}}{3 N^2 (N+1)^4 (N+2)^4}
                \biggr) S_1
\nonumber \\ &&
                +\biggl(
                        \frac{P_{334}}{6 N^3 (N+1)^3 (N+2)^3}
                        -\frac{236 N^2+433 N+272}{12 N (N+1) (N+2)} S_2
                \biggr) S_1^2
\nonumber \\ &&
                +\biggl(
                        \frac{2 P_{316}}{3 N (N+1)^3 (N+2)^3}
                        -\frac{116 N^2+229 N+132}{3 N (N+1) (N+2)} S_1^2
\nonumber \\ &&
                        +\frac{32 N^3+72 N^2-81 N+20}{3 N (N+1) (N+2)} S_2
                        -\frac{2 P_{301} S_1}{3 N^2 (N+1)^2 (N+2)^2}
                \biggr) S_{-2}
\nonumber \\ &&
                +\biggl(
                        \frac{P_{311}}{3 N^2 (N+1)^2 (N+2)^2}
                        +\frac{2 \big(16 N^3-89 N^2-192 N-210\big)}{3 N (N+1) (N+2)} S_1
                \biggr) S_{-3}
\nonumber \\ &&
                +\biggl(
                        \frac{P_{290}}{2 N (N+1)^3 (N+2)^3}
                        -\frac{4 N^2+N+8}{4 N (N+1) (N+2)} S_1^2
                        -\frac{2 N^2+9 N-4}{4 N (N+1) (N+2)} S_2
\nonumber \\ &&
                        +\frac{P_{274} S_1}{2 N^2 (N+1) (N+2)^2}
                        -\frac{(2 N+5) S_{-2}}{(N+1) (N+2)}
                \biggr) \zeta_2
                +(-1)^N \Biggl[
                        \frac{(30-N) S_2^2}{3 N (N+1) (N+2)}
\nonumber \\ &&
                        +\frac{P_{338}}{3 N (N+1)^5 (N+2)^5}
                        +\frac{4 (N-10) S_{-2}^2}{3 N (N+1) (N+2)}
                        +\frac{10 (N+16) S_{3,1}}{3 N (N+1) (N+2)}
\nonumber \\ &&
                        -\frac{4 (3 N-5) S_{2,1,1}}{3 N (N+1) (N+2)}
                        -\frac{2 (7 N+10) S_4}{3 N (N+1) (N+2)}
                        -\frac{2 (9 N-80) S_{-4}}{3 N (N+1) (N+2)}
\nonumber \\ &&
                        +\frac{-N^3+10 N^2+20 N+12}{18 N^2 (N+1)^2 (N+2)^2} S_1^3
                        -\frac{2 \big(82 N^3+149 N^2+32 N-96\big)}{3 N^2 (N+1)^2 (N+2)^2} S_{2,1}
\nonumber \\ &&
                        +\frac{4 P_{275} S_{-2,1}}{3 N^2 (N+1)^2 (N+2)^2}
                        +\frac{P_{281} S_3}{9 N^2 (N+1)^2 (N+2)^2}
                        +\frac{P_{322} S_1^2}{6 N^3 (N+1)^3 (N+2)^3}
\nonumber \\ &&
                        +\frac{P_{329} S_2}{6 N^3 (N+1)^3 (N+2)^3}
                        +\biggl(
                                \frac{P_{337}}{3 N^2 (N+1)^4 (N+2)^4}
                                +\frac{P_{272} S_2}{6 N^2 (N+1)^2 (N+2)^2}
\nonumber \\ &&
                                -\frac{2 S_3}{3 (N+1) (N+2)}
                                +\frac{4 S_{2,1}}{3 (N+1) (N+2)}
                        \biggr) S_1
                        +\biggl(
                                \frac{4 P_{297}}{3 N (N+1)^3 (N+2)^3}
\nonumber \\ &&
                                -\frac{8 (N-15) S_2}{3 N (N+1) (N+2)}
                                -\frac{4 P_{275} S_1}{3 N^2 (N+1)^2 (N+2)^2}
                        \biggr) S_{-2}
                        -\frac{4 S_{-2,2}}{3 (N+1) (N+2)}
\nonumber \\ &&
                        +\biggl(
                                -\frac{2 P_{278}}{3 N^2 (N+1)^2 (N+2)^2}
                                +\frac{8 S_1}{3 (N+1) (N+2)}
                        \biggr) S_{-3}
                        -\frac{8 S_{-3,1}}{3 (N+1) (N+2)}
\nonumber \\ &&
                        +\biggl(
                                \frac{P_{276}}{2 N (N+1)^3 (N+2)^3}
                                +\frac{(10-N) S_2}{2 N (N+1) (N+2)}
                                +\frac{(10-N) S_{-2}}{N (N+1) (N+2)}
\nonumber \\ &&
                                +\frac{\big(-N^3+10 N^2+20 N+12\big) S_1}{2 N^2 (N+1)^2 (N+2)^2}
                        \biggr) \zeta_2
                \Biggr]
        \Biggr]
        +\frac{278 N^2+451 N+144}{6 N (N+1) (N+2)} S_{2,1,1,1}
\nonumber \\ &&
        +\frac{P_{349}}{6 N (N+1)^6 (N+2)^6}
        -\frac{216 N^3+337 N^2-79 N-494}{3 N (N+1) (N+2)} S_{4,1}
\nonumber \\ &&
        -\frac{4 N^2+N+8}{48 N (N+1) (N+2)} S_1^5
        -\frac{80 N^3+539 N^2+770 N+454}{6 N (N+1) (N+2)} S_5
\nonumber \\ &&
        -\frac{48 N^3-2 N^2-637 N-240}{6 N (N+1) (N+2)} S_{2,2,1}
        -\frac{2 \big(2 N^3-143 N^2-297 N-276\big)}{3 N (N+1) (N+2)} S_{-4,1}
\nonumber \\ &&
        +\frac{16 \big(4 N^3+21 N^2+13 N+10\big)}{3 N (N+1) (N+2)} S_{-2,-3}
        -\frac{8 \big(8 N^3+41 N^2+31 N+24\big)}{3 N (N+1) (N+2)} S_{-2,1,-2}
\nonumber \\ &&
        +\frac{2 \big(16 N^3+85 N^2+94 N+74\big)}{N (N+1) (N+2)} S_{2,-3}
        +\frac{52 N^3+400 N^2+171 N+240}{3 N (N+1) (N+2)} S_{-5}
\nonumber \\ &&
        -\frac{4 \big(19 N^3+179 N^2+228 N+187\big)}{3 N (N+1) (N+2)} S_{-3,1,1}
        -\frac{2 P_{294} S_{-2,2}}{3 N (N+1)^2 (N+2)^2}
\nonumber \\ &&
        -\frac{4 \big(28 N^3+199 N^2+221 N+170\big)}{3 N (N+1) (N+2)} S_{-2,2,1}
        +\frac{P_{298} S_1^4}{144 N^2 (N+1)^2 (N+2)^2}
\nonumber \\ &&
        -\frac{4 \big(32 N^3+253 N^2+288 N+266\big)}{3 N (N+1) (N+2)} S_{2,1,-2}
        +\frac{P_{303} S_4}{24 N^2 (N+1)^2 (N+2)^2}
\nonumber \\ &&
        +\frac{8 \big(37 N^3+219 N^2+214 N+153\big)}{3 N (N+1) (N+2)} S_{-2,1,1,1}
        +\frac{P_{305} S_2^2}{48 N^2 (N+1)^2 (N+2)^2}
\nonumber \\ &&
        -\frac{2 \big(64 N^3+426 N^2+319 N+324\big)}{3 N (N+1) (N+2)} S_{-2,3}
        +\frac{P_{309} S_{-4}}{6 N^2 (N+1)^2 (N+2)^2}
\nonumber \\ &&
        +\frac{432 N^3+918 N^2-229 N-552}{6 N (N+1) (N+2)} S_{2,3}
        +\frac{P_{312} S_{-3,1}}{6 N^2 (N+1)^2 (N+2)^2}
\nonumber \\ &&
        +\frac{480 N^3+626 N^2-437 N-1112}{6 N (N+1) (N+2)} S_{3,1,1}
        +\frac{P_{285} S_{2,1,1}}{6 N^2 (N+1)^2 (N+2)^2}
\nonumber \\ &&
        -\frac{4 P_{313} S_{-2}^2}{3 N^2 (N+1)^2 (N+2)^2}
        +\frac{P_{317} S_{-2,1,1}}{3 N^2 (N+1)^2 (N+2)^2}
        +\frac{P_{318} S_{3,1}}{6 N^2 (N+1)^2 (N+2)^2}
\nonumber \\ &&
        +\frac{P_{325} S_{-3}}{6 N (N+1)^3 (N+2)^3}
        +\frac{P_{345} S_3}{18 N^3 (N+1)^3 (N+2)^3}
        +\biggl[
                \frac{P_{348}}{6 N^2 (N+1)^5 (N+2)^5}
\nonumber \\ &&
                -\frac{404 N^2+737 N+308}{6 N (N+1) (N+2)} S_{2,1,1}
                -\frac{384 N^3+158 N^2-1001 N-1240}{6 N (N+1) (N+2)} S_{3,1}
\nonumber \\ &&
                +\frac{8 \big(8 N^3+33 N^2+23 N+8\big)}{3 N (N+1) (N+2)} S_{-2}^2
                +\frac{16 N^3+613 N^2+899 N+870}{3 N (N+1) (N+2)} S_{-3,1}
\nonumber \\ &&
                +\frac{2 \big(32 N^3+338 N^2+421 N+356\big)}{3 N (N+1) (N+2)} S_{-2,2}
                +\frac{P_{288} S_{-2,1}}{3 N (N+1)^2 (N+2)^2}
\nonumber \\ &&
                +\frac{48 N^3-233 N^2-532 N-630}{3 N (N+1) (N+2)} S_{-4}
                +\frac{P_{340} S_2}{12 N^3 (N+1)^3 (N+2)^3}
\nonumber \\ &&
                -\frac{2 \big(112 N^3+739 N^2+785 N+554\big)}{3 N (N+1) (N+2)} S_{-2,1,1}
                +\frac{P_{302} S_{2,1}}{6 N^2 (N+1)^2 (N+2)^2}
\nonumber \\ &&
                +\frac{256 N^3+1452 N^2+1801 N+408}{48 N (N+1) (N+2)} S_2^2
                +\frac{P_{304} S_3}{18 N^2 (N+1)^2 (N+2)^2}
\nonumber \\ &&
                +\frac{1536 N^3+1400 N^2-2873 N-4760}{24 N (N+1) (N+2)} S_4
                +\frac{P_{307} S_{-3}}{6 N^2 (N+1)^2 (N+2)^2}
\nonumber \\ &&
                +\biggl(
                        \frac{P_{319}}{3 N (N+1)^3 (N+2)^3}
                        +\frac{128 N^3+56 N^2-547 N-276}{6 N (N+1) (N+2)} S_2
                \biggr) S_{-2}
        \biggr] S_1
\nonumber \\ &&
        +\biggl(
                \frac{480 N^2+937 N+412}{12 N (N+1) (N+2)} S_{2,1}
                +\frac{64 N^3-463 N^2-860 N-1006}{6 N (N+1) (N+2)} S_{-3}
\nonumber \\ &&
                +\frac{P_{344}}{12 N^2 (N+1)^4 (N+2)^4}
                +\frac{64 N^3+549 N^2+682 N+466}{3 N (N+1) (N+2)} S_{-2,1}
\nonumber \\ &&
                +\frac{128 N^3-193 N^2-734 N-688}{6 N (N+1) (N+2)} S_3
                +\frac{P_{283} S_2+4 P_{284} S_{-2}}{24 N^2 (N+1)^2 (N+2)^2}
%                +\frac{P_{283} S_2}{24 N^2 (N+1)^2 (N+2)^2}
%                +\frac{P_{284} S_{-2}}{6 N^2 (N+1)^2 (N+2)^2}
        \biggr) S_1^2
\nonumber \\ &&
        +\biggl(
                -\frac{1188 N^2+2109 N+1424}{72 N (N+1) (N+2)} S_2
                -\frac{572 N^2+1063 N+668}{18 N (N+1) (N+2)} S_{-2}
\nonumber \\ &&
                +\frac{P_{333}}{36 N^3 (N+1)^3 (N+2)^3}
        \biggr) S_1^3
        +\biggl(
                -\frac{672 N^3+1215 N^2-126 N-1688}{18 N (N+1) (N+2)} S_3
\nonumber \\ &&
                +\frac{P_{341}}{12 N^2 (N+1)^4 (N+2)^4}
                +\frac{144 N^3+795 N^2+538 N+638}{6 N (N+1) (N+2)} S_{-3}
        \biggr) S_2
\nonumber \\ &&
        +\biggl(
                \frac{P_{332}}{3 N (N+1)^4 (N+2)^4}
                +\frac{8 \big(8 N^3+39 N^2+33 N+24\big)}{3 N (N+1) (N+2)} S_{-3}
\nonumber \\ &&
                +\frac{288 N^3+796 N^2-175 N-92}{9 N (N+1) (N+2)} S_3
                +\frac{P_{308} S_2}{6 N^2 (N+1)^2 (N+2)^2}
        \biggr) S_{-2}
\nonumber \\ &&
        +\biggl(
                \frac{P_{331}}{6 N^3 (N+1)^3 (N+2)^3}
                +\frac{-32 N^3-256 N^2-753 N-108}{12 N (N+1) (N+2)} S_2
\nonumber \\ &&
                +\frac{2 \big(56 N^3+504 N^2+545 N+508\big)}{3 N (N+1) (N+2)} S_{-2}
        \biggr) S_{2,1}
        +\biggl(
                \frac{P_{323}}{3 N (N+1)^3 (N+2)^3}
\nonumber \\ &&
                -\frac{112 N^3+641 N^2+504 N+546}{3 N (N+1) (N+2)} S_2
                -\frac{8 \big(8 N^3+33 N^2+23 N+8\big)}{3 N (N+1) (N+2)} S_{-2}
        \biggr) S_{-2,1}
\nonumber \\ &&
        +\biggl[
                \frac{P_{326}}{4 N (N+1)^4 (N+2)^4}
                -\frac{4 N^2+N+8}{8 N (N+1) (N+2)} S_1^3
                +\frac{20 N^2+43 N+12}{4 N (N+1) (N+2)} S_{2,1}
\nonumber \\ &&
                +\frac{4 N^3-14 N^2-41 N-40}{2 N (N+1) (N+2)} S_{-3}
                +\frac{16 N^3+13 N^2-57 N-26}{4 N (N+1) (N+2)} S_3
\nonumber \\ &&
                +\frac{P_{287} S_2+P_{293} S_1^2}{8 N^2 (N+1)^2 (N+2)^2}
%                +\frac{P_{287} S_2}{8 N^2 (N+1)^2 (N+2)^2}
%                +\frac{P_{293} S_1^2}{8 N^2 (N+1)^2 (N+2)^2}
                +\frac{4 N^3+30 N^2+37 N+24}{N (N+1) (N+2)} S_{-2,1}
\nonumber \\ &&
                +\biggl(
                        \frac{P_{335}}{4 N^3 (N+1)^3 (N+2)^3}
                        -\frac{24 N^2+57 N+8}{8 N (N+1) (N+2)} S_2
                \biggr) S_1
\nonumber \\ &&
                +\biggl(
                        \frac{P_{289}}{2 N^2 (N+1)^2 (N+2)^2}
                        -\frac{20 N^2+43 N+20}{2 N (N+1) (N+2)} S_1
                \biggr) S_{-2}
        \biggr] \zeta_2
\nonumber \\ &&
        +\biggl(
                \frac{7 P_{292}}{6 N (N+1)^3 (N+2)^3}
                +\frac{7 \big(2 N^2+9 N-4\big) S_2}{12 N (N+1) (N+2)}
                +\frac{7 \big(4 N^2+N+8\big) S_1^2}{12 N (N+1) (N+2)}
\nonumber \\ &&
                -\frac{7 P_{274} S_1}{6 N^2 (N+1) (N+2)^2}
                +\frac{7 (2 N+5) S_{-2}}{3 (N+1) (N+2)}
        \biggr) \zeta_3
        +(-1)^N \Biggl[
                \frac{4 S_{2,1,-2}}{3 (N+1) (N+2)}
\nonumber \\ &&
                +\frac{P_{346}}{6 N (N+1)^6 (N+2)^6}
                +\frac{4 S_{-2,1,-2}}{3 (N+1) (N+2)}
                +\frac{(240-31 N) S_{-5}}{3 N (N+1) (N+2)}
\nonumber \\ &&
                -\frac{20 (N-3) S_{2,2,1}}{3 N (N+1) (N+2)}
                -\frac{8 (N+5) S_{2,3}}{3 N (N+1) (N+2)}
                -\frac{7 (N+10) S_5}{6 N (N+1) (N+2)}
\nonumber \\ &&
                -\frac{2 (4 N-5) S_{2,1,1,1}}{3 N (N+1) (N+2)}
                -\frac{4 (7 N-20) S_{3,1,1}}{3 N (N+1) (N+2)}
                -\frac{N^3-10 N^2-20 N-12}{144 N^2 (N+1)^2 (N+2)^2} S_1^4
\nonumber \\ &&
                +\frac{P_{268} S_2^2+16 P_{269} S_{3,1}+32 P_{275} S_{-2,2}+64 P_{275} S_{-3,1}+16 P_{277} S_{2,1,1}+2 P_{282} S_4}{48 N^2 (N+1)^2 (N+2)^2}
%                +\frac{P_{268} S_2^2}{48 N^2 (N+1)^2 (N+2)^2}
%                +\frac{P_{269} S_{3,1}}{3 N^2 (N+1)^2 (N+2)^2}
%                +\frac{2 P_{275} S_{-2,2}}{3 N^2 (N+1)^2 (N+2)^2}
%                +\frac{4 P_{275} S_{-3,1}}{3 N^2 (N+1)^2 (N+2)^2}
%                +\frac{P_{277} S_{2,1,1}}{3 N^2 (N+1)^2 (N+2)^2}
%                +\frac{P_{282} S_4}{24 N^2 (N+1)^2 (N+2)^2}
\nonumber \\ &&
                -\frac{2 P_{291} S_{-2,1}}{3 N (N+1)^3 (N+2)^3}
                +\frac{P_{322} S_1^3+2 P_{320} S_3+12 P_{327} S_{2,1}}{36 N^3 (N+1)^3 (N+2)^3}
%                +\frac{P_{320} S_3}{18 N^3 (N+1)^3 (N+2)^3}
%                +\frac{P_{322} S_1^3}{36 N^3 (N+1)^3 (N+2)^3}
%                +\frac{P_{327} S_{2,1}}{3 N^3 (N+1)^3 (N+2)^3}
                +\frac{(N+40) S_{4,1}}{N (N+1) (N+2)}
\nonumber \\ &&
                +\biggl(
                        \frac{P_{321} S_2}{12 N^3 (N+1)^3 (N+2)^3}
                        +\frac{2 S_{2,1,1}+3 S_2^2}{3 (N+1) (N+2)}
%                        +\frac{2 S_{2,1,1}}{3 (N+1) (N+2)}
%                        +\frac{S_2^2}{(N+1) (N+2)}
                        +\frac{6 P_{273} S_{2,1}+P_{279} S_3}{18 N^2 (N+1)^2 (N+2)^2}
%                        +\frac{P_{273} S_{2,1}}{3 N^2 (N+1)^2 (N+2)^2}
%                        +\frac{P_{279} S_3}{18 N^2 (N+1)^2 (N+2)^2}
\nonumber \\ &&
                        +\frac{P_{347}}{6 N^2 (N+1)^5 (N+2)^5}
                        +\frac{16 S_{3,1}-2 S_4}{3 (N+1) (N+2)}
%                        -\frac{2 S_4}{3 (N+1) (N+2)}
%                        +\frac{16 S_{3,1}}{3 (N+1) (N+2)}
                \biggr) S_1
                +\biggl(
                        \frac{P_{337}}{12 N^2 (N+1)^4 (N+2)^4}
\nonumber \\ &&
                        +\frac{-7 N^3+34 N^2+76 N+52}{8 N^2 (N+1)^2 (N+2)^2} S_2
                \biggr) S_1^2
                +\biggl(
                        \frac{P_{330}}{12 N^2 (N+1)^4 (N+2)^4}
\nonumber \\ &&
                        +\frac{5 (N+10) S_3}{3 N (N+1) (N+2)}
                        -\frac{2 (3 N-20) S_{2,1}}{3 N (N+1) (N+2)}
                        -\frac{4 S_{-2,1}}{(N+1) (N+2)}
                \biggr) S_2
\nonumber \\ &&
                +\biggl[
                        -\frac{2 P_{328}}{3 N (N+1)^4 (N+2)^4}
                        +\frac{100 S_3}{3 N (N+1) (N+2)}
                        -\frac{2 P_{280} S_2}{3 N^2 (N+1)^2 (N+2)^2}
\nonumber \\ &&
                        +\biggl(
                                \frac{2 P_{291}}{3 N (N+1)^3 (N+2)^3}
                                +\frac{4 S_2}{(N+1) (N+2)}
                        \biggr) S_1
                        +\frac{4 S_{-2,1}}{3 (N+1) (N+2)}
                \biggr] S_{-2}
\nonumber \\ &&
                +\biggl(
                        \frac{4 \big(3 N^3+48 N^2+95 N+48\big)}{3 N (N+1)^2 (N+2)^2}
                        -\frac{4 S_1}{3 (N+1) (N+2)}
                \biggr) S_{-2}^2
\nonumber \\ &&
                -\frac{2 S_{2,-3}}{3 (N+1) (N+2)}
                +\biggl(
                        \frac{P_{299}}{3 N (N+1)^3 (N+2)^3}
                        +\frac{(3 N-20) \left(4 S_{-2}-6 S_2\right)}{3 N (N+1) (N+2)}
%                        +\frac{4 (3 N-20) S_{-2}}{3 N (N+1) (N+2)}
%                        -\frac{S_2}{N (N+1) (N+2)} (2 (3 N-20))
\nonumber \\ &&
                        -\frac{4 P_{275} S_1}{3 N^2 (N+1)^2 (N+2)^2}
                \biggr) S_{-3}
                -\frac{10 S_{-2,3}}{3 (N+1) (N+2)}
                +\biggl(
                        \frac{P_{270}}{3 N^2 (N+1)^2 (N+2)^2}
\nonumber \\ &&
                        +\frac{16 S_1}{3 (N+1) (N+2)}
                \biggr) S_{-4}
                -\frac{8 \left(S_{-2,-3}+2 S_{-4,1}\right)}{3 (N+1) (N+2)}
%                -\frac{8 S_{-2,-3}}{3 (N+1) (N+2)}
%                -\frac{16 S_{-4,1}}{3 (N+1) (N+2)}
                +\biggl(
                        \frac{P_{310}}{4 N (N+1)^4 (N+2)^4}
\nonumber \\ &&
                        +\frac{(10-3 N) S_{2,1}}{2 N (N+1) (N+2)}
                        +\frac{(20-3 N) S_{-3}}{2 N (N+1) (N+2)}
                        -\frac{(N+10) S_3+4 N S_{-2,1}}{4 N (N+1) (N+2)}
%                        -\frac{(N+10) S_3}{4 N (N+1) (N+2)}
%                        -\frac{S_{-2,1}}{(N+1) (N+2)}
\nonumber \\ &&
                        +\frac{-N^3+10 N^2+20 N+12}{8 N^2 (N+1)^2 (N+2)^2} S_1^2
                        +\frac{P_{271} S_2}{8 N^2 (N+1)^2 (N+2)^2}
                        +\frac{S_1 \left(S_2+2S_{-2}\right)}{2 (N+1) (N+2)}   
%                        +\frac{S_1 S_2}{2 (N+1) (N+2)}   
%                        +\frac{S_1 S_{-2}}{(N+1) (N+2)}
\nonumber \\ &&                      
                        +\frac{P_{322} S_1}{4 N^3 (N+1)^3 (N+2)^3}                      
                        -\frac{3 N^3+48 N^2+95 N+48}{N (N+1)^2 (N+2)^2} S_{-2}
                \biggr) \zeta_2
\nonumber \\ &&             
                +\biggl(
                        -\frac{7 P_{276}}{6 N (N+1)^3 (N+2)^3}
                        +\frac{7 (N-10)}{6 N (N+1) (N+2)} \left(S_2+2 S_{-2}\right)
%                        +\frac{7 (N-10) S_2}{6 N (N+1) (N+2)}
%                        +\frac{7 (N-10) S_{-2}}{3 N (N+1) (N+2)}
\nonumber \\ &&  
                        +\frac{7 \big(N^3-10 N^2-20 N-12\big)}{6 N^2 (N+1)^2 (N+2)^2} S_1
                \biggr) \zeta_3
        \Biggr]
\Biggr\},
\end{eqnarray}
%-------------------------------------------------------------------------------------------------------------------------------
with the polynomials $P_i$
%-------------------------------------------------------------------------------------------------------------------------------
\begin{eqnarray}
P_{268}    &=& -192 N^4-2617 N^3-4838 N^2-2252 N+556 \\
P_{269}    &=& -87 N^4-889 N^3-1703 N^2-956 N-160 \\
P_{270}    &=& -45 N^4-761 N^3-1506 N^2-748 N+32 \\
P_{271}    &=& -12 N^4-213 N^3-414 N^2-204 N+76 \\
P_{272}    &=& -12 N^4-49 N^3+46 N^2+148 N+28 \\
P_{273}    &=& -6 N^4-8 N^3-7 N^2-16 N-64 \\
P_{274}    &=& 2 N^4+4 N^3+5 N^2+8 N+20 \\
P_{275}    &=& 3 N^4+7 N^3+14 N^2+20 N+32 \\
P_{276}    &=& 8 N^4+31 N^3+69 N^2+85 N+40 \\
P_{277} &=& 9 N^4-73 N^3-149 N^2-20 N+192 \\
P_{278} &=& 9 N^4+185 N^3+366 N^2+172 N-32 \\
P_{279} &=& 18 N^4-73 N^3+424 N^2+980 N+852 \\
P_{280} &=& 21 N^4+295 N^3+584 N^2+308 N+32 \\
P_{281} &=& 36 N^4+359 N^3+574 N^2+188 N-276 \\
P_{282} &=& 240 N^4+1303 N^3+2006 N^2+1076 N+268 \\
P_{283} &=& -788 N^5-4029 N^4-5508 N^3-1203 N^2+916 N+356 \\
P_{284} &=& -401 N^5-1935 N^4-2634 N^3-575 N^2+520 N+160 \\
P_{285} &=& -247 N^5-1505 N^4-2844 N^3-2119 N^2-1024 N-288 \\
P_{286} &=& -168 N^5-893 N^4-1520 N^3-463 N^2+724 N+356 \\
P_{287} &=& -40 N^5-225 N^4-432 N^3-267 N^2-12 N-12 \\
P_{288} &=& -32 N^5+429 N^4+2894 N^3+4043 N^2+732 N-408 \\
P_{289} &=& -15 N^5-81 N^4-184 N^3-123 N^2+40 N+32 \\
P_{290} &=& -3 N^5-27 N^4-100 N^3-162 N^2-125 N-40 \\
P_{291} &=& 3 N^5+18 N^4+61 N^3+152 N^2+368 N+344 \\
P_{292} &=& 3 N^5+27 N^4+100 N^3+162 N^2+125 N+40 \\
P_{293} &=& 6 N^5+3 N^4+10 N^3-29 N^2-92 N-12 \\
P_{294} &=& 8 N^5-122 N^4-814 N^3-976 N^2+75 N+172 \\
P_{295} &=& 14 N^5-3 N^4+12 N^3-113 N^2-332 N-76 \\
P_{296} &=& 15 N^5+81 N^4+184 N^3+123 N^2-40 N-32 \\
P_{297} &=& 16 N^5+176 N^4+618 N^3+1095 N^2+1046 N+432 \\
P_{298} &=& 30 N^5-15 N^4+16 N^3-281 N^2-812 N-204 \\
P_{299} &=& 61 N^5+686 N^4+2411 N^3+4228 N^2+3816 N+1384 \\
P_{300} &=& 77 N^5+427 N^4+828 N^3+641 N^2+272 N+96 \\
P_{301} &=& 83 N^5+397 N^4+654 N^3+345 N^2-8 N-32 \\
P_{302} &=& 383 N^5+1995 N^4+3226 N^3+1231 N^2-464 N-96 \\
P_{303} &=& -384 N^6-1190 N^5+1499 N^4+10446 N^3+15921 N^2+11572 N+4084 \\
P_{304} &=& -192 N^6-1614 N^5-5772 N^4-6671 N^3+2380 N^2+7108 N+3276 \\
P_{305} &=& -64 N^6+410 N^5+3337 N^4+7036 N^3+4207 N^2-748 N-748 \\
P_{306} &=& -48 N^6-346 N^5-969 N^4-1398 N^3+64 N^2+1756 N+692 \\
P_{307} &=& -32 N^6-499 N^5-2286 N^4-1209 N^3+2768 N^2+1416 N+384 \\
P_{308} &=& -32 N^6-227 N^5-371 N^4-920 N^3-443 N^2+1416 N+672 \\
P_{309} &=& -24 N^6-309 N^5-1070 N^4-21 N^3+2448 N^2+1736 N+448 \\
P_{310} &=& -16 N^6-44 N^5-82 N^4-267 N^3-631 N^2-759 N-360 \\
P_{311} &=& -8 N^6-117 N^5-476 N^4-319 N^3+580 N^2+592 N+192 \\
P_{312} &=& -8 N^6+527 N^5+2918 N^4+2755 N^3-1346 N^2-896 N-64 \\
P_{313} &=& 4 N^6-12 N^5-163 N^4-275 N^3-55 N^2+76 N+16 \\
P_{314} &=& 8 N^6-91 N^5-656 N^4-1053 N^3-488 N^2-176 N-64 \\
P_{315} &=& 16 N^6+44 N^5+82 N^4+267 N^3+631 N^2+759 N+360 \\
P_{316} &=& 35 N^6+148 N^5+71 N^4-1024 N^3-2538 N^2-1979 N-304 \\
P_{317} &=& 56 N^6-449 N^5-3594 N^4-5053 N^3-746 N^2+480 N-64 \\
P_{318} &=& 96 N^6+318 N^5+223 N^4-547 N^3-465 N^2+608 N-96 \\
P_{319} &=& 175 N^6+596 N^5-355 N^4-4336 N^3-6938 N^2-3787 N-320 \\
P_{320} &=& -362 N^7-2382 N^6-6019 N^5-6081 N^4+2449 N^3+9412 N^2
\nonumber\\ &&
+5592 N+1792 \\
P_{321} &=& -92 N^7-640 N^6-2095 N^5-3307 N^4-3091 N^3-4068 N^2
\nonumber\\ &&
-4504 N-1408 \\
P_{322} &=& -8 N^7-60 N^6-199 N^5-315 N^4-347 N^3-452 N^2-408 N-128 \\
P_{323} &=& 8 N^7-155 N^6-562 N^5+1357 N^4+6808 N^3+7236 N^2+1030 N-848 \\
P_{324} &=& 8 N^7+60 N^6+199 N^5+315 N^4+347 N^3+452 N^2+408 N+128 \\
P_{325} &=& 8 N^7+245 N^6+926 N^5-403 N^4-5232 N^3-6100 N^2-1542 N-16 \\
P_{326} &=& 24 N^7+243 N^6+1035 N^5+2234 N^4+2666 N^3+1946 N^2+1039 N+360 \\
P_{327} &=& 42 N^7+405 N^6+1280 N^5+1975 N^4+923 N^3-756 N^2-600 N-192 \\
P_{328} &=& 56 N^7+715 N^6+3716 N^5+10551 N^4+17234 N^3+15380 N^2
\nonumber\\ &&
+6132 N+392 \\
P_{329} &=& 88 N^7+1088 N^6+3597 N^5+5521 N^4+3861 N^3+972 N^2+392 N+128 \\
P_{330} &=& -248 N^8-3054 N^7-12658 N^6-27179 N^5-32542 N^4-22013 N^3-10253 N^2
\nonumber\\ &&
-3104 N+1360 \\
P_{331} &=& -189 N^8-798 N^7-589 N^6+1752 N^5+2424 N^4-185 N^3-136 N^2
\nonumber\\ &&
+1200 N+384 \\
P_{332} &=& -75 N^8-507 N^7-1503 N^6-2176 N^5-3706 N^4-12514 N^3-23231 N^2
\nonumber\\ &&
-17535 N-3632 \\
P_{333} &=& -28 N^8+2 N^7+417 N^6+1170 N^5-223 N^4-3850 N^3-3068 N^2
\nonumber\\ &&
+408 N+128 \\
P_{334} &=& -12 N^8-8 N^7+109 N^6+224 N^5-511 N^4-1674 N^3-924 N^2
\nonumber\\ &&
+408 N+128 \\
P_{335} &=& -4 N^8-13 N^7-45 N^6-249 N^5-655 N^4-586 N^3+148 N^2
\nonumber\\ &&
+408 N+128 \\
P_{336} &=& 4 N^8+13 N^7+45 N^6+249 N^5+655 N^4+586 N^3-148 N^2-408 N-128 \\
P_{337} &=& 8 N^8-58 N^7-930 N^6-4187 N^5-8890 N^4-8957 N^3-2589 N^2
\nonumber\\ &&
+1792 N+912 \\
P_{338} &=& 28 N^8+195 N^7+1330 N^6+6350 N^5+18084 N^4+30519 N^3+29316 N^2
\nonumber\\ &&
+14057 N+2264 \\
P_{339} &=& 90 N^8+518 N^7+731 N^6-1230 N^5-4059 N^4-3492 N^3-1196 N^2
\nonumber\\ &&
-392 N-128 \\
P_{340} &=& 354 N^8+1566 N^7+551 N^6-9330 N^5-20779 N^4-14512 N^3+1188 N^2
\nonumber\\ &&
+4504 N+1408 \\
P_{341} &=& -190 N^9-1511 N^8-5060 N^7-8405 N^6-6724 N^5-276 N^4+6566 N^3
\nonumber\\ &&
+8006 N^2+2200 N-1360 \\
P_{342} &=& -111 N^9-1415 N^8-7943 N^7-25380 N^6-51689 N^5-71636 N^4-69419 N^3
\nonumber\\ &&
-45233 N^2-16865 N-2264 \\
P_{343} &=& 8 N^9+62 N^8+469 N^7+2486 N^6+6861 N^5+8678 N^4+2220 N^3-5096 N^2
\nonumber\\ &&
-4296 N-912 \\
P_{344} &=& 24 N^9+157 N^8+1404 N^7+9657 N^6+35070 N^5+65248 N^4+58028 N^3
\nonumber\\ &&
+16152 N^2-4872 N-912 \\
P_{345} &=& 48 N^9+1028 N^8+5069 N^7+9105 N^6-3099 N^5-29245 N^4-32125 N^3
\nonumber\\ &&
-13748 N^2-5592 N-1792 \\
P_{346} &=& -52 N^{10}-557 N^9-4308 N^8-22323 N^7-74274 N^6-157854 N^5-211722 N^4
\nonumber\\ &&
-176525 N^3-94687 N^2-39471 N-12200 \\
P_{347} &=& -8 N^{10}+150 N^9+1746 N^8+6790 N^7+10917 N^6+311 N^5-22106 N^4
\nonumber\\ &&
-21218 N^3+9451 N^2+22384 N+8032 \\
P_{348} &=& -16 N^{11}-292 N^{10}-3230 N^9-20121 N^8-74103 N^7-173208 N^6
\nonumber\\ &&
-275499 N^5-318539 N^4-274948 N^3-165232 N^2-57448 N-8032 \\
P_{349} &=& 420 N^{11}+6567 N^{10}+46152 N^9+191204 N^8+519087 N^7+968914 N^6
\nonumber\\ &&
+1262014 N^5+1132386 N^4+672272 N^3+250266 N^2+61079 N+12200.
\end{eqnarray}
%-------------------------------------------------------------------------------------------------------------------------------

%>>D11:
Finally, diagram~11 is obtained by
%---------------------------------------------------------------------------------------------------------------------------------
\begin{eqnarray}
%% Present Notation: D11
%%%texparser:LHS:testD12Phys%%%
%D_{12} &=&
D_{11} &=&
\textcolor{blue}{C_A^2 T_F} \Biggl\{
        \frac{1}{\varepsilon^3} \Biggl[
                \frac{16 \big(2 N^2+3 N+6\big) S_2}{3 N (N+1) (N+2)}
                -\frac{32 (N+4) S_{-2}}{3 N (N+1) (N+2)}
                -\frac{16 \big(2 N^2+N+6\big) S_1^2}{3 N (N+1) (N+2)}
\nonumber \\ &&
                -\frac{8 P_{399}}{3 N^3 (N+1)^3 (N+2)^3}
                +\frac{16 P_{353} S_1}{3 N^2 (N+1)^2 (N+2)^2}
                +(-1)^N \Biggl(
                        \frac{320}{3} S_{-3}
                        -128 S_{-2,1}
\nonumber \\ &&
                        +\frac{16 P_{356}}{3 N^3 (N+1)^3 (N+2)^3}
                \Biggr)
        \Biggr]
        +\frac{1}{\varepsilon^2} \Biggl[
                \frac{8 P_{407}}{3 N^2 (N+1)^4 (N+2)^4}
                -\frac{8 \big(2 N^2+N+6\big) S_1^3}{3 N (N+1) (N+2)}
\nonumber \\ &&
                +\frac{4 P_{358} S_1^2}{3 N^2 (N+1)^2 (N+2)^2}
                +\frac{8 \big(2 N^2-19 N-16\big) S_{-3}}{3 N (N+1) (N+2)}
                +\frac{16 \big(4 N^2+9 N+12\big) S_{2,1}}{3 N (N+1) (N+2)}
\nonumber \\ &&
                +\frac{16 \big(2 N^2+13 N+32\big) S_{-2,1}}{3 N (N+1) (N+2)}
                +\frac{4 P_{363} S_2}{3 N^2 (N+1)^2 (N+2)^2}
                +(-1)^N \Biggl(
                        32 S_4
                        -\frac{64}{3} S_{-2}^2
\nonumber \\ &&
                        -\frac{8 P_{393}}{3 N^2 (N+1)^4 (N+2)^4}
                        +\frac{4 N^2+7 N+2}{(N+1) (N+2)} \biggl(
                                -\frac{40}{3} S_{-3}
                                +16 S_{-2,1}
                        \biggr)
                        +\frac{128}{3} S_{-4}
\nonumber \\ &&
                        +\frac{64}{3} S_{-2,2}
                        +\frac{32}{3} S_{-3,1}
                        -64 S_{-2,1,1}
                \Biggr)
                +\biggl(
                        -\frac{16 \big(2 N^2+13 N+32\big) S_1}{3 N (N+1) (N+2)}
\nonumber \\ &&
                        +\frac{16 P_{359}}{3 N^2 (N+1)^2 (N+2)^2}
                \biggr) S_{-2}
                -\frac{16 (4 N+13) S_3}{3 (N+1) (N+2)}
                +\biggl(
                        -\frac{8 P_{386}}{3 N (N+1)^3 (N+2)^3}
\nonumber \\ &&
                        +\frac{16 \big(2 N^2-2 N-15\big) S_2}{3 N (N+1) (N+2)}
                \biggr) S_1
        \Biggr]
        +\frac{1}{\varepsilon} \Biggl[
                \frac{16 (2 N+3) S_{2,1,1}}{3 (N+1) (N+2)}
                -\frac{4 P_{412}}{3 N^2 (N+1)^5 (N+2)^5}
\nonumber \\ &&
                -\frac{7 \big(2 N^2+N+6\big) S_1^4}{9 N (N+1) (N+2)}
                +\frac{16 \big(2 N^2+19 N+36\big) S_{-2,2}}{3 N (N+1) (N+2)}
                +\frac{4 \big(6 N^2-59 N-56\big) S_{-4}}{3 N (N+1) (N+2)}
\nonumber \\ &&
                -\frac{8 \big(6 N^2-N+60\big) S_{3,1}}{3 N (N+1) (N+2)}
                +\frac{4 \big(6 N^2+93 N+152\big) S_{-3,1}}{3 N (N+1) (N+2)}
                +\frac{\big(10 N^2-33 N+78\big) S_2^2}{3 N (N+1) (N+2)}
\nonumber \\ &&
                +\frac{2 \big(16 N^2-49 N+286\big) S_4}{3 N (N+1) (N+2)}
                -\frac{8 P_{351} S_{-2,1}}{3 N (N+1)^2 (N+2)^2}
                +\frac{2 P_{361} S_1^3}{9 N^2 (N+1)^2 (N+2)^2}
\nonumber \\ &&
                -\frac{4 P_{362} S_{2,1}}{3 N^2 (N+1)^2 (N+2)^2}
                +\frac{4 P_{372} S_3}{9 N^2 (N+1)^2 (N+2)^2}
                -\frac{2 P_{404} S_2}{3 N^3 (N+1)^3 (N+2)^3}
\nonumber \\ &&
                -\frac{8 \big(10 N^2+59 N+136\big) S_{-2,1,1}}{3 N (N+1) (N+2)}
                +\biggl(
                        \frac{4 P_{401}}{3 N (N+1)^4 (N+2)^4}
                        -\frac{16 (2 N-7) S_{2,1}}{3 N (N+2)}
\nonumber \\ &&
                        +\frac{8 \big(10 N^2+59 N+136\big) S_{-2,1}}{3 N (N+1) (N+2)}
                        +\frac{8 \big(22 N^2-43 N-90\big) S_3}{9 N (N+1) (N+2)}
\nonumber \\ &&
                        +\frac{2 P_{368} S_2}{3 N^2 (N+1)^2 (N+2)^2}
                \biggr) S_1
                +\biggl(
                        \frac{8 P_{351} S_1}{3 N (N+1)^2 (N+2)^2}
                        -\frac{8 P_{389}}{3 N (N+1)^3 (N+2)^3}
\nonumber \\ &&
                        +\frac{4 \big(2 N^2-29 N-56\big) S_2}{3 N (N+1) (N+2)}
                        -\frac{4 \big(10 N^2+59 N+136\big) S_1^2}{3 N (N+1) (N+2)}
                \biggr) S_{-2}
\nonumber \\ &&
                +\biggl(
                        -\frac{2 P_{402}}{3 N^3 (N+1)^3 (N+2)^3}
                        -\frac{4 \big(6 N^2+27 N+79\big) S_2}{3 N (N+1) (N+2)}
                \biggr) S_1^2
\nonumber \\ &&
                +\biggl(
                        \frac{4 P_{369}}{3 N^2 (N+1)^2 (N+2)^2}
                        -\frac{4 \big(6 N^2+93 N+152\big) S_1}{3 N (N+1) (N+2)}
                \biggr) S_{-3}
\nonumber \\ &&
                +\biggl(
                        \frac{P_{397}}{N^3 (N+1)^3 (N+2)^3}
                        -\frac{4 (N+4) S_{-2}}{N (N+1) (N+2)}
                        -\frac{2 \big(2 N^2+N+6\big) S_1^2}{N (N+1) (N+2)}
\nonumber \\ &&
                        +\frac{2 \big(2 N^2+3 N+6\big) S_2}{N (N+1) (N+2)}
                        +\frac{2 P_{353} S_1}{N^2 (N+1)^2 (N+2)^2}
                \biggr) \zeta_2
\nonumber \\ &&
                +(-1)^N \Biggl[
                        \frac{4 P_{409}}{3 N^2 (N+1)^5 (N+2)^5}
                        -\frac{8 P_{350} S_3}{N (N+1)^2 (N+2)^2}
                        -\frac{8 P_{384} S_{-2,1}}{N^2 (N+1)^2 (N+2)^2}
\nonumber \\ &&
                        +\frac{4 N^2+7 N+2}{(N+1) (N+2)} \biggl(
                                -4 S_4
                                +\frac{8}{3} S_{-2}^2
                                -\frac{16}{3} S_{-4}
                                -\frac{8}{3} S_{-2,2}
                                -\frac{4}{3} S_{-3,1}
                                +8 S_{-2,1,1}
                        \biggr)
\nonumber \\ &&
                        +\biggl(
                                \frac{8 P_{356}}{N^3 (N+1)^3 (N+2)^3}
                                -\frac{352}{3} S_{-2,1}
                        \biggr) S_2
                        +\frac{304}{3} S_5
                        +\biggl(
                                \frac{4 P_{388}}{3 N^2 (N+1)^2 (N+2)^2}
\nonumber \\ &&
                                +\frac{272}{3} S_2
                                +\frac{64}{3} S_{-2}
                        \biggr) S_{-3}
                        +\frac{304}{3} S_{-5}
                        -\frac{512}{3} S_{-2} S_{2,1}
                        -32 S_{2,3}
                        -64 S_{2,-3}
                        +\frac{448}{3} S_{-2,3}
\nonumber \\ &&
                        -\frac{128}{3} S_{-2,-3}
                        -160 S_{-4,1}
                        +\frac{512}{3} S_{2,1,-2}
                        -\frac{128}{3} S_{-2,1,-2}
                        +\frac{32}{3} S_{-2,2,1}
                        +\frac{16}{3} S_{-3,1,1}
\nonumber \\ &&
                        -32 S_{-2,1,1,1}
                        +\biggl(
                                \frac{2 P_{356}}{N^3 (N+1)^3 (N+2)^3}
                                +40 S_{-3}
                                -48 S_{-2,1}
                        \biggr) \zeta_2
                \Biggr]
        \Biggr]
\nonumber \\ &&
        +\frac{2 P_{416}}{3 N^2 (N+1)^6 (N+2)^6}
        +\frac{\big(-2 N^2-N-6\big) S_1^5}{6 N (N+1) (N+2)}
        +\frac{16 \big(N^2+3 N+6\big) S_{2,1,1,1}}{3 N (N+1) (N+2)}
\nonumber \\ &&
        -\frac{8 \big(N^2+6 N-21\big) S_{2,3}}{N (N+1) (N+2)}
        +\frac{4 \big(2 N^2-5 N+24\big) S_{-5}}{N (N+1) (N+2)}
        +\frac{8 \big(4 N^2-17 N-4\big) S_{-2,-3}}{3 N (N+1) (N+2)}
\nonumber \\ &&
        +\frac{4 \big(4 N^2-5 N+12\big) S_{-2,1,-2}}{N (N+1) (N+2)}
        -\frac{8 \big(10 N^2+71 N+144\big) S_{-2,2,1}}{3 N (N+1) (N+2)}
\nonumber \\ &&
        -\frac{4 \big(10 N^2+123 N+304\big) S_{2,1,-2}}{3 N (N+1) (N+2)}
        -\frac{4 \big(11 N^2+30 N-16\big) S_5}{N (N+1) (N+2)}
\nonumber \\ &&
        +\frac{4 \big(14 N^2+75 N+276\big) S_{3,1,1}}{3 N (N+1) (N+2)}
        +\frac{4 \big(20 N^2+33 N-266\big) S_{4,1}}{3 N (N+1) (N+2)}
\nonumber \\ &&
        +\frac{2 \big(22 N^2+305 N+520\big) S_{-4,1}}{3 N (N+1) (N+2)}
        +\frac{4 \big(23 N^2+53 N+90\big) S_{2,2,1}}{3 N (N+1) (N+2)}
\nonumber \\ &&
        +\frac{4 \big(30 N^2+9 N-40\big) S_{-2,3}}{3 N (N+1) (N+2)}
        +\frac{2 \big(38 N^2+93 N+240\big) S_{2,-3}}{3 N (N+1) (N+2)}
\nonumber \\ &&
        +\frac{4 \big(38 N^2+217 N+488\big) S_{-2,1,1,1}}{3 N (N+1) (N+2)}
        -\frac{2 \big(42 N^2+351 N+664\big) S_{-3,1,1}}{3 N (N+1) (N+2)}
\nonumber \\ &&
        -\frac{16 P_{360} S_{-2,2}}{3 N^2 (N+1)^2 (N+2)^2}
        +\frac{P_{364} S_1^4}{36 N^2 (N+1)^2 (N+2)^2}
        -\frac{4 P_{366} S_{3,1}}{3 N^2 (N+1)^2 (N+2)^2}
\nonumber \\ &&
        +\frac{2 P_{367} S_{2,1,1}}{3 N^2 (N+1)^2 (N+2)^2}
        +\frac{4 P_{370} S_{-2,1,1}}{3 N^2 (N+1)^2 (N+2)^2}
        -\frac{2 P_{371} S_{-3,1}}{3 N^2 (N+1)^2 (N+2)^2}
\nonumber \\ &&
        +\frac{P_{376} S_2^2}{12 N^2 (N+1)^2 (N+2)^2}
        +\frac{P_{378} S_4}{6 N^2 (N+1)^2 (N+2)^2}
        +\frac{4 P_{390} S_{-2,1}}{3 N (N+1)^3 (N+2)^3}
\nonumber \\ &&
        +\frac{2 P_{403} S_{2,1}}{3 N^3 (N+1)^3 (N+2)^3}
        -\frac{2 P_{406} S_3}{9 N^3 (N+1)^3 (N+2)^3}
        +\biggl(
                -\frac{2 P_{411}}{3 N (N+1)^5 (N+2)^5}
\nonumber \\ &&
                +\frac{4 \big(2 N^2-23 N-46\big) S_{2,1,1}}{3 N (N+1) (N+2)}
                +\frac{\big(6 N^2-31 N-78\big) S_2^2}{2 N (N+1) (N+2)}
                -\frac{2 P_{365} S_{2,1}}{3 N^2 (N+1)^2 (N+2)^2}
\nonumber \\ &&
                +\frac{8 \big(10 N^2+71 N+144\big) S_{-2,2}}{3 N (N+1) (N+2)}
                -\frac{4 \big(30 N^2+9 N+148\big) S_{3,1}}{3 N (N+1) (N+2)}
\nonumber \\ &&
                -\frac{4 \big(38 N^2+217 N+488\big) S_{-2,1,1}}{3 N (N+1) (N+2)}
                +\frac{2 \big(42 N^2+351 N+664\big) S_{-3,1}}{3 N (N+1) (N+2)}
\nonumber \\ &&
                +\frac{\big(136 N^2-85 N+98\big) S_4}{3 N (N+1) (N+2)}
                -\frac{4 P_{370} S_{-2,1}}{3 N^2 (N+1)^2 (N+2)^2}
                +\frac{2 P_{377} S_3}{9 N^2 (N+1)^2 (N+2)^2}
\nonumber \\ &&
                +\frac{P_{395} S_2}{3 N^3 (N+1)^3 (N+2)^3}
        \biggr) S_1
        +\biggl(
                -\frac{2 \big(10 N^2+107 N+226\big) S_3}{3 N (N+1) (N+2)}
\nonumber \\ &&
                -\frac{4 \big(4 N^2-25 N-41\big) S_{2,1}}{3 N (N+1) (N+2)}
                +\frac{2 \big(38 N^2+217 N+488\big) S_{-2,1}}{3 N (N+1) (N+2)}
\nonumber \\ &&
                +\frac{P_{373} S_2}{6 N^2 (N+1)^2 (N+2)^2}
                +\frac{P_{414}}{3 N^4 (N+1)^4 (N+2)^4}
        \biggr) S_1^2
\nonumber \\ &&
        +\biggl(
                \frac{P_{396}}{9 N^3 (N+1)^3 (N+2)^3}
                -\frac{2 \big(28 N^2+110 N+279\big) S_2}{9 N (N+1) (N+2)}
        \biggr) S_1^3
\nonumber \\ &&
        +\biggl(
                \frac{P_{415}}{3 N^4 (N+1)^4 (N+2)^4}
                -\frac{2 \big(14 N^2+61 N+104\big) S_{-2,1}}{3 N (N+1) (N+2)}
\nonumber \\ &&
                +\frac{4 \big(23 N^2+42 N-1\big) S_{2,1}}{3 N (N+1) (N+2)}
                -\frac{4 \big(79 N^2+257 N+441\big) S_3}{9 N (N+1) (N+2)}
        \biggr) S_2
\nonumber \\ &&
        +\biggl(
                \frac{4 P_{405}}{3 N (N+1)^4 (N+2)^4}
                -\frac{4 \big(4 N^2-5 N+12\big) S_{-2,1}}{N (N+1) (N+2)}
                +\frac{2 P_{370} S_1^2}{3 N^2 (N+1)^2 (N+2)^2}
\nonumber \\ &&
                +\frac{8 \big(10 N^2+71 N+144\big) S_{2,1}}{3 N (N+1) (N+2)}
                -\frac{2 \big(38 N^2+217 N+488\big) S_1^3}{9 N (N+1) (N+2)}
\nonumber \\ &&
                -\frac{8 \big(64 N^2+131 N+220\big) S_3}{9 N (N+1) (N+2)}
                +\frac{2 P_{374} S_2}{3 N^2 (N+1)^2 (N+2)^2}
\nonumber \\ &&
                +\biggl(
                        -\frac{4 P_{390}}{3 N (N+1)^3 (N+2)^3}
                        -\frac{2 \big(26 N^2+223 N+472\big) S_2}{3 N (N+1) (N+2)}
                \biggr) S_1
        \biggr) S_{-2}
\nonumber \\ &&
        +\biggl(
                -\frac{2 P_{391}}{3 N (N+1)^3 (N+2)^3}
                +\frac{\big(-42 N^2-351 N-664\big) S_1^2}{3 N (N+1) (N+2)}
\nonumber \\ &&
                +\frac{\big(66 N^2+123 N+472\big) S_2}{3 N (N+1) (N+2)}
                +\frac{2 P_{371} S_1}{3 N^2 (N+1)^2 (N+2)^2}
        \biggr) S_{-3}
\nonumber \\ &&
        +\biggl(
                \frac{2 P_{375}}{3 N^2 (N+1)^2 (N+2)^2}
                -\frac{2 \big(22 N^2+305 N+520\big) S_1}{3 N (N+1) (N+2)}
        \biggr) S_{-4}
\nonumber \\ &&
        +\biggl(
                \frac{P_{407}}{N^2 (N+1)^4 (N+2)^4}
                +\frac{2 \big(2 N^2+13 N+32\big) S_{-2,1}}{N (N+1) (N+2)}
                +\frac{2 \big(4 N^2+9 N+12\big) S_{2,1}}{N (N+1) (N+2)}
\nonumber \\ &&
                +\frac{P_{358} S_1^2}{2 N^2 (N+1)^2 (N+2)^2}
                +\frac{P_{363} S_2}{2 N^2 (N+1)^2 (N+2)^2}
                +\frac{\big(-2 N^2-N-6\big) S_1^3}{N (N+1) (N+2)}
\nonumber \\ &&
                +\frac{\big(2 N^2-19 N-16\big) S_{-3}}{N (N+1) (N+2)}
                +\biggl(
                        \frac{P_{379}}{N (N+1)^3 (N+2)^3}
                        +\frac{2 \big(2 N^2-2 N-15\big) S_2}{N (N+1) (N+2)}
                \biggr) S_1
\nonumber \\ &&
                -\frac{2 (4 N+13) S_3}{(N+1) (N+2)}
                +\biggl(
                        \frac{2 P_{359}}{N^2 (N+1)^2 (N+2)^2}
                        -\frac{2 \big(2 N^2+13 N+32\big) S_1}{N (N+1) (N+2)}
                \biggr) S_{-2}
        \biggr) \zeta_2
\nonumber \\ &&
        +\biggl(
                \frac{P_{398}}{3 N^3 (N+1)^3 (N+2)^3}
                -\frac{4 (23 N+44) S_{-2}}{3 N (N+1) (N+2)}
                +\frac{2 \big(2 N^2+N+6\big) S_1^2}{3 N (N+1) (N+2)}
\nonumber \\ &&
                -\frac{2 \big(2 N^2+27 N+54\big) S_2}{3 N (N+1) (N+2)}
                -\frac{2 P_{352} S_1}{3 N^2 (N+1)^2 (N+2)^2}
        \biggr) \zeta_3
\nonumber \\ &&
        +(-1)^N \Biggl[
                \frac{2 P_{381} S_4}{N^2 (N+1)^2 (N+2)^2}
                +\frac{2 P_{380} S_{-3,1}}{3 N^2 (N+1)^2 (N+2)^2}
                -\frac{8 P_{350} S_{3,1}}{N (N+1)^2 (N+2)^2}
\nonumber \\ &&
                -\frac{2 P_{413}}{3 N^2 (N+1)^6 (N+2)^6}
                +\frac{4 P_{383} S_{-2,2}}{3 N^2 (N+1)^2 (N+2)^2}
                -\frac{4 P_{385} S_{-2,1,1}}{N^2 (N+1)^2 (N+2)^2}
\nonumber \\ &&
                +\frac{4 N^2+7 N+2}{(N+1) (N+2)} \biggl(
                        -\frac{38}{3} S_5
                        +\biggl(
                                -\frac{34}{3} S_2
                                -\frac{8}{3} S_{-2}
                        \biggr) S_{-3}
                        -\frac{38}{3} S_{-5}
                        +\frac{64}{3} S_{-2} S_{2,1}
\nonumber \\ &&
                        +4 S_{2,3}
                        +8 S_{2,-3}
                        +\frac{44}{3} S_2 S_{-2,1}
                        -\frac{56}{3} S_{-2,3}
                        +\frac{16}{3} S_{-2,-3}
                        +20 S_{-4,1}
                        -\frac{64}{3} S_{2,1,-2}
\nonumber \\ &&
                        +\frac{16}{3} S_{-2,1,-2}
                        -\frac{4}{3} S_{-2,2,1}
                        -\frac{2}{3} S_{-3,1,1}
                        +4 S_{-2,1,1,1}
                        +\biggl(
                                -5 S_{-3}
                                +6 S_{-2,1}
                        \biggr) \zeta_2
                \biggr)
\nonumber \\ &&
                +\biggl(
                        \frac{8 P_{350} S_3}{N (N+1)^2 (N+2)^2}
                        +\frac{4 P_{357} S_{-2,1}}{N^2 (N+1)^2 (N+2)^2}
                \biggr) S_1
                +\biggl(
                        32 S_4
                        +\frac{128}{3} S_{-2,2}
\nonumber \\ &&
                        -\frac{4 P_{393}}{N^2 (N+1)^4 (N+2)^4}
                        +\frac{56}{3} S_{-3,1}
                        -112 S_{-2,1,1}
                \biggr) S_2
                +\biggl(
                        \frac{4 P_{400}}{N^3 (N+1)^3 (N+2)^3}
\nonumber \\ &&
                        +32 S_{2,1}
                        -\frac{608}{3} S_{-2,1}
                \biggr) S_3
                -24 S_3^2
                +\frac{256}{3} S_6
                +\biggl(
                        \frac{160}{3} S_4
                        +\frac{368}{3} S_{3,1}
                        +32 S_{-2,2}
\nonumber \\ &&
                        +64 S_{-3,1}
                        -\frac{128}{3} S_{-2,1,1}
                \biggr) S_{-2}
                +\biggl(
                        -\frac{4 P_{382}}{3 N^2 (N+1)^2 (N+2)^2}
                        -\frac{128}{3} S_2
                \biggr) S_{-2}^2
\nonumber \\ &&
                +\biggl(
                        -\frac{2 P_{410}}{3 N^3 (N+1)^3 (N+2)^3}
                        +\frac{2 P_{354} S_1}{N^2 (N+1)^2 (N+2)^2}
                        +168 S_3
                        +8 S_{2,1}
\nonumber \\ &&
                        -64 S_{-2,1}
                \biggr) S_{-3}
                +\frac{112}{3} S_{-3}^2
                +\biggl(
                        \frac{2 P_{387}}{3 N^2 (N+1)^2 (N+2)^2}
                        +\frac{184}{3} S_2
                        -96 S_{-2}
                \biggr) S_{-4}
\nonumber \\ &&
                +\frac{832}{3} S_{-6}
                +16 S_{4,2}
                -\frac{160}{3} S_{4,-2}
                +32 S_{5,1}
                +\frac{64}{3} S_{-2,1}^2
                -56 S_{-3,3}
                +\frac{8}{3} S_{-4,2}
\nonumber \\ &&
                +\biggl(
                        \frac{4 P_{408}}{N^3 (N+1)^3 (N+2)^3}
                        +16 S_{2,1}
                \biggr) S_{-2,1}
                +64 S_{-4,-2}
                +\frac{352}{3} S_{-5,1}
                -32 S_{2,3,1}
\nonumber \\ &&
                -\frac{64}{3} S_{2,-3,1}
                -\frac{368}{3} S_{3,1,-2}
                -32 S_{3,2,1}
                -16 S_{-2,2,2}
                -\frac{32}{3} S_{-2,2,-2}
                -\frac{464}{3} S_{-2,3,1}
\nonumber \\ &&
                -\frac{128}{3} S_{-3,1,-2}
                -\frac{32}{3} S_{-3,2,1}
                -16 S_{-4,1,1}
                -64 S_{2,-2,1,1}
                +\frac{128}{3} S_{-2,1,1,2}
\nonumber \\ &&
                -\frac{176}{3} S_{-2,2,1,1}
                +\frac{8}{3} S_{-3,1,1,1}
                -16 S_{-2,1,1,1,1}
                +\biggl(
                        \frac{P_{392}}{N^2 (N+1)^4 (N+2)^4}
\nonumber \\ &&
                        +12 S_4
                        -8 S_{-2}^2
                        +16 S_{-4}
                        +8 S_{-2,2}
                        +4 S_{-3,1}
                        -24 S_{-2,1,1}
                \biggr) \zeta_2
\nonumber \\ &&
                +\biggl(
                        -\frac{2 P_{394}}{3 N^3 (N+1)^3 (N+2)^3}
                        -\frac{4 P_{355} S_1}{N^2 (N+1)^2 (N+2)^2}
                        +32 S_3
                        -\frac{40}{3} S_{-3}
\nonumber \\ &&
                        -16 S_{2,1}
                        +16 S_{-2,1}
                \biggr) \zeta_3
        \Biggr]
\Biggr\},
\end{eqnarray}
%---------------------------------------------------------------------------------------------------------------------------------
with
%---------------------------------------------------------------------------------------------------------------------------------
\begin{eqnarray}
P_{350}    &=& 4 N^4+22 N^3+41 N^2+29 N+8 \\
P_{351}    &=& 24 N^4+223 N^3+591 N^2+367 N-120 \\
P_{352}    &=& 2 N^5+6 N^4+20 N^3+157 N^2+424 N+224 \\
P_{353}    &=& 2 N^5+18 N^4+50 N^3+37 N^2+28 N+32 \\
P_{354}    &=& 4 N^5-59 N^4-247 N^3-412 N^2-298 N-144 \\
P_{355}    &=& 4 N^5+13 N^4+9 N^3-20 N^2-26 N-16 \\
P_{356}    &=& 4 N^5+29 N^4+64 N^3+80 N^2+84 N+32 \\
P_{357}    &=& 4 N^5+85 N^4+265 N^3+372 N^2+246 N+112 \\
P_{358}    &=& 6 N^5+30 N^4+82 N^3-55 N^2-188 N-16 \\
P_{359} &=& 10 N^5+61 N^4+134 N^3+80 N^2-4 N+16 \\
P_{360} &=& 12 N^5+127 N^4+330 N^3+190 N^2-80 N-8 \\
P_{361} &=& 14 N^5+54 N^4+146 N^3-239 N^2-620 N-112 \\
P_{362} &=& 14 N^5+66 N^4+75 N^3-147 N^2-318 N-128 \\
P_{363} &=& 14 N^5+88 N^4+138 N^3+19 N^2+20 N+64 \\
P_{364} &=& 30 N^5+102 N^4+274 N^3-607 N^2-1484 N-304 \\
P_{365} &=& 30 N^5+362 N^4+951 N^3+839 N^2+126 N-48 \\
P_{366} &=& 34 N^5+32 N^4-480 N^3-769 N^2-184 N-80 \\
P_{367} &=& 34 N^5+204 N^4+459 N^3+285 N^2-270 N-160 \\
P_{368} &=& 50 N^5+500 N^4+1490 N^3+1425 N^2+624 N+400 \\
P_{369} &=& 56 N^5+389 N^4+757 N^3+325 N^2-72 N+64 \\
P_{370} &=& 60 N^5+749 N^4+2151 N^3+1355 N^2-576 N-96 \\
P_{371} &=& 132 N^5+1283 N^4+3129 N^3+1685 N^2-704 N-32 \\
P_{372} &=& 134 N^5+621 N^4+356 N^3-1340 N^2-1250 N-160 \\
P_{373} &=& 138 N^5+1600 N^4+4738 N^3+3033 N^2-1152 N+16 \\
P_{374} &=& 156 N^5+999 N^4+2097 N^3+1125 N^2-112 N+224 \\
P_{375} &=& 188 N^5+1271 N^4+2437 N^3+877 N^2-512 N+128 \\
P_{376} &=& 190 N^5+1230 N^4+1210 N^3-739 N^2+1396 N+1648 \\
P_{377} &=& 354 N^5+2856 N^4+6826 N^3+5297 N^2+2104 N+1520 \\
P_{378} &=& 642 N^5+2296 N^4-1342 N^3-7981 N^2-2976 N+448 \\
P_{379} &=& -5 N^6-41 N^5-62 N^4+218 N^3+554 N^2+244 N-96 \\
P_{380} &=& 4 N^6-17 N^5+240 N^4+1087 N^3+1864 N^2+1322 N+656 \\
P_{381} &=& 4 N^6+3 N^5-55 N^4-148 N^3-196 N^2-168 N-64 \\
P_{382} &=& 4 N^6+19 N^5+33 N^4+16 N^3-80 N^2-136 N-64 \\
P_{383} &=& 4 N^6+19 N^5+249 N^4+784 N^3+1096 N^2+680 N+320 \\
P_{384} &=& 4 N^6+23 N^5+82 N^4+153 N^3+96 N^2-26 N-16 \\
P_{385} &=& 4 N^6+31 N^5+252 N^4+683 N^3+840 N^2+466 N+208 \\
P_{386} &=& 5 N^6+41 N^5+62 N^4-218 N^3-554 N^2-244 N+96 \\
P_{387} &=& 16 N^6+40 N^5-93 N^4-401 N^3-728 N^2-718 N-304 \\
P_{388} &=& 20 N^6+83 N^5+234 N^4+437 N^3+248 N^2-194 N-80 \\
P_{389} &=& 26 N^6+206 N^5+546 N^4+435 N^3-15 N^2+270 N+416 \\
P_{390} &=& 72 N^6+660 N^5+1616 N^4+202 N^3-2488 N^2-631 N+1408 \\
P_{391} &=& 136 N^6+1192 N^5+3220 N^4+3034 N^3+1492 N^2+3043 N+2560 \\
P_{392} &=& -6 N^7-68 N^6-314 N^5-703 N^4-925 N^3-1002 N^2-832 N-256 \\
P_{393} &=& 6 N^7+68 N^6+314 N^5+703 N^4+925 N^3+1002 N^2+832 N+256 \\
P_{394} &=& 54 N^7+396 N^6+1240 N^5+2279 N^4+2914 N^3+2744 N^2+1644 N+416 \\
P_{395} &=& -152 N^8-1408 N^7-3809 N^6-1733 N^5+4194 N^4+3032 N^3+304 N^2
\nonumber \\ &&
+1472 N+256 \\
P_{396} &=& -32 N^8-236 N^7-137 N^6+2945 N^5+6572 N^4+2680 N^3-2136 N^2
\nonumber \\ &&
-672 N-192 \\
P_{397} &=& -N^8-3 N^7+3 N^6+7 N^5-44 N^4-124 N^3-160 N^2-168 N-64 \\
P_{398} &=& N^8-9 N^7-255 N^6-1327 N^5-2932 N^4-2864 N^3-512 N^2
\nonumber\\ &&
+984 N+448 \\
P_{399} &=& N^8+3 N^7-3 N^6-7 N^5+44 N^4+124 N^3+160 N^2+168 N+64 \\
P_{400} &=& 12 N^8+103 N^7+347 N^6+562 N^5+444 N^4+183 N^3+176 N^2
\nonumber\\ &&
+252 N+96 \\
P_{401} &=& 12 N^8+106 N^7+203 N^6-512 N^5-2048 N^4-2412 N^3-2668 N^2
\nonumber \\ &&
-4256 N-2800 \\
P_{402} &=& 14 N^8+106 N^7+87 N^6-1127 N^5-2560 N^4-1056 N^3+776 N^2
\nonumber\\ &&
+224 N+64 \\
P_{403} &=& 37 N^8+287 N^7+701 N^6+308 N^5-521 N^4+555 N^3+1684 N^2
\nonumber\\ &&
+624 N+96 \\
P_{404} &=& 42 N^8+330 N^7+789 N^6+295 N^5-970 N^4-656 N^3+672 N^2
\nonumber\\ &&
+1008 N+384 \\
P_{405} &=& 58 N^8+598 N^7+2573 N^6+6669 N^5+13304 N^4+19661 N^3+15835 N^2
\nonumber \\ &&
+2802 N-2304 \\
P_{406} &=& 308 N^8+2321 N^7+5204 N^6+1705 N^5-4382 N^4+2216 N^3+10704 N^2
\nonumber \\ &&
+7848 N+2304 \\
P_{407} &=& 2 N^9+13 N^8+30 N^7+56 N^6+214 N^5+566 N^4+847 N^3+986 N^2
\nonumber \\ &&
+832 N+256 \\
P_{408} &=& 4 N^9+43 N^8+220 N^7+613 N^6+889 N^5+449 N^4-300 N^3-428 N^2
\nonumber \\ &&
-176 N-32 \\
P_{409} &=& 14 N^9+158 N^8+653 N^7+972 N^6-439 N^5-3213 N^4-5644 N^3-7788 N^2
\nonumber \\ &&
-6720 N-2048 \\
P_{410} &=& 20 N^9+119 N^8+348 N^7+817 N^6+1661 N^5+2157 N^4+1860 N^3+1284 N^2
\nonumber \\ &&
+528 N+96 \\
P_{411} &=& 28 N^{10}+293 N^9+817 N^8-1908 N^7-19268 N^6-64344 N^5-126146 N^4
\nonumber \\ &&
-148587 N^3-88356 N^2-8012 N+10944 \\
P_{412} &=& 6 N^{11}+58 N^{10}+224 N^9+437 N^8+356 N^7-637 N^6-2926 N^5-5169 N^4
\nonumber \\ &&
-6432 N^3-7916 N^2-6720 N-2048 \\
P_{413} &=& 30 N^{11}+408 N^{10}+2436 N^9+9180 N^8+27724 N^7+69523 N^6+128540 N^5
\nonumber \\ &&
+163980 N^4+153568 N^3+118816 N^2+66880 N+16384 \\
P_{414} &=& 32 N^{11}+294 N^{10}+530 N^9-1855 N^8-7195 N^7-10521 N^6-14976 N^5
\nonumber \\ &&
-19216 N^4-7232 N^3+6000 N^2+2880 N+384 \\
P_{415} &=& 100 N^{11}+984 N^{10}+3700 N^9+6789 N^8+6993 N^7+6027 N^6+9288 N^5
\nonumber \\ &&
+18344 N^4+22704 N^3+12720 N^2+2880 N+384 \\
P_{416} &=& 16 N^{13}+204 N^{12}+1099 N^{11}+3170 N^{10}+4865 N^9+3051 N^8+3872 N^7
\nonumber \\ &&
+31054 N^6+92314 N^5+143392 N^4+147016 N^3+117920 N^2+66880 N
\nonumber \\ &&
+16384.
\end{eqnarray}
%---------------------------------------------------------------------------------------------------------------------------------
As for the diagrams $D_1$ and $D_5-D_{10}$ it can be thoroughly expressed by nested harmonic sums.

%\input{D5phys.tex}
%\fi

\vspace{5mm}
\noindent
{\bf Acknowledgment.}~
We would like to thank A.~Hasselhuhn, C.G.~Raab, and F.~Wi\ss{}brock for discussions and M.~Steinhauser for the possibility to use
the package {\tt MATAD3.0}. The graphs have been drawn using {\tt Axodraw}~\cite{Vermaseren:1994je}. This work was supported in part 
by the Austrian Science Fund (FWF) grants P20347-N18 and SFB F50 (F5009-N15) and the European Commission through contract 
PITN-GA-2012-316704 ({HIGGSTOOLS}).
%--------------------------------------------------------------------------------------------------

%-----------------------------------------------------------------------------
\end{document}